THE ANDRZEJ SOŁTAN INSTITUTE FOR NUCLEAR STUDIES
INSTYTUT PROBLEMÓW JĄDROWYCH im. ANDRZEJA SOŁTANA
Otwock-Świerk, 05-400, Poland

# Ionization Cluster Size Distributions Created by Low Energy Electrons and Alpha Particles in Nanometric Track Segment in Gases

***Aliaksandr Bantsar***

*Thesis submitted to the Andrzej Sołtan Institute for Nuclear Studies in partial fulfilment of the requirement for the Ph.D. degree in Physics performed under supervision of Prof. DSc Marian Jaskóła.*

# *Dedication*

Pracę dedykuję moim Rodzicom – Marii i Tadeuszowi,
którzy wierzyli we mnie i dawali mi siły do jej zrealizowania.

Посвящается моим Родителям Марии и Тадеушу, которые своей верой
и поддержкой вдохновили меня и дали силы написать настоящую работу.

I dedicate this work to my Parents (Maria and Tadeusz)
who believed in me and gave me the power to finish this work.

## *Acknowledgments*

I am very grateful to my co-workers: Elżbieta Jaworska, Adam Dudziński and Jacek Kula for their technical assistance, interesting observations and comments.

I'd like to thank dr Nicholas Keeley for a critical reading of my dissertation and for constructive suggestions concerning the English.

I would like to thank dr Bernd Grosswendt for helpful comments and valuable discussions.

I am grateful in a special way to dr Stanisław Pszona. It is not possible to fully appreciate his participation, as it was a collaborative continuation of his work starting in the early 1970s.

I would especially like to thank to Prof. DSc Marian Jaskóła for his persevering continued interest and for his assistance in finishing this work.

This work has been partially supported by Grants **N N202 288738**, N N401 216134 and Contract Nr. 569/N-COST/2009/0 of the Ministry of Science and Education of Poland.

# Abstract

The interaction of ionizing radiation with nanometric targets is a field of interest for many branches of science such as: radiology, oncology, radiation protection and nanoelectronics. A new experimental technique known as nanodosimetry has been developed for the qualitative as well as quantitative description of these types of interactions.

The work presented here is a contribution to this development, namely by further improvement of the new experimental technique called the Jet Counter, originally developed at the Andrzej Sołtan Institute for Nuclear Studies. The Jet Counter is a unique device in the world for studying the interaction of low energy electrons with nanometer targets in the range 2-10 *nm* (in unit density).

The basic characteristics of the experimental device called the Jet Counter with the emphasis on my own contributions, namely the piezoelectric valve, single ion counting efficiencies and automation of the acquisition system are presented.

The basic experimental result is the frequency distribution of ionization cluster size produced by ionizing particles in a gaseous (nitrogen or propane) nanometric track segment.

The measurements were performed at the Jet Counter facility using the single-ion-counting method. The frequencies were measured based on counting the number of ionizations in coincidence with the ionizing particle after passing through the simulated nanometric volume.

The first experimental data on the frequency distribution of ionization cluster size produced by low energy "single" electrons (100 *eV* – 2000 *eV*) in target cylinders of nitrogen ($N_2$) 0.34 $\mu g/cm^{2.}$ in diameter are presented.

New experimental data on the frequency distribution of ionization cluster size produced by 3.8 *MeV* $\alpha$-particles in a target cylinder of nitrogen ($N_2$) and propane gas ($C_3H_8$), ranging from 0.1 to 0.5 $\mu g/cm^2$ in diameter are presented.

Experimental results are compared with Monte Carlo simulations. A Bayesian analysis is applied for convoluting the measured spectra to the true cluster size distributions.

New quantities characterizing the interaction of ionizing radiation with the nanometre level are proposed, namely – $P_1$ (the probability of forming a cluster size $\nu = 1$), $M_1$ (mean cluster size – the first moment of the distribution), $F_2$ (the sum distribution function of forming an ionization cluster size $\nu \geq 2$). It has been shown that these quantities may substitute for the traditional (macro) dosimetric quantities.

Summarizing, the Jet Counter is the first and unique measuring facility based on single-ion counting which can be used to investigate ionization-cluster formation in nanometer target volumes (up to a few µg/cm2) for single ionizing particles. New nanometric radiation quantities are proposed in the hope that they will be of use in the practice of targeted radiotherapy.

# *Streszczenie*

Oddziaływanie promieniowania jonizującego ze strukturami nanometrowymi jest przedmiotem zainteresowania wielu dyscyplin nauki i techniki takich jak radiobiologia, onkologia, ochrona przed promieniowaniem czy też nanoelektronika. Aby opisać jakościowo jak i ilościowo takie oddziaływania rozwinęła się nowa technika eksperymentalna (często zwana nanodozymetrią).

Przedstawiona rozprawa doktorska pod tytułem **“Rozkłady klastrów jonizacyjnych tworzonych przez nisko-energetyczne elektrony i cząstki alfa na nanometrowym odcinku toru w gazach”**, jest wkładem w rozwój techniki eksperymentalnej w oparciu o tzw. Jet Counter. Jest to technika rozwinięta całkowicie w Polsce i jak dotychczas jest unikalną w skali światowej, w zakresie eksperymentów z oddziaływaniem nisko-energetycznych elektronów z symulowanymi strukturami nanometrowymi w przedziale od 2 do 10 *nm* (w skali gęstości jednostkowych). W pracy przedstawiono podstawowe cechy stanowiska eksperymentalnego Jet Counter jak też uwypuklono prace, dzięki którym stanowisko to udoskonalono. Dotyczy to szczególnie prac nad stabilnością zaworu piezoelektrycznego, wydajności liczenia pojedynczych jonów oraz automatyzacji systemu akwizycji danych.

Podstawową informacją eksperymentalną uzyskiwaną jest widmo częstości tworzenia klastrów jonizacyjnych na nanometrowym odcinku toru cząstki jonizującej w gazie (azocie lub propanie). Otrzymane widmo klastrów jest mierzone techniką śledzenia pojedynczych ”przelotów” cząstki naładowanej przez strukturę nanometrową. W pracy opisano wyniki doświadczeń dla cząstek alfa oraz (przede wszystkim) nisko-energetycznych elektronów w przedziale energii 100 *eV* do 2000 *eV.* Doświadczenia dla nisko-energetycznych elektronów przeprowadzono dla azotowej struktury nanometrowej o grubości 3.4 *nm*. A dla cząstek alfa – dla azotowych i propanowych struktur nanometrowych w przedziale 0.1 – 5 *nm*.

Wyniki pomiarów widm analizowano metodą Bayesa celem ich transformacji na 100% wydajność detektora jonów. Wyniki porównano z obliczeniami metodą Monte Carlo uzyskując dobrą zgodność rezultatów. W pracy przedstawiono też propozycję nowych wielkości, które opisują zjawiska oddziaływania na poziomie nanometrów – są to: $P_1$ – prawdopodobieństwo tworzenia jednej jonizacji, $M_1$ – pierwszy moment rozkładu oraz $F_2$ – kumulanta rozkładu dla klastrów większych od 2. Wykazano, że parametry te mogą zastąpić dotychczas używane parametry oparte o makrodozymetrię.

Konkludując – wykazano, że Jet Counter jest unikalnym stanowiskiem pomiarowym pozwalającym na uzyskiwanie wyników oddziaływania cząstek naładowanych w nanometrowym odcinku toru cząstki w gazach. Zaproponowano nowy system wielkości do opisu w/w oddziaływań. W związku z rozwojem radioterapii celowanej wyrażam nadzieję, że technika opisana w pracy znajdzie wkrótce szersze praktyczne zastosowania.

# *Preface*

The present paper is a summary of a series of experimental investigations in the field of nanodosimetry. The detailed results of these investigations are described in the following publications:

1. **A.Bantsar**, B.Grosswendt, J.Kula and S.Pszona "C*lusters of ionisation in nanometre targets for propane – experiments with a jet counter*" Radiat. Prot. Dosim. **110** 845-850 (2004)
2. **A.Bantsar**, B.Grosswendt and S.Pszona "*Formation of ion clusters by low-energy electrons in nanometric targets: experiment and monte carlo simulation*", Radiat. Prot. Dosim. **122** 82-85 (2006)
3. S.Pszona, **A.Bantsar** and H. Nikjoo "*Ionization cluster size distribution for alpha particles: experiment and modeling*" Radiat. Prot. Dosim. **122** 28-31(2006)
4. B.Grosswendt, S.Pszona and **A.Bantsar** "*New descriptors of radiation quality based on nanodosimetry, a first approach*" Radiat. Prot. Dosim. **126** 432-444 (2007)
5. S.Pszona, **A.Bantsar** and J.Kula *"Charge cluster distribution in nanosites traversed by a single ionizing particle – an experimental approach"* Nucl. Instr. and Meth. **B266** 4911-4915 (2008)
6. **A.Bantsar,** B.Grosswendt, S.Pszona, J.Kula *"Single track nanodosimetry of low electron energy"* Nucl. Instr. and Meth. in Phys. Res. **A599** 270-274 (2009)

The main results of the above mentioned papers are included along with several as yet unpublished results and some details, mostly of an experimental nature, which were not given in the original papers. Together, these form the basis of the present dissertation. The experiments were carried out at the Andrzej Sołtan Institute for Nuclear Studies at Świerk in the years 2001-2010 using the "Jet Counter" facility constructed by dr S. Pszona's group.

# *Table of contents*

# 1 Introduction

The interaction of ionizing radiation with living tissues is an object of investigation which started together with the discovery of radiation. Since the beginning the problem of understanding the nature of such interactions can be divided (in a crude simplification) into physical and biological phases. The physical phase initiates all events, which is very important for the understanding and quantification of the biological phases. Both disciplines work together. Targets inside living tissues can only be defined based on a knowledge of radiobiology. Physics for its part provides the necessary parameters that characterize such interactions at the needed level of tissue organization.

At present it is generally accepted that the initiation of radiation damage to genes or cells is the result of the spatial distribution of inelastic interactions of single ionizing particles within the DNA (deoxyribonucleic acid) molecule or in its neighborhood and is, in consequence, determined by the stochastics of particle interactions in the volume — a few nanometres in size (comparable to the DNA size).

Here, so-called clustered damage in segments of the DNA is of particular importance, as pointed out by Goodhead [1]. This clustered damage, which may lead to mutagenic, genotoxic or other potential lethal lesions such as single or double strand breaks (SSB or DSB), can be assumed to be caused by a combination of primary or secondary particle interaction processes in the DNA and of successive reactions of damaged sites with reactive species (for instance, OH radicals) produced by ionizing particles within the neighborhood of the DNA.

An encouraging starting point to tackle these challenges is, for instance, the finding of Brenner and Ward [2] that the yields of clusters of multiple ionizations produced by ionizing radiation of different quality within sites 2 to 3 *nm* in size correlate well with yields observed for double strand breaks. Watt [3], who analyzed the best parameters which take into account the influence of the quality of the radiation on the radiobiological effectiveness, found that the maximum effectiveness is reached for a mean free path for primary ionization equal to 2 *nm*.

On account of the complexity of radiation-induced damage and the almost insuperable difficulties for its detailed experimental investigation, our present knowledge on this topic almost exclusively stems from Monte Carlo simulations based on more or less highly sophisticated models of DNA as well as on cross section sets for water vapour or liquid water. For an overview of the computational modeling of DNA damage see for instance the articles by Nikjoo *et al.* [4] and Friedland *et al.* [5].

The essential results of such simulations are the yields of single- or double-strand breaks (SSB or DSB) in the DNA and also, in part, the distribution of DNA fragments. Typically, for all of these data the radiation damage strongly depends on the radiation quality and cannot be described satisfactorily by macroscopic quantities, which, like absorbed dose, take into account neither the track structure of the ionizing particles nor the structure of the radio-sensitive sub-cellular targets.

In view of this fact, one of the aims of current nanodosimetry is to develop an experimental procedure and a method that can be applied to measuring quantities also valid in sub-cellular structures in the determination of the radiation induced frequency distribution of ionization cluster size (number of ionizations per primary particle) in liquid water as a substitute for sub-cellular material, in volumes that are comparable in size with those of the most probable radio-sensitive volumes of biological systems (segments of the DNA 2.2 *nm*, chromosome 11 *nm*, nucleosomes in 30 *nm* chromatin fibers, Figure 1). Such frequency distributions are, in large part, governed by the same basic physical interaction data as those that can be expected if charged particles interact, for instance, with DNA segments. In consequence, the frequency distributions of ionization cluster size in nanometric volumes of liquid water (nanodosimetry) can also be used for the definition of new descriptors of radiation quality.

In nanoelectronic elements radiation effects may be manifested by a large diversity of secondary effects, so-called single-event effects, SEE, depending on the hit region and on the type of interacting charged particles. The are several reports [6, 7, 8] on the formation of single-event upsets in microelectronic devices caused by charged particles (cosmic rays, particles emitted in solar events, $\alpha$-particles from materials contaminated with natural radionuclides). As the size of the elements of electronic circuits is constantly decreasing (currently approaching a few nanometres), their capacity also decreases, as does the charge necessary to manifest an SEE. It is only a matter of time before elementary circuits of RAM will attain a few nanometers in size. At these nanometre sizes, single-event effects are of particular importance.

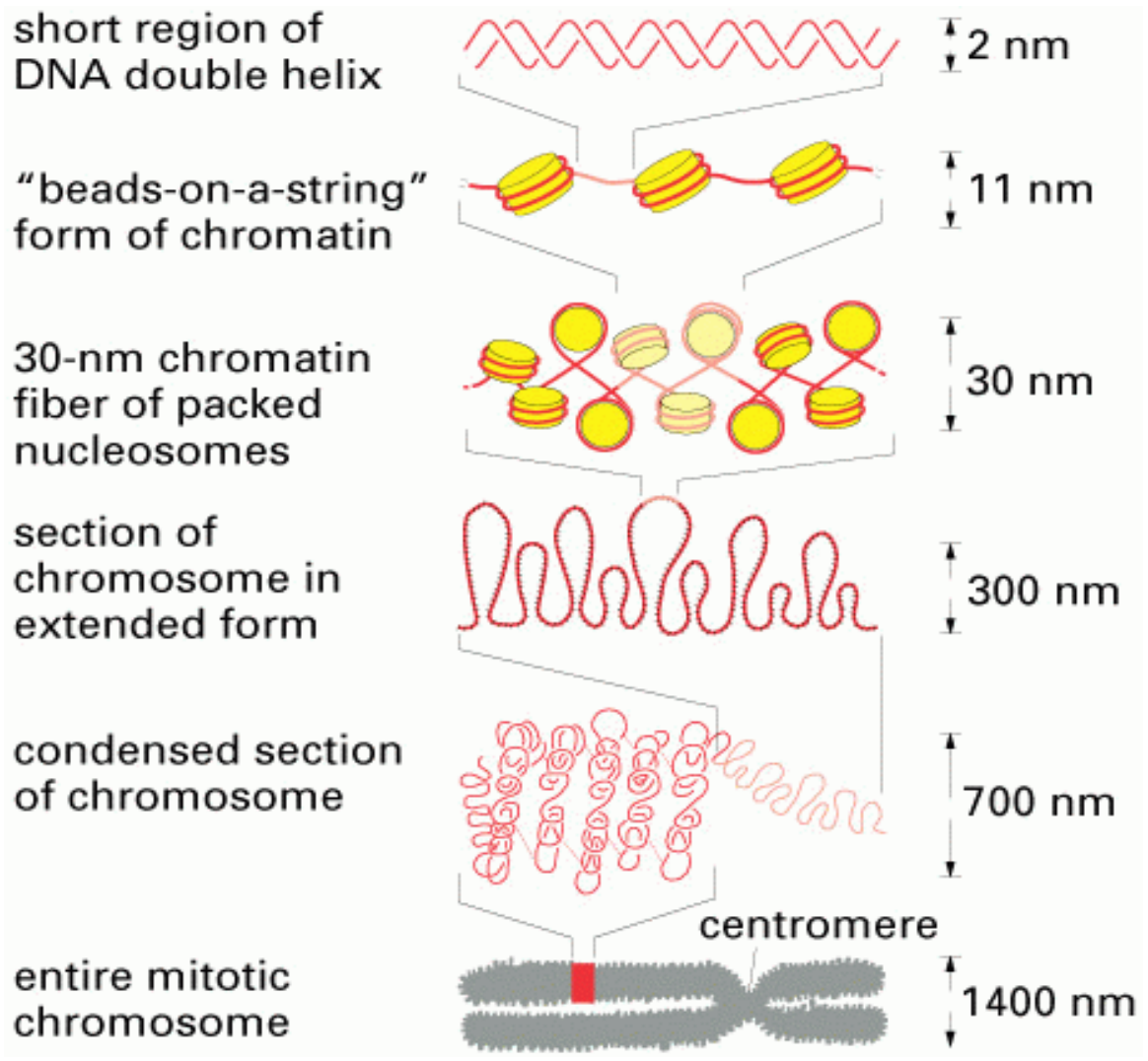

*Figure 1: Schematic view of DNA package in chromosome Ref.[9].*

Nowadays, absorbed dose, LET (linear energy transfer) or LET100 are commonly used as macroscopic quantities to characterize radiation fields. In many cases these quantities are still good parameters [10, 11], but if the size of the radio-sensitive element of a target is a few *nm* they do not work in principle. It is, therefore, one of the challenges of current radiation physics to define more appropriate physical quantities for the description of radiation on the nanometer scale, which:

- are based on particle interactions in nanometric sites and thus may serve as a tool for an adequate description of the induction of radiobiological effects due to particle interactions in sub-cellular structures or the description of single-event effects in nanoelectronics. The price which possibly has to be paid for this might be the loss of a correlation to quantities like absorbed dose, as was pointed out by Amols *et al*. [12];
- and are easily measurable.

Summarizing, the aims of this Ph.D. thesis are:

- to give an overview of track formation by charged particles;
- to give a short overview of recent measuring devices for ionization cluster measurements;
- to present new experiments on the formation of ionization clusters by $\alpha$-particles and low energy electrons in molecular nitrogen ($N_2$) and propane gas ($C_3H_8$);
- to summarize the main principle of ionization cluster-size formation in nanometric targets with new nanometric quantities, namely – $P_1$ (frequency required to create one ionization), $M_1$ (mean cluster size – first moment of the distribution), $F_2$ (frequency required to create two or more ionization);
- to apply the general principles of cluster-size formation to develop methods that can be applied in order to relate the results of gaseous measurements to radiobiology, or in principle to measurements for liquid water or other materials (for example – Si in nanoelectronic devices);
- to propose a tentative procedure of how to apply cluster-size distributions for nanometric targets to radiobiological or nanoelectronic effects.

# 2 Track formation – state of the art

## 2.1 Interaction of charged particles with matter

### 2.1.1 Heavy charged particles

The passing of a charged particle through matter is characterized by a loss of energy and a deflection from its original direction. These effects are the result of two processes: inelastic collisions with atomic electrons of the material and elastic scattering from the nuclei.

So, a complete description of a charged particle track in matter should consist of elastic and inelastic scattering in terms of a specification of the cross-section including:

- “total elastic”,
- “total inelastic”,
- “total ionization”,
- “total excitation”,
- “single differential” cross-section (representing the energy loss in each encounter),
- “double differential” cross-section (giving the probability of energy loss and angle of deviation of the outgoing particle),
- “charge-transfer” cross-section between the incoming particle and the atomic electrons,
- “multiple ionization” cross-section,
- “dissociation” cross-section and stopping power and so on.

Not all the necessary data are available for materials of interest (radiation biology), but some of the processes like emission of Cherenkov radiation, nuclear reactions and Bremsstrahlung are rare in comparison to the atomic collision processes and can be neglected in this treatment.

So, the energy loss of a heavy charged particle in matter is caused mainly by inelastic collisions in which energy is transferred from the particle to the atom causing its ionization or/and excitation.

The inelastic collisions are statistical in nature. However, the fluctuations in total energy loss per macroscopic path length are small and the process can be described by the average energy loss per unit path length. This macroscopic quantity is often called the stopping power and may be calculated with the original Bethe-Bloch formula (Bethe [13], Bloch [14]) improved by including a hard-collision term and a shell correction:

$$-\frac{dE}{dx} = 2\pi N_A r_e^2 m_e c\rho \frac{Z}{A}\frac{z^2}{\beta^2}\left[\ln\frac{2m_e\gamma^2 v^2 E_{\max}}{I^2} - 2\beta^2 - \delta(\rho) - 2\frac{C}{Z}\right] \tag{1}$$

where the additional terms within the square brackets ($C/Z$ and $\delta(\rho)$) take into account electron shell or density-related effects [15]. An additional relativistic effect may be added [16, 17].

Where: $Z$ – atomic number, $z$ – charge of incident particle in units of $e$, $v$ – speed of incident particle, $r_e = 2.8179402894(58) \cdot 10^{-13}\ cm$ – is the classical electron radius, $m_e$ – electron mass, $A$ – atomic mass, $\rho$ – medium density, $I$ – mean excitation potential, $\beta = v/c$, $\gamma = 1/\sqrt{1-\beta^2}$, $c$ – velocity of light, $N_A$ – the Avogadro number, $\delta(\rho)$ – correction term for the density effect in condensed media.

Note that the only medium-related parameters in this formula are the electron density, $Z\rho/A$, and the mean excitation potential, $I$. In effect the projectile is traversing a (nearly) free-electron gas.

The maximum allowable energy transfer $E_{\max}$ between a projectile of mass $M >> m_e$ and an electron emitted in an ionization event, derived from relativistic two-body kinematics, is as follows:

$$E_{\max} \approx \frac{2m_e c^2 \beta^2}{1-\beta^2} \qquad \textbf{(2)}$$

(e.g. in the case of an $\alpha$-particle, $E_{\max}$ is a few *keV*).

Approximately, the probability of generating an electron of energy $E$ (within a track segment of length $l$) is given by:

$$p(E)dE = \frac{4\pi N_A e^4}{m_e c^2} \frac{z^2}{\beta^2} \frac{Z}{A} \frac{\rho l}{E^2} \qquad \textbf{(3)}$$

So, while most ionization electrons generated by the projectile will have energy below the ionization threshold of the medium (typically 10 *eV*), there is a finite probability of generating higher energy electrons. Such electrons, termed δ-electrons, will create further ionization in the medium and transport energy away from the main projectile track.

The consequence of this is that when studying radiation effects on the cell as a whole (microdosimetry) or on the mammalian body (dosimetry) the radiation can be envisioned as a field of “uniform rays”. When looking at the DNA scale, on the other hand, this approximation breaks down and we see a stochastic distribution of ionization clusters.

### 2.1.2 Electrons

The specific ionization by electrons at higher energies (beta particles) is roughly 2 or 3 orders of magnitude smaller than for $\alpha$-particles. Consequently their path length in matter is longer in the same proportion.

As an energetic electron traverses matter, it interacts with it through Coulomb interactions with atomic orbital electrons (ionization, excitation) and atomic nuclei. Through these collisions the electrons may lose their kinetic energy (collision and radiative losses) or change their direction of travel (scattering).

The total energy loss of electrons is composed of two parts:

$$\left(\frac{dE}{dx}\right)_{tot} = \left(\frac{dE}{dx}\right)_{coll} + \left(\frac{dE}{dx}\right)_{rad} \tag{4}$$

The treatment of the energy loss by collisions for incoming electrons follows the same line as for massive charged particles. Nevertheless, the Bethe-Bloch formula must be modified somewhat for two reasons: the small mass of the electron and the fact that for electrons the collisions are between identical particles, so that the calculation must take into account their indistinguishability. Also, the maximum allowable energy transfer becomes $E_{\max} = T_e / 2$ where $T_e$ is the kinetic energy of the incident electron. So, the Bethe-Bloch formula then becomes:

$$-\frac{dE}{dx} = 2\pi N_A r_e^2 m_e c^2 \rho \frac{Z}{A} \frac{1}{\beta^2} \left[ \ln \frac{\tau^2(\tau+2)}{2(I/m_e c^2)^2} + (1-\beta^2) + \frac{\frac{\tau^2}{8} - (2r+1)\ln 2}{(\tau+1)^2} - \delta(\rho) - 2\frac{C}{Z} \right] \tag{5}$$

where $\tau$ is the kinetic energy of the electron in units of $m_e c^2$.

The remaining quantities are as described previously in (Eq.1).

The collisions between the incident electron and an orbital electron or nucleus of an atom may be elastic or inelastic. In an elastic collision the electron is deflected from its original path but no energy loss occurs, while in an inelastic collision the electron is deflected from its original path and some of its energy is transferred to an orbital electron (ionization or excitation) or emitted in the form of Bremsstrahlung (radiation loss).

Electrons, however, since they have the same mass as orbital electrons in matter, are easily deflected during collision. For this reason the electrons follow a tortuous path as they pass through absorbing media. The range of penetration of electrons in matter is therefore substantially less than their full path length. The energy absorption from electrons depends mainly on the number of absorbing electrons in the path of the electrons, or on the areal density of electrons in the absorber.

When an electron passes close to a nucleus in matter, the strong attractive Coulomb force causes the electron to deviate sharply from its original path. The change in direction is due to radial acceleration, and the electron, in accordance with the classical theory of physics, loses energy by electromagnetic radiation, which is called Bremsstrahlung. The likelihood of Bremsstrahlung production increases with the atomic number of the absorber and with the electron energy. In tissue equivalent material (low *Z*), the production of Bremsstrahlung is only relevant for the dosimetry of high energy electrons.

Energy loss due to Bremsstrahlung is an important process for higher energy electrons (> 10 *MeV*) and depends strongly on the absorbing material. For each material we can define a critical energy $E_c$ at which the radiation loss is equal to collision loss.

$$\left(\frac{dE}{dx}\right)_{rad} = \left(\frac{dE}{dx}\right)_{coll} \quad \text{for} \quad E = E_c \tag{6}$$

This critical energy may by estimated by the approximate formula of Bethe and Heitler:

$$E_c \approx \frac{1600 m_e c^2}{Z} \quad (7)$$

## 2.1.3 Particle tracks

A complete description of the track of an electron or charged particle would include the spatial coordinates of every interaction with the medium, the characteristics (energy, excited state etc.) of the projectiles after the collision, the characteristics of the target after the collision, and the energy, direction and other characteristics of any ejected secondary particles. This is a lot of information even for a single particle after being completely slowed down. Of course, not all events are known as we are limited by currently available cross-section data. The only method of calculating individual tracks is the Monte Carlo method based on event by event simulation. In chapter 4 a Monte Carlo model for the simulation of electron and $\alpha$-particle tracks in gaseous media (propane and nitrogen) is presented which was developed and adopted for the needs of the experiment with the Jet Counter.

Figure 2 shows the simulated ionization component of particle track segments for 4 different particles with the same speed in liquid water. As can be seen, these particles have different ionization densities. If we compare it with the size of DNA (critical target for living cells), 60 *MeV* carbon ions always produce ionization in this volume but 5 *MeV* protons and 2.72 *keV* electrons – only sometimes. It is very interesting that 5 *MeV* protons have an ionization yield almost the same as 2.72 *keV* electrons and as a result these two particles may have identical radio biological effectiveness.

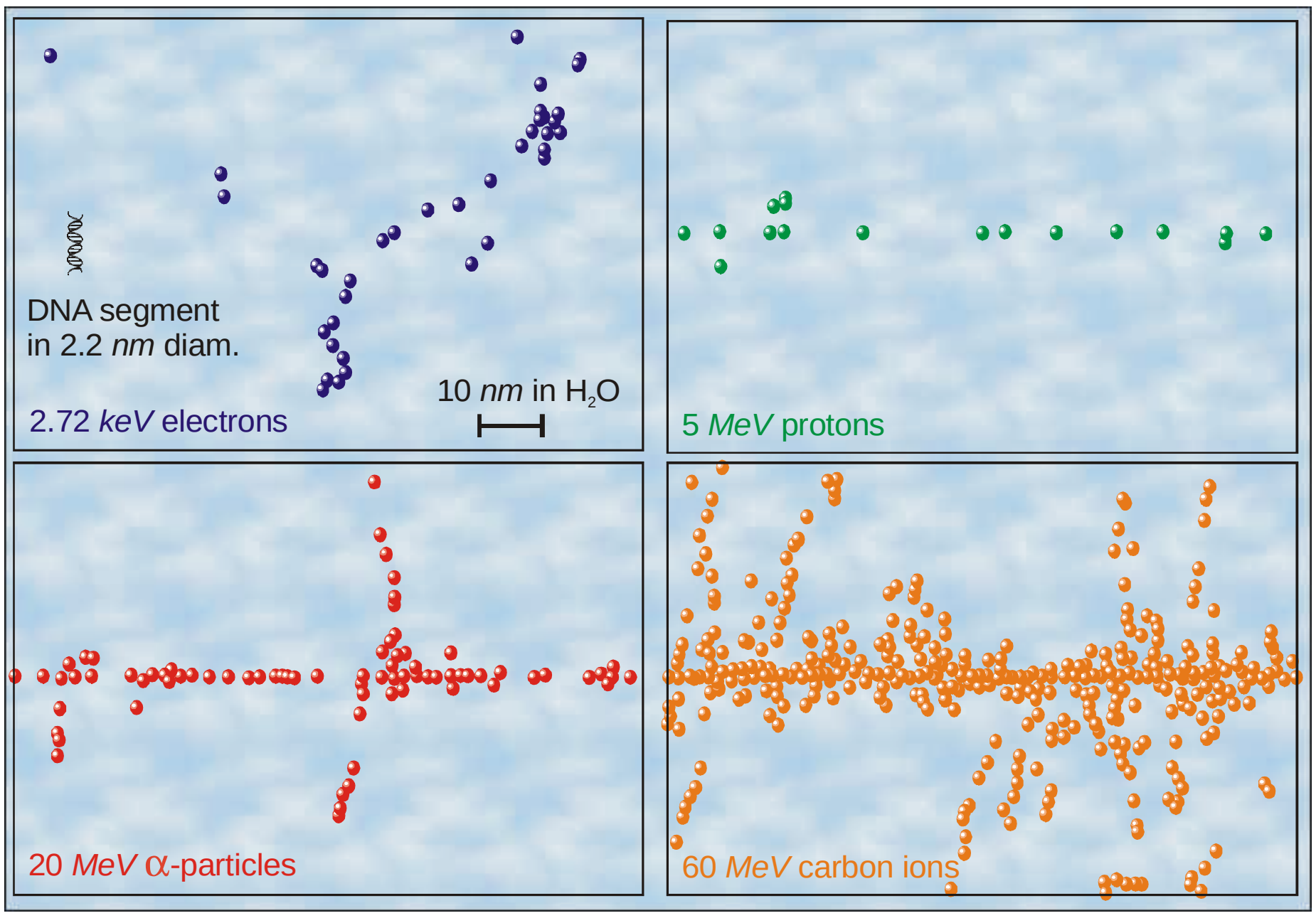

*Figure 2: Particle track segments due to ionization, 100 nm in length in liquid water. Courtesy: Bernd Grosswendt [18]. Modified.*

## 2.2 Physical description of a charged particle track

The track of a charged particle as obtained from Monte Carlo simulations is up to now the basic tool for interpretation of observed phenomena in radiation chemistry, radiobiology, nanoelectronics *etc*. Recently, the situation seems to have changed owing to the development of new experimental techniques. Therefore, it seems reasonable to assume that both approaches are perspective for many applications.

### 2.2.1 Dosimetry quantities for characterization of particle tracks (classical approach)

#### 2.2.1.1 Stopping power and ranges

By *stopping power* $S$ (Eq.8) we mean the average energy loss of the particle per unit length, measured for example in *MeV/cm*. The mass stopping power $S/\rho$ (Eq.9) is the stopping power divided by the density $\rho$ of the substance and may be expressed in units like *MeV/(mg/cm²)*.

$$S = -\frac{dE}{dx} \tag{8}$$

$$\frac{S}{\rho} = \frac{dE}{\rho dx} \tag{9}$$

The Bethe-Bloch formula (Eq.1) may be used to calculate the mass stopping power for charged particles.

The *range* of the charged particle, $R$, is defined as the expectation value of the path length that it follows until it comes to rest (Attix [19]). The Continuous Slowing Down Approximation Range, $R_{CSDA}$, is defined as

$$R_{CSDA} = \int_0^{E_0} \left( \frac{dE}{\rho dx} \right)^{-1} dE \tag{10}$$

#### 2.2.1.2 Linear energy transfer (LET)

*Linear energy transfer* (LET), $L_\infty$ is a measure of the **energy transferred to the material** as an ionizing particle travels through it and is defined as

$$L_\infty = \frac{dE}{dx} \tag{11}$$

The definition of LET can also be extended to indirectly ionizing particles such as photons or neutrons. In this case the term LET concerns the spectrum of LET of the secondary particles involved in the interactions with the medium of interest.

The major limitations of $L_\infty$ for characterization of a radiation field are as follow:

- $L_\infty$ does not account for the distribution of energy loss and energy deposition in a thin layer or for small targets. This distribution becomes significant if the particle's free path is comparable with the thickness of the layer or the dimensions of the target.
- Heavy charged particles with the same LET but different charge and velocity produce different δ-ray spectra. In particular, the response of detectors (biological cell, DNA, TLDs) irradiated with heavy charged particles of identical LET can differ significantly.

The collision stopping power is sometimes used as a synonym for LET, $L_\infty$. The infinity subscript denotes that the total energy of all δ-rays resulting from hard collisions was accounted for when calculating the stopping power. Radiative energy losses are not included in the LET because these photons do not contribute to the energy deposition in the vicinity of the heavy charged particle track.

The energy restricted LET, $L_\Delta$, only includes contributions from those δ-rays whose initial energies are lower than the cut-off energy $\Delta$. The cut-off energy is typically in the range of 100 *eV*. $L_\Delta$ is applied in some radiobiological models to calculate the energy deposition in small targets irradiated with energetic ions.

The range restricted LET $L_r$, is defined as that part of the total energy loss $dE/dx$ which is deposited within a cylinder of radius $r$ and length $dx$.

The values of stopping power and restricted LET for particles and media of interest to radiation protection and medical physics can be obtained in [20, 21, 22, 23].

### 2.2.1.3 Radial dose distribution

The *radial dose distribution*, $D(r)$, around the ion path is defined as the average energy deposited in a cylinder with radius between $r$ and $r+dr$, normalized to its mass. $D(r)$ is of principal importance in track structure theory (Katz [24]; Waligórski [25]; Horowitz [26]; Geiss *et al.* [27]; Paganetti *et al.* [28]) in predicting the response of physical detectors and biological systems. The general representation of $D(r)$ is

$$D(r) \approx \frac{Z^{*2}}{(v/c)^2 r^2} \tag{12}$$

$Z^*$ – is the particle's "effective charge" described by a formula which accounts for charge-pickup at low ion velocities,

$v$ – is the particle's velocity,

$c$ – is the speed of light in vacuum,

$r$ – is defined between $r_{min}$ (about 0.1 *nm*) and $r_{max}$ which is determined by the maximum range of δ-electrons and depends on the ion's velocity. The integral of $D(r)$ per unit path should in general be equal to the LET of the particle.

The main disadvantage of $D(r)$ for modeling the response of physical detectors (biological cells, DNA, TLD) is that it cannot be derived for photons and electrons, i.e. modeling of detector response to these low-LET particles is not possible. Also, it is still a macroscopic parameter applied to the nanometer scale.

### 2.2.2 Microdosimetric quantities: energy imparted, lineal energy, specific energy

When an ionizing particle interacts with matter it transfers its energy to the medium in the form of ionization and excitation. The elementary quantity used in microdosimetry to describe the energy transfer in a single interaction is the *energy deposited,* $\varepsilon_i$ , which is defined as ([29]):

$$\varepsilon_i = T_{in} - T_{out} + Q_{\Delta m} \qquad \textbf{(13)}$$

Where: $T_{in}$ – is the energy of the incident ionizing particle,

$T_{out}$ – is the sum of the energies of all ionizing particles leaving the interaction,

$Q_{\Delta m}$ – is the change of the rest mass of the atom and all particles involved in the interaction.

The energy deposited is a stochastic quantity, i.e., it is subject to random fluctuation for a given incident particle energy. The contribution from all energy depositions $\varepsilon_i$ in volume $V$ is called the *energy imparted,* $\varepsilon = \sum_i \varepsilon_i$ . The energy imparted can be expressed in the form of the number of ionizations, $j$, which occurin the volume. If $W$ is the average energy produced in the medium per ionization event, then

$$\varepsilon = jW \qquad \textbf{(14)}$$

Heavy charged particles crossing the volume of interest are characterized by the so-called *lineal energy,* $y$ , which is defined as the actual energy deposited $\varepsilon$ in the volume, divided by the mean chord length $\bar{l}$ of this volume:

$$y = \frac{\varepsilon}{\bar{l}} \qquad \textbf{(15)}$$

The lineal energy is a stochastic quantity, usually expressed in units of *keV/μm*.

The related quantity, *specific energy* (imparted), $z$ , is defined as the energy deposited per mass of the volume, $m$ , and is expressed in grays (1 *Gy* = 1 *J/kg*),

$$z = \frac{\varepsilon}{m} \qquad \textbf{(16)}$$

The mean values of $y$ and $z$ are microdosimetric analogues of LET, $L$, and dose, $D$. For spherical targets of unit density ($\rho = 1 \; g/cm^3$) the relationship between specific and lineal energy is

$$z[Gy] = 0.204 \frac{y[keV/\mu m]}{d^2[\mu m]} \qquad \textbf{(17)}$$

Where: $d$ is the sphere diameter.

### 2.2.3 Nanodosimetric quantities: $\nu$, $P_1$, $M_1$ and $F_2$

Along with the development of experimental nanodosimetry (see chapter 2.6) the necessity of defining new quantities is evident. The first concept of what has to be measured was given by Pszona: "*It is of special interest to obtain experimental data on the frequency of occurrence of various numbers of ions when an ionizing particle is traversing a tissue domain of the order of a nanometer*" [30]. Later on the name cluster size was introduced, and the frequency distribution of cluster size has become the main experimental data which can be obtained from nanodosimetric devices.

Therefore the following quantities can be defined:

A *cluster* is a spatially correlated group of ionizations, created by a charged particle in a nanometric target volume. In DNA studies a typical shape dimension and spatial location of the target volume relative to a particle track ranges between 1 and 10 *nm*. The cluster approach has been applied to study the relation of DNA strand breaks by charged particles (Grosswendt *et al.* [31], Garty *et al.* [32], Michalik [33]).

As was shown above, macroscopic definitions (like absorbed dose) of the radiation field are not valid directly for the description of nanometric targets (like DNA). The nanodosimetric concept presented here is based on the track structure of the ionizing particles and its statistical characteristics.

Let $P_\nu(Q;d)$ be the probability that exactly $\nu$ ionizations are produced within a specified cylindrical target volume by a single primary particle of radiation quality $Q$ (particle type, energy, …), passing the volume at a distance $d$ to its main axis (see Figure 3, left). In the present work, the Jet Counter simulates a nanometric-sized volume with incident particles at $d = 0$ (see Figure 3, right).

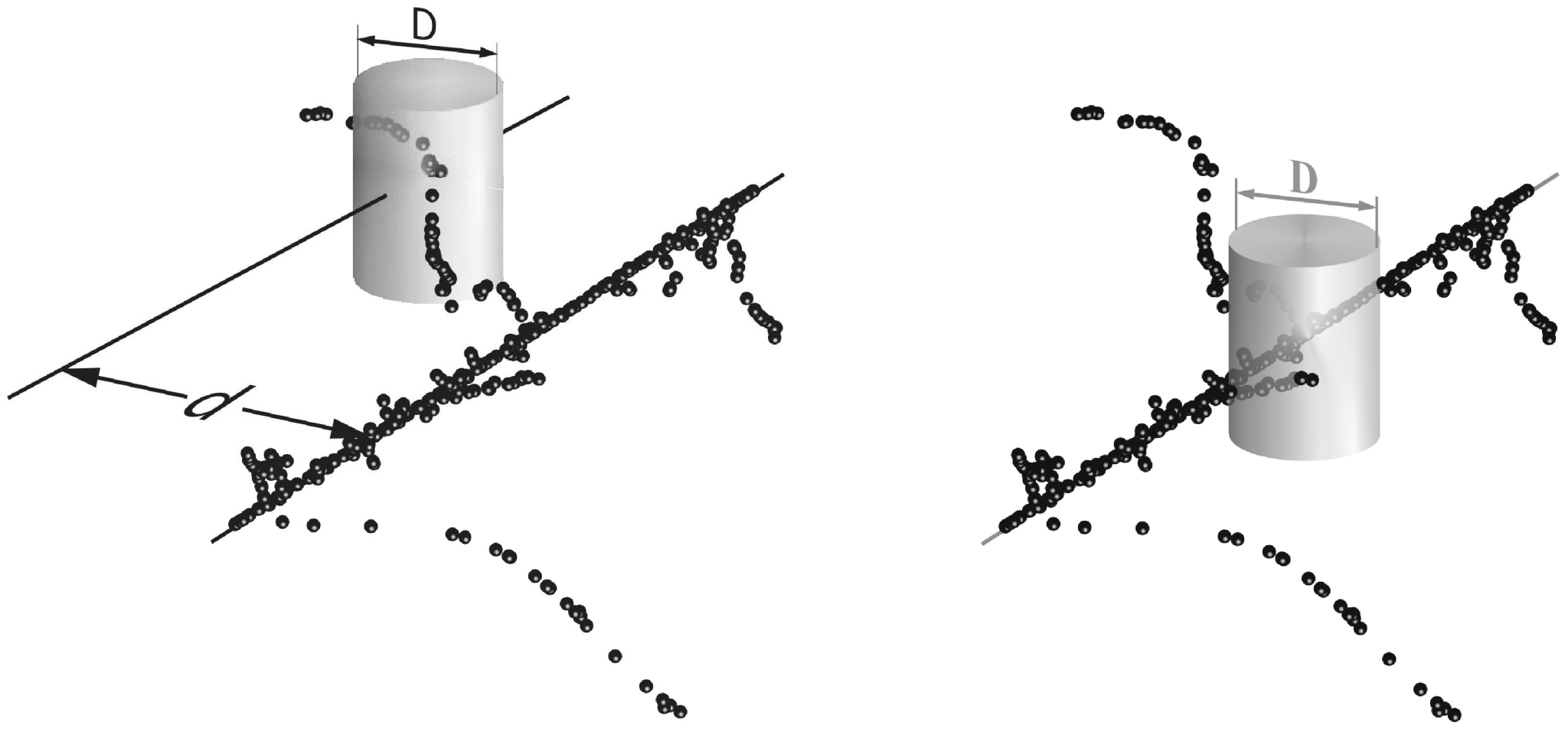

*Figure 3: Ionization cluster-size formation by a primary particle passing a specified cylindrical target volume of diameter $D$: left is at a distance $d$ from the cylinder's main axis and right with $d = 0$. The particle track segment shown is represented by its ionization component. Ref.[34]*

The number of ionizations produced by a primary particle (including its secondary electrons) within the target volume is called the *cluster size* $\nu$ in what follows. It is a stochastic quantity, and represents the result of a superposition of the ionization component of the particle track structure and the geometric characteristics of the target volume. The distribution of the probability $P_\nu(Q;d)$ with respect to formation of cluster size $\nu$ therefore describes the stochastic nature of the conjunction of track structure and target volume.

In this sense, the moments $M_\xi(Q;d)$ of the distribution are the characteristics of special aspects of conjunction and may strongly depend on the type and energy of the primary particle, on the area density of the target volume, on the shape and size of the target volume, and on the distance $d$ as follows:

$$M_\xi(Q;d) = \sum_{\nu=0}^{\infty} \nu^\xi P_\nu(Q;d) \text{ with } \sum_{\nu=0}^{\infty} P_\nu(Q;d) = 1 \qquad \textbf{\textit{(18)}}$$

The first and second moments $M_1(Q;d)$ and $M_2(Q;d)$ are of particular interest as $M_1(Q;d)$ stands for the mean cluster size, and $M_2(Q;d)$ is needed to characterize the fluctuation of cluster size formation which is commonly expressed by the variance $M_2(Q;d) - M_1(Q;d)^2$ [35].

The cumulative frequency (probability), $F_k(Q;d)$, of a cluster size $\nu \geq k$,

$$F_k(Q;d) = \sum_{\nu=k}^{\infty} P_\nu(Q;d) \qquad \text{for } k = 1,2,3,4... \qquad \textbf{\textit{(19)}}$$

From the point of view of radiation biology, if the nanometer target (defined previously) is equivalent to DNA, the hypotheses (see chapter 7) are:

1. The probability $P_1(Q;d)$ to create cluster size $\nu = 1$ should be proportional to the probability of SSB (single strand break) in DNA (one ionization is needed to create one SSB),
2. The frequency $F_2(Q;d) = \sum_{\nu=2}^{\infty} P_\nu(Q;d)$ to create $\nu \geq 2$ should be proportional to the probability of DSB (double strand break) formation in DNA (at least two ionizations are needed to break the helix in DNA).

To study this question systematically, the probability $P_\nu(Q;d)$ of cluster size $\nu$ due to ionizing particles at radiation quality $Q$ must be determined in cylindrical target volumes of liquid water 2.2-2.4 *nm* in diameter and *3.4 nm* in height, for different kinds of ionizing particles at various energies, and compared with the probability of strand-break formation afterwards.

In the special case of a macroscopic target volume, where the initial particle energy $T$ is completely absorbed, $M_1(Q;d)$ is equal to the mean number $N(T)$ of ion pairs formed, which is conventionally expressed by $N(T) = T/W(T)$, where $W(T)$ is the so-called *W*-value defined as the mean energy expended per ion pair formed upon the complete degradation of a charged particle [36].

So, as a result, to characterize a particle track on a specific nanometric scale, the frequency distribution of cluster-size is used. Certainly, for application purposes it is more convenient to use statistical parameters, derived from the frequency distribution of the cluster-size for a given nanomeric scale, like $P_1(Q;d)$, $M_1$, (Eq.18) and $F_2$ (Eq.19) or a combination of them.

Following the analysis of a compound Poisson process by De Nardo *et al.* [34] the mean cluster size $M_1(Q;d)$ and the variance are factorized in two terms:

$$M_1(Q,d) = \kappa(Q) \times m_1(Q,d) \qquad \textbf{(20)}$$

$$M_2(Q;d) - M_1(Q;d)^2 = \kappa(Q) \times m_2(Q,d) \qquad \textbf{(21)}$$

Here, $\kappa(Q)$ is the mean number of primary ionizations produced by a primary particle when crossing the interacting volume of the Jet Counter. It is equal to the ratio $(D\rho)/(\lambda\rho)_{ion}$ of the mass per area of the Jet Counter's diameter to the mean free ionization path length of the primary particles. The variables $m_1(Q,d)$ and $m_2(Q,d)$ describe the contribution of secondary electrons to the cluster size.

One of the most important ratios $m_2(Q,d)/m_1(Q,d)$, derived using (Eq.20) and (Eq.21), is:

$$\frac{M_2(Q,d)}{M_1(Q,d)} - M_1(Q,d) = \frac{m_2(Q,d)}{m_1(Q,d)} \qquad \textbf{(22)}$$

From the mathematical point of view, (Eq.22) is formally equivalent to the expressions used in microdosimetry to relate the ratio of the variance of specific energy $z$ to the absorbed dose, in the case of the dose-dependent microdosimetric distribution $f(z;d)$, with the single-event quantity dose-mean specific energy per event $\bar{z}_d$ [a detailed discussion of these quantities is given by Kellerer [37] and Kellerer and Chmelevsky [38].

So, as a result, to characterize a particle track on a specific nanometric scale the frequency distribution of cluster-size is used. Certainly, for application purposes it is more convenient to use statistical parameters derived from the frequency distribution of cluster-size like $P_1(Q;d)$, $M_1$, $M_2$ (Eq.18) and $F_2$ (Eq.19) or combinations of them.

More details about the concept of cluster size formation in nanometric targets with main dependencies are in De Nardo *et al.* [34] .

As mentioned before, it is the aim of experimental nanodosimetry not only to measure the frequency distributions of ionization cluster size in arbitrary nanometric gaseous volumes but also to determine the frequency distributions of ionization clusters $P_\nu(T)$, which would be measured in a nanometric volume of liquid water as a substitute for sub-cellular structures. For this purpose, a procedure must be applied experimentally that ensures that ionization cluster-size frequencies measured in a gas are equivalent to cluster-size frequencies for a nanometric liquid water target of specified dimensions. One requirement, which must be fulfilled to reach such a material equivalence is that, at least, the mean cluster size of the measured $P_\nu(T)$ distributions are approximately the same as those to be expected in liquid water. This leads to the following scaling rule that relates the mass per area $(D\rho)^{(gas)}$ of the diameter of a gaseous measuring volume to the mass per area $(D\rho)^{(water)}$ of the diameter of a liquid water cylinder for which the ionization cluster-size distribution is to be determined according to the paper of Grosswendt [39]:

$$(D\rho)^{(gas)} = \frac{(D\rho)^{(water)}}{\varepsilon} \times \frac{(\lambda\rho)_{ion}^{(gas)}(T)}{(\lambda\rho)_{ion}^{(water)}(T)} \times K_{el} \qquad \textbf{(23)}$$

Here: $\varepsilon$ – is the experimental efficiency of ion detection and counting,

$(\lambda\rho)_{ion}^{(gas)}(T)$ or $(\lambda\rho)_{ion}^{(water)}(T)$ – represents the mass per area of the mean free-path length of a particle at energy $T$ with respect to ionization in the gaseous system or in liquid water,

$D$ – is diameter of the medium cylinder,

$\rho$ – is the medium density,

$K_{el} = \frac{m_1^{water}(T, d\rho)}{m_1^{gas}(T, d\rho)}$ is a correction factor, the moment ratio, which takes into account the different interaction properties of secondary electrons in the two media.

Unfortunately, (Eq.23) can only be used if the first moments of the single-ionization distributions in gas and in liquid water are known, for instance, by a series of Monte Carlo calculations. It can, however, be directly applied if the contribution of secondary electrons to the cluster-size distributions in both media is negligible and $K_{el} = 1$. This leads to a very simple scaling procedure to reach material equivalence, which is based only on the primary ionization:

$$(D\rho)^{(gas)} = \frac{(D\rho)^{(water)}}{\varepsilon} \times \frac{(\lambda\rho)_{ion}^{(gas)}(T)}{(\lambda\rho)_{ion}^{(water)}(T)} \qquad \textbf{(24)}$$

Detailed discussions of equations (Eq.23) and (Eq.24) are presented by Grosswendt in references [39, 40, 41].

The mean free path lengths $(\lambda\rho)_{ion}^{(nitrogen)}(T)$ and $(\lambda\rho)_{ion}^{(propane)}(T)$ of $\alpha$-particles with respect to ionization in nitrogen or propane gas are presented in Figure 4 as a function of energy $T$, in comparison with those in liquid water.

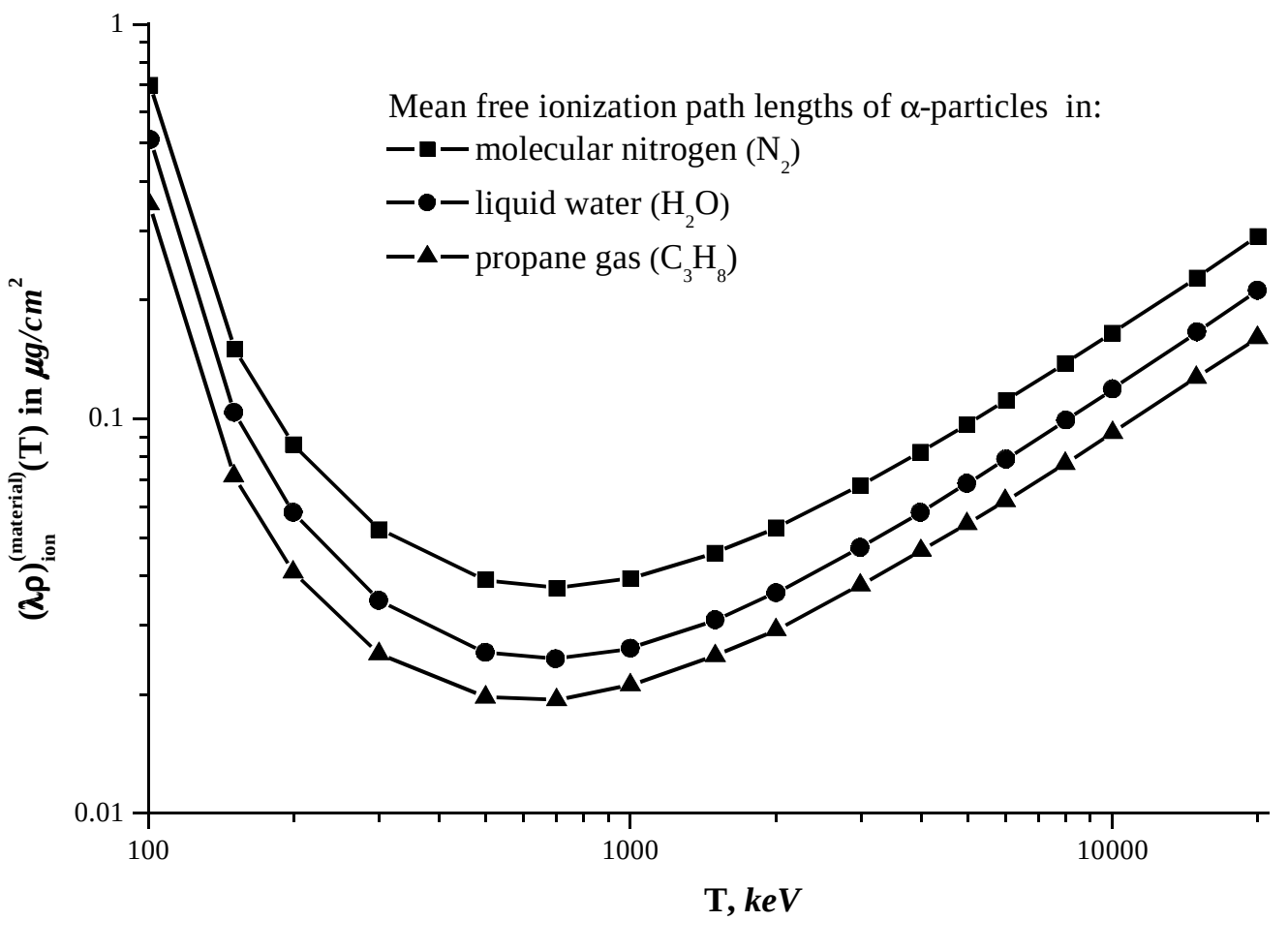

*Figure 4: Mean free ionization path lengths $(\lambda\rho)_{ion}^{(material)}(T)$ of $\alpha$-particles in nitrogen (■), liquid water (●) or propane (▲) as a function of the particle energy T.*

At first glance, it can be seen from the figure that $(\lambda\rho)_{ion}^{(nitrogen)}(T)$ is always greater than $(\lambda\rho)_{ion}^{(water)}(T)$, whereas $(\lambda\rho)_{ion}^{(propane)}(T)$ is always smaller then $(\lambda\rho)_{ion}^{(water)}(T)$. In consequence, the mass per area of the equivalent diameter in nitrogen is greater than that specified for water, whereas in propane it is always smaller.

For example, a 2.2 *nm* diameter helix of DNA corresponds to a water cylinder with a diameter equal to 0.22 $\mu g/cm^2$ (water is the equivalent material). Using equation (Eq.24), for 3.8 $MeV$ $\alpha$-particles with an experimental efficiency of ion detection and counting $\varepsilon = 100\,\%$, the equivalent diameters for molecular nitrogen and propane gas are 0.32 $\mu g/cm^2$ and 0.18 $\mu g/cm^2$ respectively.

At first glance, this procedure seems to be of doubtful value, since it is hardly conceivable that gaseous systems well suited for proportional counter experiments show the same mechanisms of radiation interaction as sub-cellular material. This argument is generally true for excitation processes that strongly depend on the target species, but it is not so serious from the point of view of ionization cluster-size formation, because the energy distribution of secondary electrons set in motion by impact ionization does not strongly depend on the type of target molecules.

## 2.3 Radiation effects at DNA levels

Radiation Causes Ionizations of:

- ATOMS which may affect
  - MOLECULES which may affect
    - CELLS which may affect
      - TISSUES which may affect
        - ORGANS which may affect
          - THE WHOLE BODY.

Although we tend to think of biological effects in terms of the effect of radiation on living cells, in actuality, ionizing radiation, by definition, interacts only with atoms by a process called ionization. Thus, all biological damage effects begin as a consequence of the interaction of radiation with the atoms forming the cells. As a result, radiation effects on humans proceed from the lowest to the highest levels as noted in the above list.

All subsequent biological effects can be traced back to the interaction of radiation with atoms. Radiation-sensitive targets in the human (mammalian) cell are concentrated in the cellular nucleus and, in particular, the deleterious effects of ionizing radiation are known to arise from radiation damage to the DNA molecule (about 2.2 *nm* in diameter) either by direct ionization or indirectly via the action of hydroxyl radicals which are produced during water radiolysis (Maurer *et al.* [42]).

If radiation interacts with the atoms of the DNA molecule, or some other cellular component critical to the survival of the cell, it is referred to as a direct effect. Such an interaction may affect the ability of the cell to reproduce and, thus, survive. If enough atoms are affected such that the chromosomes do not replicate properly, or if there is significant alteration in the information carried by the DNA molecule, then the cell may be destroyed by “direct” interference with its life-sustaining system.

If a cell is exposed to radiation, the probability of the radiation interacting with the DNA molecule is very small since these critical components make up such a small part of the cell. However, each cell, just as is the case for the human body, is mostly water (about 70 %). Therefore, there is a much higher probability of radiation interacting with the water that makes up most of the cell's volume. When radiation interacts with water it produces fragments such as hydrogen (H) and hydroxyls (OH). These fragments may recombine or may interact with other fragments or ions to form compounds, such as water, which would not harm the cell. However, they could combine to form toxic substances, such as hydrogen peroxide ($H_2O_2$), which can contribute to the destruction of the cell – an "indirect" effect.

Von Sonntag [43] has estimated that direct effects contribute about 40 % to cellular DNA damage, while the effects of water radicals amount to about 60 %. A paper by Krisch *et al.* [44] on the production of strand breaks in DNA initiated by $OH^-$ radical attack has the direct effects contribution at 50 %.

The typical damages of DNA by "direct" and "indirect" ionization effects are:

- Single strand break (SSB) – a break in the double-stranded DNA in which only one of the two strands has been cleaved; both strands have not separated from each other. This type of damage is rather easy to repair, as the opposite DNA strand remains intact.
- Double strand break (DSB) – a break in the double-stranded DNA in which both strands have been cleaved; however, the two strands have not separated from each other. This type of damage is unrepairable or very difficult to repair and as a result the DNA is nonfunctional and the cell in consequence dies.

So, the target for ionization radiation in a cell is the DNA which has nanometer dimensions. The aim of current radiation physics is to give adequate characteristics of ionizing radiation on a scale comparable with DNA. These characteristics should be measurable for application purposes and correlated with SSB, DSB and other types of DNA damage.

## *2.4 Radiation effects in microelectronic devices*

In many cases, microelectronic devices are present in high radiation environments, far above the exposures typically encountered by any biological system (e.g. high energy physics, space). Even natural radiation or technological contamination of electronic components by natural radionuclides may cause radiation effects. The difference only is in the scale of these effects.

In micro and nano-electronic elements radiation effects may be manifested by a large diversity of secondary effects, so-called single-event effects (SEE), depending on the hit region and on the type of interacting charged particles.

A single event upset (SEU) is a change of state caused by ions or electro-magnetic radiation striking a sensitive node in a micro-electronic device, such as a microprocessor, semiconductor memory, or power transistor. The state change is a result of the free charge created by ionization in or close to an important node of a logic element (e.g. a memory "bit"). The error in the device output or operation caused as a result of the strike is called an SEU or a soft error. The SEU itself is not considered permanently damaging to the transistor's or circuit's functionality, unlike the case of a single event latchup (SEL), single event gate rupture (SEGR), or single event burnout (SEB). These are all examples of a general class of radiation effects in electronic devices called single event effects (SEE).

There are several reports [6, 7, 8] on the formation of a single-event upset (SEU) in microelectronic devices caused by charged particles (cosmic rays, particles emitted in solar events, $\alpha$-particles from materials contaminated with natural radionuclides).

As the size of the elements of electronic circuits is constantly decreasing (currently approaching a few nanometres, see Figure 5), their capacity also decreases as does the charge necessary to manifest an SEE. It is only a matter of time until the elementary circuits of RAMs attain a few nanometers in size. At these nanometre sizes, single-event effects are of particular importance.

As, at present, no such single nanometer-sized electronic structures exist, the only way to study the charge generation issue is by mathematical or experimental simulation. Experimental approaches have recently become available for studying such topics. In these experiments, the nanometre-sized electronic structure is replaced by a nanometre-sized gas target. Here, nitrogen and propane appear to be the most convenient media. One experimental approach was successfully applied in the micro-dosimetry of devices, as described extensively by Bradley *et al.* [45].

So, one of the challenges of current nano-dosimetry is to experimentally simulate nanometric semiconductor elements in a radiation field.

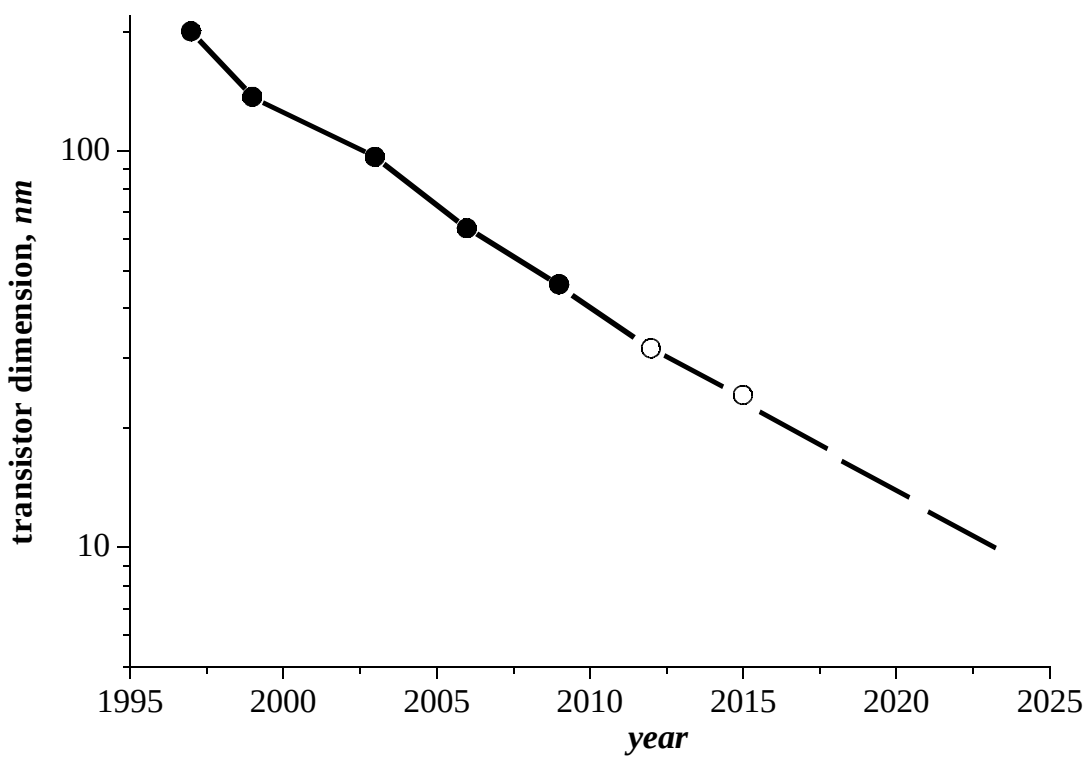

*Figure 5: Transistor dimensions in electronics versus time: (●) – current, (○) – prognosis.*

## 2.5 Remarks on the weaknesses of the absorbed-dose concept in the case of nanometric targets

(after B. Grosswendt [18])

As mentioned above, current radiation-therapy treatment planning is conventionally based on the assumption that the interaction of ionizing particles in matter and, in consequence, also the inducing of radiation effects, like damage to living cells or to cellular substructures, can be satisfactorily described by the absorbed dose. This assumption seems to be, at least, questionable if one bears in mind that the absorbed dose considers neither the atomic structure of matter nor the stochastic nature of particle interactions and the track structure of ionizing particles. This raises the question of the validity of the absorbed dose to describe radiation damage to sub-cellular structures with sizes of a few nanometres.

To study the impact of the target size on the validity of absorbed-dose quantities let us consider a small but still macroscopic piece of matter of volume $V$ and mass $m$ which is irradiated by a number $n$ of ionizing particles. If we additionally assume that the energy deposition in the target volume is homogeneously distributed, the absorbed dose $D$ at a point within $V$ is:

$$D = \frac{\tilde{\varepsilon}}{m} = \frac{\sum_{j=1}^{n} \sum_{i} \varepsilon_i^{(j)}}{m} \tag{25}$$

Where: $\tilde{\varepsilon}$ – is the mean energy imparted to the piece of matter which, by definition, is equal to the sum of all the energy deposits in the target volume $V$,

$\varepsilon_i^{(j)}$ – represents the *i*-th energy deposited within the target volume $V$ due to an inelastic interaction of the *j*-th particle.

In view of the fact that the absorbed dose is defined as a point quantity, the major prerequisite for the validity of (Eq.25) is the supposed homogeneous distribution of energy deposits within the target volume.

But, considering the fact that energy deposition by ionizing particles is determined by a series of discrete inelastic interaction processes, it is obvious that a homogeneous distribution of energy deposition within $V$ can only be reached in the case of:

- a very large number $n$ of primary particles entering the target volume
- or in radiation fields of very high particle fluence.

It can never be fulfilled in radiation fields of low intensity like those, for instance, which are usually of interest in radiation protection.

If the condition of a homogeneous distribution of energy deposits is not fulfilled, the definition of the absorbed dose according to (Eq.25) meets its limits. To demonstrate this, let us skip the assumption of a homogeneous distribution of energy deposits, and take into account the track structure of the ionizing particles in matter (see the left-hand side of Figure 6 for $n=1$). If we do this, (Eq.25) can no longer be interpreted as the absorbed dose at a point, but only as the mean absorbed dose within the target volume which, of course, may depend on $V$.

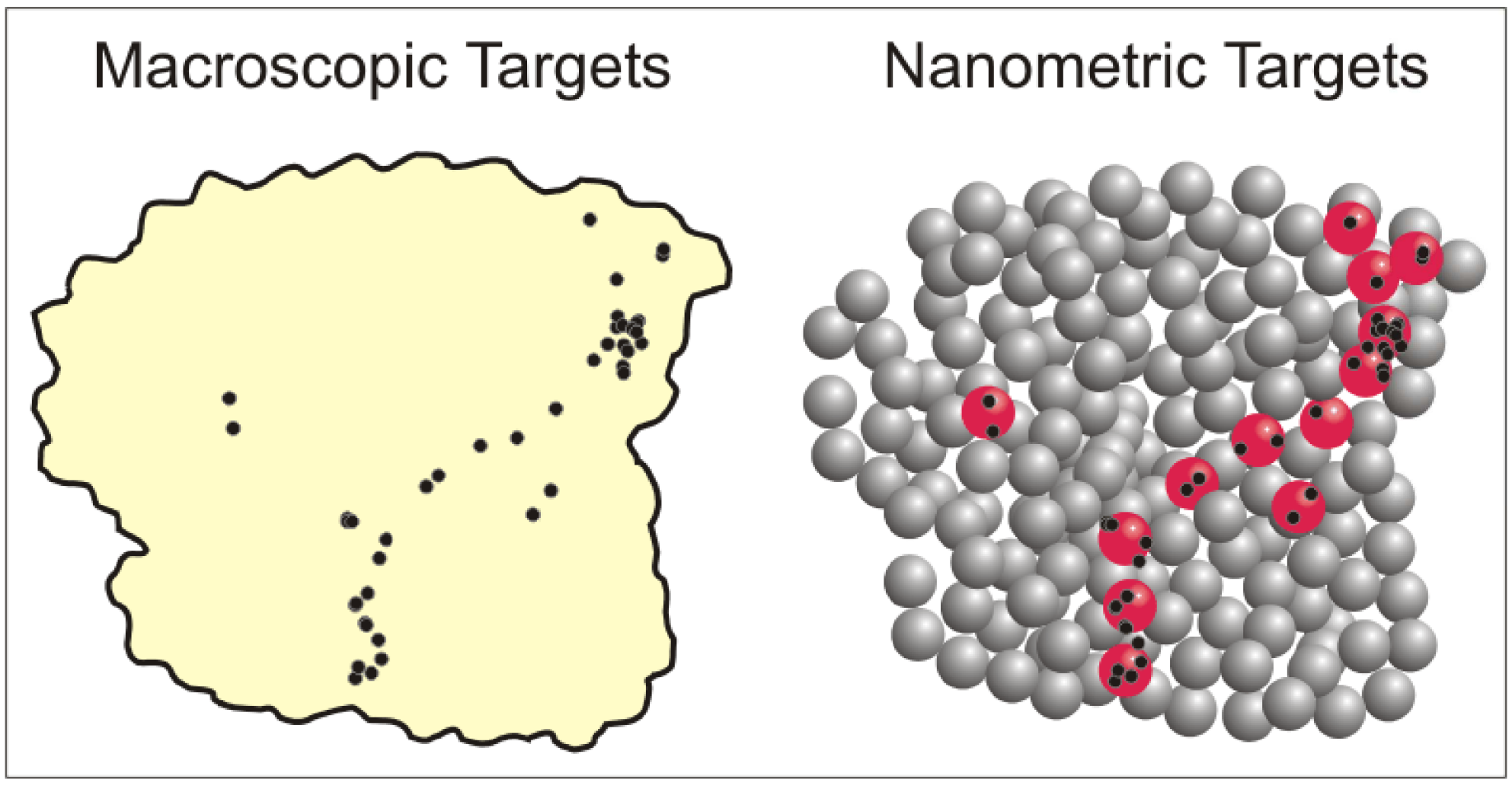

*Figure 6: Schematic view of a particle track in a homogeneous macroscopic target of volume V (left hand side) and of the same track in a collection of N nanometric targets of volume V/N (right hand side) Ref.[18].*

Suppose now that the irradiated target volume $V$ is composed of $N$ radiation sensitive sub-volumes of mass $\Delta m = m / N$, and ask for the mean absorbed dose in those sub-volumes of $V$ which receive at least one energy deposit due to an inelastic particle interaction (see the red spheres in the right-hand part of Figure 6). If the number of sub-volumes which receive an energy deposit is indicated by $n_{hit}$, the microscopic absorbed dose $D_{micros}$ and its relation to the macroscopic absorbed dose $D$ is given by (Eq.26).

$$D_{micros} = \frac{\bar{\varepsilon}}{n_{hit} \times \Delta m} = \frac{\sum_{j=1}^{n} \sum_{i} \varepsilon_i^{(j)}}{n_{hit} \times \Delta m} = \frac{N}{n_{hit}} \times \frac{\sum_{j=1}^{n} \sum_{i} \varepsilon_i^{(j)}}{m} = \frac{N}{n_{hit}} \times D \qquad (26)$$

In view of the fact that the number $n_{hit}$ of hit sub-volumes is always less than or equal to the total number $N$ of sub-volumes of $V$, the microscopic absorbed dose $D_{micros}$ is generally greater than or equal to $D$, independently of the number of sub-volumes.

To give an impression of the large differences which may exist between $D_{micros}$ and $D$, in particular in the field of radio-biology, let us suppose that the irradiated $V$ is given by the nucleus of a living human cell, and the sub-volumes by DNA segments of 10 base pairs (2.3 *nm* in diameter, 3.4 *nm* in length), which, in radio-biology, are generally accepted to represent the most important radiation sensitive volumes of a living cell. Taking into account a total length of the DNA of 1.5 *m* in each human cell, the number $N$ of radiation sensitive sub-volumes in a cell nucleus is about $4.5 \cdot 10^8$. This number is generally greater by orders of magnitude than the number of hit sub-volumes apart from the application of extremely strong radiation fields. **In consequence, it can almost never be expected in practice that quantities based on a macroscopic absorbed-dose concept are very well suited to a detailed description of radiation damage to target volumes comparable in size to those of short DNA segments.**

Typical examples in which metrological problems can be expected if absorbed-dose quantities are applied are:

- in a microbeam facility for radiobiology – if single cells are irradiated with ionizing particles using, for instance, a microbeam facility,
- in hadron therapy in the region of the spread-out Bragg peak, or in radio-nuclide therapy if radioactive nuclei are inserted directly into the tumour cells.
- in Boron-neutron-capture therapy (BNCT) [46].

In all of these examples, the absorbed dose applied to the tumour volume can not be a representative quantity to directly characterize what really happens in the radiation-sensitive nanometric volumes of the irradiated cells. So, it's clear that new quantities are needed.

## 2.6 Experimental set ups for track characterization on the nanometric scale

One of the aims of current nanodosimetry is to develop experimental procedures. A method of measuring the relevant quantities that is applicable to sub-cellular structures is the determination of the radiation induced frequency distribution of the ionization cluster size (number of ionizations per primary particle) in liquid water. Liquid water is used as a substitute for sub-cellular material in volumes that are comparable in size with those of the most probable radio-sensitive volumes of biological systems (DNA segments 2.2-2.4 *nm*, nucleosomes 11 *nm*, chromatin fibers 30 *nm*, see Figure 1). The idea of new descriptors of radiation action at the nanometric level, namely the frequency distribution of the creation of ionization clusters, was first proposed by Pszona [30]. Such frequency distributions are, to a large extent, governed by the same basic physical interaction data as those that can be expected if charged particles interact, for instance, with DNA segments. Consequently, frequency distributions of ionization cluster size in nanometric volumes of liquid water (nanodosimetry) can also be used for the definition of new descriptors of radiation quality.

***The idea of experimental nanodosimetry, similar to microdosimetry is to assume that the size of a gas-filled measuring volume must have the same mass per area as the size of the simulated target of liquid water, as a substitute for a sub-cellular biological target.***

In order to simulate, for instance, a cylindrical volume of liquid water, 10 *nm* in diameter, by a gaseous volume, 1 *cm* in diameter, the gas area density must be 1 $\mu g/cm^2$, which corresponds to a gas pressure of the order of 86 *Pa* in molecular nitrogen at room temperature.

A proposed measuring device (see Figure 7) for determining ionization cluster-size distributions in 'nanometric' gaseous target volumes consists of

- a low-pressure interaction chamber with target volume,
- an electrode system to create an extracting field and extract ions or low-energy electrons from the interaction chamber,
- an evacuated drift column that includes at its end a single particle detector.

Charged particles $Q$ enter the interaction chamber, penetrate or pass through a walled (wall-less) target volume of definite shape and size and reach a trigger detector. Positive ions ⊕ or low energy electrons $e^-$ induced by each primary particle $Q$, including its secondaries, within the target volume are extracted from the interaction chamber into an evacuated drift chamber and counted by an ion or electron counter, which can detect single particles. In measuring devices, the frequency distribution $P_\nu(T)$ of ionization cluster size, $n$, is measured for a great number of primary particles $Q$ at a specified energy $T$ by counting the number of ionizations (or low energy electrons) caused within the sensitive measuring volume by each single primary particle, including its secondaries.

A fundamental difference between ion-counting and electron-counting is the fact that radiation-induced electrons (δ-electrons) have a wide range of kinetic energies (up to a few *keV*) and cannot be thermalized within a small gas target. Ions, on the other hand, have much lower initial kinetic energy. As a result the track image obtained using an ion-counting device will reflect the place where the ionizations took place whereas an image obtained using an electron based device will reflect the location of the electrons after they have thermalized. The former is, of course, the more interesting as the damage is formed at the place of ionization. The use of an electron based device will therefore tend to shift the location of the measured ionizations from the track core to the δ-electron track ends, resulting in reduced efficiency in imaging the track core and over-estimation of the ionization density at the δ-electron track ends.

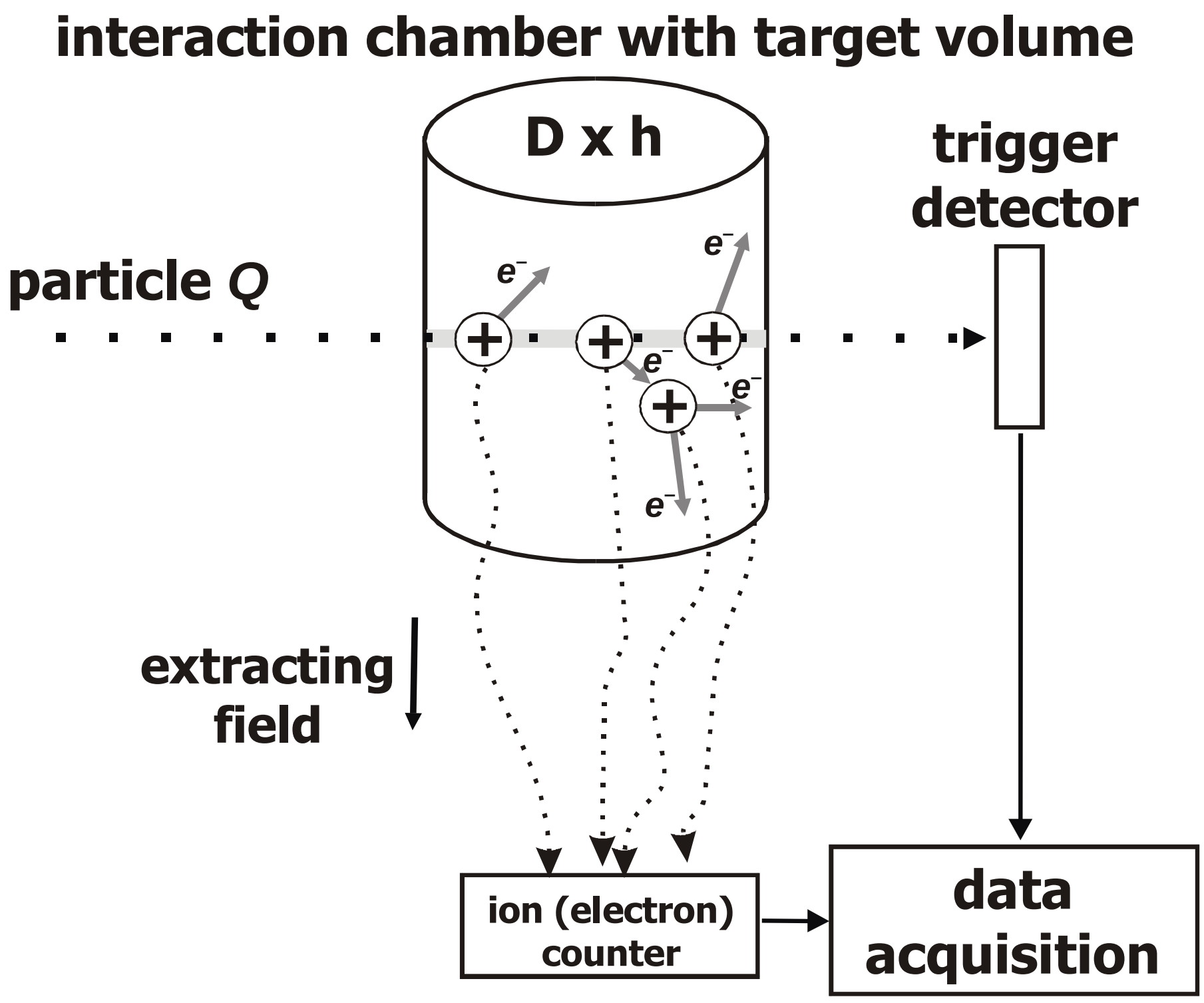

*Figure 7: Schematic view of a typical ion(electron)-counting measuring device which can be applied to determining ionization cluster-size distributions.*

Applying this method, it is assumed that:

1. the interaction mechanisms of ionizing radiation in the counter gas are similar to those in biological tissue,
2. the cross sections, the kind and number of interactions, and the most important energy loss channels are almost independent of the gas used,
3. the particle tracks are not noticeably disturbed by any component of the measuring device.

The final results are, of course, not unique for different measuring devices but also depend, apart from the particle energy, on the irradiation geometry, the type and pressure of the measuring gas as well as on the detection and counting efficiency of the measuring device, which ranges between about 20 to 60 %.

### 2.6.1 Ion-counting measuring devices

There are three ion-counting nanometric measuring devices and only two of them are currently in use. The first one is **a track ion counter,** ***developed*** by Pszona [30] in 1976. ***The first results (in the world) obtained were measurements of the cluster size spectra for*** $\alpha$***-particles for 0.15 nm sites in nitrogen.*** Later, the nanodosimeter at the Weizmann Institute of Science was created by Shchemelinin *at al.* [47] ***as a further*** development of the track ion counter idea. The final device is the Jet Counter facility at the Andrzej Sołtan Institute for Nuclear Studies (SINS) built by the group of S.Pszona [48] in 1994, ***which will be used as the basic instrument in this work.***

#### 2.6.1.1 The Track Ion Counter

The first measuring device was proposed by Pszona [30] in 1976 and previously announced in 1973 [49]. A schematic diagram of the Track Ion Counter (TIC) is shown in Figure 8. Positive ions, produced in a cylindrical volume above an orifice (F) by charged particles traversing the volume move in a constant electric field. Some of these ions pass through the orifice (F) and are subsequently accelerated and detected by a «spiratron» electron multiplier. The upper part of the counter, above the diaphragm with the orifice (F), contains an ion source and a Faraday cup. The Faraday cup is inserted in order to estimate the yield of the ion source.

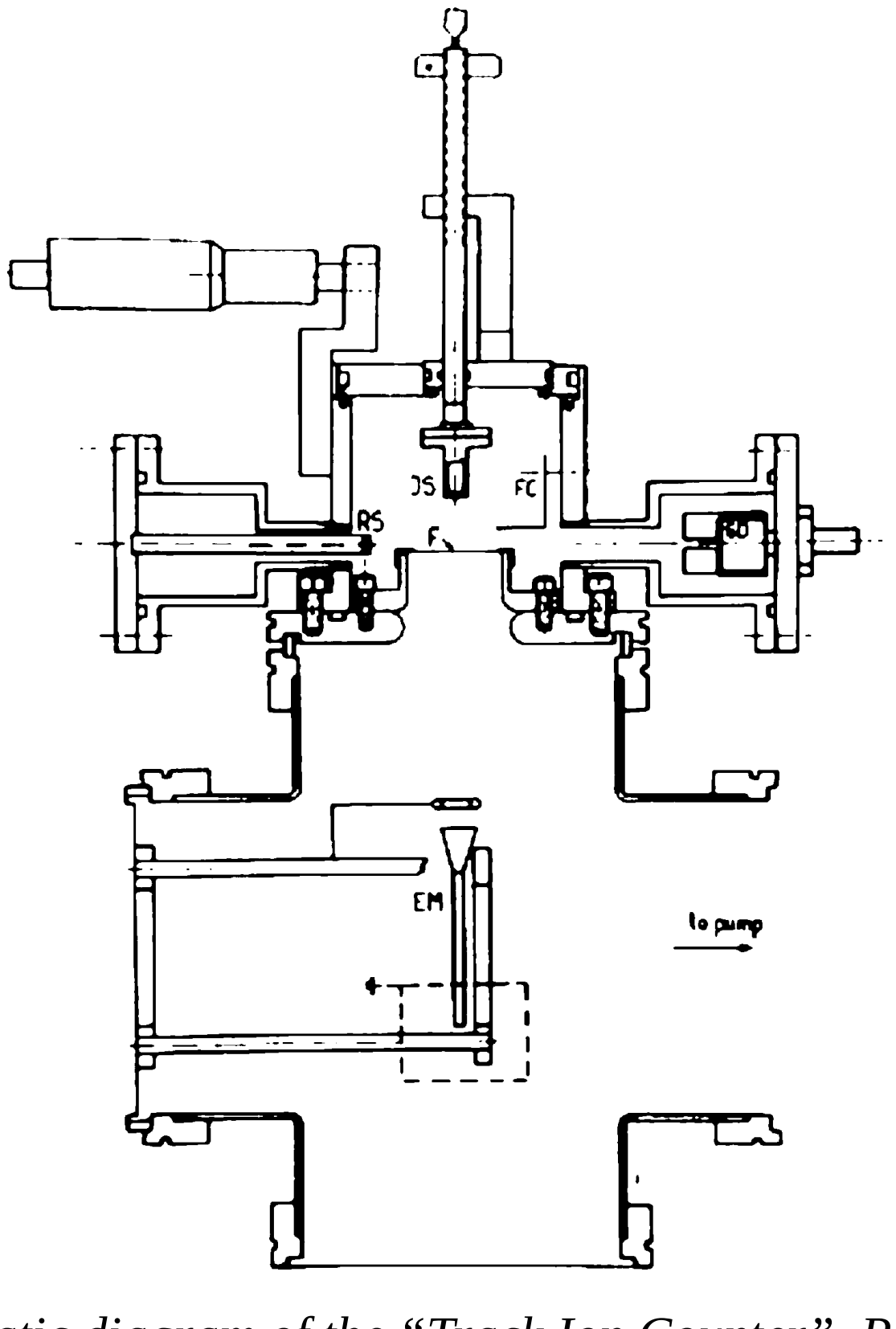

*Figure 8: Schematic diagram of the "Track Ion Counter". Ref. [30]*

The lower part of the device consists of a «spiratron» 4219 Bendix electron multiplier. The ion source may be moved to any position to allow the estimation of ion detection efficiency by the spiratron. The track ion counter is connected to a PMC-10C oil diffusion pump with a pumping speed of 5000 *l/s*. The gas flow through the orifice (F) (0.5 *mm* in diameter) determines the pressure in the upper and lower regions of the device.

The results of preliminary measurements were the frequency of various numbers of ions ($N_2$, $H_2$, $CH_4$ and $CO_2$) created within a gas domain which corresponds to a cylinder of tissue of 0.15 *nm* dia and 7.6 *nm* height. The gas domain was irradiated by $\alpha$-particles from $^{241}$Am.

It is interesting to note that the track ion counter achieved an efficiency in the counting of positive ions of about 45 % ± 5 %.

#### *2.6.1.2 The nanodosimeter at the Weizmann Institute of Science*

The ion-counting nanodosimeter (ND) developed at the Weizmann Institute of Science in 1996 (Shchemelinin *et al.* [47, 50], Garty *et al.* [51] and in its extended version, Bashkirov *et al.* [52]) is presented in Figure 9. It is in fact based on the idea of "The Track Ion Counter" Ref.[30], being at the same time an improved version. This device is currently situated at Physikalisch-Technische Bundesanstalt (PTB), Germany.

The ion-counting ND (see Figure 9, Ref.[53]) consists of a large gas-filled ionization volume (IV), traversed by a radiation field. Radiation-induced ions formed within a small subsection of this volume (termed the sensitive volume – SV) are extracted into vacuum, detected and counted.

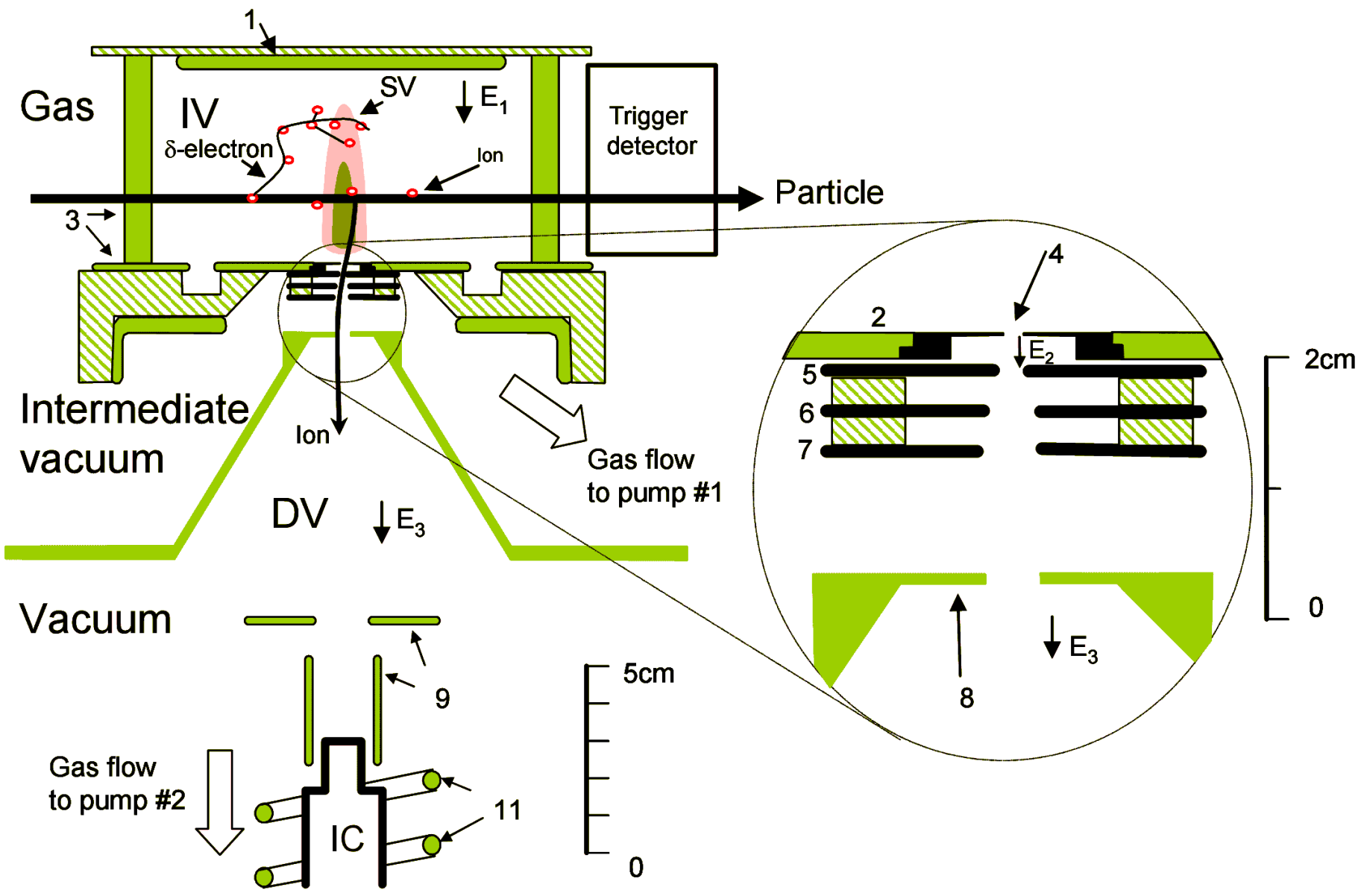

*Figure 9: A detailed diagram of the ion counting nanodosimeter device. In the ionization volume (IV), the anode (1), cathode (2) and field shaping electrodes (3) determine the extraction field $E_1$. Ions created within the sensitive volume (SV), are extracted via the aperture (4) into the intermediate vacuum region, they are focused via the electrodes $A_1$ (5), $A_2$ (6), $A_3$ (7) and $A_4$ (8) into the detection volume (DV). The ions are then accelerated and focused by the electrodes (9) into the ion counter (IC). A helical coil (11) protects the ion counter from discharges. Note that the SV and δ-electrons are schematic representations and not to scale. Ref.[53]*

A detailed scheme of the ion counting ND is shown in Figure 9. The charged particle beam traverses a gas-filled interaction volume (IV) and reaches a trigger detector. Ions induced within a wall-less region, denoted the "sensitive volume" (SV, within the IV) are extracted into the vacuum-operated detection volume (DV) and are detected by an ion counter (IC). The pressure difference between the IV and the DV is maintained by a differential pumping system.

The data acquisition system registers the arrival time of the ions at the counter with respect to the trigger. These data are used afterwards to determine the frequency of the ionization cluster size.

The size and shape of the wall-less detection volume are determined by the extraction efficiency of ions through the aperture, and depend on gas density, aperture diameter and the electric fields above and below the ion extraction aperture. Consequently, the sensitive volume is represented by a map of tapered, cylindrically symmetrical volume-contours representing equal ion-extraction efficiencies. These maps were determined by calculations based on the electric field geometry of the measuring device and on measured ion-transport parameters. The size of the sensitive volume, at unit density, is parameterized by the 50 % contour of the ion-extraction efficiency.

Devices based on differential pumping systems (TIC and ND) have a limited application in simulating nanometre volumes (not more than a few *nm* in the unit density scale) due to the limits derived from the pumping speed of the vacuum system (the vacuum pressure near the ion detector should not increase above $10^{-3}$ *hPa*).

#### 2.6.1.3 The Jet Counter at the Andrzej Sołtan Institute for Nuclear Studies (SINS)

The Jet Counter, JC, at the Andrzej Sołtan Institute for Nuclear Studies was developed by the group of S.Pszona [48, 54] in 1994. It can be applied to experiments for measuring the frequency distribution of ionization cluster size produced not only by $\alpha$-particles but also (uniquely) by low-energy electrons.

A schematic view of the Jet Counter is given in Figure 10. The gas cavity that simulates a nanometric volume at unit density is obtained by pulse expansion of a gas (for instance, nitrogen) into an interaction chamber, leading to a pulsed jet of gas molecules. The ions induced in the gas cavity by primary particles are extracted into a vacuum and detected by an ion counter as a function of their arrival time with respect to a trigger signal provided by a primary-particle detector, at least for $\alpha$-particles.

However, because of the strong scattering behaviour of electrons in the case of impact ionization or elastic scattering, the primary-particle detector cannot be used for triggering the time-of-arrival measurements when electrons are being used as primary particles. To solve this problem, mono-energetic electrons were produced by a pulsed electron gun, which was operated at a low electron beam current and emitted primary electrons into the interaction chamber during time periods of the order of 1 $\mu s$. This time window is used, afterwords, to trigger the time-of-arrival measurements.

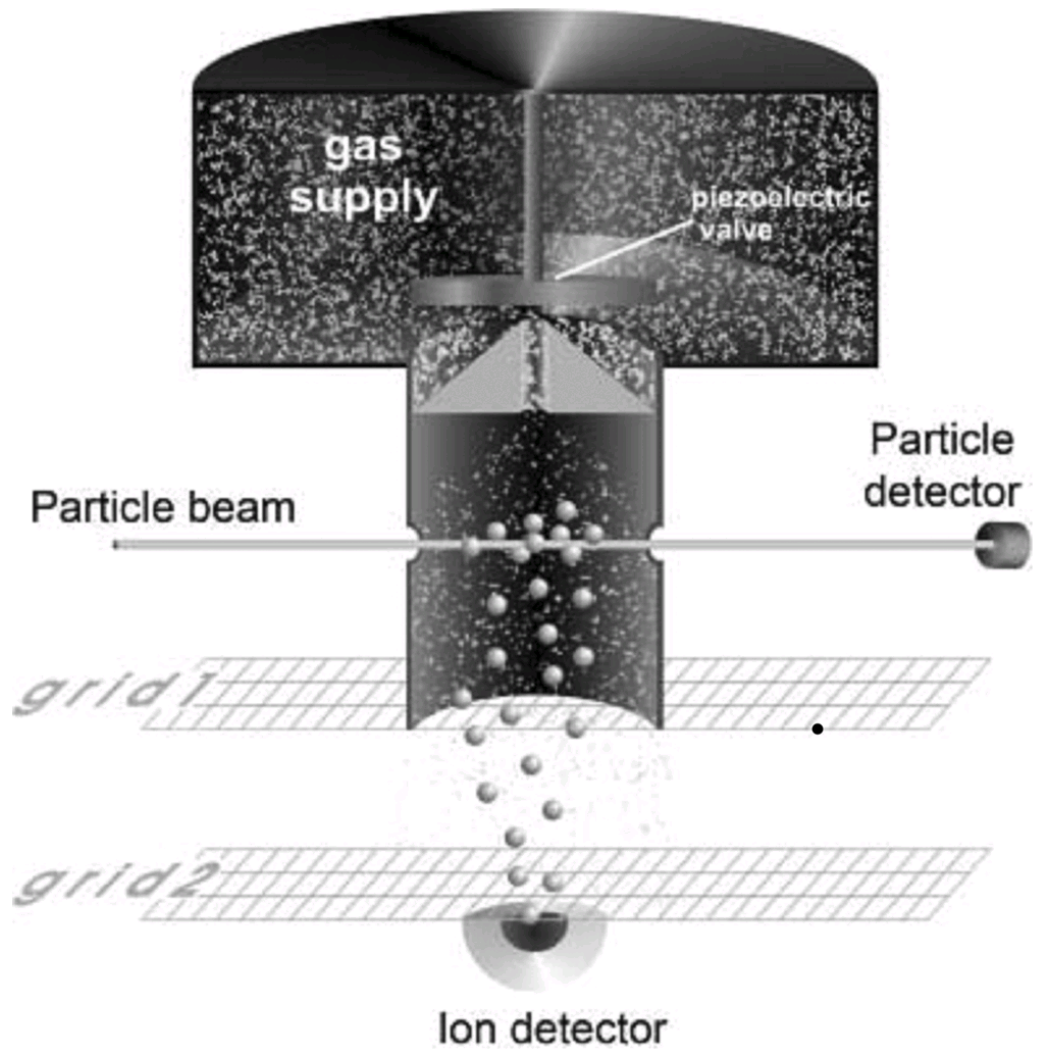

*Figure 10: Schematic view of the Jet Counter used for measuring the ionization cluster-size distributions of α particles and low-energy electrons in 'nanometric' volumes of gas media. The ions produced within the interaction chamber by the projectiles are guided to an ion detector, where they are counted using time-of-arrival techniques. For α particles, the time-of-arrival measurement is triggered by the signal of a silicon detector if a single particle penetrates through the target volume and is detected; for electrons it is triggered by a signal derived from the period of time in which primary electrons are injected into the interaction chamber by an electron gun.*

Experiments with the Jet Counter cover the density region up to 1–2 $\mu g/cm^2$ with a detection efficiency of about 30–40 %.

For more details see chapter 3.

### 2.6.2 Electron-counting measuring devices

#### 2.6.2.1 The Track-nanodosimetric counter of Laboratori Nazionali di Legnaro

The track-nanodosimeter counter at Laboratori Nazionali di Legnaro (De Nardo *et al*. [55]) is the only one at present based on the principle of single-electron detection and counts single electrons at very low gas pressure without being affected by gas gain fluctuation. In addition, it is also possible to measure the probability distribution of ionization cluster-size formation in nanometric volumes as a function of the distance from the centre line of a primary particle beam.

The detector consists essentially of an electron collector and a single electron counter (SEC). A schematic diagram of the apparatus is presented in Figure 11. The electron collector is a system of electrodes enclosing an almost wall-less cylindrical volume whose height equals its diameter. Electrons created inside this volume, the sensitive volume (SV) of the counter, are transferred into the drift column of the SEC and are detected one by one, using a multi-step avalanche chamber (MSAC). Two collimators positioned in front of a solid-state detector (SSD) define the $\alpha$-particle track, with respect to which the detector can be moved (using a screw).

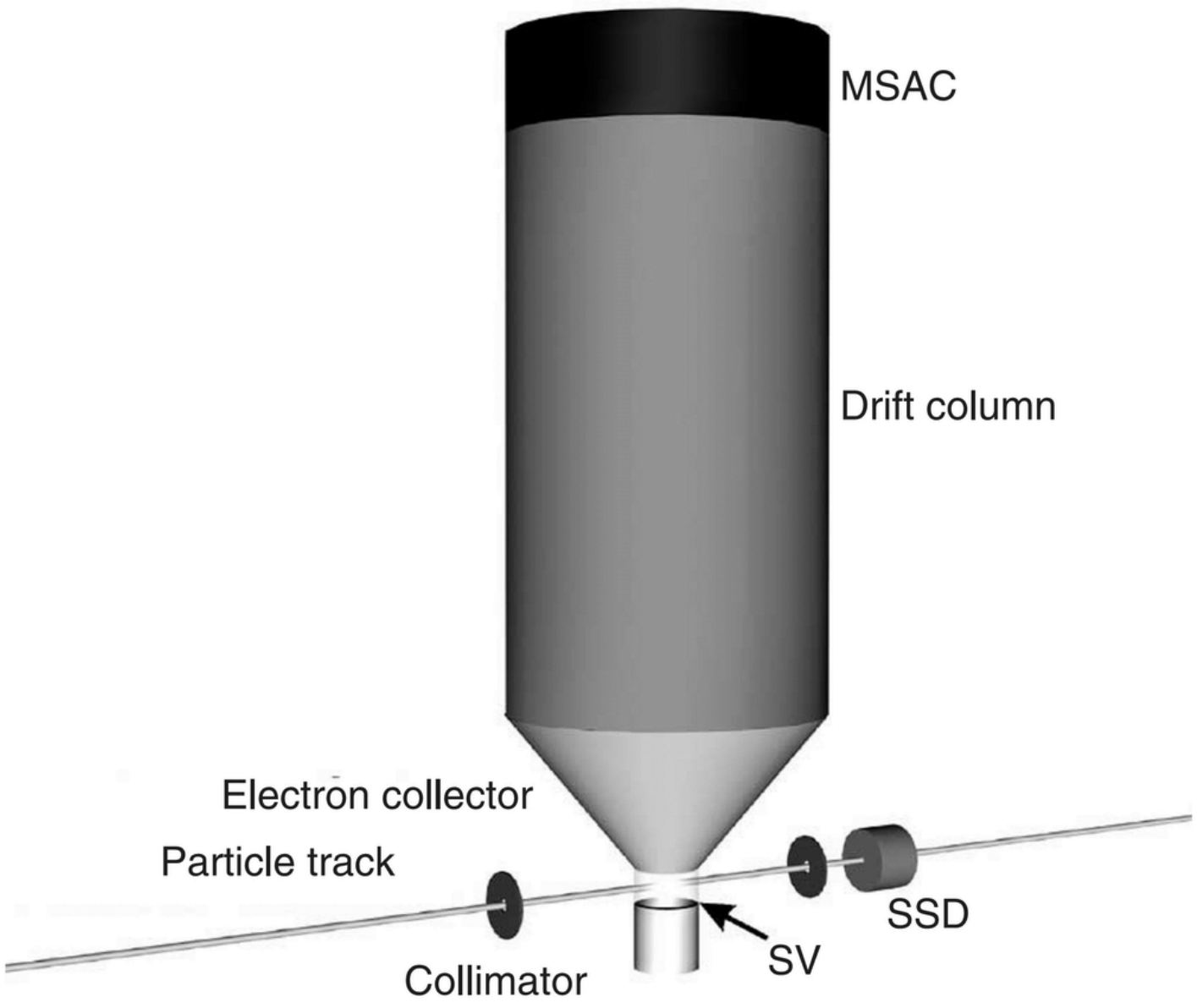

*Figure 11: Schematic representation of the experimental set-up for track-nanodosimetry measurements (not to scale). See text for explanation. Ref.[55].*

The SSD signal triggers the counter acquisition system to start counting single electrons produced by particles in the SV. As the whole detection system is immersed in the counting gas at a given pressure, there are some limitations in measuring the gas density. The pressure inside the whole system defines the simulated nanometer size and the pressure should be such as to allow for good operation of the MSAC. Usually, measurements have been performed at a few hundred *Pa* of propane, which corresponds to at least 20–30 *nm* with efficiencies of about 20 *%*.

# 3 The Jet Counter facility

## 3.1 Simulation of nanometer size

A gas cavity which simulates a nanometric volume (at unit density) is created by the method explained in Figure 12. A simulated nanometer size is obtained by pulse expansion of gas (producing a nitrogen or propane jet) to the volume of the interaction chamber (IC). The IC volume has a cylindrical form, (diameter 10 *mm*; height 10 *mm*) with walls of 1 $mg/cm^2$ Mylar. The gas jet is created by a pulse operated valve PZ, which injects gas from the volume R into the IC (over a valve, and through a nozzle with an orifice 1 mm in diameter). The lower part of the IC, the cavity between the grid S and the edge of the IC, represents the site of simulated nanometer size (SNS), i.e. the cavity from which ions are extracted and collected with known efficiency. The SNS cavity is shown schematically in Figure 12 (B) as an enlargement. It forms a cylinder with height equal to its diameter, shown in Figure 12 (C). A dynamic vacuum condition, for the proper working of the ion detector AF180H and electron detector CH1, is provided by a 500 *l/s* turbomolecular pump. During gas injection the gas pressure inside the Jet Counter device increases up to $10^{-3}$ *hPa* and before the next gas injection the vacuum is recovered down to $10^{-6}$ *hPa*. The volume of the Jet Counter chamber is about 20 $dm^3$ (litres).

The instantaneous density of gas flowing through the IC (as well as its maximum which constitutes the SNS) was determined by measurement of the transmission rate of electrons with known energy and known total scattering cross section [56]. For details see chapter 3.4.4. Figure 13 shows the time dependence of the transmission rate of a 1 *keV* electron beam penetrating through a propane jet. As can be seen from the figure, the area of maximum instantaneous gas density (the range of gas densities at lowest transmission rate) exists for about 200 $\mu s$. This period of time is used later as a time window for measuring the ion species (nitrogen or propane ions) which are produced by charged particles which penetrate through the SNS. The position of the electron gun (EG) and the electron detector (CH1), with respect to the alpha source and Si detector is shown in Figure 12 (D).

## 3.2 Method for measuring ion cluster size spectra created by α-particles

The experiments with $\alpha$-particles were carried out using the apparatus presented in Figure 12. A collimated (4.6 *MeV*) $\alpha$-particle beam from a ($^{241}$Am) radioactive source (Amersham gold-plated type AMM2) penetrates through the Mylar wall in the IC and is degraded to 3.8 *MeV.* Afterwards, the $\alpha$-particles intersect the SNS chamber (along its diameter) at half its height and are registered by a Si detector. The ions created by a single $\alpha$-particle along its path (as well as by delta electrons) within the SNS are removed by the electric field of the grid (G) and then guided through a second grid G1 to the ion detector AF180H.

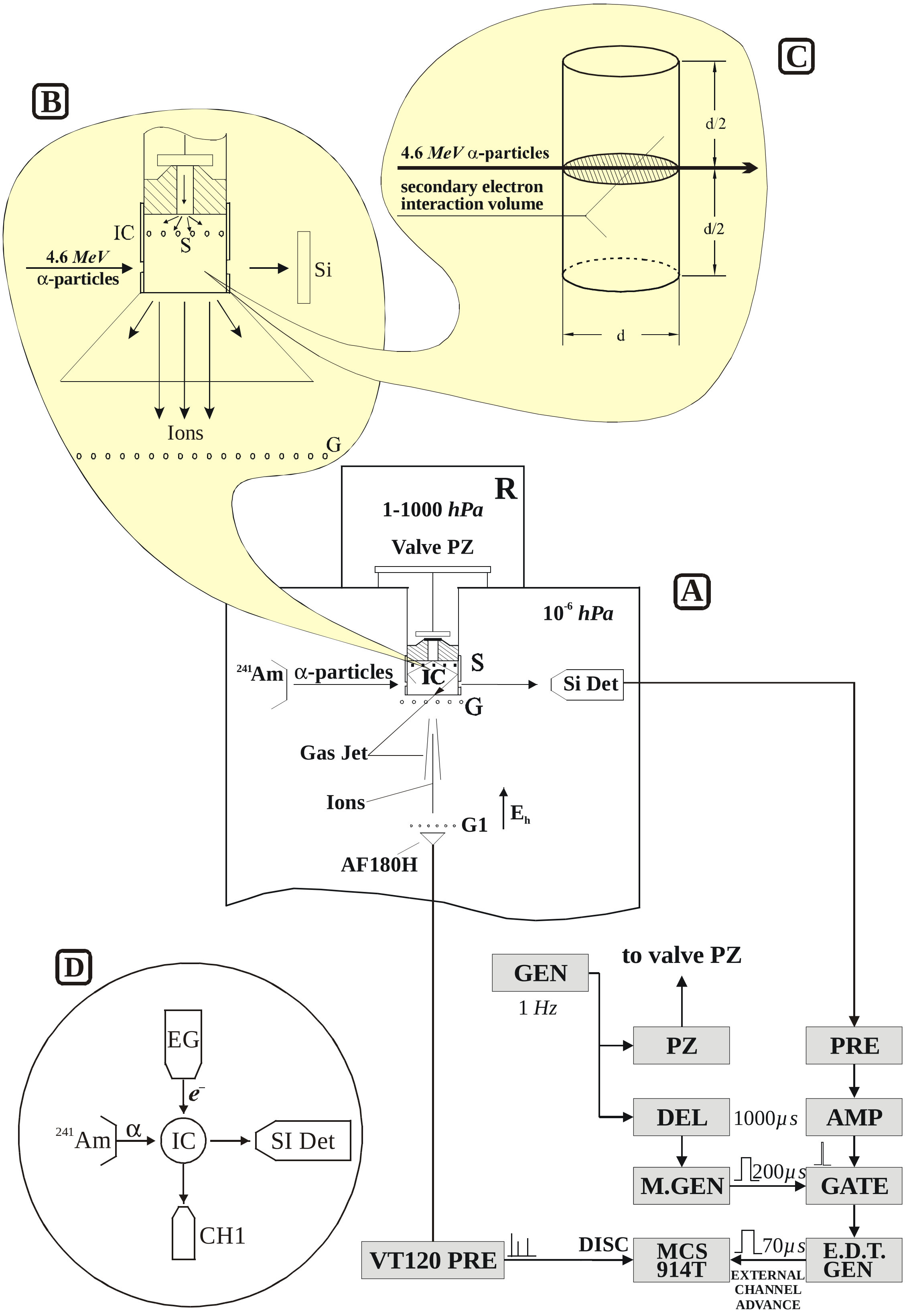

*Figure 12: Schematic view of the experimental set-up of the Jet Counter. Configuration for experiments with α-particles. IC – interaction chamber, S – grounded grid to shape IC, GEN – generator, EG – electron gun, BEAM.CH. – beam chopper, PZ – piezoelectric valve, CH1 – electron detector, G – extracting grid, G1 – accelerating grid, AF180H – ion detector.*

A typical time-of-flight spectrum of nitrogen ions registered by the ion detector is presented in Figure 14. It shows a single pronounced peak at 40 *μs*, with a rather broad time spread of up to 100 *μs*. It should be pointed out that the voltages on grids G and G1 were optimized to provide maximum resolution between successive ions.

The pulses from this detector are amplified by a fast preamplifier, VT120, and enter the 10 ns resolution multiscaler, MCS (914T Ortec). The associated electronic set-up as well as the timing chart of the experiment for registering signal clusters is shown in Figure 12 (A). The channel on the MCS is advanced after each pulse from the solid state Si detector. In this way a signal cluster spectrum for a given thickness of the gas target (SNS) is recorded. From the recorded data, the signal cluster distribution as a function of cluster size is derived. One of the important features of this method is the ability to measure the 'zero size' event rate. The measured spectra are de-convoluted to the true number of ions spectra.

The frequency distribution spectra in nitrogen and propane corresponding to different nanometre sizes ranging from 0.1 to 0.5 *μg/cm²* were measured and the results are presented in chapter 6.2 and 6.3.

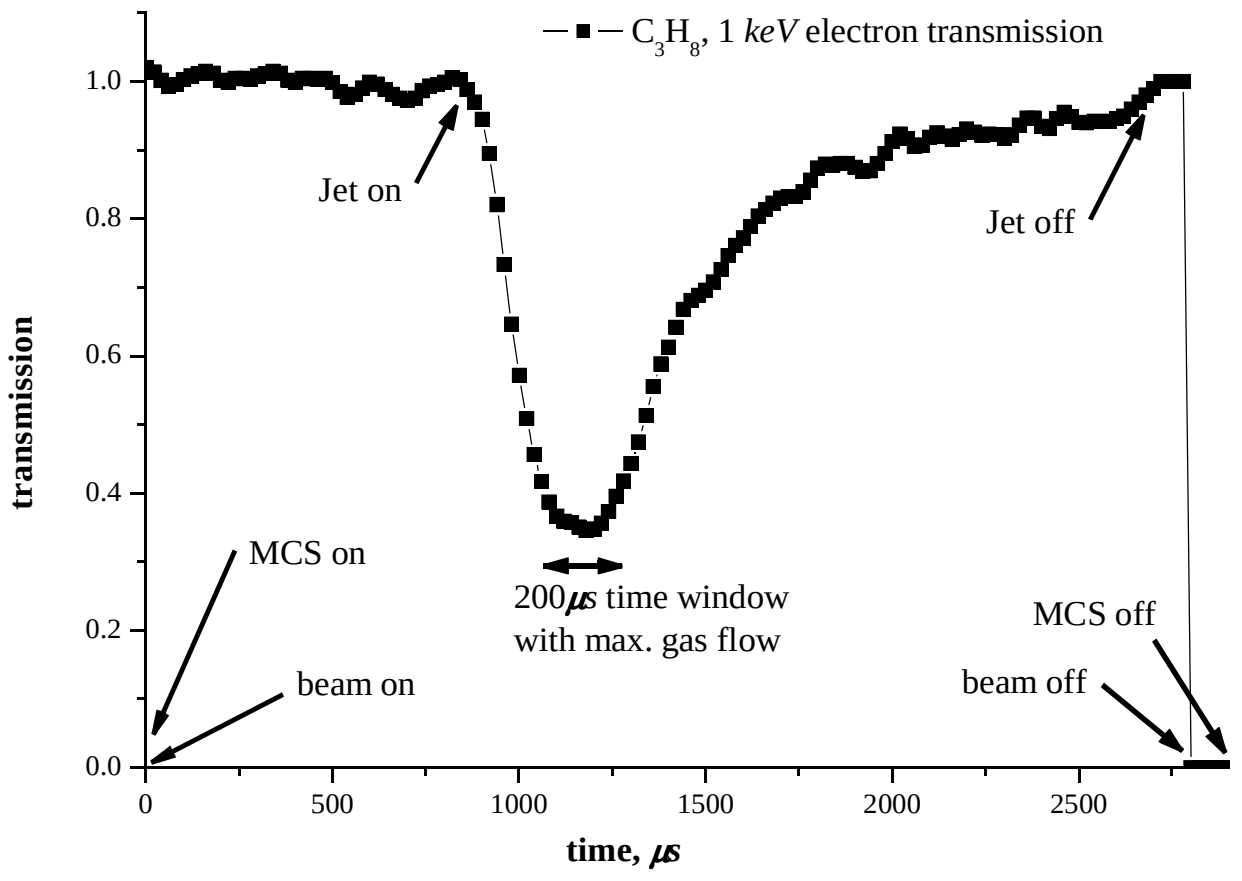

*Figure 13: Transmission of 1 keV electrons through propane jets, timing chart, time window 200 μs (as gate on).*

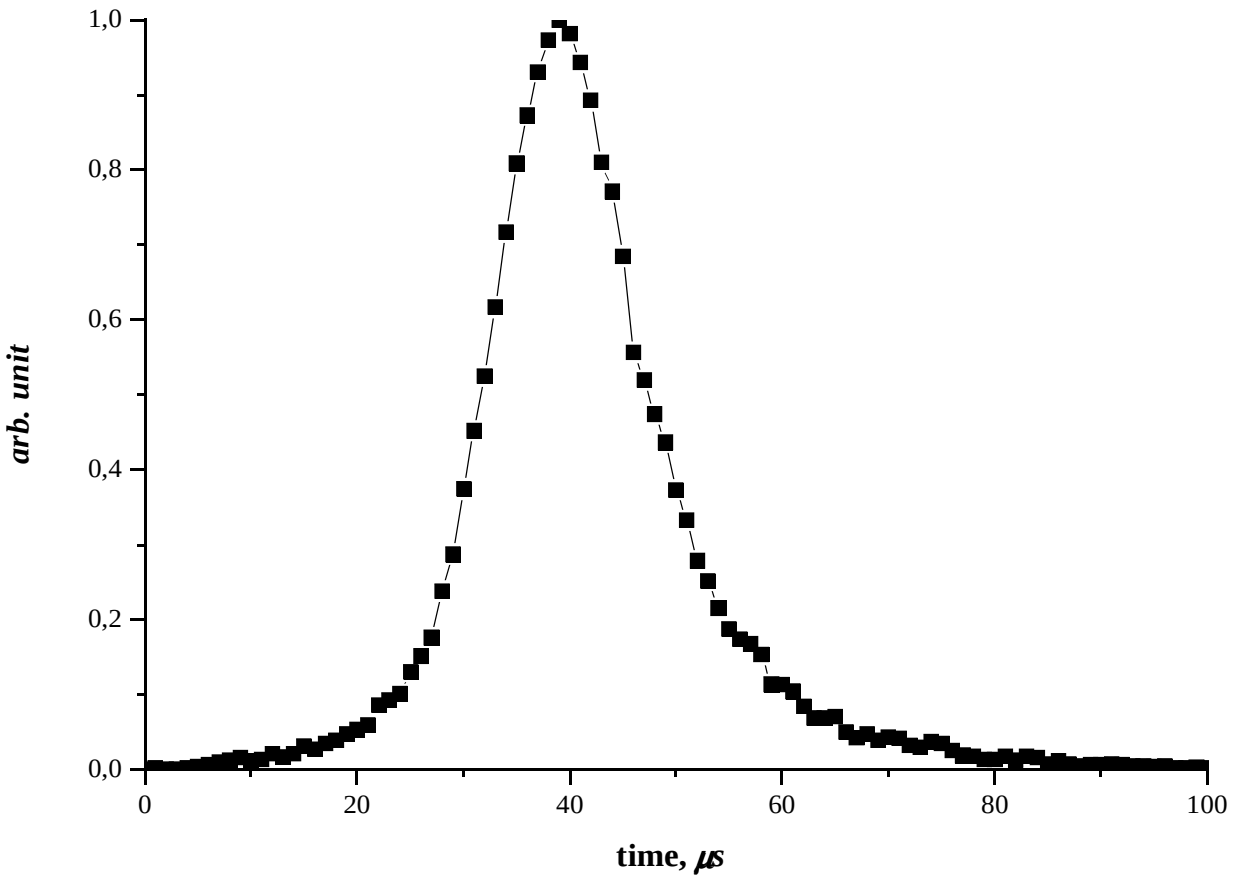

*Figure 14: Time of flight spectrum of nitrogen ions.*

## 3.3 Method for measuring ion cluster size spectra created by “single” electrons

The experiments with single mono-energetic low-energy electrons were carried out using the Jet Counter presented in Figure 15. The electron gun, which was controlled by a beam chopper, was operated at a current which led, during the time window, to a mean rate of 1 electron per 4 $\mu s$ chopper pulse. As the Jet Counter works in pulse mode, the single electrons are generated at the moment of maximum gas density in the IC. As in the transmission experiments (see Figure 13), the electrons entered the IC through an orifice in the Mylar wall where they interacted with the gas jet (alongside of its diameter) at half its height and were registered by the electron detector CH1.

The ions (cluster of ions) created by each single electron (or by its secondaries) within the SNS, were removed by the electric field of the grid (G) and guided through G1 to an AF180H detector.

A typical time-of-flight spectrum of nitrogen ions registered by the ion detector is the same as in the experiment with $\alpha$-particles presented in Figure 14.

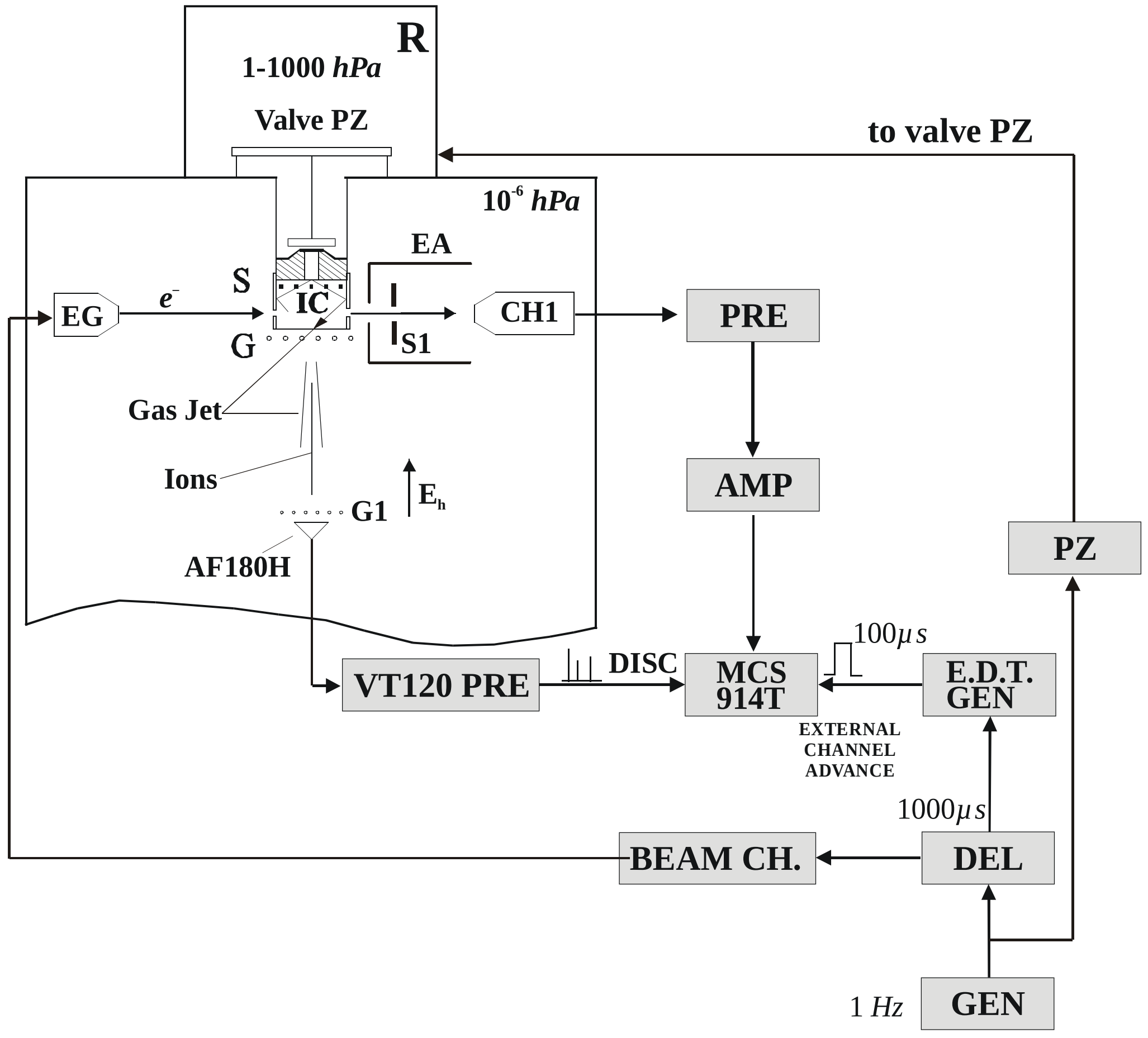

*Figure 15: Schematic view of the measuring set-up of the Jet Counter. Configuration for the experiments with a single electron. GEN – generator, EG – electron gun, EA – electron analyzer, S1 – grid, BEAM.CH. – beam chopper, PZ – piezoelectric valve, CH1 – electron detector, EA – electron analyzer, AF180H – ion detector, IC – interaction chamber.*

The pulses from the AF180H detector were amplified by the fast preamplifier VT 120 and entered a 10 *ns* resolution multiscaler, MCS, type 914T ORTEC. The associated electronics set up, as well as the timing chart of the experiment for registering the signal clusters is shown in Figure 15.

The following sequences of steering pulses were found to be optimal: At the beginning, a GEN pulse with a repetition rate of 1 Hz started the driver of the PZ valve. Since this valve needs around 1000 $\mu s$ to open completely (see Figure 13), a time delay of 1000 $\mu s$ was needed to match the electron beam to the maximum jet density. This was done by the trigger DEL, which in turn opens the electron gun by the beam chopper, for a time period of about 4 $\mu s$. At each electron energy, the mean electron beam intensity was fixed at 1 electron per 4 $\mu s$ chopper pulse. At the same time, an external dwell time generator E.D.T. GEN gave a 100 $\mu s$ pulse for steering the gate of the multiscaler to register all counts arriving at the ion detector as well as to switch up the multiscaler to the next channel. To derive one cluster-size distribution, around 10000 electron track passages were analyzed.

The frequency distribution spectra for 100 *eV*, 200 *eV*, 300 *eV*, 500 *eV*, 1000 *eV* and 2000 *eV* electrons in nitrogen for 0.34 $\mu g/cm^2$ target thickness were measured and the results are presented in chapter 6.1.

## 3.4 Essential parts of the Jet Counter facility

In the current chapter a detailed overview of the essential parts of the Jet Counter facility that influence the stability and quality of the measurements is given. Most of these data have never been published but are of importance for explaining the mode of operation of the Jet Counter device. This chapter gives details about the work have been done to improve the original Jet Counter set up and the quality (reproducibility) of the measured results.

### 3.4.1 Electron gun and beam chopper

The reasons for installing an electron gun (EG) in the present experiment are as follows: first, for defining the gas-target thickness via transmission measurements (see chapter 3.4.4) and second as a source of “single” electrons in the main experiment with electrons (see chapter 3.3).

The EG is a device that produces mono-energetic electrons. A schematic view and photo of the EG (model EQ22 made by Specs) is presented in Figure 16. The EG may produce electrons in the energy range from 50 *eV* to 5 *keV* (but the lowest energy for good operation is above 300 *eV*). By using a power supply (PS) (PU-EQ22/35 made by Specs), the EG may generate stabilized current from 1 *nA* — 100 $\mu A$. For currents lower than 1 *nA* the stabilization does not work and the currents may change unpredictably.

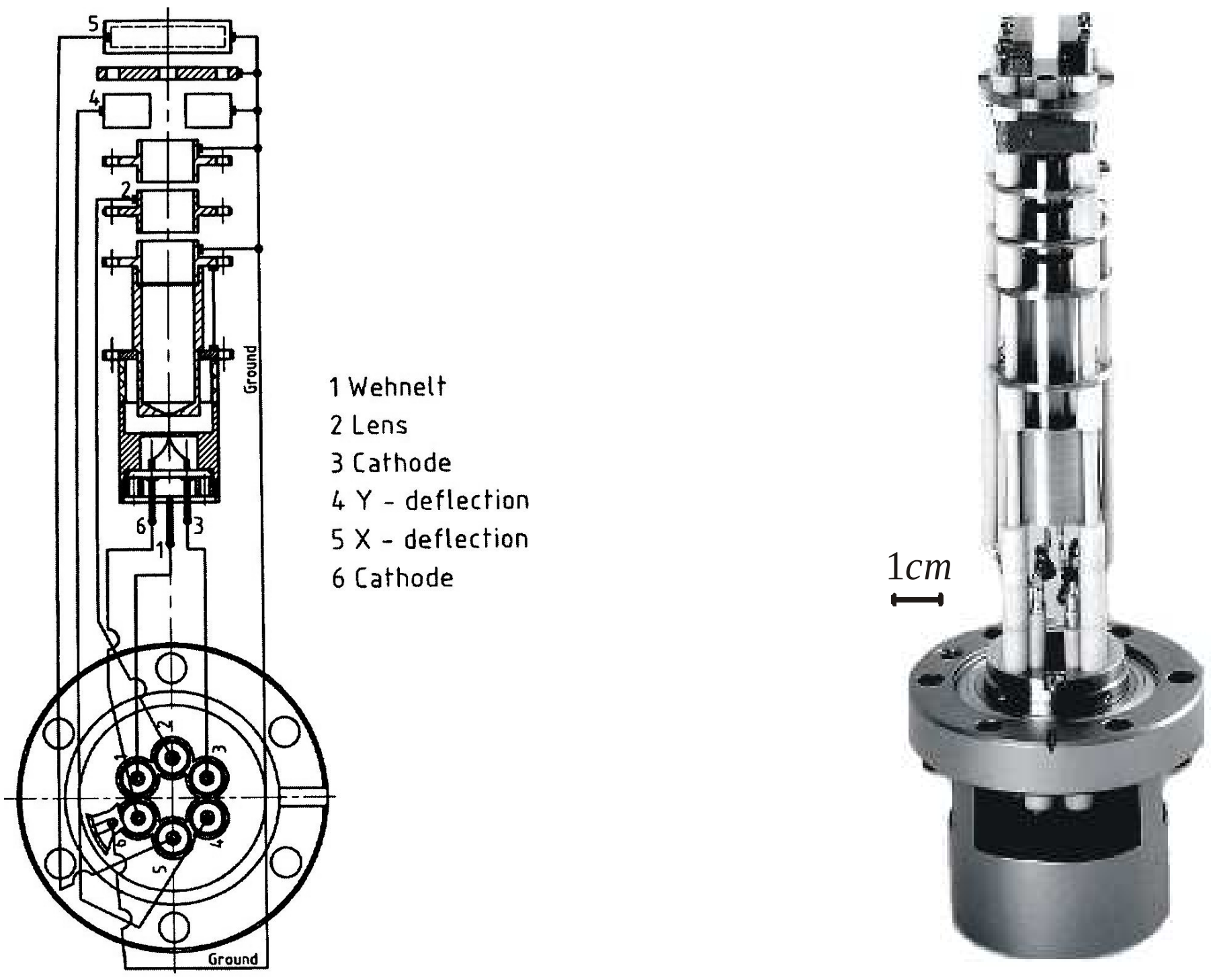

*Figure 16: Schematic view (left) and photo (right) of the electron gun. Ref.[57]*

For the producing of short pulses of electrons in the range 0.8-15 $\mu s$ and 0.5-7 *ms* by the EG a beam chopper (BC) (Beam Chopper 9 99 800 made by Specs) was used. The BC was connected between the EG and PS as presented in Figure 17. When the EG does not generate electrons by applying a closing voltage on the Wehnelt electrode (extracting grid), the BC generates a short opening impulse on the Wehnelt electrode and in consequence the EG produces a short electron impulse, equal in duration to the opening impulse on the Wehnelt electrode. (The intensity of the electron pulses is determined by the closing voltage on the Wehnelt electrode).

The EG is opened by a short opening impulse of about 20 *V* on the Wehnelt electrode which is generated by the BC.

It was discovered that the electron pulses produced by the electron gun (indicate on the evident) undesirable oscillations, presented by curve A of Figure 18. These oscillations complicated the electron transmissions measurements. This effect was reduced to an acceptable level of about ±3 *%* (see curve B of Figure 18) by introducing an additional capacitor C1 in the PS PU-EQ 22/35 (see Figure 19).

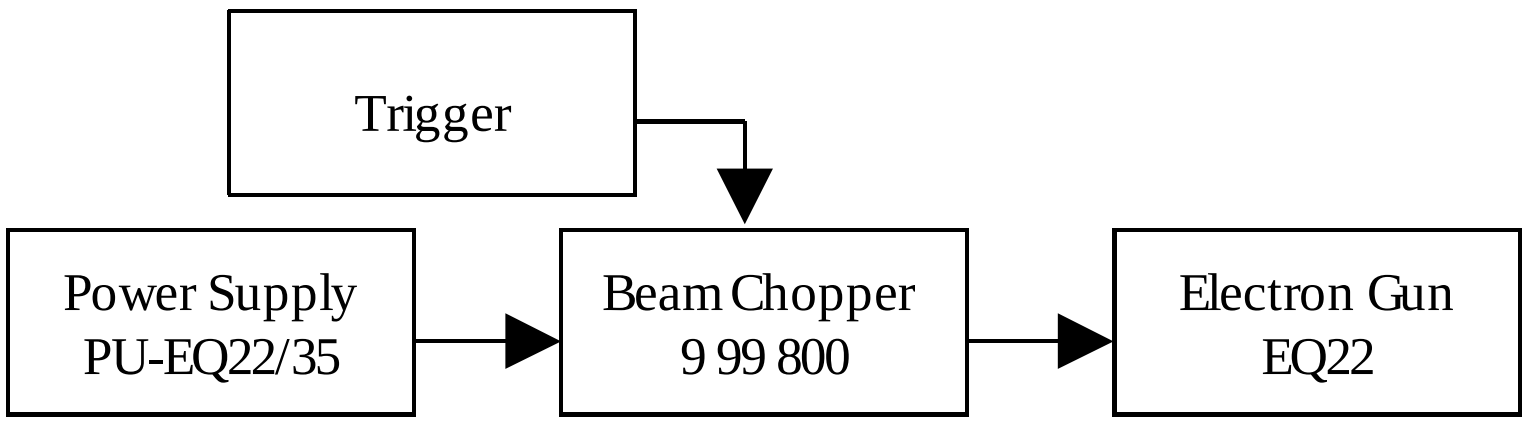

*Figure 17: Beam Chopper connection diagram.*

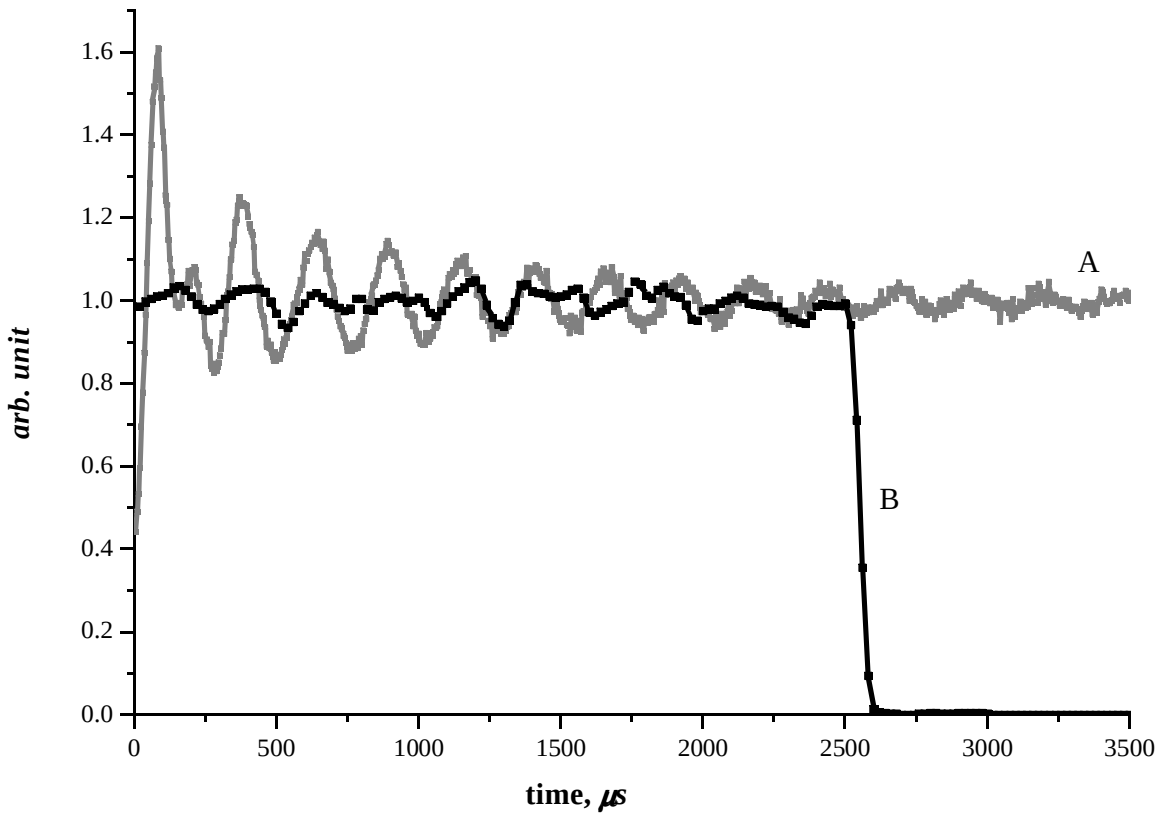

*Figure 18: Electron impulse generated by the EG as seen by the channeltron detector without capacitor (A) and with capacitor C1 (B).*

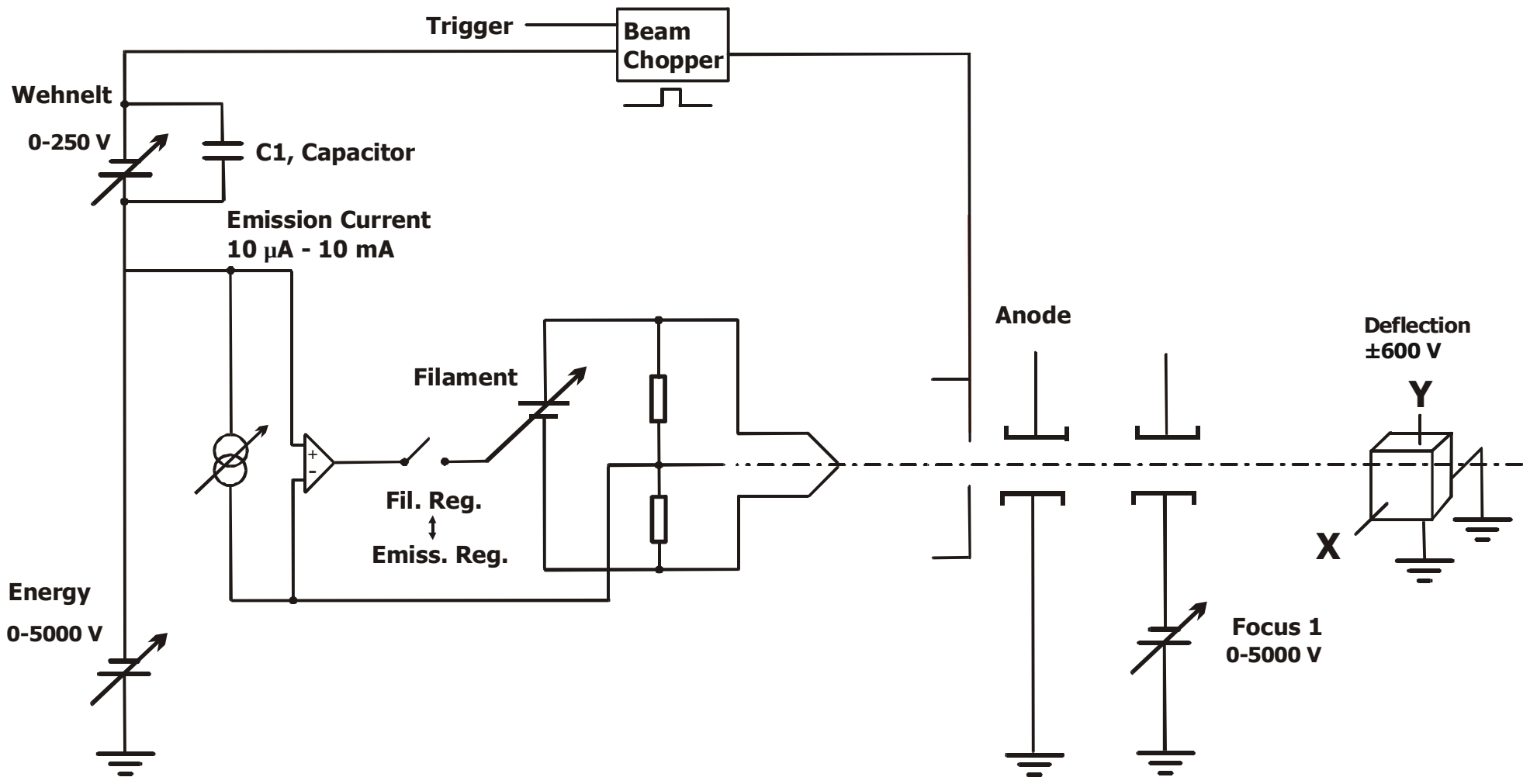

*Figure 19: Overview of the voltages generated by the Power Supply PU-EQ 22/35 and Beam Chopper. Ref.[58]. Modified.*

### 3.4.2 Electron gun as a source of a “single” electron beam

In experiments planned with electrons a single electron beam ought to be applied.

The standard electron gun EG with controlled filament heating power has been checked as a “single” electron beam source. For this purpose an electron gun (EG) with beam chopper (BC) was used (see Figure 17). Using an external trigger, the BC generates a short (1-100 $\mu s$) opening voltage impulse on the Wehnelt electrode in the EG. As a result, the EG emits electrons. The quality of such electron beams was measured. It was found that the EG produces electrons with a Poisson-like distribution. The number of emitted electrons is given by the formula (Eq.27)

$$p_n = \frac{N_{mean}^{n}}{n!} e^{-N_{mean}} \qquad (27)$$

where $p_n$ is the probability of emitting $n$ electrons with a mean number of emitted electrons $N_{mean}$. The electron distribution was studied using different filament currents and duration of extracting voltage on the Wehnelt electrode.

The measured frequency distributions of the number of emitted electrons for widths of the extracting impulse of $1\,\mu s$, $10\,\mu s$ and $100\,\mu s$ are compared with the Poisson distributions and are presented in Figure 20 A,B,C. Satisfactory agreement with the Poisson distribution was obtained. However, for mean intensities higher than 1 electron per $1\,\mu s$ some differences were observed. These differences can be explained by the effect of pile up. Figure 21 presents the frequency distribution of the number of emitted electrons for the same extracting voltage on the Wehnelt electrode and for different widths of the extracting impulse. The results of the measurements are compared with Poisson distributions taking the measured $N_{mean}$ value. Very good agreement, for a wide range of widths of the extracting voltage and low intensity (less than 1 electron per $1\,\mu s$), was found.

The mean number of emitted electrons $N_{mean}$ is linearly dependent on the width of the extracting voltage pulse on the Wehnelt electrode (see Figure 22), generated by the beam chopper. As a conclusion one can say that an adjustment to the required $N_{mean}$ can be achieved by: the proper choice of the voltage on the Wehnelt electrode and of the width of the extracting voltage impulse in the beam chopper.

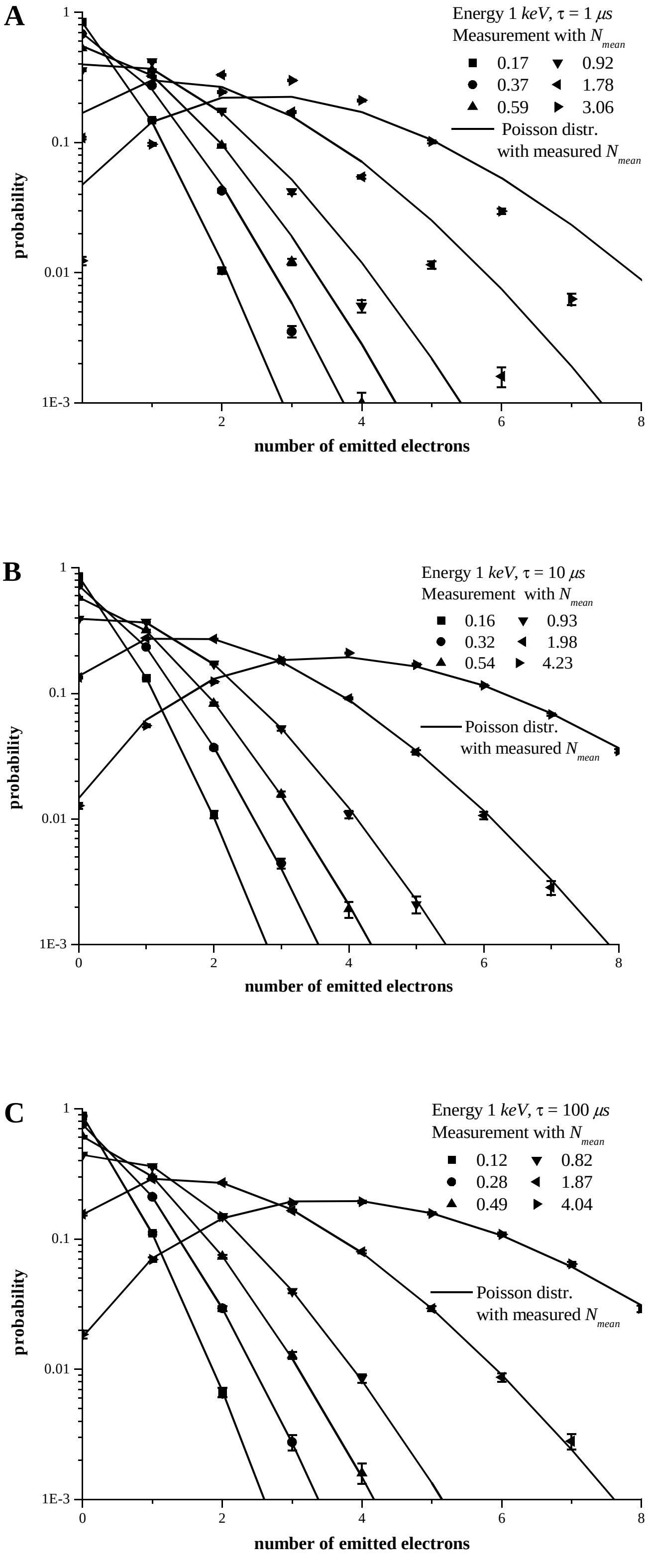

*Figure 20: Frequency distribution of the number of emitted electrons for three widths of the extracting impulses: A – 1 μs, B – 10 μs and C – 100 μs. The data are compared with Poisson distributions (solid lines) with the measured mean number of emitted electrons* $N_{mean}$*. Experimental results with statistical uncertainties.*

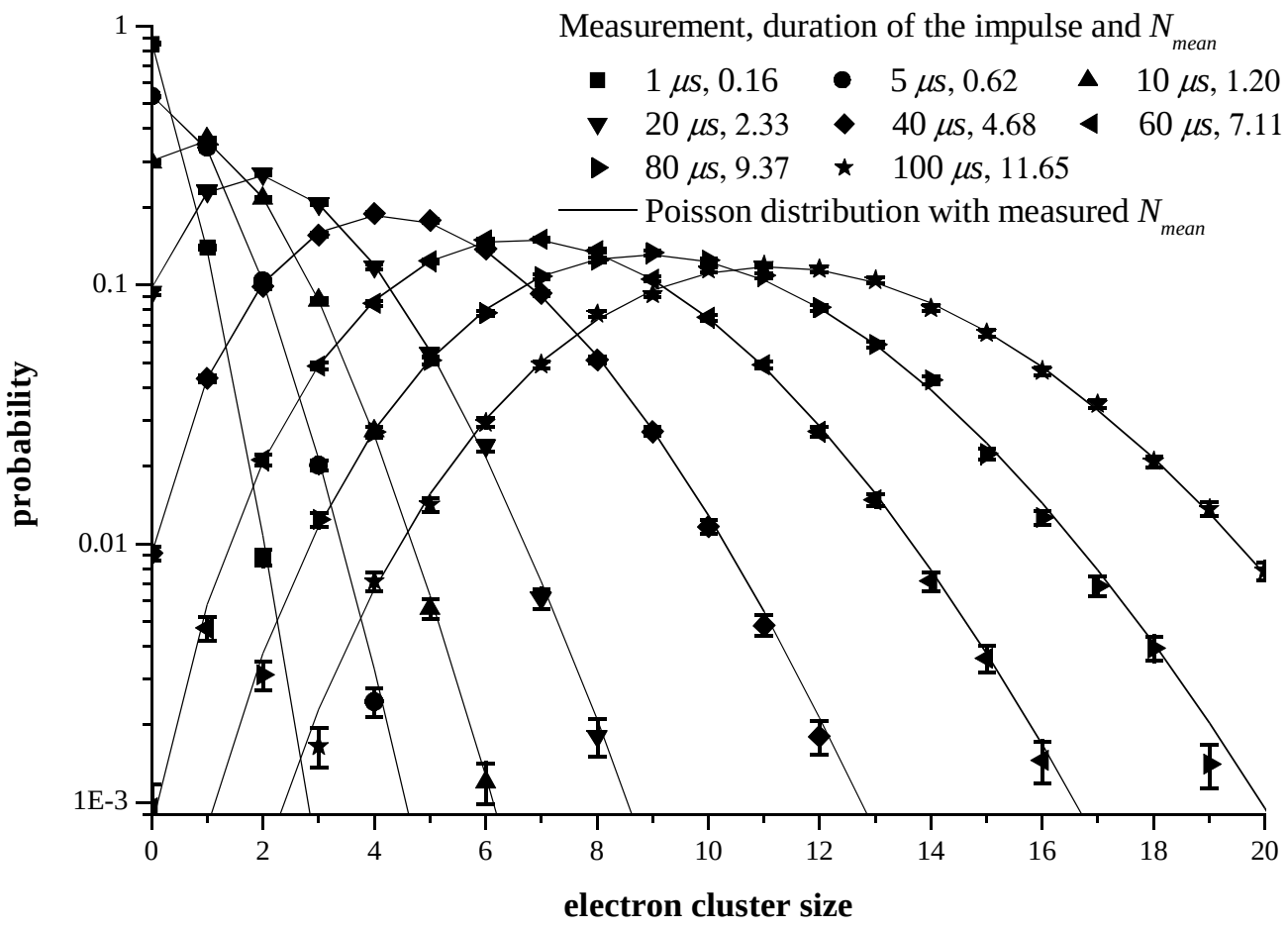

*Figure 21: Frequency distribution of the number of emitted electrons for the same (constant) extracting voltage and different widths of the extracting impulse. The data are compared with Poisson distributions (solid lines) with the measured mean number of emitted electrons* $N_{mean}$. *Experimental results with statistical uncertainties.*

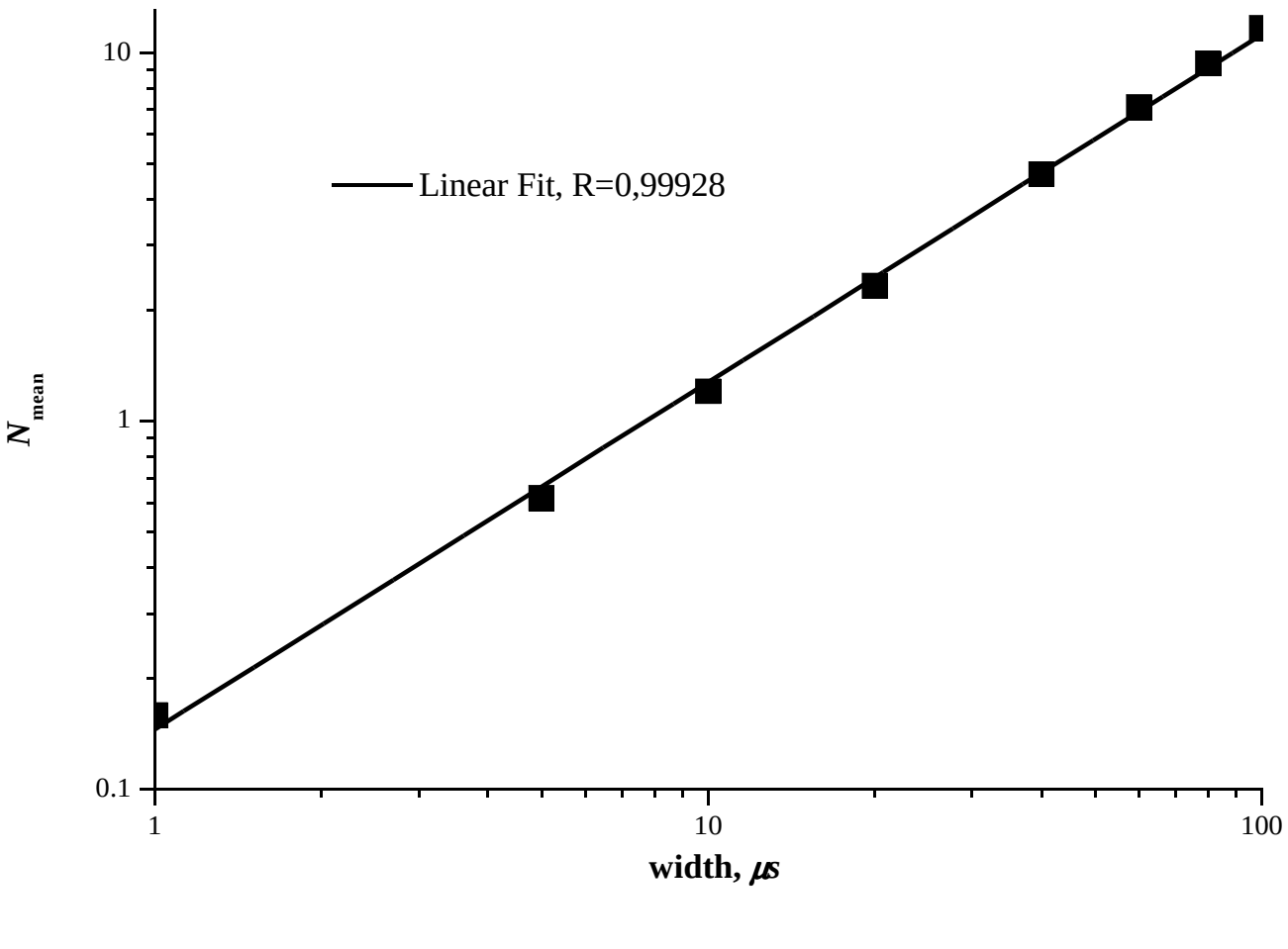

*Figure 22: Mean number of emitted electrons* $N_{mean}$ *for the same extracting voltage as a function of the width of the extracting impulse. The data are compared to a linear fit. Statistical uncertainties are smaller than the symbol size.*

### 3.4.3 Piezoelectric valve and its characteristics

A piezoelectric valve was selected for pulse injection of gas into the interaction volume. It should exhibit very fast operation with good stability and reproducibility. It works on the principle of the piezoelectric effect: "Piezoelectric materials show the converse phenomena in which the material is subject to mechanical stress when an electric field is applied to it" [59].

To achieve good operation conditions, the automatic gas control system was experimentally verified. First, the gas pressure in reservoir (R) at the point behind the valve and the voltage amplitude to open the valve were checked. After some hundreds of test measurements it was decided that such control was not enough, as the gas density changed too much during the measuring period (20 *hours*). Second, taking into account that the valve should always inject the same amount of gas and the repetition is constant, it was decided to try to regulate the pressure behind the valve in such a way as to have constant gas flow leakage in the reservoir behind the valve (in units of *Pa l/s*). This system appears to be better than the previous one. The details of the gas control are as follows: the gas-flow control system is presented in Figure 23. The MKS PR-4000 power supply/readout unit (PR4000) with Baratron Type 122B pressure transducer (PT) may stabilize the pressure in the reservoir (R) with a hysteresis of 50 *Pa* by the electro-magnetic valve (V1). The personal computer (PC) connected to the PR-4000 via RS232 logs the actual pressure every second. The PC calculates the actual gas-flow (leakage from the gas reservoir R) using the logged values of the pressure and a back loop to the PR-4000. Through manipulation of the valve opening the desired gas flow leakage was maintained.

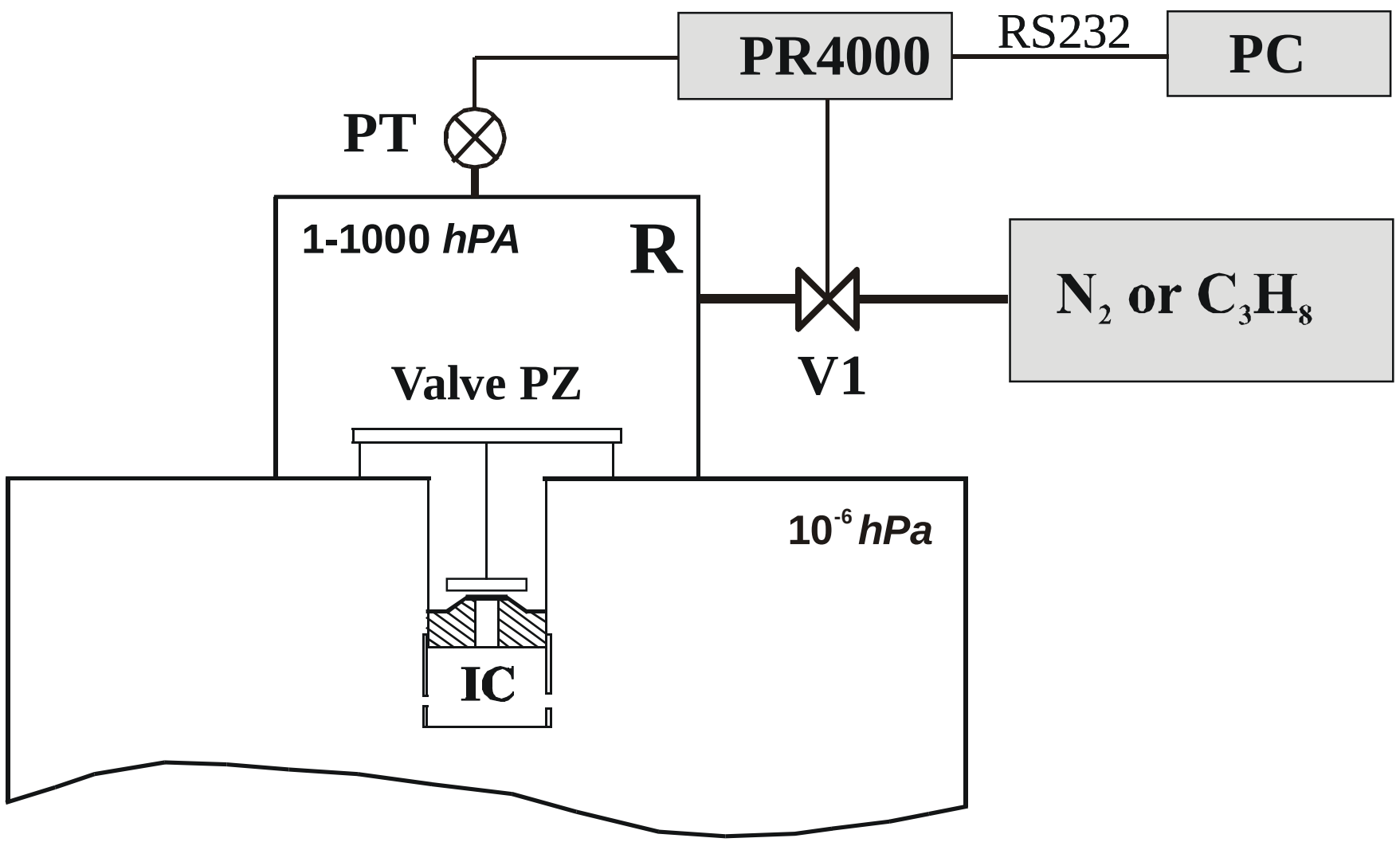

*Figure 23: Schematic view of the gas-flow control system. PT – Baratron Type 122B pressure transducer, PR4000 – MKS PR-4000 power supply/readout unit, PC – personal computer, R – gas reservoir with a volume of about 1420 $cm^3$, Valve PZ – piezoelectric valve, V1 – electromagnetic valve, IC – interaction chamber, $N_2$ or $C_3H_8$ bottle.*

A comparison of these two methods for the stabilization of the gas pressure (for target thicknesses at which the measurements were performed) is presented in Figure 24. As may be seen from Figure 24, the mean value of the area gas density in the IC is 0.342 *μg/cm²* with a standard deviation (std) of 0.027 *μg/cm²* or about 8 *%* of the absolute value of the gas density for the first method of gas pressure control. For the gas-flow stabilization system, the area density in the IC is 0.523 *μg/cm²* with a standard deviation of 0.009 *μg/cm²* (3 times lower than with the previous stabilization method) that is about 2 *%* of the absolute value of the gas density (4 times lower than with the previous stabilization). One can conclude that the best solution to achieve gas density stability to about 2 *%* during more than 200 hours is by gas flow control.

The thickness of the gas targets was determined using the electron transmission method described in chapter 3.4.4.

The gas flow leakage stabilization system was selected as better and was used in the present measurement.

“For example, 49.6 *μg* of nitrogen gas ($1.06 \cdot 10^{18}$ *molecules*) are in one valve gas impulse with 0.523 *μg/cm²* of gas target density”.

Summarizing: the uncertainties in the gas target density are caused by the temperature of the gas in the reservoir R (1.6 *%*), uncertainties in gas flow stabilization (1.5 *%*), temperature drift of the MKS PR-4000 and Baratron Type 122B, accuracy measurements of the MKS PR-4000 (±14 *Pa*) and uncertainties of the total electron scattering cross sections which are used in the transmission measurement (5 *%*).

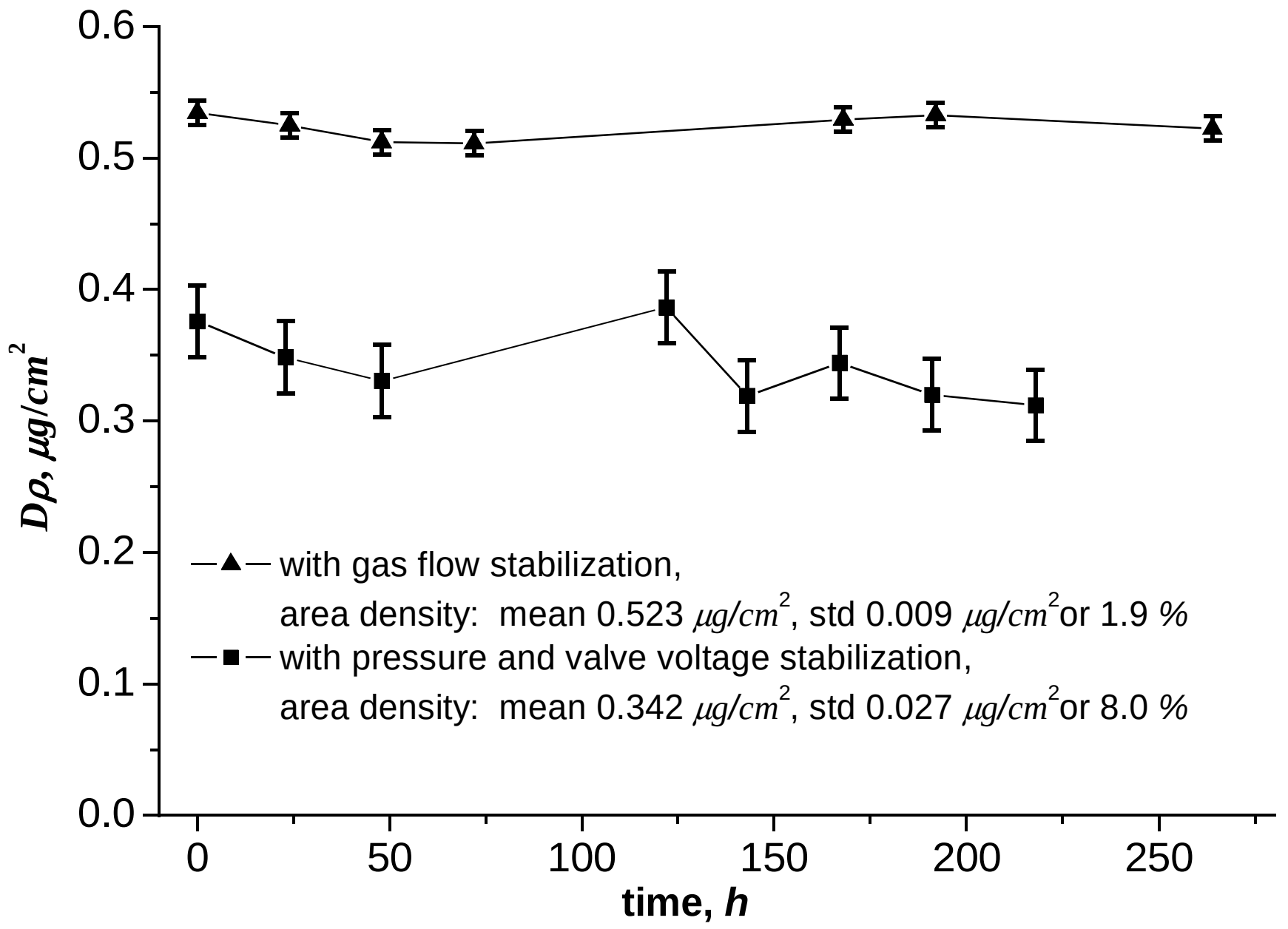

*Figure 24: Long time gas area density Dρ stability versus time for two types of stabilization systems.*

### 3.4.4 Nanometer gas volume definition by transmission measurements

It has been assumed that the ionization processes at nanometre level can be modelled in a gas cavity with an appropriate size. A gas cavity which simulates a nanometre size volume (in unit density) was created as shown in Figure 12. A simulated nanometre size (SNS) is obtained by injecting every second a 1 *ms* pulse of gas (nitrogen or propane) into the interaction chamber (IC). The gas target density in this chamber is controlled by the gas pressure in the gas reservoir R and by the voltage applied to the piezoelectric valve (PZ) described in chapter 3.4.3.

A schematic view of the set-up for the transmission measurement is presented in Figure 25. The piezoelectric valve is opened with one second repetition. By the same trigger, the electron gun (EG) generates a 2.5 *ms* electron impulse that passes through the gas jet and is detected by an electron detector channeltron (CH1). A cumulative transmission curve (see Figure 13 or Figure 26) after many repetitions (a few hundreds) was obtained. The retarding field of the electron analyzer (EA) was used to protect the electron detector (CH1) from scattered electrons (only the primary beam attenuated by the gas target reached the CH1). The retarding field was produced by the grid S1 and the voltage applied to the entrance of the channeltron CH1.

The electron transmission through the analyzer for different electron energies is presented in Figure 27.

The transmission of mono-energetic electrons through the electron analyzer for different voltages on grid S1 is presented in Figure 28. The energy resolution of the electron analyzer (EA), presented in Figure 29 and 30, was estimated to be about 4 *%* and is sufficient for attenuation measurements.

When the transmission is at its minimum, the gas density shows a maximum. To derive the gas area density a formula (Eq.28) based on Beer's law [60] is used.

$$\frac{(D\cdot\rho)_{meas}}{\mu g/cm^2} = A\cdot\frac{ln(t(T))}{\sigma_{tot}(T)/cm^2} \qquad \textbf{(28)}$$

Where: $T$ – energy of the mono-energetic electrons,

$t$ – transmission of the mono-energetic electrons,

$\sigma_{tot}$ – total scattering cross section for electrons, see Figure 31 with tabulated values in Table 1 (Appendix C), Ref.[56]

$A = -M/N_A$ is equal to -4.65174·$10^{-17}$ $\mu g$ for molecular nitrogen, and -7.32168·$10^{-17}$ $\mu g$ for propane gas,

$M$ – molar mass of molecular nitrogen or propane gas,

$N_A$ – Avogadro's constant is equal to 6.02214179(30)·$10^{23}$ $mol^{-1}$.

The nitrogen and propane area density vs. electron transmission for 200 *eV*, 300 *eV*, 500 *eV*, 1000 *eV* and 2000 *eV* is given in Figure 32 with tabulated values in Tables 2 and 3 (Appendix C).

The importance of using the electron analyzer (EA) for electron transmission is illustrated in Figure 26. As may be seen, the response curve without the EA has a higher minimum (0.135 – which corresponds to 0.425 $\mu g/cm^2$ of $N_2$) than with the EA (0.085 – 0.523 *μg/cm²* of $N_2$), this means that we detect not only the primary beam but also scattered electrons. As a result, if the EA is not used there is an error in the estimation of the gas volume of the order of 0.425 *μg/cm²* via 0.523 *μg/cm²* with the EA. Therefore, it is extremely important to use the EA for a precise estimation of the gas target thickness.

The electron transmission with energies of 200 *eV*, 300 *eV*, 500 *eV* and 1000 *eV* through the nitrogen gas jet for the same gas density is shown in Figure 33. The calculated target thickness (at the minima of transmission) using equation (Eq.28) and the maximum of the gas density for different electron energy transmissions is presented in Figure 34. Mean area density is about 0.340 *μg/cm²* for 4 different energies with a standard deviation of 0.0082 *μg/cm²* or 2.5 *%*.

Usually in the experiments for gas density measurement 1000 *eV* electrons are used. In the main experiment with "single" electrons the density is calculated at the time of the gas maximum. But in experiments with $\alpha$-particles the density is calculated as a mean value during 200 *μs* coincidence time. In the case presented in Figure 33 the mean density for 1000 *eV* is 0.377 *μg/cm²* with a maximum of 0.402 *μg/cm²* and a standard deviation of 0.015 *μg/cm²* or 3.9 *%*.

All experiments proved that transmission measurements are a perfect instrument for nanometer gas volumes definition.

For the channeltron and its characteristic see chapter 3.4.5.

The reproducibility of the stability of the gas jet is described in chapter 3.4.3.

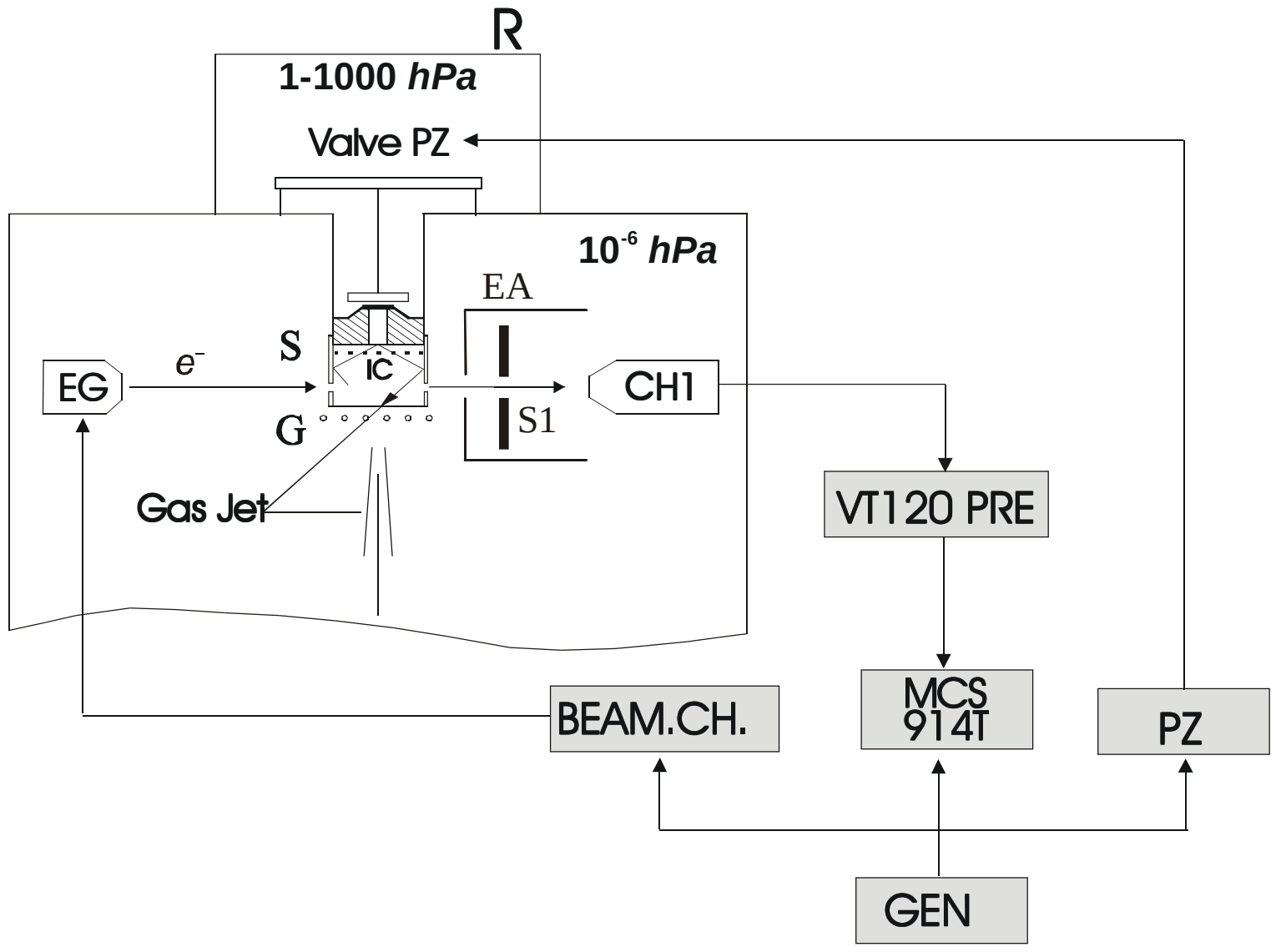

*Figure 25: Schematic view of the set-up for the transmission measurement. EA — electron analyzer, S1 – grid, CH1 – channeltron; others as in Figure 12.*

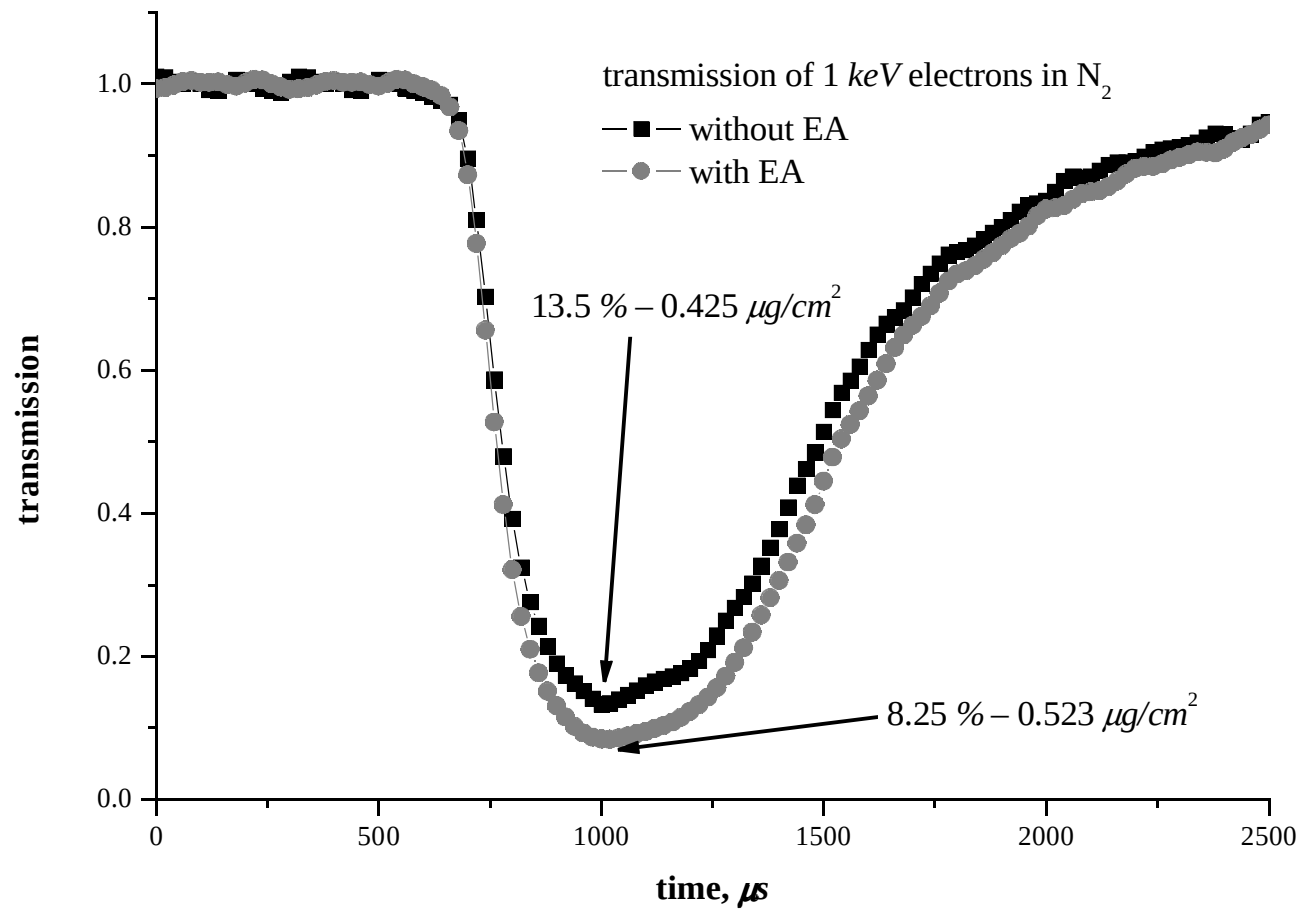

*Figure 26: Transmission of 1 keV mono-energetic electrons through nitrogen jets with (●) and without (■) the electron analyzer (EA).*

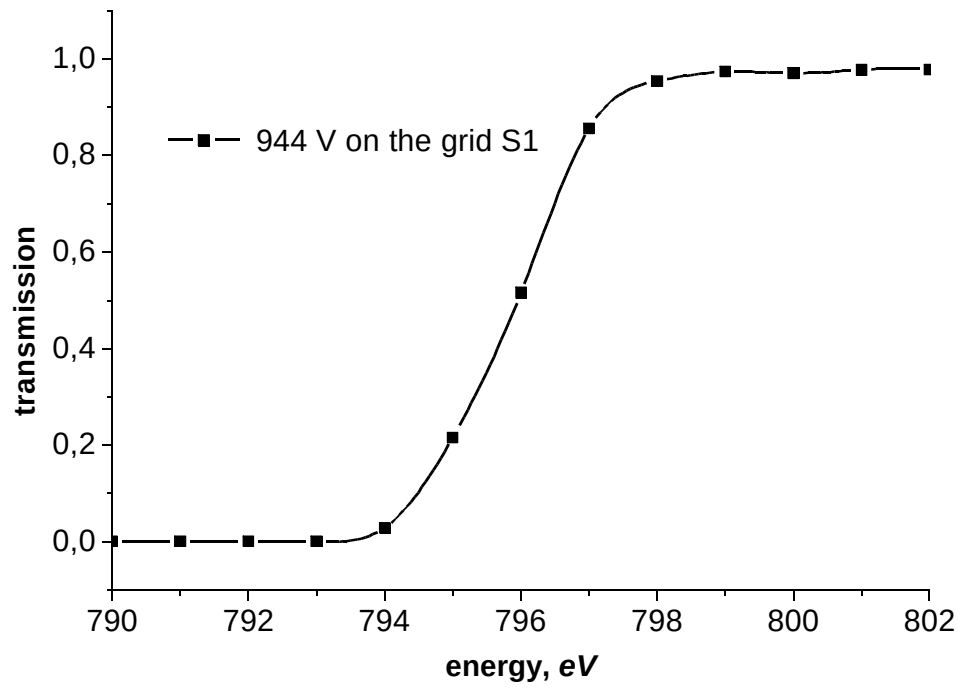

*Figure 27: Electron transmission through the analyzer vs. electron energy for grid S1 voltage equal to 944 V (vacuum $10^{-6}$ hPa).*

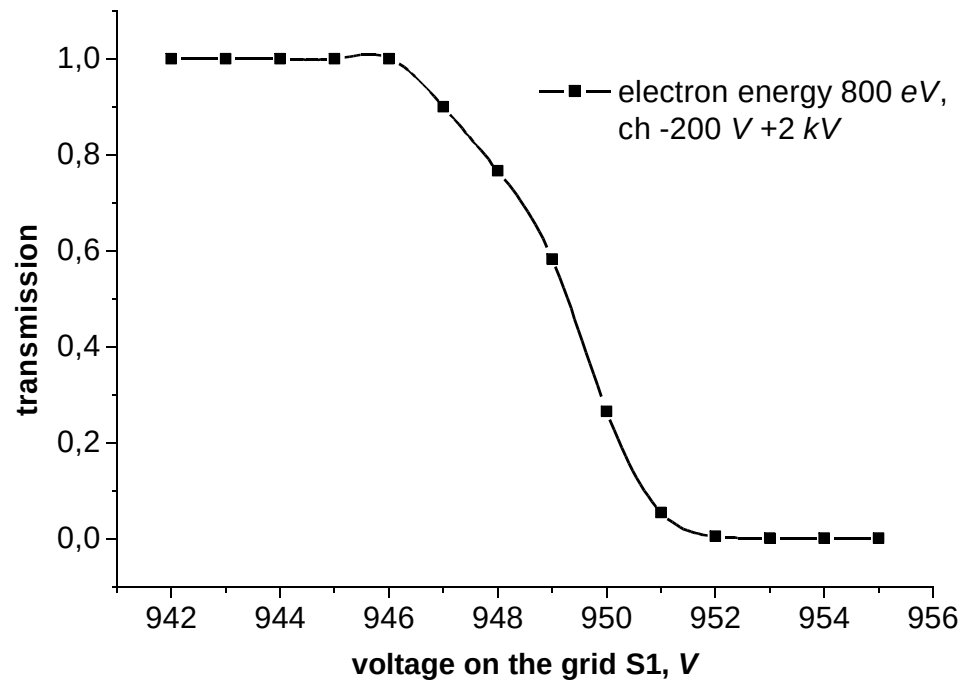

*Figure 28: Transmission of mono-energetic electrons of 800 eV through the electron analyzer for different voltages on grid S1 (vacuum $10^{-6}$ hPa).*

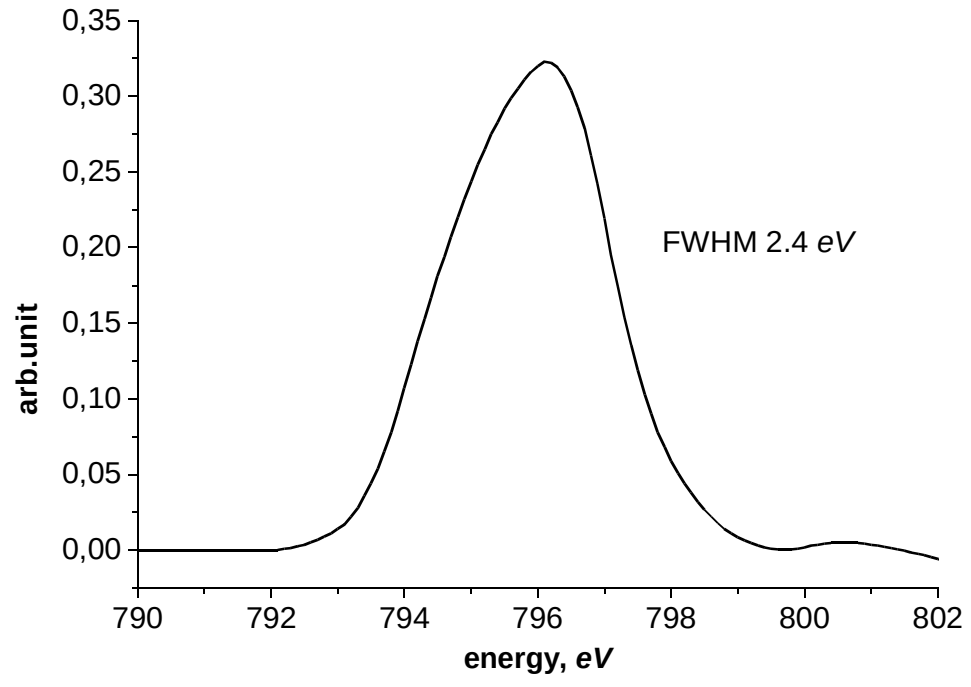

*Figure 29: Energy resolution of the electron analyzer versus electron energy and constant voltage on the grid S1.*

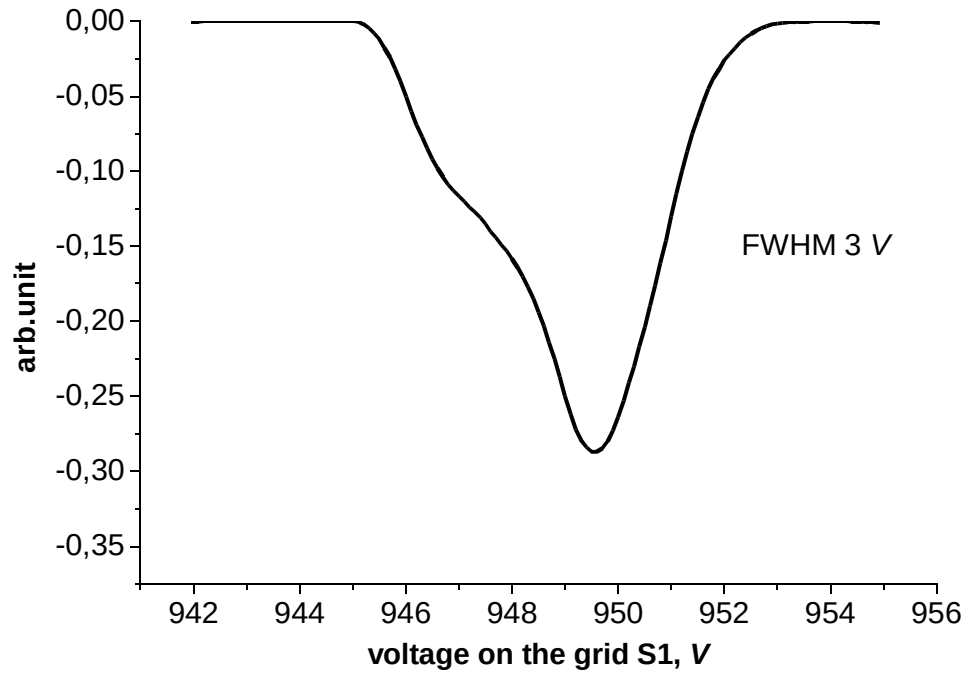

*Figure 30: Energy resolution of the electron analyzer for constant electron energy versus voltage on the grid S1.*

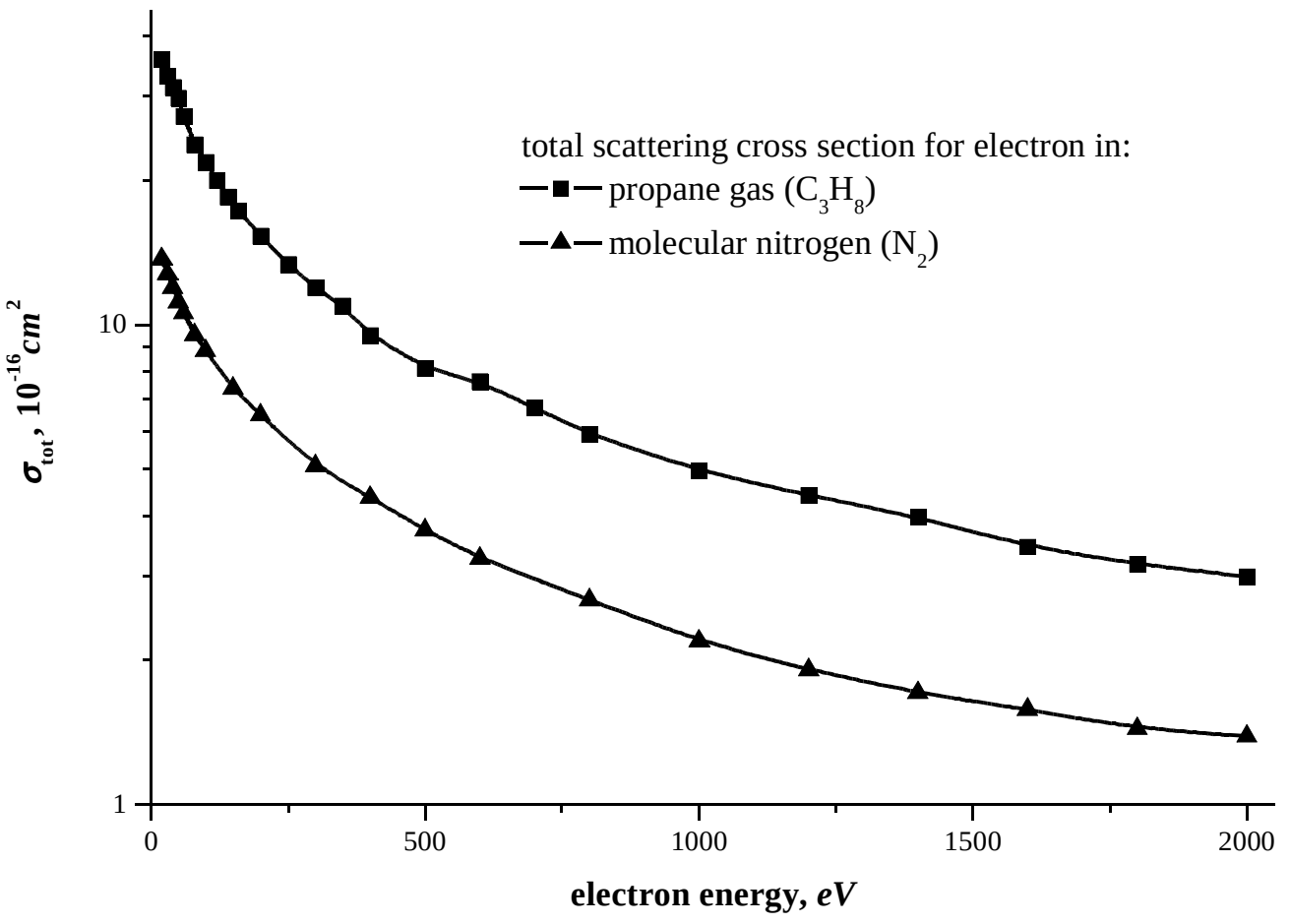

*Figure 31: Fig. Total scattering cross section vs. electron energy in: (■) – propane gas ($C_3H_8$) and (▲) molecular nitrogen ($N_2$), Ref.[56].*

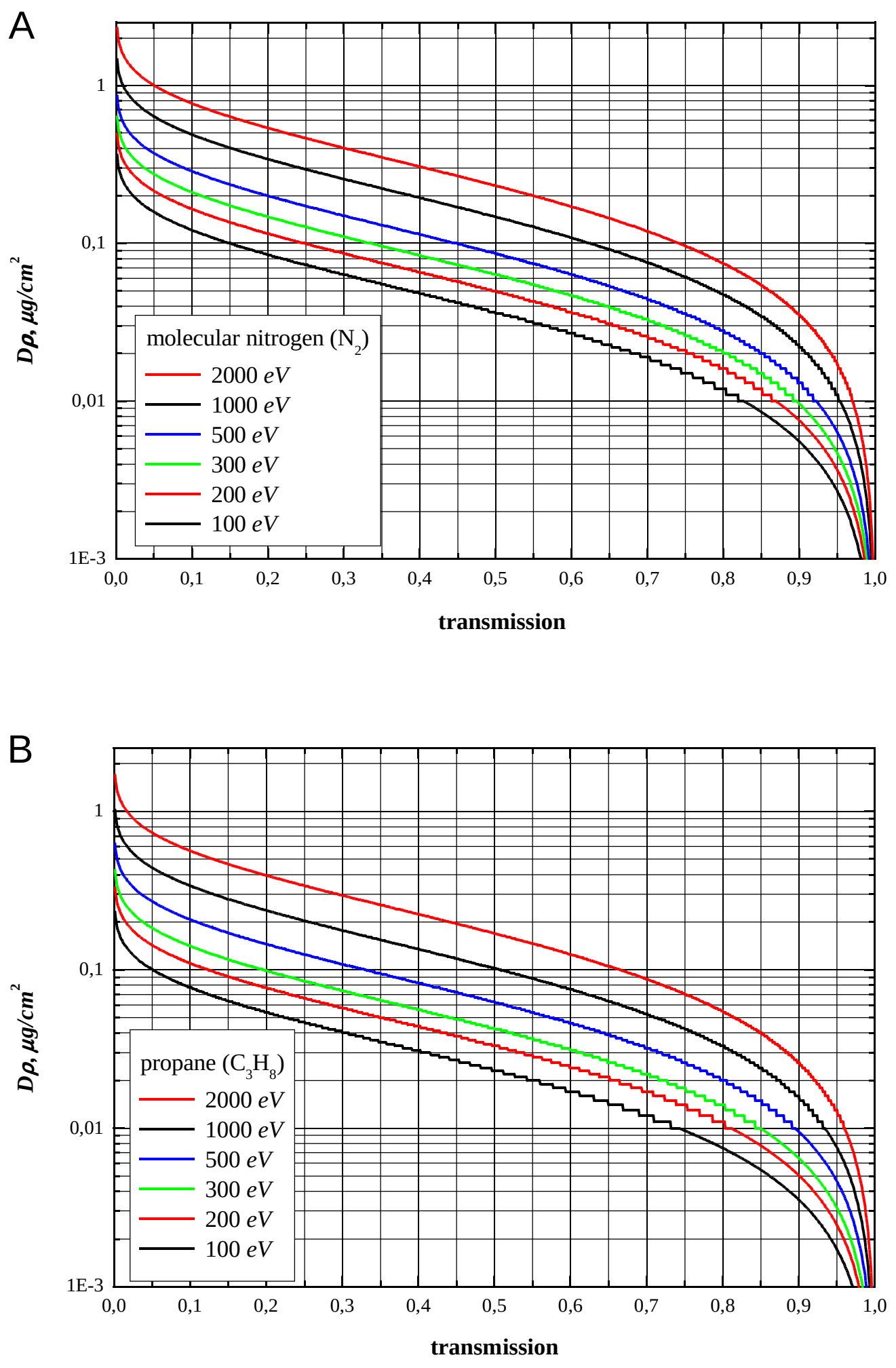

*Figure 32: Nitrogen (A) and propane (B) area density $D\rho$ vs. electron transmission for 200 eV, 300 eV, 500 eV, 1000 eV and 2000 eV.*

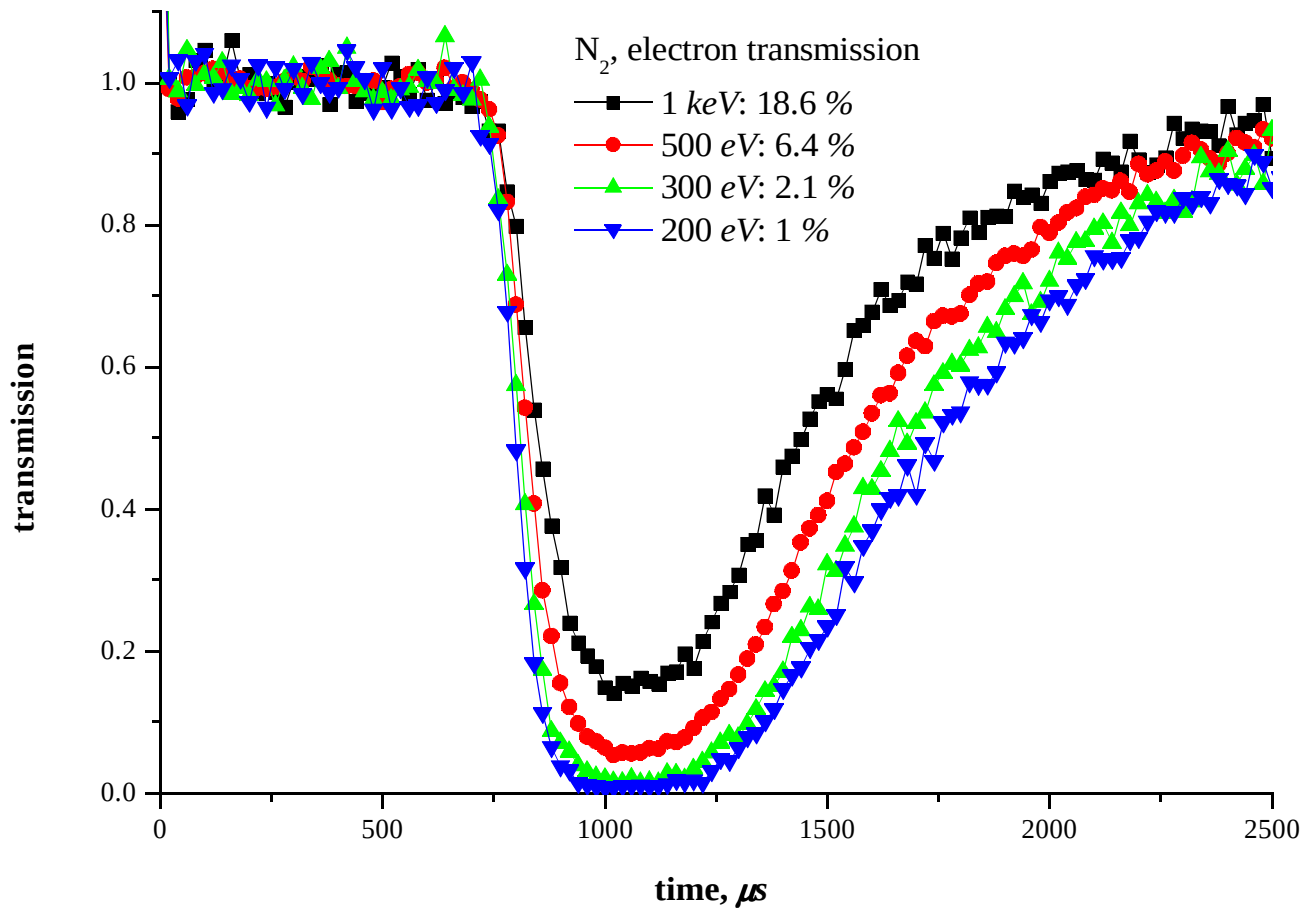

*Figure 33: Transmission of electrons at (■) 1 keV, (●) 500 eV, (▲) 300 eV and (▼) 200 eV through the nitrogen gas jet.*

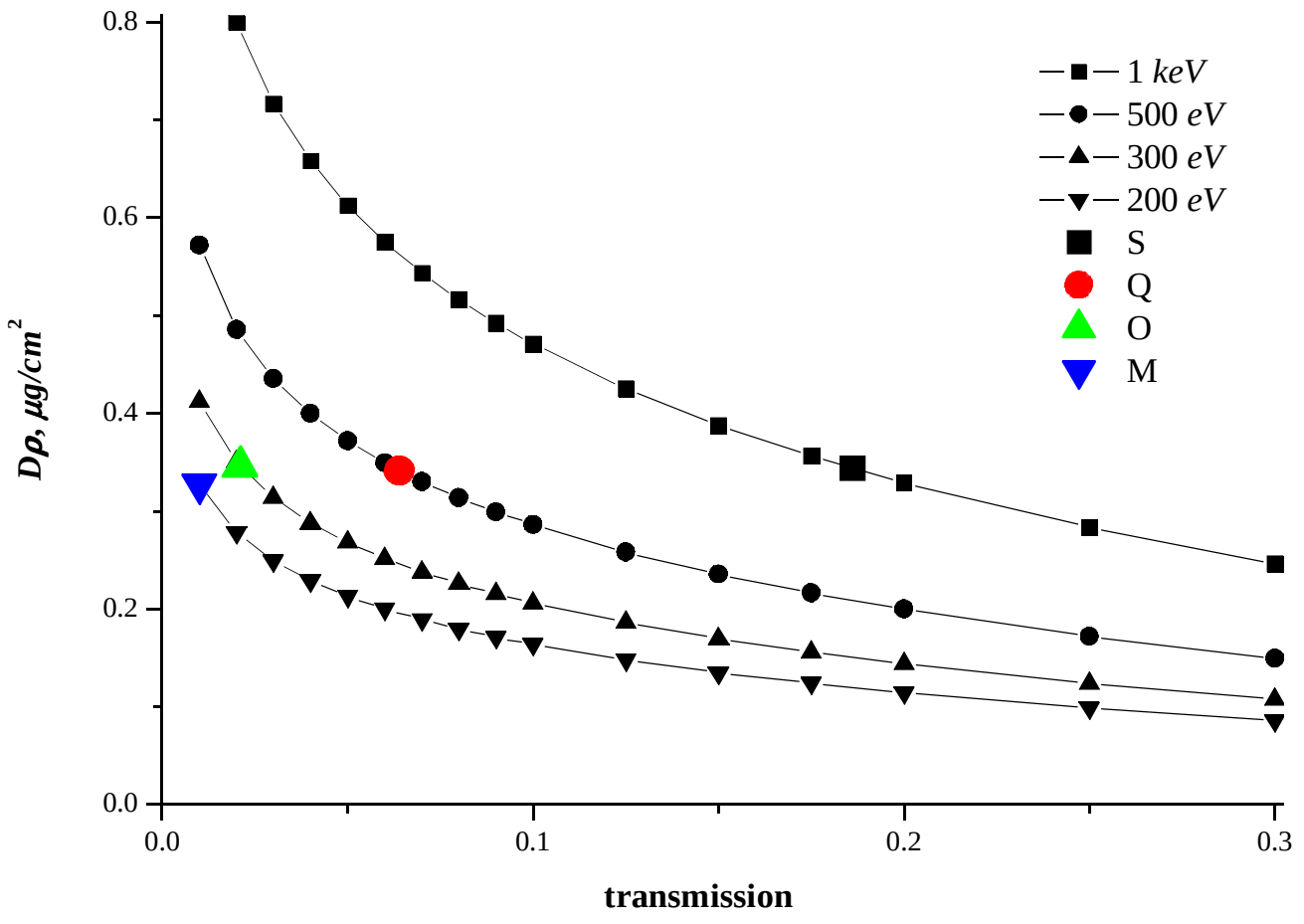

*Figure 34: Calculated nitrogen area density Dρ vs. electron transmission for (■) 1 keV, (●) 500 eV, (▲) 300 eV and (▼) 200 eV. Points M, O, Q and S are measured transmissions for different energies, but for the same gas area density.*

Summarizing: the errors in gas density measurements are of the order 8 % and mostly depend on knowledge of $\sigma_{tot}$ – total elastic scattering cross section for electrons (about 5 %), uncertainty of the transmission method measurements (about 5 %). The statistics of the transmission measurements give only 0.2 %.

### 3.4.5 Electron multipliers; their efficiency for ion and electron detection

#### 3.4.5.1 Electron multipliers

Two types of electron multipliers for electron and ion detection were used in the present work, namely; – a Continous Dynode Electron Multiplier (CDEM, see Figure 35, [61]) type Philips Channeltron X719BL and a Discrete Dynode Electron Multiplier (DDEM, see Figure 36) type ETP AF180H. Depending on the application, one or the other is preferable. Taking into account their main characteristics such as counting efficiency for charged particles, operating voltage, counting rate, dark count rate, resistance, size *etc.* a CDEM for electron detection and a DDEM for ion detection was used.

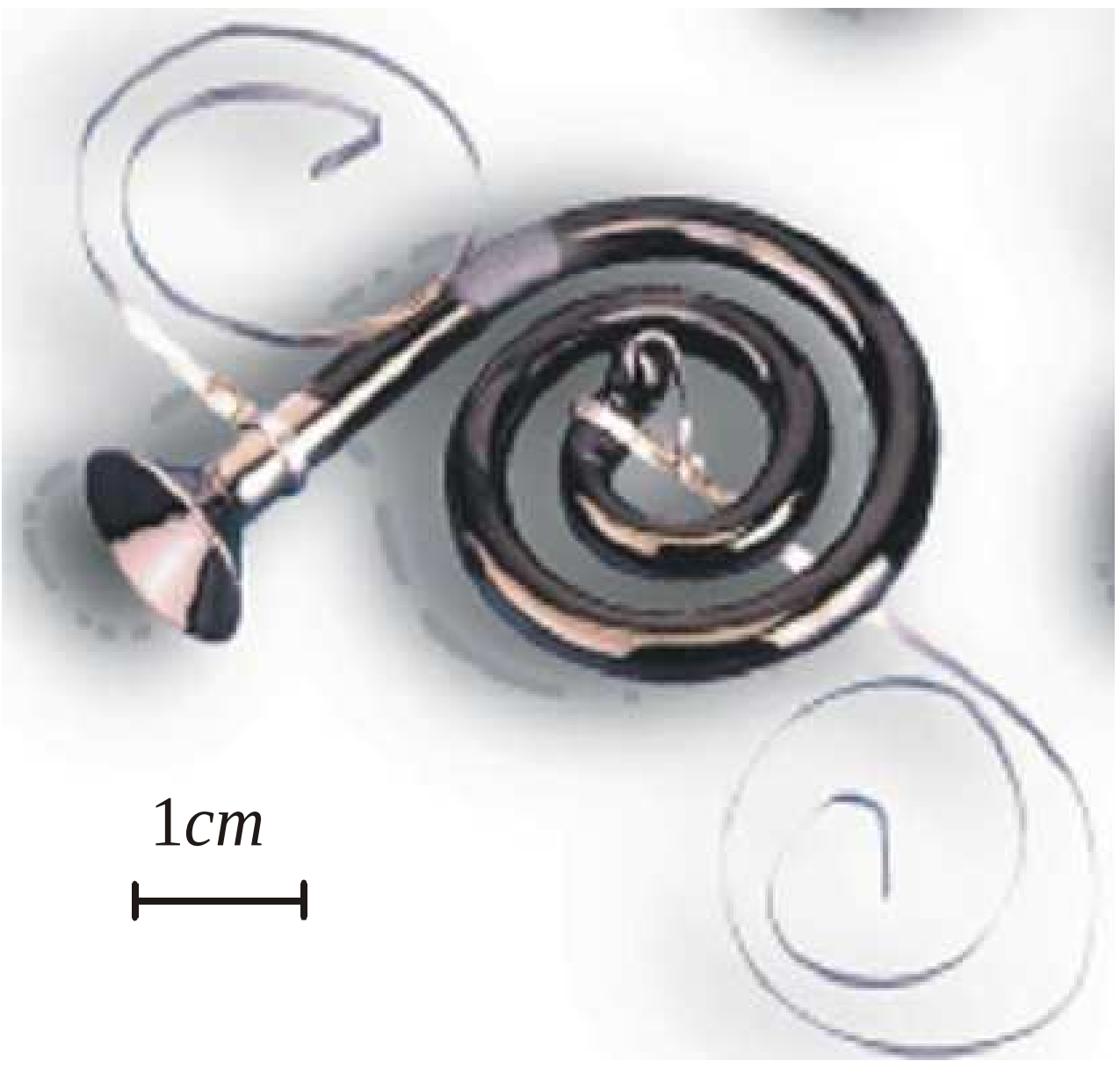

*Figure 35: Photo of a CDEM type X719BL detector.*

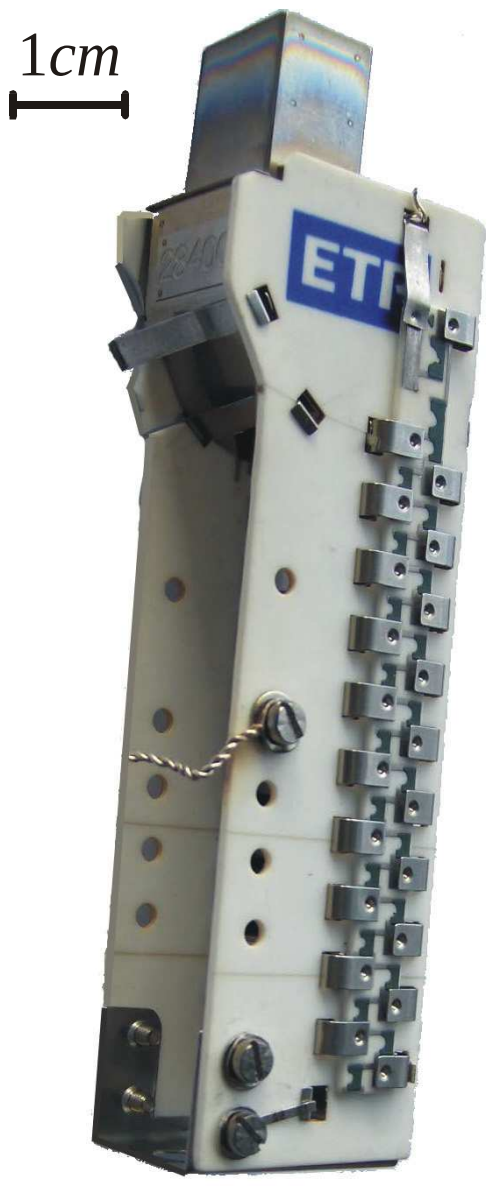

*Figure 36: Photo of a DDEM type AF180H detector.*

A CDEM X719BL was chosen for electrons due to the small mounting place needed. A DDEM AF180H was chosen for ion detection due to its ability to withstand the high counting rates.

The counting efficiency of the particle detectors is a very important parameter for our studies. The efficiency of the electron detector (CDEM) is needed to evaluate the true mean number of electrons in cluster size measurements with “single” electrons. The efficiency of the ion detector (DDEM) is needed to reconstruct the measured cluster size distributions for 100 % detection efficiency.

The efficiency of the CDEM and the DDEM must be precisely measured for each new copy of these detectors (one of disadvantages of these detectors) due to different efficiency characteristics for the same type of detectors.

### 3.4.5.2 Continuous dynode electron multiplier – Philips Channeltron X719BL

The signal impulse shape after the VT120A preamplifier is presented in Figure 37. Full width at half maximum (FWHM) is about 15 *ns*. The counting rate is about $10^4$ *cps* and the dark count rate 1-2 *cps*. The saturation of counting rate versus voltage on the detector is shown in Figure 38. The pulse counting pulse height distribution for different voltages on the detector is in Figure 39.

The efficiency of the detector for a given charged particle is the ratio of the number of counted pulses to primary particle flux. The result is a percentage of counted particles.

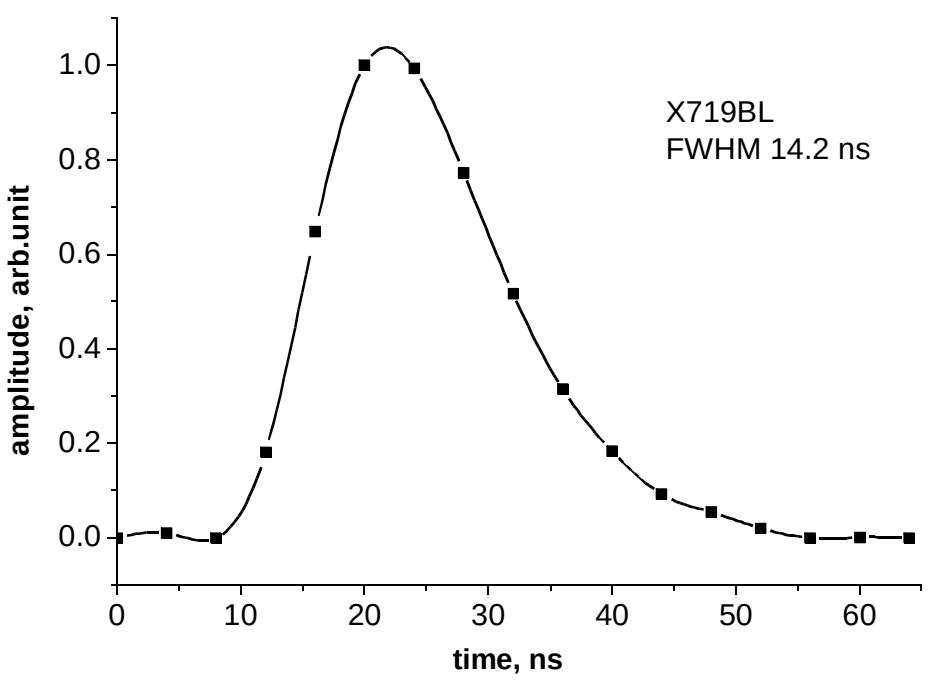

*Figure 37: X719BL signal impulse shape after VT120A preamplifier, digitized with a Lecroy 9354 Osciloscope.*

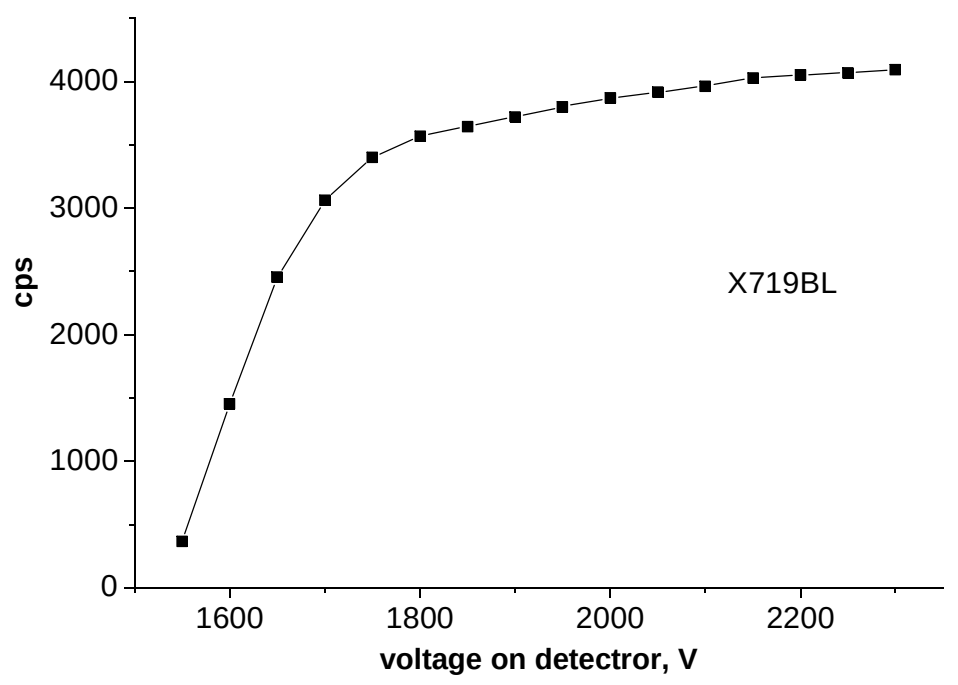

*Figure 38: X719BL counting rate versus detector voltage.*

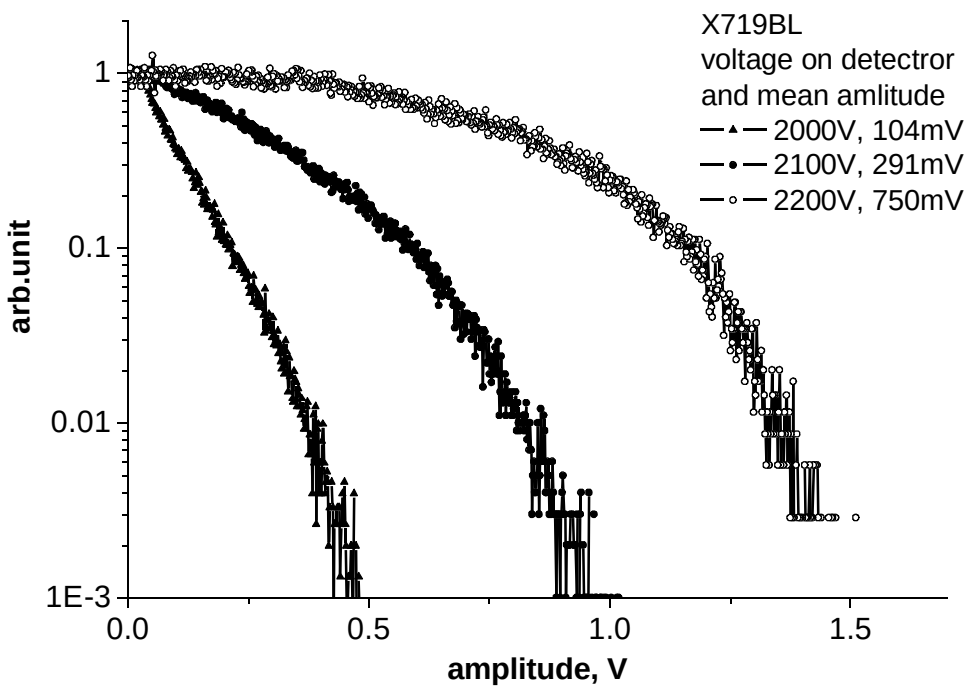

*Figure 39: Pulse counting pulse height distribution X719BL CEM for different detector voltages.*

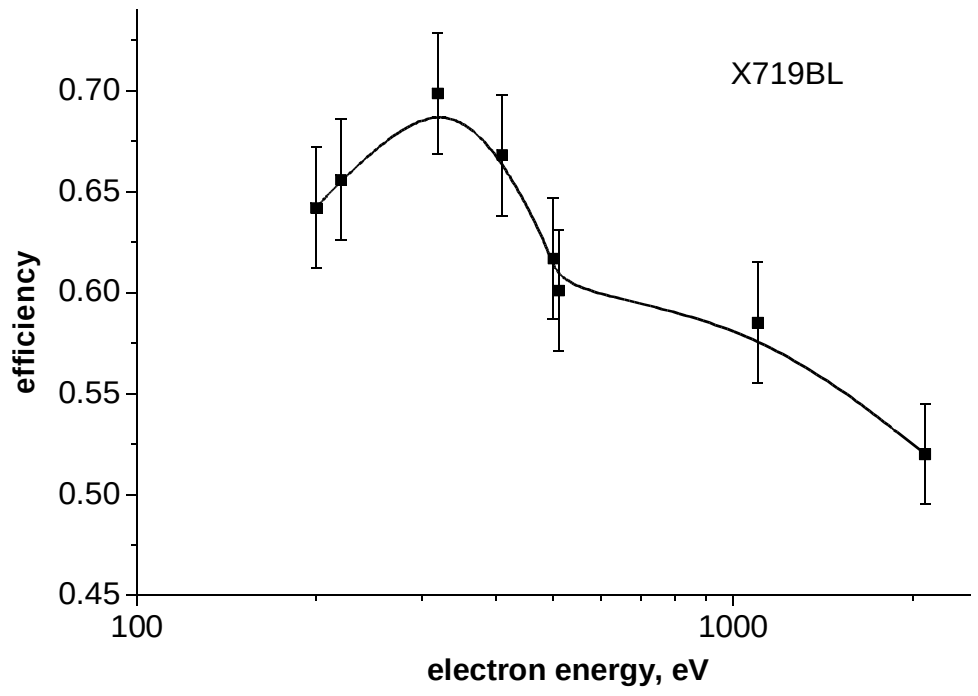

*Figure 40: X719BL Efficiency map for electrons versus electron energy (2200 V on detector, 5 mV discriminator level)*

The efficiency measurement for the CDEM was made in the mounting place in the Jet Counter (see Figure 12 and Figure 15) with the same cables, preamplifier and counter (Turbo MCS 911). The CDEM has a low count rate and direct measurements of electron counts and current measurements at the same time are not possible. The impulse method for efficiency measurement proposed by Pszona [62] was used. The gun generates a 1 $\mu s$ electron impulse (a few electrons in one impulse) with 1 *Hz* repetition. These electrons are counted by the CDEM. Then the same CDEM is connected to an electrometer acting as a Faraday Cup which measures the current from the gun. The gun generates the same impulses with the same intensity but with a repetition of about 500 *kHz*. It is possible to evaluate the true number of electrons in a single impulse from the gun, and to calculate the CDEM efficiency as a function of electron energy. For proper current measurements the voltage combination in the analyzer and in the entrance of the CDEM were taken into account to prevent secondary electron emission from the surface of the CDEM. As a final result, the efficiency map for electrons as a function of electron energy, 2200 *V* on detector, 5 *mV* discriminator level, is shown in Figure 40. The maximum efficiency for electrons is near 300 *eV*, corresponding to the maximum of the secondary electron emission from the surface of the CDEM where the amplification has a maximum value.

### 3.4.5.3 Discrete dynode electron multiplier – ETP AF180H

The signal impulse shape after the VT120A preamplifier is presented in Figure 41. The FWHM is about 5 *ns*. The counting rate is much better than in a CDEM and is about $10^6$ *cps* with a dark count rate of 0.3-0.5 *cps*. The pulse counting pulse height distribution for different voltages on the detector is shown in Figure 42.

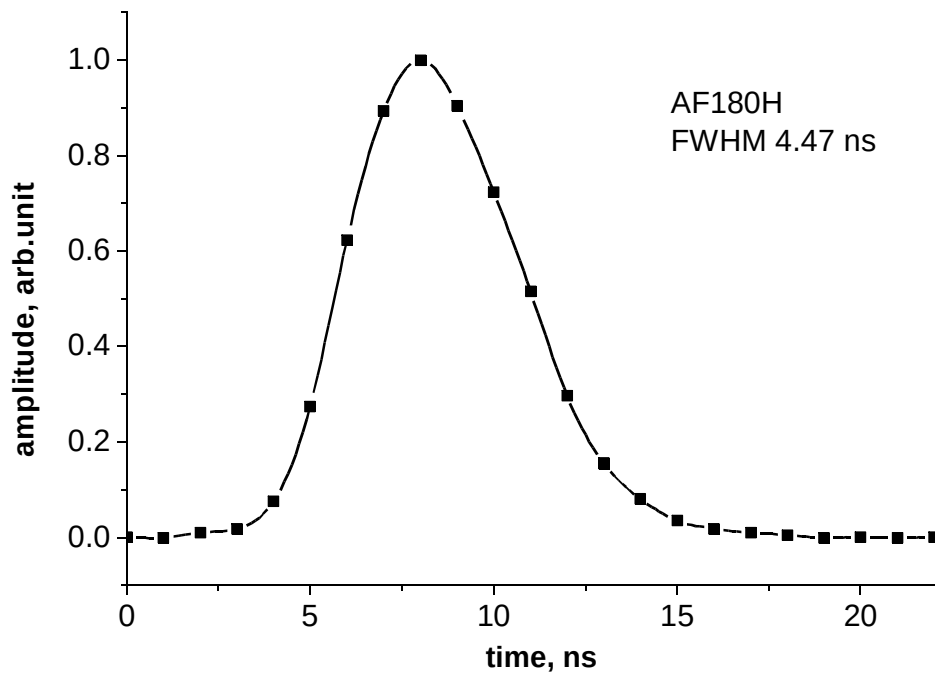

*Figure 41: AF180H signal impulse shape after the VT120A preamplifier, digitized with a Lecroy 9354 Osciloscope.*

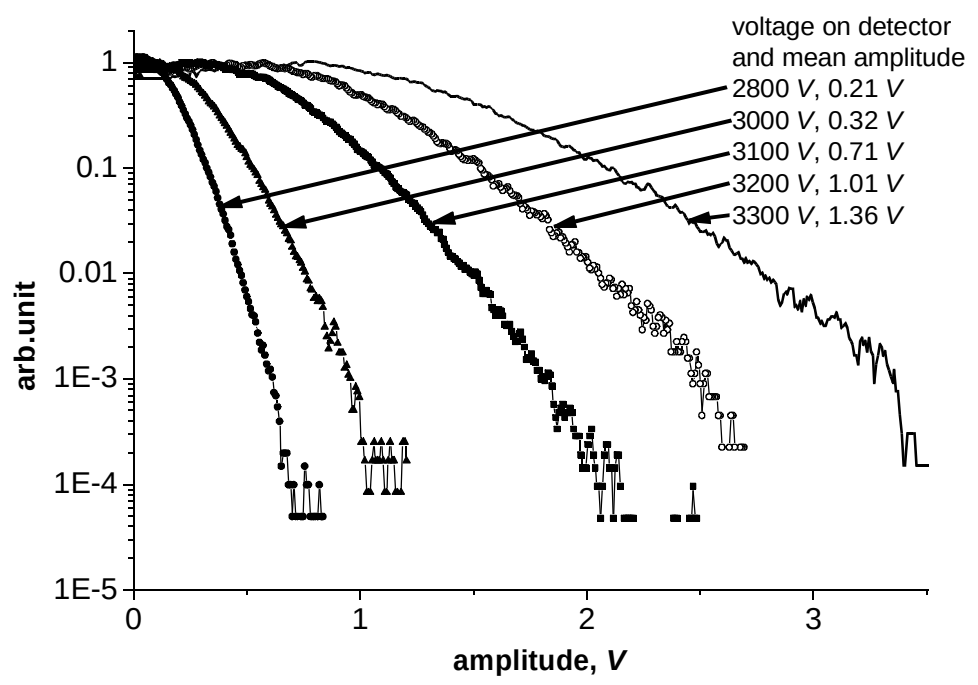

*Figure 42: Pulse counting pulse height distribution AF180H DDEM versus detector voltage.*

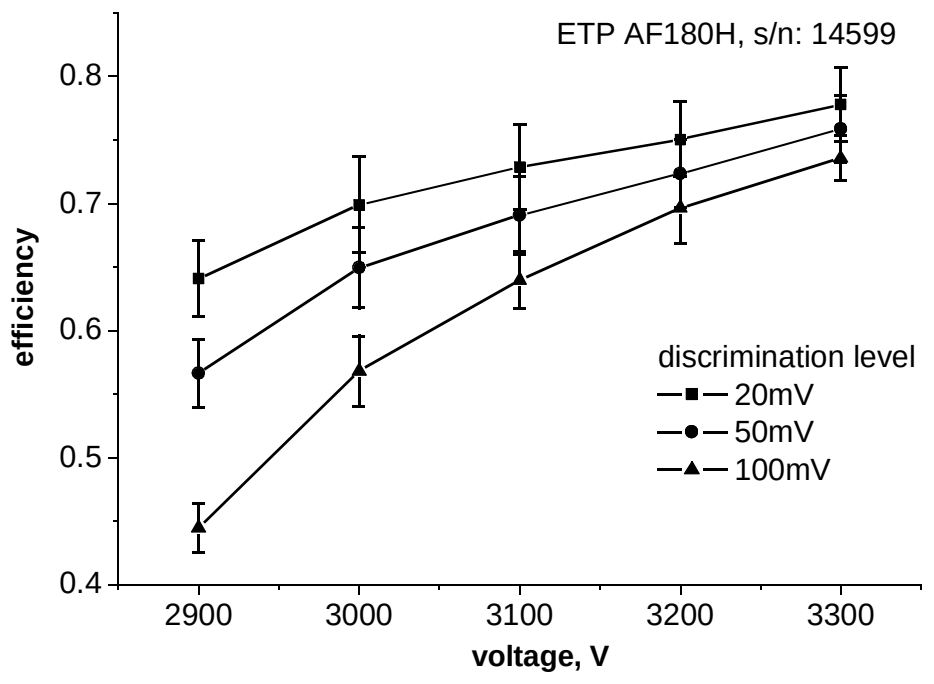

*Figure 43: ETP AF180H s/n: 14599 Efficiency map for $N_2^+$ ions versus detector voltage and discrimination level in the counting device (Ortec, Turbo MCS 914T).*

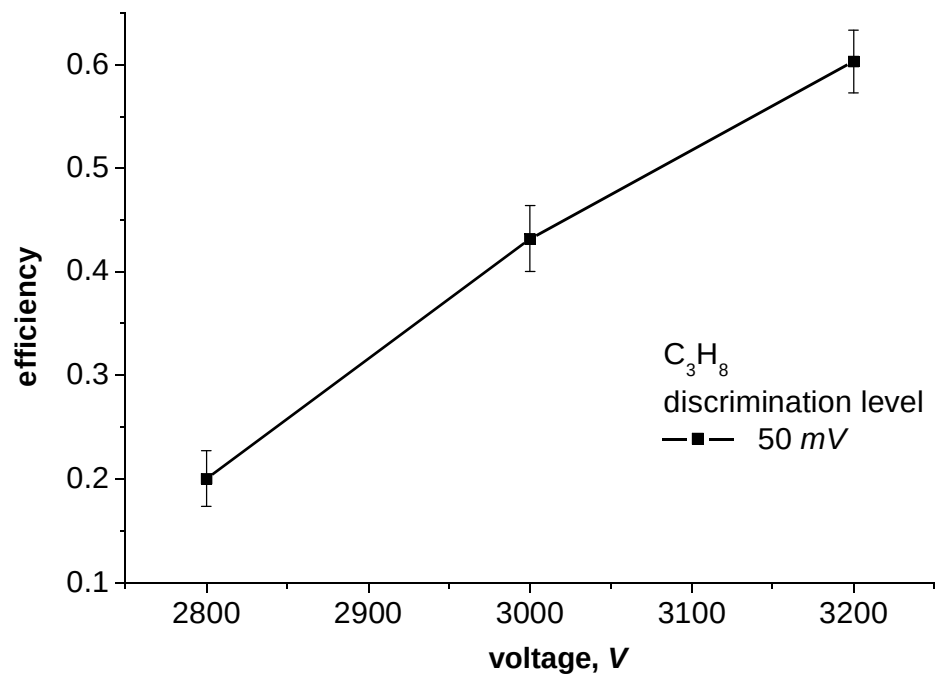

*Figure 44: ETP AF180H s/n: 14599 Efficiency map for $C_3H_8$ ions versus detector voltage and discrimination level in the counting device (Ortec, Turbo MCS 914T).*

As with the CDEM, the efficiency measurements for the DDEM were made in the Jet Counter at its operational position (see Figures 12 and 15). As the DDEM has a sufficient counting rate (about $10^6$ cps), direct comparison of ion current and ion counts are possible. The source of ions is the Jet Counter. The gas is continuously injected into the IC, at the same time as the gun emits electrons and ionizes the gas. All ions produced in the IC are removed by the electric field of the grid (G) and then guided through G1 to the AF180H detector. The ion intensity is regulated by the electron gun and the flow of the gas injected into the IC. At a specific ion flux, the ion current (DDEM is connected as a Faraday Cup) and the ion count rate are measured. The efficiency is calculated. During the measurement the vacuum is about $10^{-4}$ *hPa*. The advantage of this method is that the efficiency is measured for the same ions (ion type and ion energy) and also with the same detection electronics. The final results of the efficiency map for 3 *keV* propane and nitrogen ions versus detector voltage are presented in Figure 43 and Figure 44 respectively.

### 3.4.6 Ion detection in the Jet Counter and it's efficiency

The efficiency of ion detection is the basic parameter that influences the shape of the signal spectra and must be known with relatively good precision.

The efficiency of the ion detection $\varepsilon$ consists of:

- efficiency of ion extraction from the IC $\varepsilon_{ext}$ ,
- efficiency of ion guiding to ion detector by electrostatic field $\varepsilon_{guid}$
- efficiency of ion detector $\varepsilon_{ion}$ .

$$\varepsilon = \varepsilon_{ext} \cdot \varepsilon_{guid} \cdot \varepsilon_{ion} \qquad (29)$$

The component $\varepsilon_{ext} \cdot \varepsilon_{guid}$ was studied with SIMION 3D version 6.0 [63]. This is a program for simulation of electrostatic lens analysis with the possibility of observing the traveling path of ions in a simulated electrostatic field. In the simulations the real geometry of the Jet Counter was taken into account with applied voltages on the extracting grids. The starting points of ions in the IC were homogeneously placed within the IC. The interaction of ions with the neutral gas was not taken into account. The result of the simulation is that $\varepsilon_{ext} \cdot \varepsilon_{guid}$ is about 80 %.

A typical electric field in the Jet Counter is presented in Figure 45 with equipotential lines and ion track lines (from the points of ion creation in the interaction chamber to the ion detector AF180H). In simulations, the optimal voltages on grids S1.. S3 were -10 *V*, -30 *V* and -130 *V*. The voltage on the entrance to the ion detector was -3100 *V*. Experiments showed that the voltage on S1 must be 0 *V*, S2 = -30 *V*, S3 = -130 *V*. The applications of any voltage on *S1* decreases the collection efficiency of ions at the ion detector AF180H.

In principle, there is also a marked probability of ion loss due to charge exchange because of collisions between ions and neutral molecules of the gas ($N_2$ or $C_3H_8$), particularly inside the IC volume, and due to the recombination process because of the low ion extraction field strength applied.

Unfortunately, there is not yet a direct method for the evaluation of the efficiency $\varepsilon_{ext} \cdot \varepsilon_{guid}$ .

For the efficiency of the ion detector $\varepsilon_{ion}$ for different type of ions, see chapter 3.4.5.

Finally, the overall efficiency $\varepsilon$ was estimated to be about 40 % with an uncertainty between 5 % and 10 %.

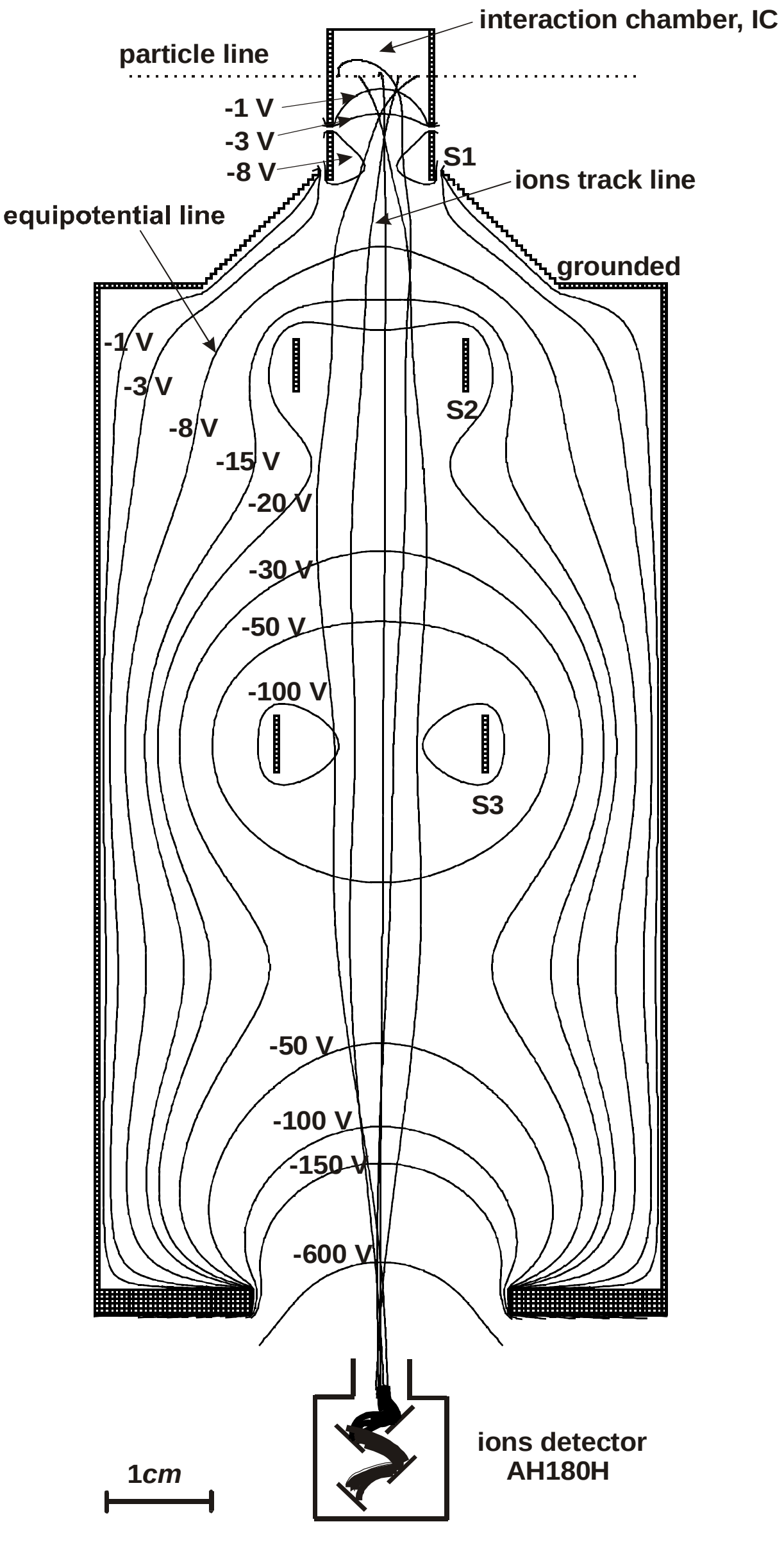

*Figure 45: Typical electric field in the Jet Counter. S1..S3 – extracting grids.*

**Influence of the efficiency on the measurement of ion cluster size distributions**.

The ability of a detector (Jet Counter) to register all ions in a cluster depends on its efficiency to register single ions. If $\varepsilon$ is the counting efficiency for single ions, the probability to count $\mu$ out of $\nu$ ions is given by the binomial probability according to

$$B(\mu,\nu,\varepsilon) = \binom{\nu}{\mu} \varepsilon^{\mu} (1-\varepsilon)^{\nu-\mu} \qquad \textbf{(30)}$$

As a consequence of (Eq.30), the probability of counting all $\nu$ ions is given by the following expression:

$$B(\nu;\nu;\varepsilon) = \varepsilon^{\nu} \qquad \textbf{(31)}$$

Registration of all ions in each cluster, therefore, needs an overall efficiency $\varepsilon$ close to 100 %. Since $\varepsilon$ was estimated to be about 40 %, the measured signal cluster size differs in shape. Consequently, a de-convolution procedure must be applied to the measured distribution at least in principle (see chapter 5).

### 3.4.7 α-particle source $^{241}$Am

An $^{241}$Am radioactive source of $\alpha$-particles (Amersham gold-plated type AMM2) was used. The energy of the $\alpha$-particles which passed through a 1 *mg/cm*$^2$ thick Mylar wall in the IC was degraded to 3.8 *MeV*.

The $\alpha$-particles were detected by a Si surface barrier detector (produced by W.Czarnacki in SINS). The calibration of the Si detector was performed using an Amersham mixed alpha spectrometric source: $^{239}$Pu (5.155 *MeV*), $^{241}$Am (5.486 *MeV*) and $^{244}$Cm (5.805 *MeV*).

The energy spectra of $\alpha$-particles from the $^{241}$Am–source presented in Figure 46 demonstrate a change in the energy resolution ($\Delta E$) from 331 *keV* to 875 *keV* versus the Mylar foil thickness.

The energy resolution ($\Delta E$) is sufficient for our experiment.

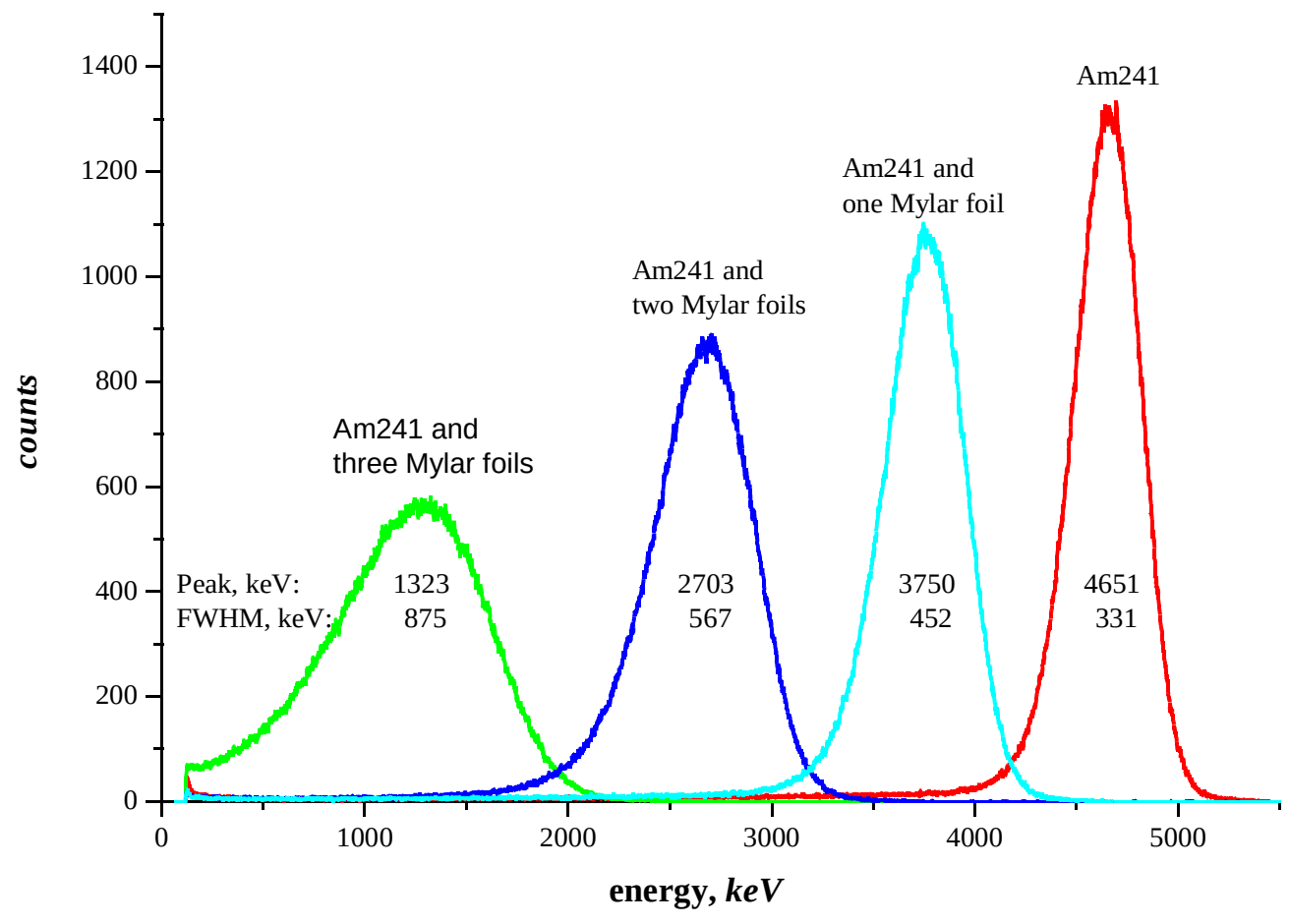

*Figure 46: Spectra of $^{241}$Am alone and degraded after penetrating through Mylar foils.*

## 3.5 Conclusions

It can be stated that the Jet Counter, which simulates a geometric structure of nanometre size at unit density by an injection of a pulsed gas beam of nitrogen into an IC, is well suited for studying the formation of ionisation clusters not only by $\alpha$-particles but also by primary electrons.

It should be mentioned that the Jet Counter facility has well studied characteristics:

- very well defined simulated nanometer size by transmission measurements of mono-energetic electrons (see chapter 3.4.4). This was achieved by the application of an electron analyzer;
- very good stability of the simulated nanometer size. This is the result of the application of the piezoelectric valve and gas-flow stabilization system (see chapter 3.4.3);
- a source of “single” electrons for cluster size distribution measurements created by low energy electrons was defined and studied (see chapter 3.4.2);
- the detection efficiency of the ion and electron detectors has been previously studied (see chapter 3.4.5). These values are very useful for the definition of the mean number of electrons in measurements with “single” electrons and for the shape of the cluster size distributions in all experiments.

The Jet Counter has a rather high efficiency for the detection of single ions and represents the first measuring device based on single-ion counting which can be used to investigate ionization cluster formation in target volumes 0.9–5 *nm* in diameter at unit density. Such target volumes are comparable in size to sub-cellular structures like segments of DNA or nucleosomes.

# 4 Monte Carlo model of ionization cluster formation in molecular nitrogen and propane gas

In order to simulate the ionization cluster distribution produced by $\alpha$-particles and electrons a track-structure code developed by Grosswendt [34, 64] was used. This code was adapted to the needs of our experiment by taking into account the geometry of the Jet Counter and its specific properties.

The ionization cluster distribution produced by 3.8 *MeV* $\alpha$-particles in the Jet Counter was calculated by simulation of the ionization pattern of track segments in cylindrical volumes of diameter $d$ and height $h$, of a mass per area between 0.092 $\mu g/cm^2$ and 0.538 $\mu g/cm^2$ in the case of perpendicular particle incidence at half the cylinder's normal height.

To calculate the ionization-yield distribution caused by $\alpha$-particles at energies of a few *MeV* during their penetration through layers of nitrogen or propane of small mass per area, we assumed that:

1. the energy loss due to impact ionization or excitation along short track segments does not appreciably change the initial particle energy,
2. the influence of elastic scattering on the particle energy and flight direction can also be neglected in the case of short track segments,
3. the primary particle energy is high enough to allow charge changing processes to be neglected.

The first assumption can be justified by the electronic stopping powers of $\alpha$-particles in nitrogen (propane). In the case of 3.8 *MeV* $\alpha$-particles in nitrogen (propane), for instance, the total mass stopping power is 933.0 (1304.8) $eV{\cdot}cm^2/\mu g$ (see ICRU [21]), which leads to an energy loss of less than 1 % for a penetration through a 1 $\mu g/cm^2$ layer of nitrogen (propane).

The validity of the second assumption is obvious from the detour factor of 3.8 *MeV* $\alpha$-particles which is equal to 0.9878 (0.9951) in nitrogen (propane) (see ICRU 21) thus demonstrating that the particles' projected range is almost equal to the continuous slowing-down range .

The third assumption can be justified by the results of Grosswendt and Baek [65] and by Baek and Grosswendt [66, 67] with regard to the influence of charge changing processes of protons on their $W$-value.

In view of these facts, the structure of $\alpha$-particle track segments at an energy of a few *MeV* is almost exclusively based on their ionization cross section, on the spectral and angular distribution of secondary electrons produced by impact ionization, and on the properties of secondary electron degradation. The main steps of the simulation of their track segments are therefore:

1. the determination of the distance to the successive point of ionization impact interaction,
2. the determination of the energy and direction of the secondary electron set in motion, and
3. the simulation of electron transport.

For the latter purpose, the Monte Carlo model must be able to follow electron histories down to the ionization threshold energy of 15.58 *eV* in the case of nitrogen and 11.08 *eV* in the case of propane, taking into account elastic electron scattering, impact ionization and the reliable excitation processes influencing electron degradation.

## 4.1 Simulation of the primary ionization pattern of α-particle track segments

Within the framework of the present work, the traveling distance $(\lambda\rho)_\alpha$ between two successive interaction points of $\alpha$-particles at energy $T_\alpha$ can be calculated in the conventional way according to (Eq.32) if it is assumed that the ideal gas law is valid for nitrogen and propane:

$$\frac{(\lambda\rho)_\alpha}{\mu g / cm^2} = A \cdot \frac{\ln \xi}{\sigma_{ion}(T_\alpha) / cm^2} \qquad \textbf{\textit{(32)}}$$

Where $\sigma_{ion}(T_\alpha)$ is the ionization cross section of $\alpha$-particles at energy $T_\alpha$,

$\xi$ – a random number uniformly distributed between 0 and 1,

$A$ – the same as in (Eq.28).

Since no comprehensive sets of experimental ionization cross sections of $\alpha$-particles at energies of a few *MeV* in nitrogen and propane are available, the treatment of direct ionization by $\alpha$-particles was based on the Hansen-Kocbach-Stolterfoht (HKS) model published by ICRU [68]. This semi-empirical model includes the single-differential cross section of charged particles with respect to the energy of secondary electrons set in motion by the ionization process, and the double-differential cross section with respect to the energy and the emission angle of secondary electrons, without the use of any empirical parameters. The only parameters which must be known for its application are the binding energies $B_k$ and occupation numbers $N_k$ of electrons in all subshells $k$ of weakly bound electrons of the target system. These data were taken from the publication by Hwang *et al.* [69] assuming four orbitals of outer or weakly-bound valence electrons in nitrogen and 10 orbitals of outer or weakly-bound valence electrons in propane. The ionization cross section $\sigma_{ion}(T_\alpha)$ was calculated by an integration of single-differential cross sections for specified subshells $k$, followed by a summation over all subshells. Since the HKS model has so far been tested only for a few target systems, the ionization cross sections derived from the model were compared with the results of the semi-empirical model of Rudd *et al.* [70] for protons. To calculate the cross sections of $\alpha$-particles at energy $T_\alpha$ from the proton data, the Rudd model was applied at a proton energy of $T_p = (m_p / m_\alpha) \cdot T_\alpha$ ($m_p$ and $m_\alpha$ are the proton and $\alpha$-particle masses, respectively) and multiplied by a factor of 4 to take the charge of the projectile into account . At $T_\alpha$ = 4.6 *MeV*, the ionization cross section in nitrogen based on the HKS model is equal to $4.74 \cdot 10^{-16}$ $cm^2$, and that derived from the Rudd model is equal to $5.31 \cdot 10^{-16}$ $cm^2$. This means that the latter value is about 10 *%* greater than the former. As, however, the proton cross sections of the Rudd model are assumed to be affected by an uncertainty of about 10 *%*, the agreement between both models is very satisfactory and confirms the applicability of the HKS model, at least, for $\alpha$-particles at an energy of a few *MeV* in nitrogen.

To simulate the secondary electron distribution produced by impact ionization of the $\alpha$-particles, the partial single-differential cross sections of the HKS model for the four subshells $k$ specified by Hwang *et al.* [69] were applied. After determination of the secondary electron energy, the polar angle $\theta$ of the electron's flight direction relative to that of the $\alpha$-particle was sampled using the double-differential cross section of the HKS model at a specified electron energy as the probability density, after normalization to its integral over $\cos(\theta)$ within the limits $-1 \leq \cos(\theta) \leq 1$. The azimuthal angle of the electron direction was assumed to be uniformly distributed between $0$ and $2\pi$.

## 4.2 Simulation of the ionization pattern produced by secondary electrons

The contribution of secondary electrons to the ionization pattern of α-particles was calculated by simulating their histories in nitrogen (propane) from one interaction point to an other, taking into account elastic electron scattering, a series of different excitation processes and impact ionization. At each point of interaction, the electron's flight direction in the case of elastic scattering or its energy loss and flight direction in the case of inelastic scattering was calculated, supplemented by the energies and flight directions of the secondary particles set in motion by the scattering process. The main steps taken to follow the histories of electrons through nitrogen (propane), therefore, were:

1. the determination of the distance to the subsequent point of interaction,
2. the determination of the type of interaction the electron will suffer at this point, and
3. the sampling of the energy loss and the new flight direction caused by the selected interaction process, possibly supplemented by the energies and flight directions of secondary particles, if liberated. As external electro-magnetic fields were not taken into account, it was assumed that the electrons travel along straight lines which connect successive interaction points.

If we assume that the target molecules can be treated as independent points homogeneously distributed in space, the traveling length $(\lambda\rho)_{el}$ of an electron at energy $T$ between two successive interaction points is governed by an exponential probability density and can be sampled in the conventional way using (Eq.33).

$$\frac{(\lambda\rho)_{el}}{\mu g / cm^2} = A \frac{\ln \xi}{\sigma_{tot}(T) / cm^2} \qquad \textbf{(33)}$$

Here, $\xi$ – is again a pseudo-random number uniformly distributed between 0 and 1,

$A$ – the same as in (Eq.28) and

$\sigma_{tot}(T)$ – is the total scattering cross section of an electron at energy $T$ given by the following equation:

$$\sigma_{tot}(T) = \sigma_{el}(T) + \sum_j \sigma_{exc}^{(j)}(T) + \sum_k \sigma_{ion}^{(k)}(T) \qquad \textbf{(34)}$$

Where $\sigma_{el}(T)$ – is the integrated elastic scattering cross section of an electron,

$\sigma_{exc}^{(j)}(T)$ – is the cross section for the excitation of an electron to state $j$,

$\sigma_{ion}^{(k)}(T)$ – is the integrated partial cross section of impact ionization of an electron for a state of threshold energy $I_k$.

The summation over $j$ and $k$ includes all significant excitation and ionization processes.

The type of event an electron suffers at each interaction point is sampled from the discrete probability densities $p_\nu(T)$ of the interaction effects taken into account in the calculation. These probability densities were set equal to the ratios of cross sections with respect to a specified interaction process of type $\nu$ to the cross section of total electron scattering at energy $T$.

In the case of an elastic interaction, the polar angle of the electron's flight direction after scattering relative to its initial direction was determined on the basis of the differential elastic cross section, assuming in addition that the azimuthal scattering angle is uniformly distributed between $0$ and $2\pi$. If excitation to a particular state $j$ is selected, the initial electron energy must be reduced by the excitation energy required for the process assuming, however, that the electron direction remains unchanged.
In the case of impact ionization (only single ionization is taken into account), a secondary electron is liberated which may be able to contribute to the energy transport and which must, therefore, be treated in the same way as the primary electron. For this purpose, not only the energy loss and the direction of the initial electron after impact ionization must be determined but also the energy and direction of the secondaries.

The complete history of a primary electron is simulated as long as its energy has been degraded to a value smaller than the ionization threshold energy of 15.58 *eV* in nitrogen and 11.08 *eV* in propane. The degradation of secondary electrons is treated like that of the primaries if their initial energy is greater than the predefined energy threshold, otherwise it is assumed that they come to rest directly at their source point. This latter assumption is also made in the case of photons emitted after excitation events, apart from the excitation of Rydberg states which are assumed to lead in part to autoionization. The formation of ionization clusters was analyzed after each ionization event, taking into account the detection efficiency of the measuring device and that of energy losses after each ionization or excitation event. More details of the treatment of electron interactions in molecular nitrogen and propane gas, in particular the description of the cross sections used for electron elastic scattering and electron impact ionization or excitation are given in Appendix A and Appendix B with tabulated values in Table 21 and Table 23 (Appendix C).

# 5 Reconstruction of cluster distributions at 100 % detection efficiency through a Bayesian analysis

## 5.1 Basic information

In all experiments, the results achieved $A(\varepsilon)$ are measured with some efficiency $\varepsilon$ < 100 %, and of course $A(\varepsilon$ = 100 %) is needed. In many cases for macroscopic parameters (e.g. current, dose, flux), to reconstruct the $A(\varepsilon$ = 100 %) value – it is enough to divide $A(\varepsilon)$ by $\varepsilon$ (all types of calibrations). In single event measurements, such as frequency measurement of cluster size distributions, this simple trick is not applicable.

One possible way to reconstruct the cluster distribution at 100 % detection efficiency is to apply a Bayesian analysis [35].

*Bernoulli Trials:* Repeated independent trials with only two possible outcomes for each trial and a probability of outcome which remains the same throughout the trials.

If $p$ is the probability of success and $q$ the probability of failure, $p+q=1$

*Newton's Theorem:* If we make $\nu$ Bernoulli trials with probabilities $\varepsilon$ for success and $(1-\varepsilon)$ for failure, the probability of $\mu$ successes and $(\nu-\mu)$ failures is given by (Eq.30).

If $P(\nu)$ is the real probability of ionization cluster-size formation, we can, therefore, expect in a nanodosimetric measurement at detection efficiency $\varepsilon$ an experimental distribution which is given by the Binomial Distribution:

$$\tilde{P}_{\mu}(\varepsilon)=\sum_{\nu=\mu}^{\infty}\binom{\nu}{\mu}\varepsilon^{\mu}(1-\varepsilon)^{\nu-\mu}P(\nu) \qquad \textbf{(35)}$$

Figure 47 shows the results of the application of the Binomial Distribution to a calculated Monte Carlo frequency distribution of ion cluster size spectra for a 0.2 *µg/cm*$^2$ diameter sensitive volume irradiated by 3.8 *MeV* $\alpha$-particles. The Binomial Distribution is compared with the MC with the same detection efficiency $\varepsilon$ = 30 %. It should be noted that both MC simulations contain $10^7$ events and the maximum cluster sizes are different (for $\varepsilon$ = 30 % it is 10, for $\varepsilon$ = 100 % – 21). For the Binomial Distribution case, the maximum cluster size is the same as with MC $\varepsilon$ = 100 %.

Summarizing, in a real experiment with $\varepsilon$ < 100 %, the maximum cluster size will be lower than in the true distribution with $\varepsilon$ = 100 %.

## 5.2 Unfolding procedure 1

Let $m$ be the maximum cluster size which measured at detection efficiency $\varepsilon$, then

$$\widetilde{P}_{\mu}(\varepsilon) = \sum_{\nu=\mu}^{m} \binom{\nu}{\mu} \varepsilon^{\mu} (1-\varepsilon)^{\nu-\mu} P(\nu) + \sum_{\nu=m+1}^{\infty} \binom{\nu}{\mu} \varepsilon^{\mu} (1-\varepsilon)^{\nu-\mu} P(\varepsilon) \qquad (36)$$

if the second sum is negligible compared with the last term of the first sum, we get a system of equations which may be solved starting with $\mu = m$:

$$P(m) = \frac{\widetilde{P}_m(\varepsilon)}{\varepsilon^m} \qquad (37)$$

$$P(m-1) = \frac{\widetilde{P}_{m-1}(\varepsilon) - \binom{m}{m-1} \varepsilon^{m-1} (1-\varepsilon) P(m)}{\varepsilon^{m-1}} \qquad (38)$$

and so on and so forth.

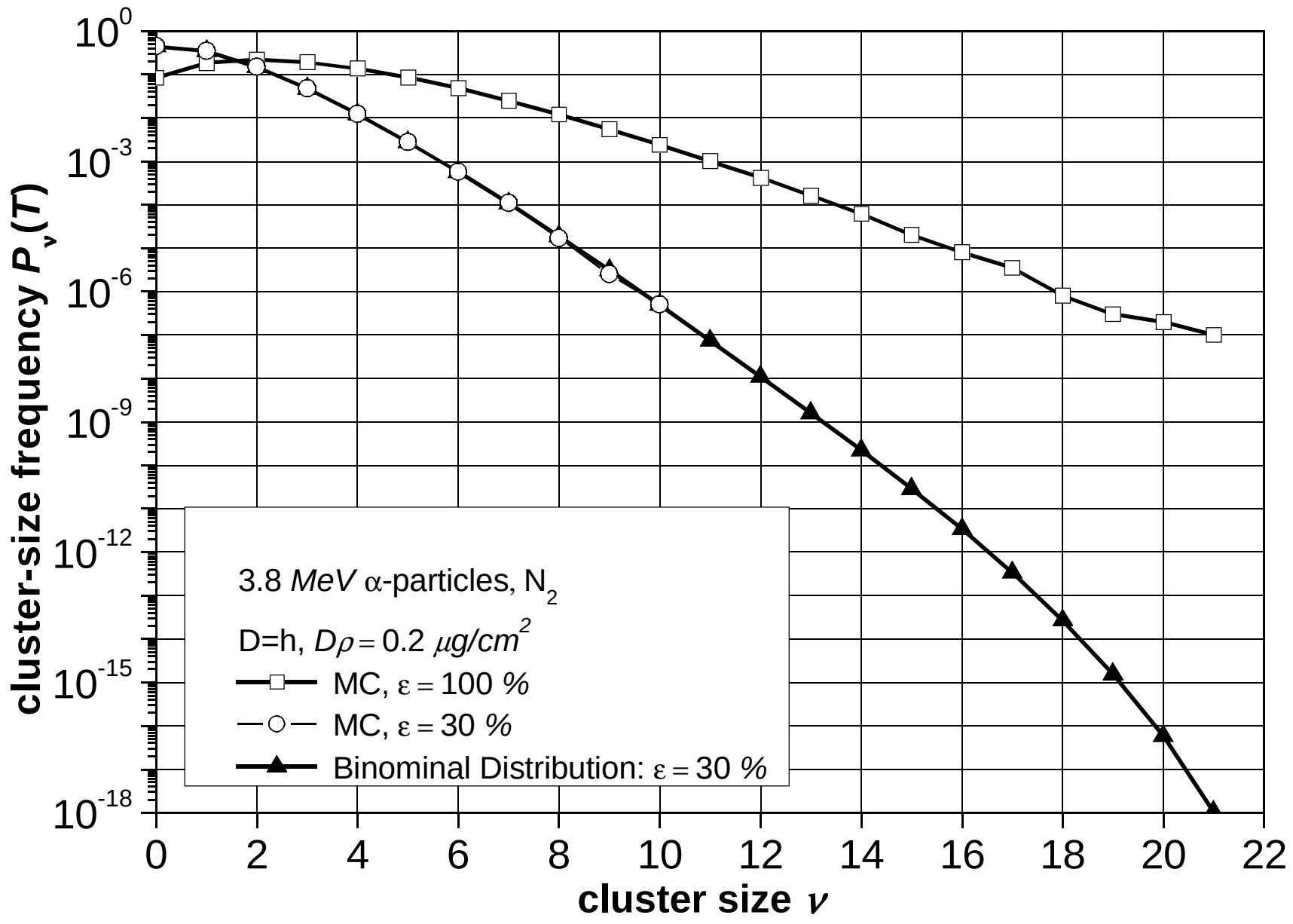

*Figure 47: Frequency distribution of ion cluster-size spectra for a 0.2 μg/cm² diameter sensitive volume irradiated by 3.8 MeV $\alpha$-particles. (□) – Monte Carlo calculations for ion detector efficiencies $\varepsilon$ = 100 %, (○) – Monte Carlo calculations with $\varepsilon$ = 30 %, (▲) – Binomial Distribution using (Eq.35) with $\varepsilon$ = 30 % of Monte Carlo calculations with $\varepsilon$ = 100 %. Molecular nitrogen.*

Unfortunately, however, this procedure is not applicable since the necessary condition is not fulfilled in general.

For example, in our experiment:

- the measured highest cluster size is lower than in the true distribution,
- the solution of (Eq.36) is very sensitive to the measured highest cluster size which has an uncertainty of 10-100 % due to low statistics.

Nevertheless, this procedure was applied in [71] as a test.

## 5.3 Unfolding procedure 2 – Bayesian unfolding algorithm

If we have a Binomial Distribution, the left part of (Eq.36), with a maximum cluster size $m$ which has been measured at detection efficiency $\varepsilon$.

Let us assume:

*Event*: $E = \{\text{a cluster size } \mu \text{ is detected at efficiency}\,;\ \mu = 0,1,2,...\}$

*Hypothesis*: $H = \{\text{a cluster size } \nu \text{ is produced}:\ \nu = 0,1,2,...\}$

Using the Bayesian theorem [35], the probability of detecting cluster size $\mu$ with efficiency $\varepsilon$ if a cluster size $\nu$ is produced:

$$P(E \cap H|\varepsilon) = P(H = \nu|E = \mu,\varepsilon) \times P(E = \mu,\varepsilon) = P(E = \mu|H = \nu,\varepsilon) \times P(H = \nu|\varepsilon) \qquad (39)$$

Result:

$$P(\nu|\mu,\varepsilon) = \frac{P(\mu|\nu,\varepsilon) \times P(\nu,\varepsilon)}{P(\mu|\varepsilon)} \qquad (40)$$

With:

$$P(\mu|\varepsilon) = \sum_{\nu'} P(\mu|\nu',\varepsilon) \times P(\nu'|\varepsilon) \qquad (41)$$

$$P(\mu|\nu,\varepsilon) = \binom{\nu}{\mu} \varepsilon^{\mu} (1-\varepsilon)^{\nu-\mu} \qquad (42)$$

Based on (Eq.39-42) the block-scheme diagram of the Bayesian unfolding algorithm is shown in Figure 48. The iteration algorithm enables control of the final result with $X^2 \leq \text{Limit}$.

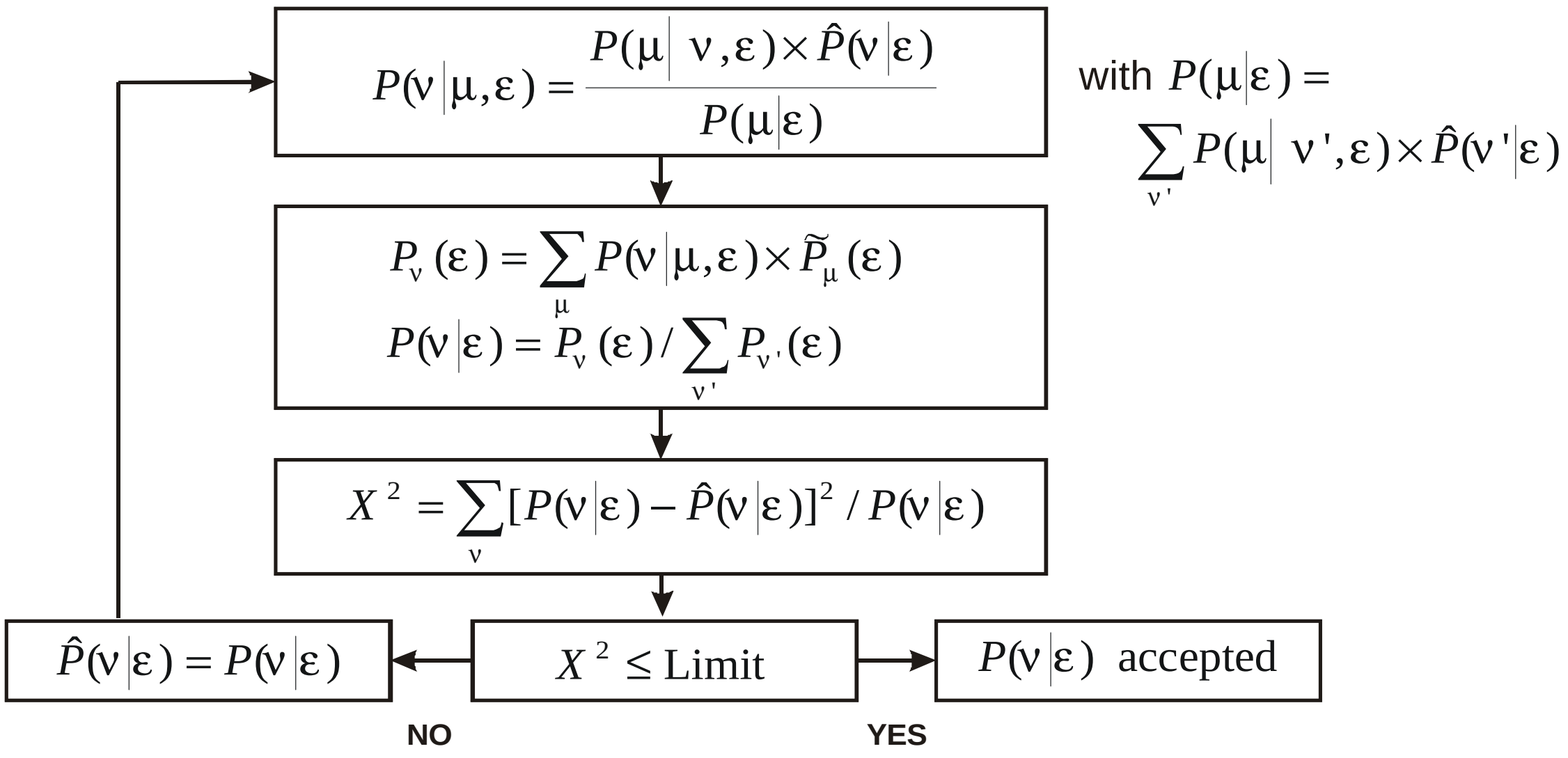

*Figure 48: The block-scheme diagram of the Bayesian unfolding algorithm.*

In the present work, the Bayesian unfolding algorithm developed by [72] was used.

Some examples of the application of this procedure are presented in Figures 49-50. Monte Carlo (MC) simulations of the frequency distribution of the ion cluster-size spectra for 0.4 and 1.6 $\mu g/cm^2$ (molecular nitrogen) diameter sensitive volumes irradiated by 3.8 *MeV* $\alpha$-particles are presented. Simulations were made in the case of ion detection efficiencies $\varepsilon$ = 30 % and $\varepsilon$ = 100 % (for comparison with unfolding) with statistics of $10^6$ events. For 0.4 $\mu g/cm^2$ the unfolding procedure works well. The consequent iterations show good agreement with MC $\varepsilon$ = 100 %. For 1.6 $\mu g/cm^2$ the unfolding procedure does not work. The consequent iterations show that the final iteration is oscillatory and completely differs from MC $\varepsilon$ = 100 %.

If we look at the frequency distributions for 0.4 and 1.6 $\mu g/cm^2$, the main difference is that the maximal cluster size in 0.4 $\mu g/cm^2$ MC $\varepsilon$ = 30 % is two times larger than the mean cluster size (or maximum in the distribution) for MC $\varepsilon$ = 100 %. So, the distribution for MC $\varepsilon$ = 30 % is represented for unfolding to 100 %.

For 1.6 $\mu g/cm^2$, *the* maximal cluster size in MC $\varepsilon$ = 30 % is comparable with the mean cluster size (or maximum in the distribution) for MC $\varepsilon$ = 100 %. As a result, there is not enough information for a proper unfolding to 100 %.

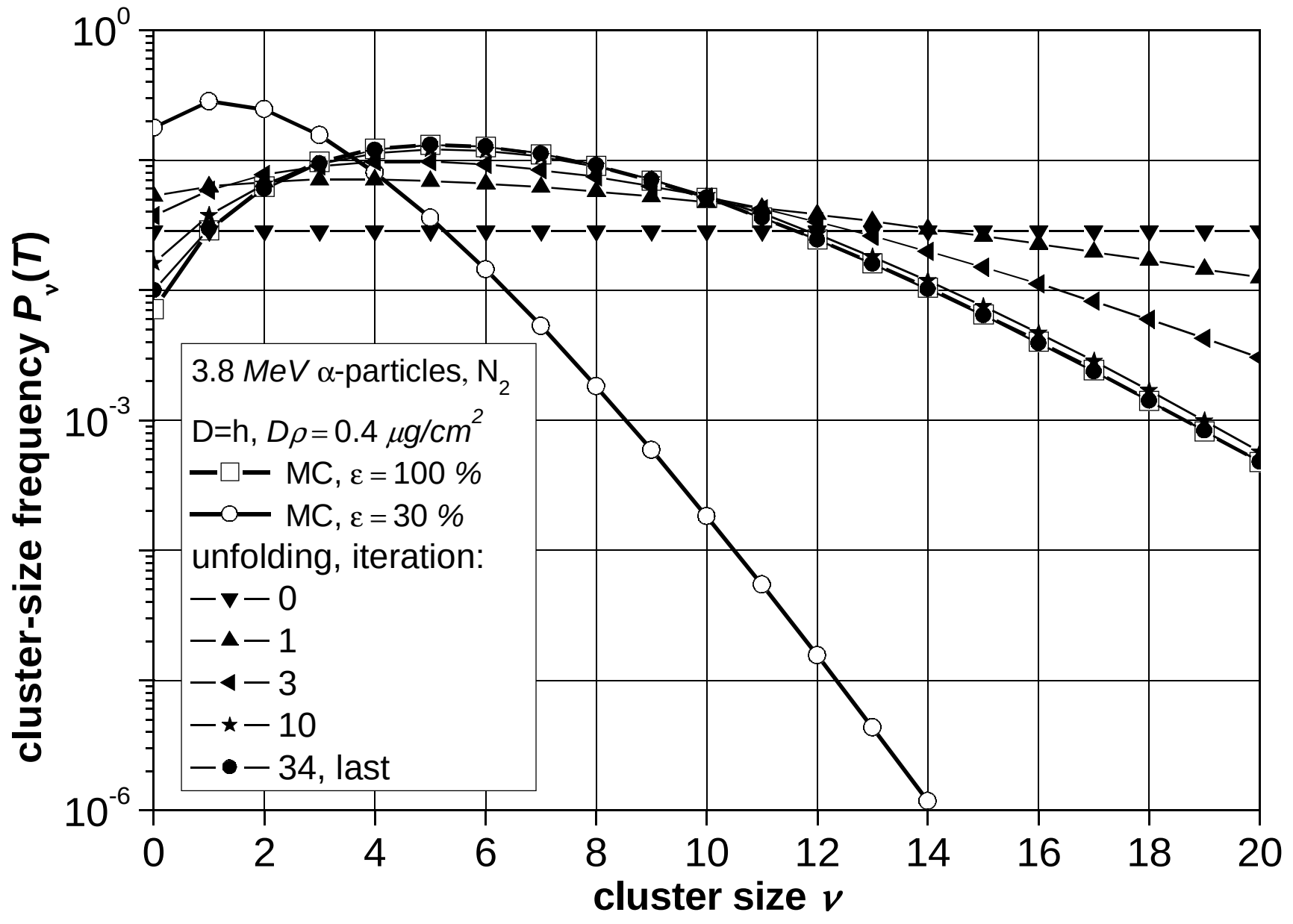

*Figure 49: Frequency distribution of ion cluster-size spectra for a 0.4 $\mu g/cm^2$ diameter sensitive volume irradiated by 3.8 MeV $\alpha$-particles. Molecular nitrogen. (□) – Monte Carlo calculations for ion detector efficiencies $\varepsilon$ = 100 %; (○) – Monte Carlo calculations with $\varepsilon$ = 30 %. Applying the Bayesian unfolding procedure with number of iteration events: (▼) – 0, (▲) – 1, (◀) – 3, (★) – 10, (●) – 34, last.*

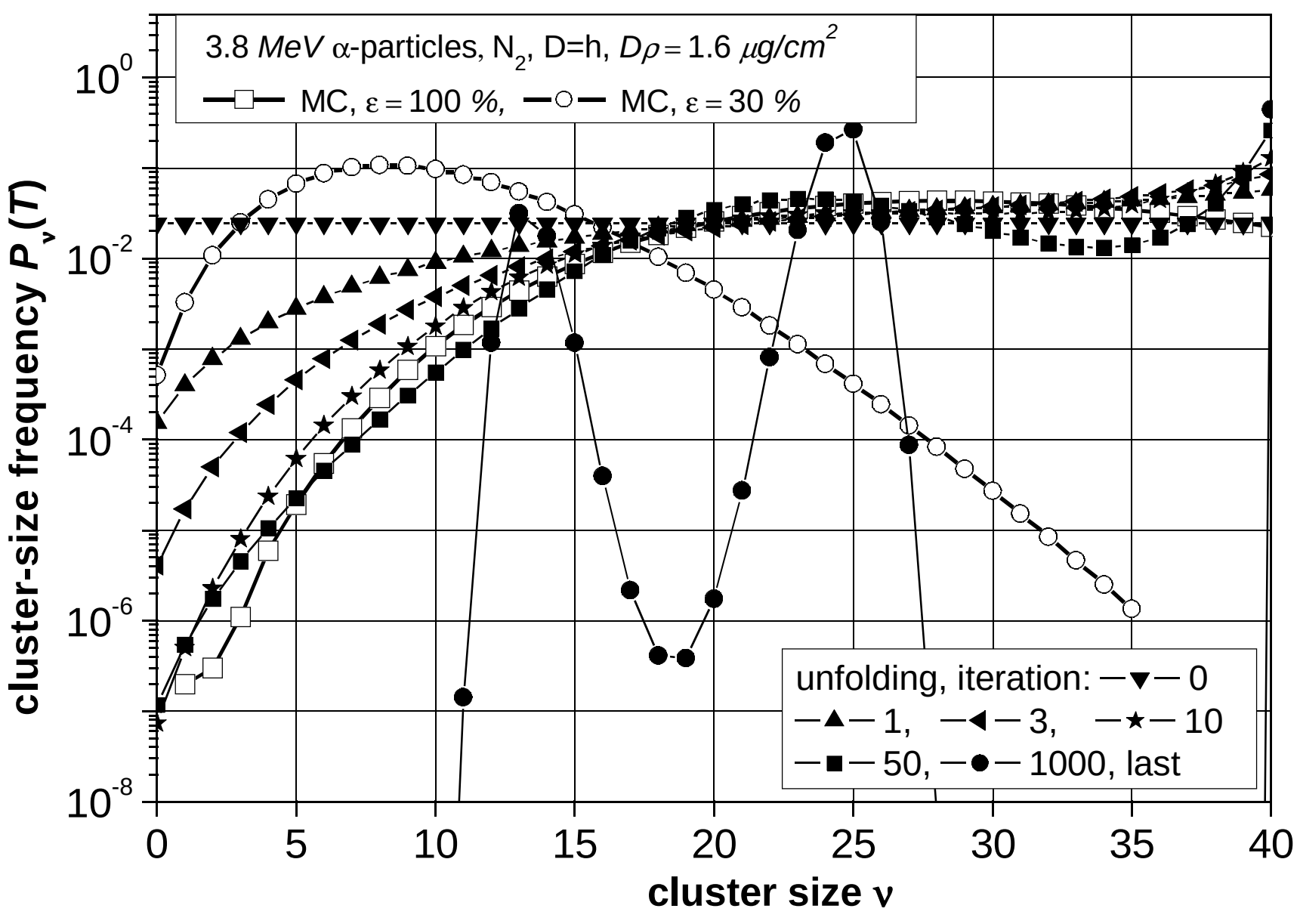

*Figure 50: Frequency distribution of ion cluster-size spectra for a 1.6 μg/cm² diameter sensitive volume irradiated by 3.8 MeV $\alpha$-particles. Molecular nitrogen. (□) – Monte Carlo calculations for ion detector efficiencies $\varepsilon$ = 100 %; (○) – Monte Carlo calculations with $\varepsilon$ = 30 %. Applying the Bayesian unfolding procedure with number of iteration events: (▼) – 0, (▲) – 1, (◀) – 3, (★) – 10, (■) – 50, (●) – 1000, last.*

## 5.4 Conclusions

- If there is a constant detection efficiency $\varepsilon$, a Bayesian unfolding procedure is easily applicable and is rather fast on a desktop computer (Intel P4-2GHz), with a CPU time of the order of 1 *s*,
- Depending on the shape of the distribution $P_\nu(\varepsilon)$, the procedure works very well when handled with care,
- One prerequisite for its application is the measurement or calculation of cluster-size distributions for values of $\mu$ as large as possible, at least at low efficiencies $\varepsilon$.

The frequencies $P_\nu(\varepsilon)$ should be measured or calculated, at least, down to $1{\cdot}10^{-6}$. For $\alpha$-particles this impossible (a very long measuring time, more than 6 months, being necessary). For electrons it works well.

All measured frequency distributions of ion cluster-size spectra presented in this work were de-convoluted to 100 % using the Bayesian unfolding procedure. The de-convolution results for electrons and $\alpha$-particles are presented in Figures 51-56, 58-64, and 66-68 with numerical values in Tables 4-19.

More information on this topic may be found in references [35, 73-76].

# 6 Experimental results

The experimental results, presented in this chapter, were obtained at the Andrzej Sołtan Institute for Nuclear Studies during the years 2001-2010. The experiments were made at the Jet Counter facility which was originally devised by dr S.Pszona's group [48, 54]. The cluster-size distributions created by low energy electrons in molecular nitrogen, as the first measurements of this quantity, were published in [77, 78 and 79]. During this period, the results for $\alpha$-particles in propane gas were also published in [71, 80 and 81]. The results for $\alpha$-particles in molecular nitrogen were not published, as similar measurements were performed previously by dr S.Pszona's group and published in [64].

## 6.1 Cluster-size distributions due to low-energy electrons in molecular nitrogen

The experiments were carried out for mono-energetic electrons at energies of 100 *eV*, 200 *eV*, 300 *eV*, 500 *eV*, 1 *keV* and 2 *keV*, which penetrate through a nitrogen cylinder, 0.34 $\mu g/cm^2$ in height and 0.34 $\mu g/cm^2$ in diameter, corresponding to a water cylinder of 0.23 $\mu g/cm^2$ x 0.23 $\mu g/cm^2$ (according to (Eq.24)). The efficiency of single ion counting by the Jet Counter was estimated to be 30 *%*.

Figures 51-56 show experimental frequency distributions of ion cluster-size spectra due to electrons at 100 *eV*, 200 *eV*, 300 *eV*, 500 *eV*, 1 *keV* and 2 *keV*, measured with the Jet Counter filled with molecular nitrogen. These data are compared with the results of a Monte Carlo simulation of the experimental arrangement, assuming a Poisson-like distribution (see Eq.27).) of the number of electrons injected into the Jet Counter’s interaction chamber by the electron gun, with a mean number of primary electrons $N_{mean}$ of 1.75 at 100 *eV*, 1.19 at 200 *eV*, 1.02 at 300 *eV*, 0.93 at 500 *eV*, 0.93 at 1 *keV* and 1.06 at 2 *keV*. The agreement between experimental and calculated frequency distributions is striking.

In the experiment, the $N_{mean}$ were set to be close to 1. In reality, experiment shows that it is almost impossible to control the real number of $N_{mean}$ that interact with the target as the low energy electrons are not well focused and are deflected by external (earth and devices) magnetic fields. So, the $N_{mean}$ used in the Monte Carlo simulations were calculated using dependencies from the properties of a compound Poisson process as described by De Nardo *et al.* [34]; if a Poisson-like distribution is assumed for the frequency distribution of the injected primary electrons:

$$M_1^{(equiv)}(T) = \varepsilon \times M_1^{(meas)}(T) \tag{43}$$

$$M_2^{(equiv)}(T) - M_1^{(equiv)}(T) = \varepsilon^2 \times [M_2^{(meas)}(T) - M_1^{(meas)}(T)] \tag{44}$$

$$M_1(T; N_{mean}) = N_{mean} \times M_1(T; \text{single}) \tag{45}$$

$$M_2(T; N_{mean}) - M_1^2(T; N_{mean}) = N_{mean} \times M_2(T; \text{single}) \tag{46}$$

So, to resolve the $N_{mean}$ Monte Carlo calculations were performed for nitrogen, in the case of a single (just one) electron and a mass per unit area of the target diameter of 0.34 *μg/cm²* with a detection efficiency of 100 *%*. The results of these Monte Carlo simulations are presented in Figures 51-56. As can be seen, the Monte Carlo simulations for 100 *%* efficiencies in the case of a single electron and electrons with $N_{mean}$ is different. In the case of a single electron, the MC ionization cluster-size distribution spectra have the maximum cluster size.

The results of measured frequency distributions of ion cluster-size spectra de-convolved to 100 *%* efficiency for all energies are also included in Figures 51-56. The agreement with the Monte Carlo simulations of the experiment with detection efficiency 100 *%* is very good, only the step near 0 to 1 cluster size shows some deviations.

Based on these distributions, the following parameters, which directly describe the radiation quality on the nanometric scale, can be derived:

- the first moment of the frequency distribution $M_1$ (Eq.18), i.e., the mean number of ions (ionizations) in a cluster for a given geometry of irradiation as well as for a given SNS;
- the cumulative frequency, $F_2$ (Eq.19) – the frequency required to create cluster-size equal to 2 or higher,
- cluster-size frequency $P_1$ - the frequency required to create cluster-size equal to 1.

The calculated $M_1$, $F_2$ and $P_1$ parameters show good agreement between the experimental results and the Monte Carlo simulations of the experiment with detection efficiency 30 *%*; de-convoluted experimental results and Monte Carlo simulations of the experiment with detection efficiency 100 *%*.

The numerical values of all ionization cluster-size distribution spectra presented in Figures 51-56 with calculated $M_1$, $F_2$ and $M_2$ (second moment – useful in (Eq.18) and (Eq.19)) are tabulated in Tables 4-9 (Appendix C).

The experimental results presented here are the first of their kind for low-energy electrons with energies ranging from 100 *eV* to 2000 *eV*.

The experiments using "single" low-energy electrons (100 *eV* – 2 *keV*) interacting with a nitrogen jet of nanometre size comparable to that of a short DNA segment show discrete frequency distributions of cluster size with extended cluster sizes. These cluster-size distributions were determined for the first time for electrons.

It has been shown (based on experimental data) that low-energy electrons interacting with a DNA-like segment are able to create single and clustered damage (assuming that SSB formation is proportional to the frequency of a single ionization while DSB formation needs at least two ionizations within a short DNA segment). In nanoelectronics they can generate charge clusters.

In view of this, the nanodosimetric quantities $M_1$, $F_2$ and $P_1$ can be used as new tools for the qualitative interpretation of observed biological endpoints due to monoenergetic electrons arising from the photoelectric effect of low-energy characteristic X rays, due to low-energy Auger electrons, and due to delta electrons of charged particle tracks.

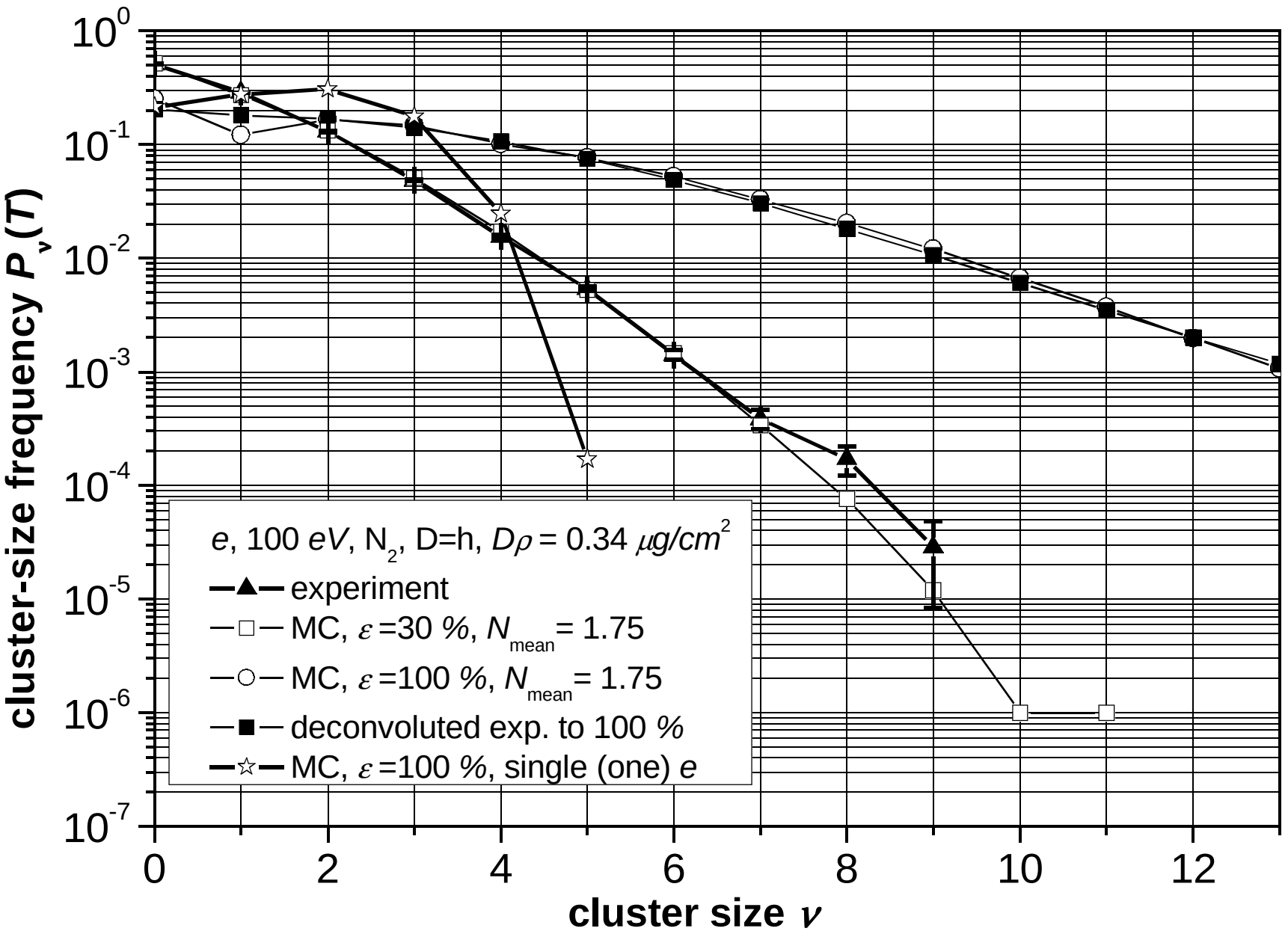

*Figure 51: Frequency distribution of ion cluster-size spectra due to 100 eV electrons in molecular nitrogen in the case of a target volume with mass per unit area of the diameter of 0.34 μg/cm²: (▲) – measurements; (□) – results of a Monte Carlo simulation of the experimental data performed with a single-ion detection efficiency of ε = 30 % and a mean number of primary electrons $N_{mean}$ = 1.75; (○) – results of a Monte Carlo simulation of the experimental data performed with ε = 100 % and $N_{mean}$ = 1.75; (■) – deconvoluted experimental results to ε = 100 % with the assumption of experimental single-ion detection efficiency ε = 30 %; (☆) – results of a Monte Carlo simulation performed for a single (just one) electron with a single-ion detection efficiency of 100 %.*

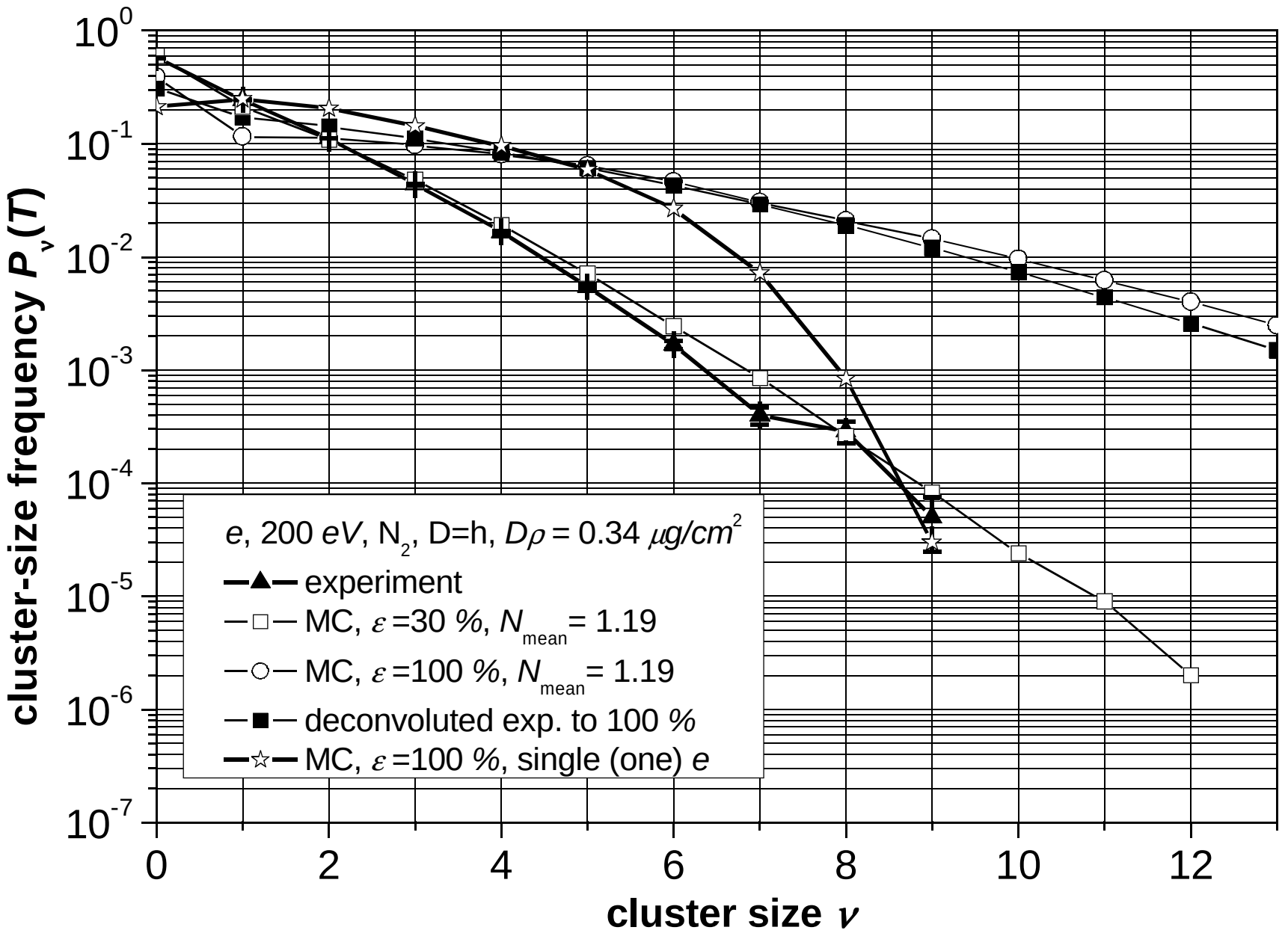

*Figure 52: The same as Figure 51 for 200 eV electrons with $N_{mean}$ = 1.19.*

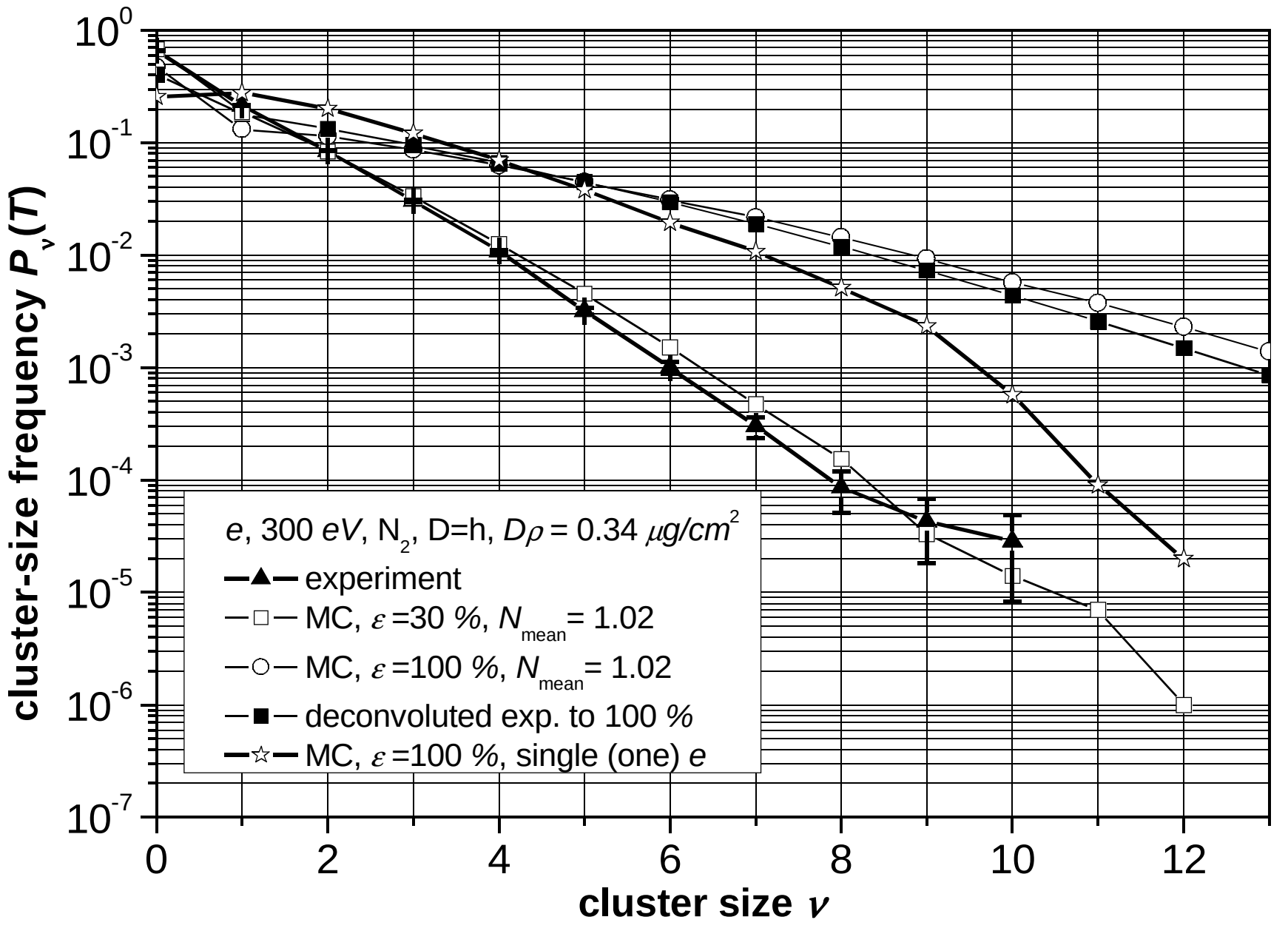

*Figure 53: The same as Figure 51 for 300 eV electrons with $N_{mean}$ = 1.02.*

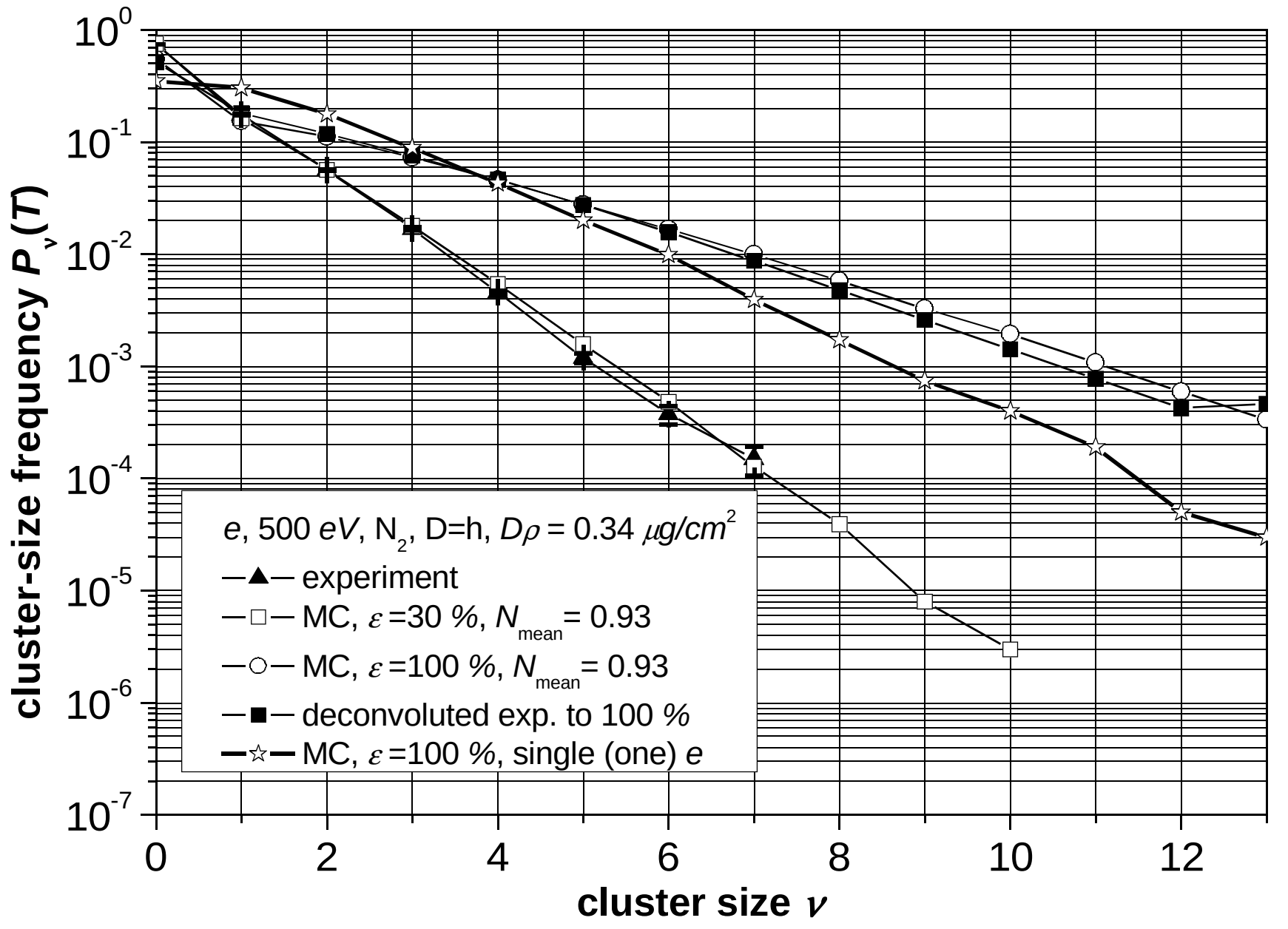

*Figure 54: The same as Figure 51 for 500 eV electrons with $N_{mean}$ = 0.93.*

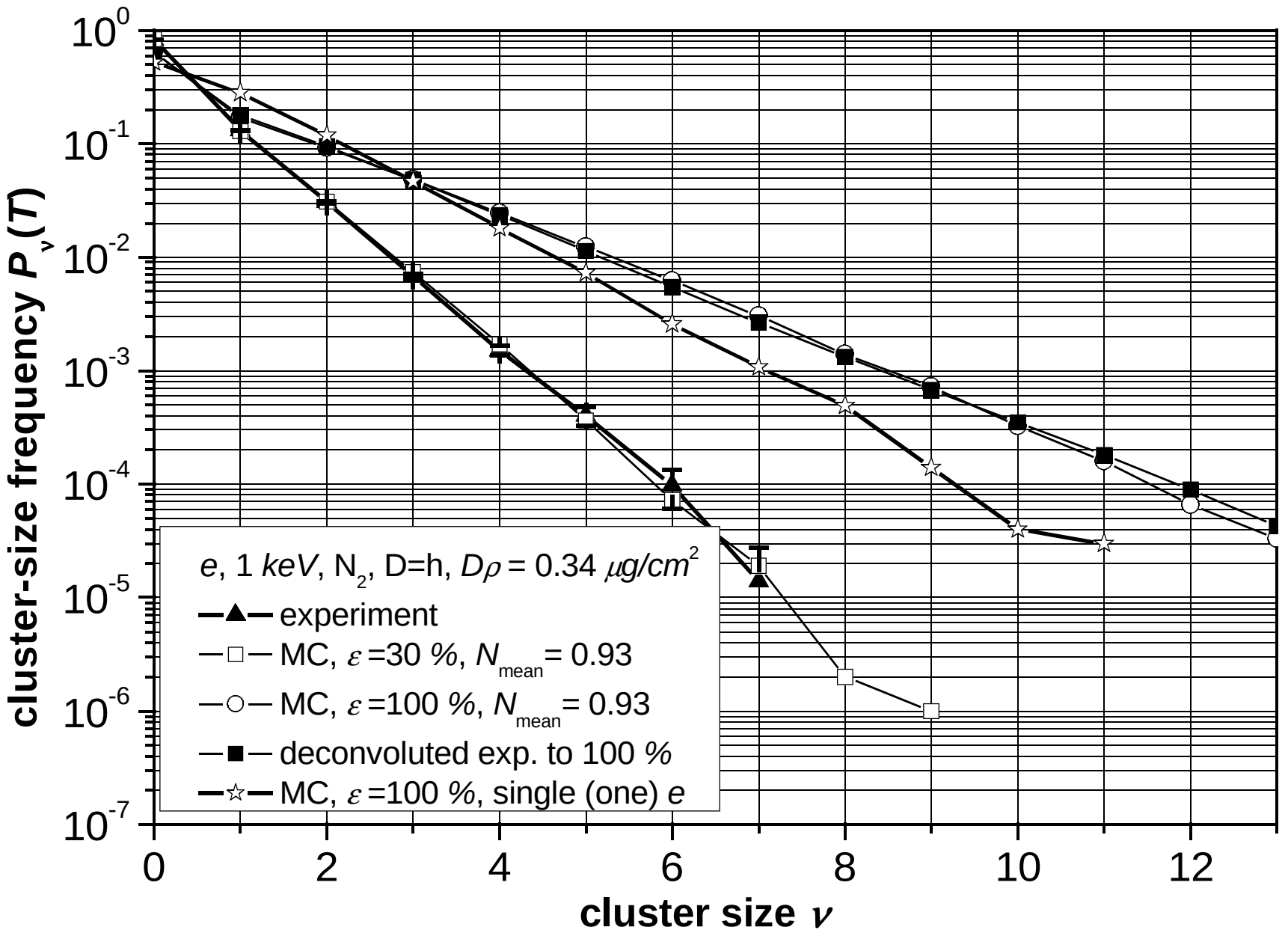

*Figure 55: The same as Figure 51 for 1 keV electrons with $N_{mean}$ = 0.93.*

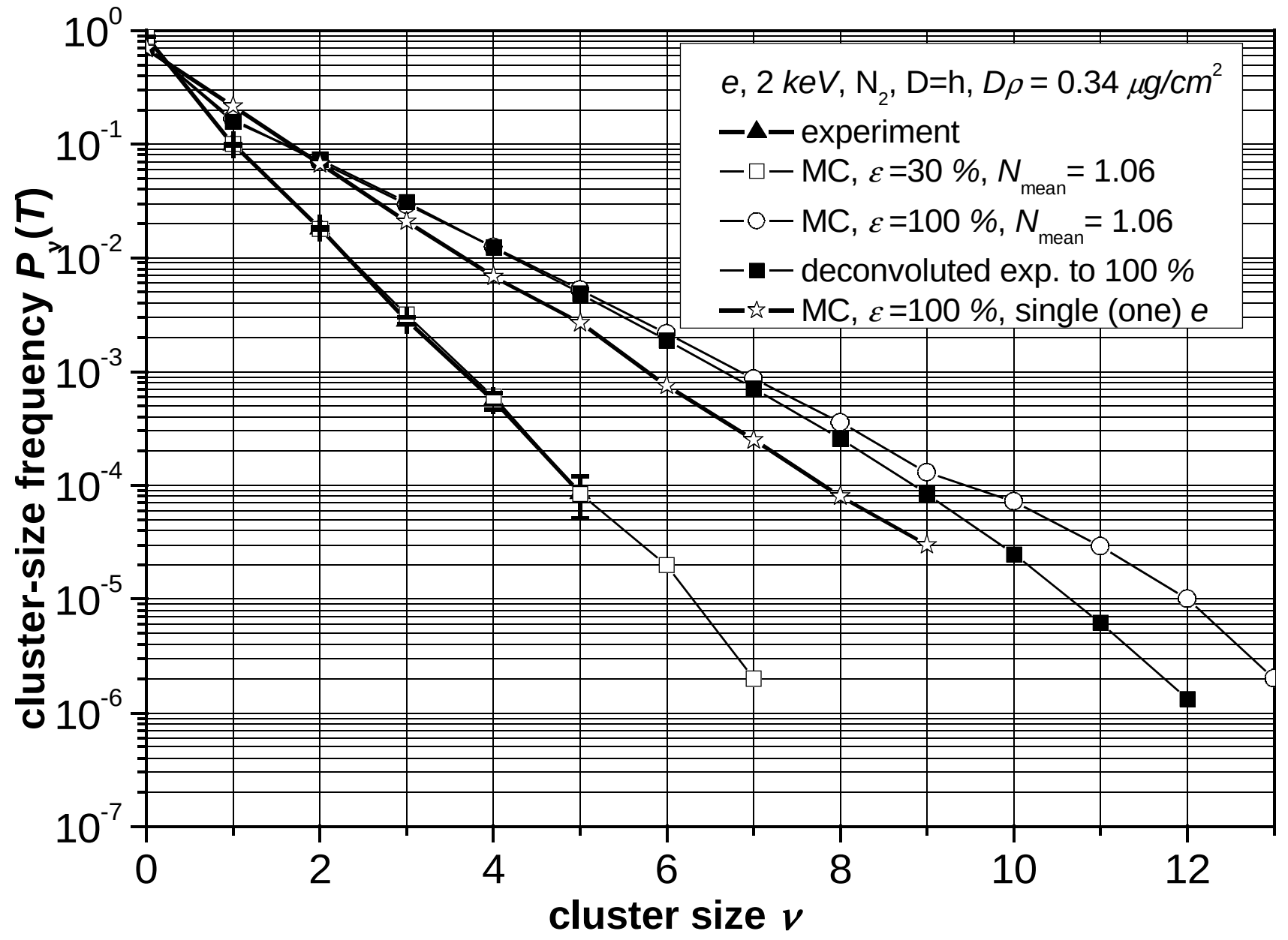

*Figure 56: The same as Figure 51 for 2 keV electrons with $N_{mean}$ = 1.06.*

## 6.2 Cluster-size distributions due to α-particles in molecular nitrogen

Figure 57 shows the calculated frequency distribution of ion cluster-size spectra $P(T_\alpha,\nu)$ just to give an impression of the formation of ionization clusters by 3.8 *MeV* $\alpha$-particles in "nanometric" sensitive volumes of nitrogen. $P(T_\alpha,\nu)$ describes the probability that exactly a cluster size $\nu$ (number of ions) is produced by a single particle track in cylindrical volumes of diameter $D$ and height $h$, at a mass per unit area of 0.092 $\mu g/cm^2$, 0.130 $\mu g/cm^2$, 0.187 $\mu g/cm^2$, 0.291 $\mu g/cm^2$, 0.354 $\mu g/cm^2$, 0.387 $\mu g/cm^2$ and 0.538 $\mu g/cm^2$ of molecular nitrogen. The ion detection efficiency is assumed to be 100 *%*.

At very small values of the mass per unit area, $P(T_\alpha,\nu)$ decreases strongly and has a maximum at $\nu = 0$. With increasing mass per unit area, the probability for $\nu = 0$ decreases and the maximum value of the cluster size distribution is shifted to higher values of $\nu$.

Figures 58-64 show measured frequency distributions of ion cluster-size spectra in the case of a mass per unit area of the simulated nanometre size (cylinder *h* = *d*) of 0.092 $\mu g/cm^2$ to 0.538 $\mu g/cm^2$ of molecular nitrogen. The measurements are compared with the corresponding results of the Monte Carlo simulation for a detection efficiency of 40 *%* for 0.092 $\mu g/cm^2$, 0.130 $\mu g/cm^2$, 0.187 $\mu g/cm^2$, 0.291 $\mu g/cm^2$, 0.354 $\mu g/cm^2$; 30 *%* for 0.387 $\mu g/cm^2$; and 25 *%* for 0.538 $\mu g/cm^2$ (it should be noted that the numerical results for the 100 *%* efficiency are those of Figure 57). As can be seen from these figures, the agreement between measured and calculated distributions is very satisfactory. It should be mentioned here that no normalization procedure was applied to the experimental data.

The detection efficiency in comparing with the Monte Carlo result was chosen to give the best fit. Nevertheless, the reasons for the discrepancy in the detection efficiency and the decreasing detection efficiencies for densities higher than 0.354 $\mu g/cm^2$, might be the increasing loss of ions by molecular processes within the interaction chamber, such as recombination and other charge-changing effects, with increasing mass per area of the sensitive target volume, or the experimental determination of the latter quantity, which is affected by an uncertainty of the order of 10 *%*. The discrepancy may also have been caused by our limited knowledge of the cross sections, in particular, that for ionization by $\alpha$-particles (see Ref. [64, 82]).

Using (Eq.24), 0.092 $\mu g/cm^2$ of molecular nitrogen corresponds to a water cylinder (*h* = *d*) of 0.065 $\mu g/cm^2$, 0.130 $\mu g/cm^2$ to 0.091 $\mu g/cm^2$, 0.187 $\mu g/cm^2$ to 0.132 $\mu g/cm^2$, 0.291 $\mu g/cm^2$ to 0.205 $\mu g/cm^2$, 0.354 $\mu g/cm^2$ to 0.249 $\mu g/cm^2$, 0.387 $\mu g/cm^2$ to 0.272 $\mu g/cm^2$, 0.538 $\mu g/cm^2$ to 0.379 $\mu g/cm^2$ respectively.

The next aspect of cluster formation by $\alpha$-particles which was investigated, was the shape of the cluster probability $P(T_\alpha,\nu)$. For this purpose, the first moment $M_1(T_\alpha)$ of the experimental distribution $P(T_\alpha,\nu)$ was calculated and used as the mean value of a Poisson-like distribution. The results are presented in Figure 65 for simulated nanometre volumes of 0.092 $\mu g/cm^2$ and 0.354 $\mu g/cm^2$ in comparison with the experimental data and those of the Monte Carlo simulation. That ionization cluster probabilities produced by heavy charged particles are governed by Poisson's law is confirmed experimentally only for the smallest sensitive volume, in the case of 3.8 *MeV* $\alpha$-particles in nitrogen. If the target volume increases, the deviation of the measured or calculated cluster-size probabilities from those of a Poisson-like distribution also increases, at least in general. For greater values of the cluster size $\nu$, the experimental cluster probabilities and those of the Monte Carlo simulation are always greater than the probabilities calculated by applying Poisson's law. This fact is due to the contribution of secondary electrons to the formation of ion cluster sizes. For more details see Ref. [64].

Also, the deconvolution procedure presented in chapter 5 was tested on the experimental measurements. The results are presented in Figures 58-64. For the smallest simulated densities with 40 % detection efficiency the agreement is rather good. Only for the highest simulated densities with decreasing detection efficiency (not well defined) is the agreement not so good but still satisfactory.

The numerical values of all frequency distributions of ion cluster-size spectra presented in Figures 58-64 with calculated $M_1$, $F_2$ (radiation descriptors on the nanometric scale) and $M_2$ (second moment – useful in (Eq.18) and (Eq.19)) values are listed in Tables 10-17.

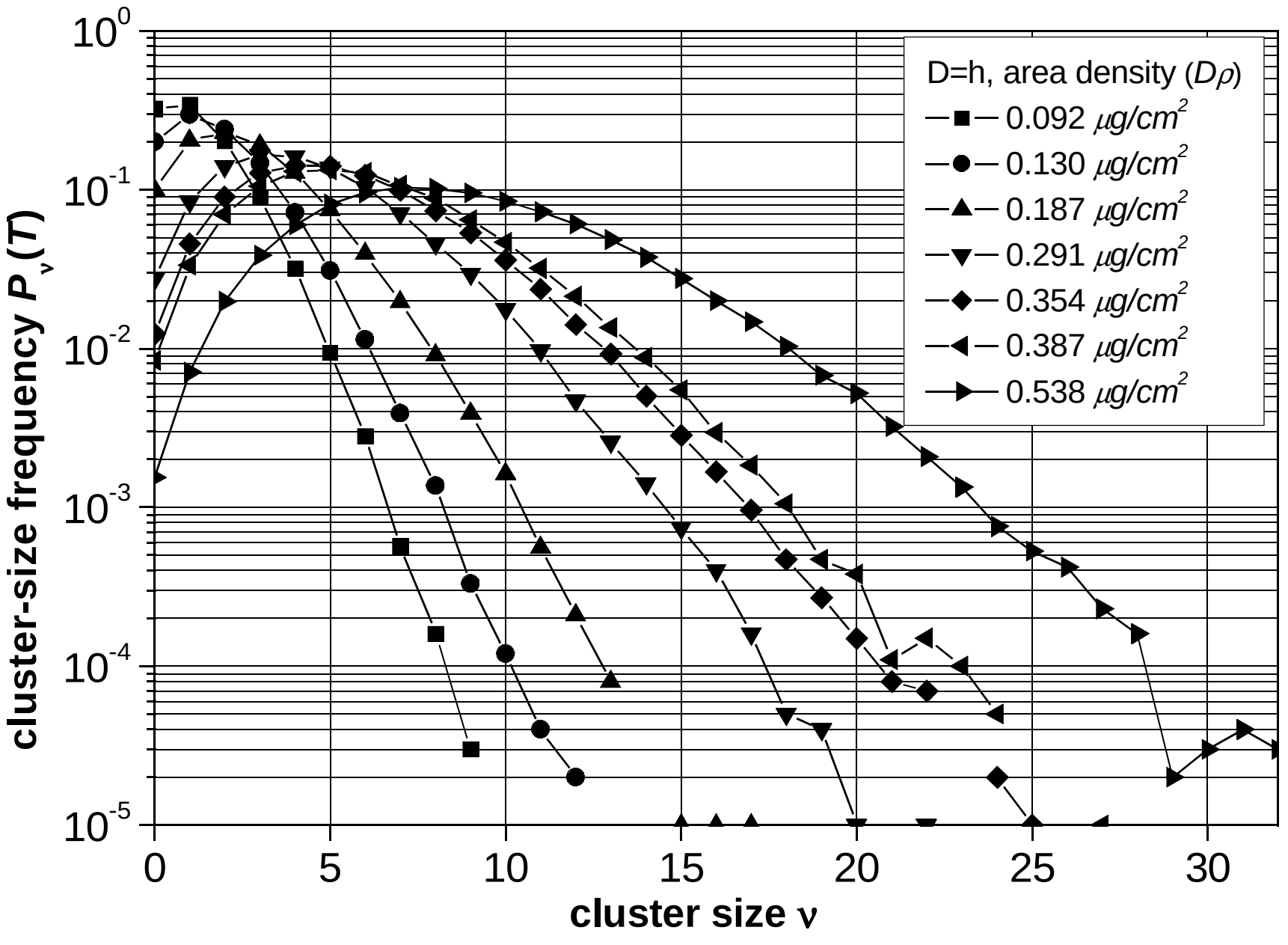

*Figure 57: Calculated frequency distribution of ion cluster-size spectra $P_\nu(T)$ with respect to ionization produced by 3.8 MeV $\alpha$-particles in molecular nitrogen upon diametrical penetration through cylinders between 0.092 μg/cm² and 0.538 μg/cm² in diameter and height. The ion detection efficiency is assumed to be $\varepsilon$ = 100 %.*

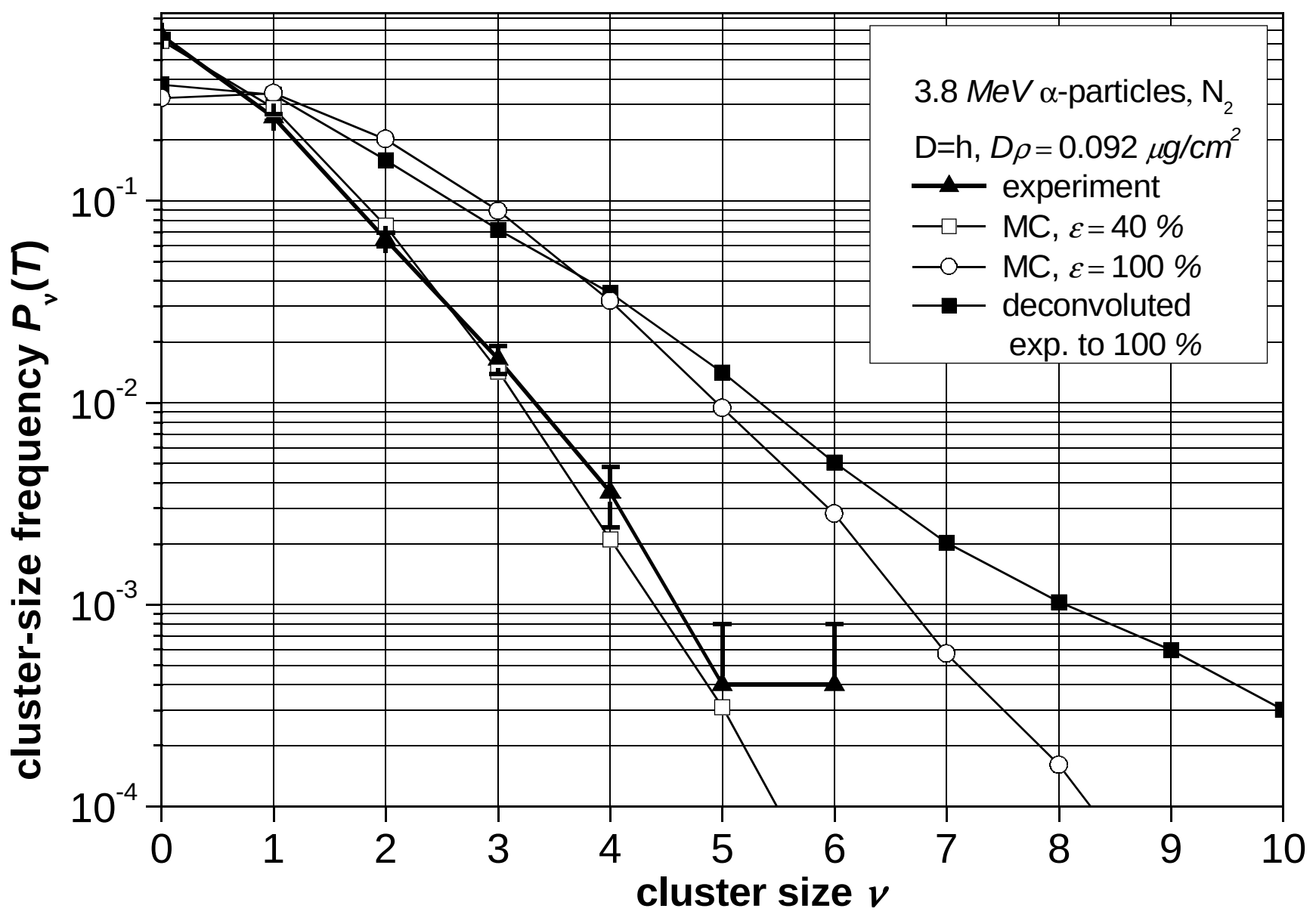

*Figure 58: Frequency distribution of ion cluster-size spectra for 0.092 μg/cm² diameters of sensitive volume irradiated by 3.8 MeV $\alpha$-particles. (▲) – experimental spectra, (□) and (○) – Monte Carlo calculations for different ion detection efficiencies $\varepsilon$ = 40 % and 100 % respectively, (■) – deconvoluted experimental results to $\varepsilon$ = 100 % with assumption of experimental single-ion detection efficiency $\varepsilon$ = 40 %. Molecular nitrogen.*

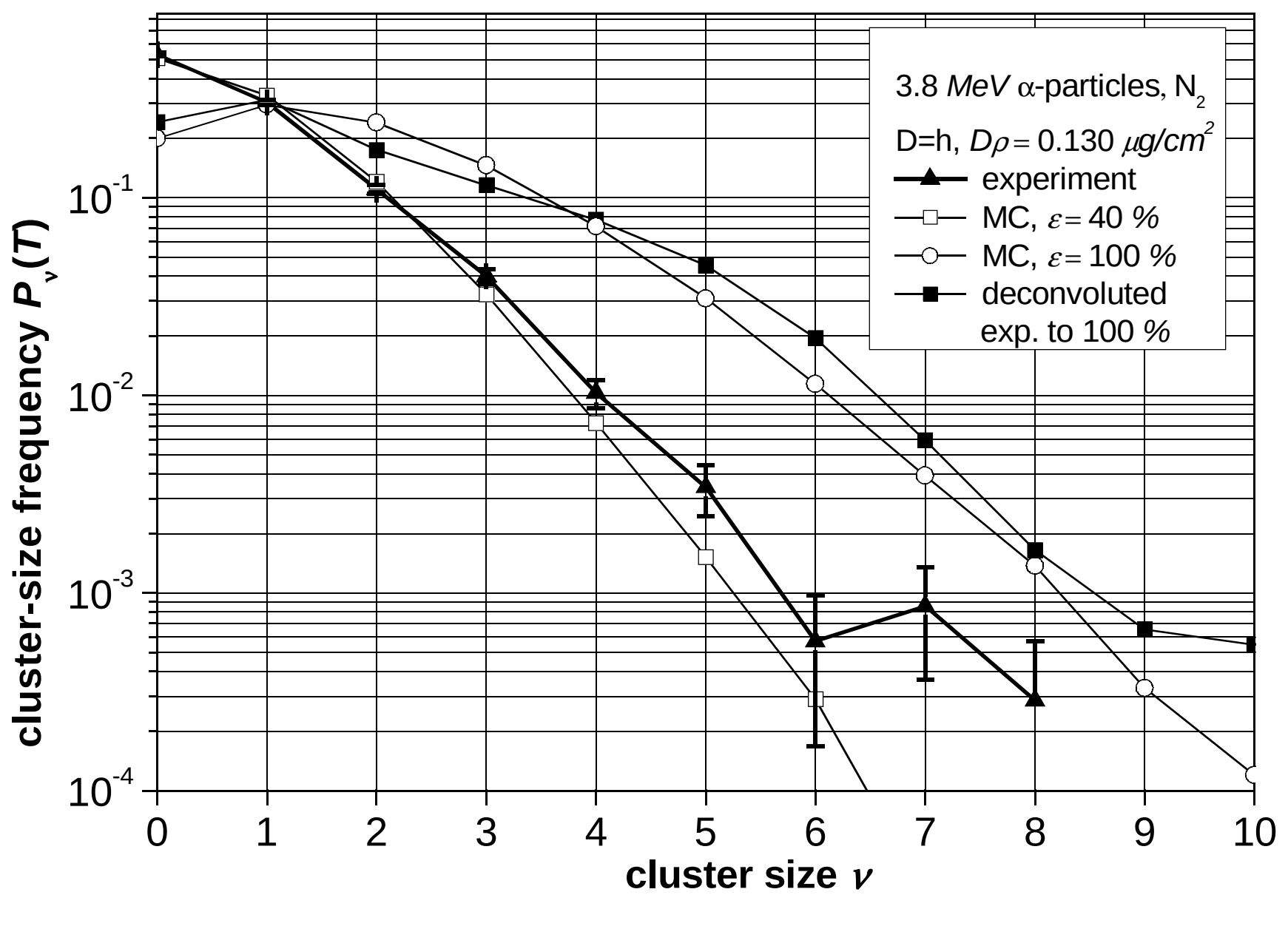

*Figure 59: The same as in Figure 58 for 0.130 μg/cm².*

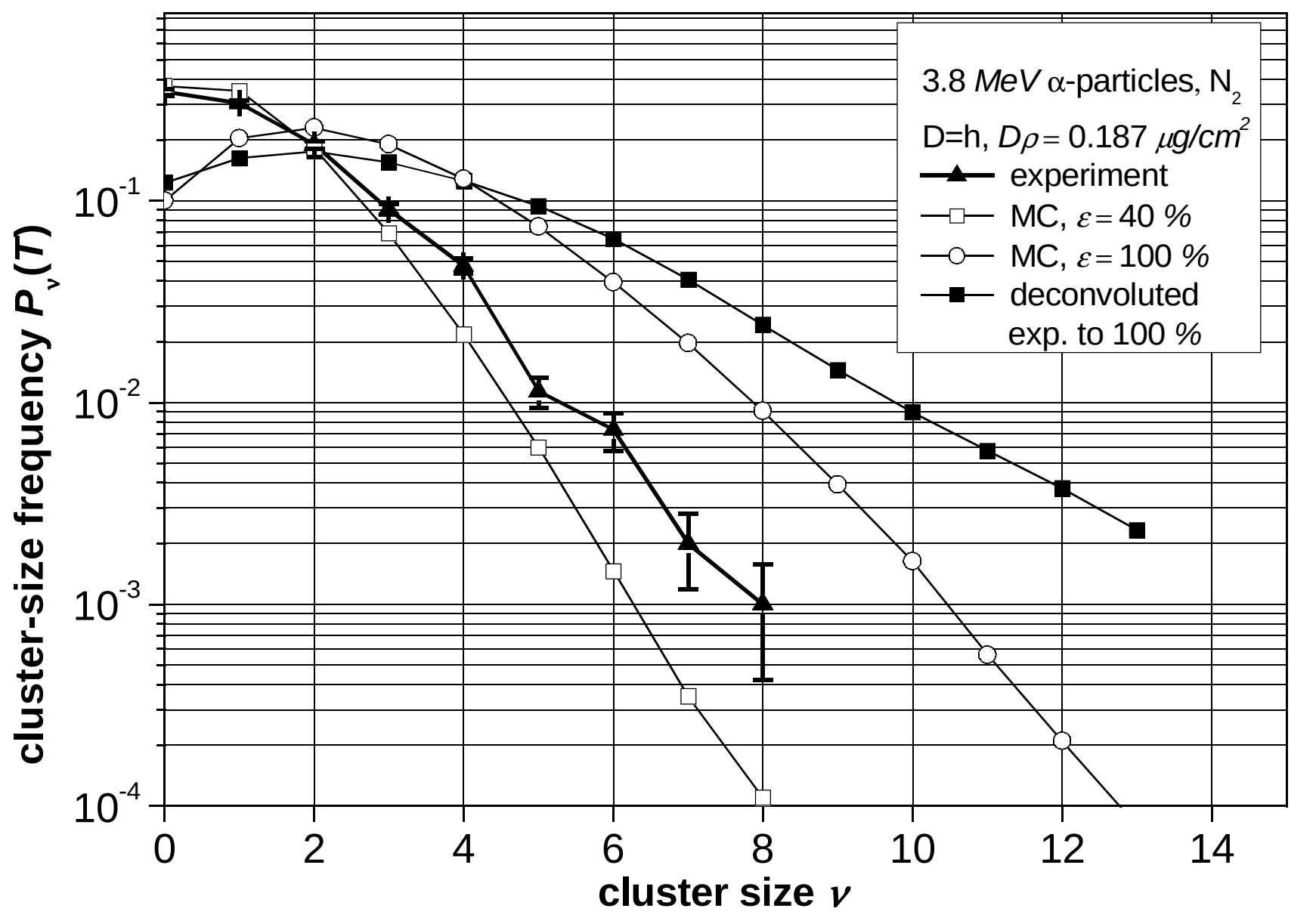

*Figure 60: The same as in Figure 58 for 0.187 μg/cm².*

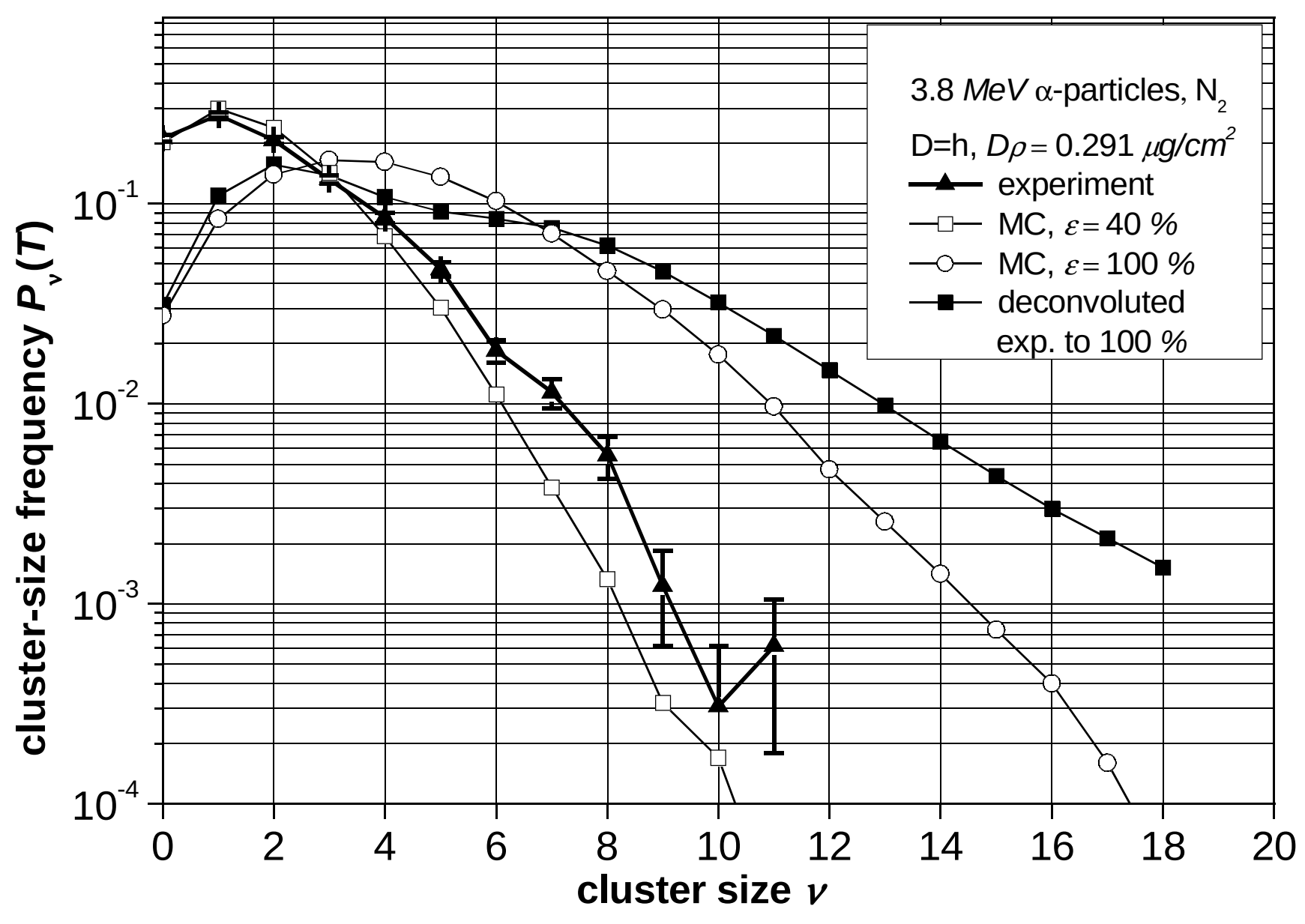

*Figure 61: The same as in Figure 58 for 0.291 μg/cm².*

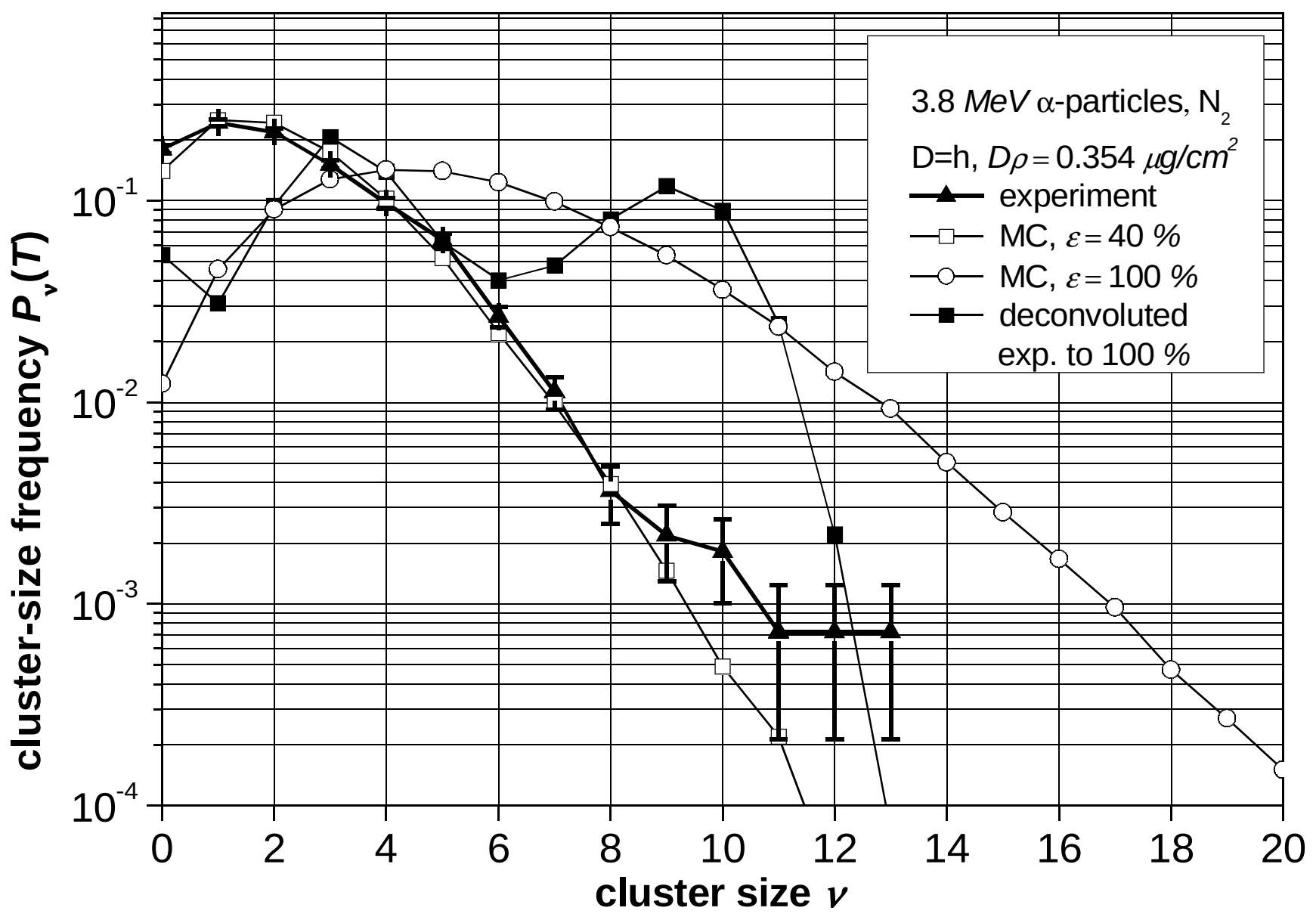

*Figure 62: The same as in Figure 58 for 0.354 μg/cm²*.

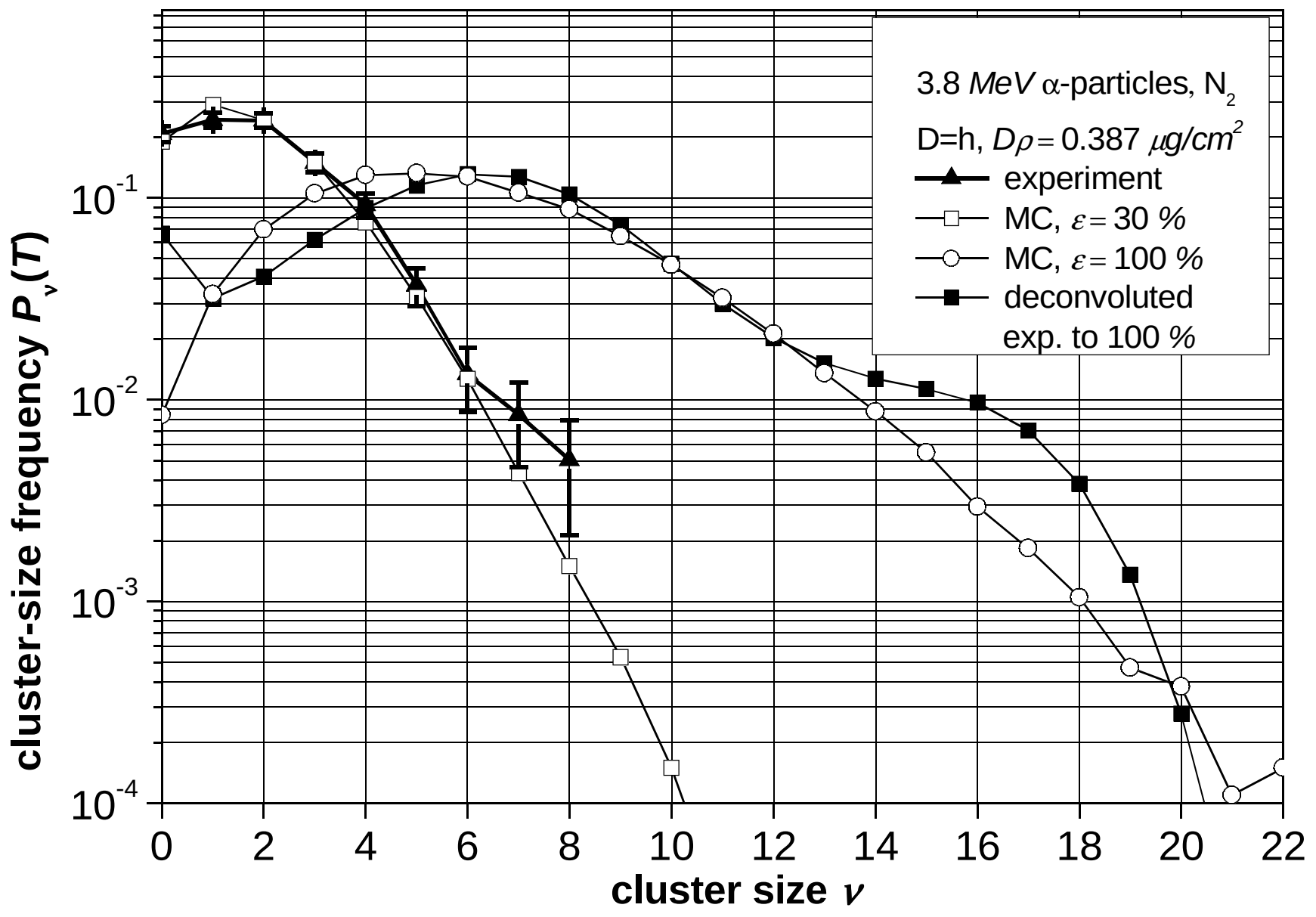

*Figure 63: The same as in Figure 58 for 0.387 μg/cm² with* $\varepsilon$ *= 30 % and 100 % respectively.*

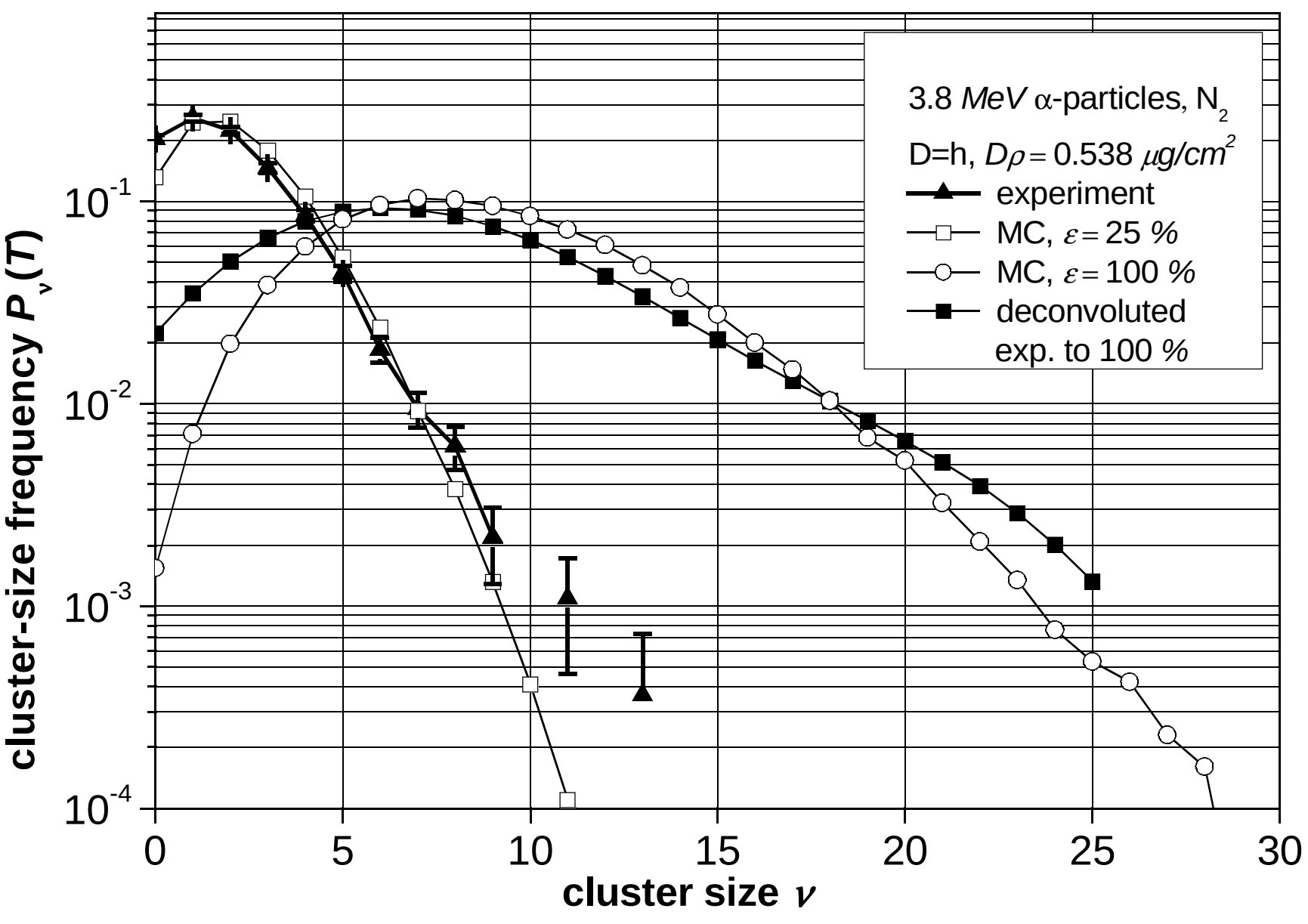

*Figure 64: The same as in Figure 58 for 0.538 μg/cm² with ε = 25 % and 100 % respectively*

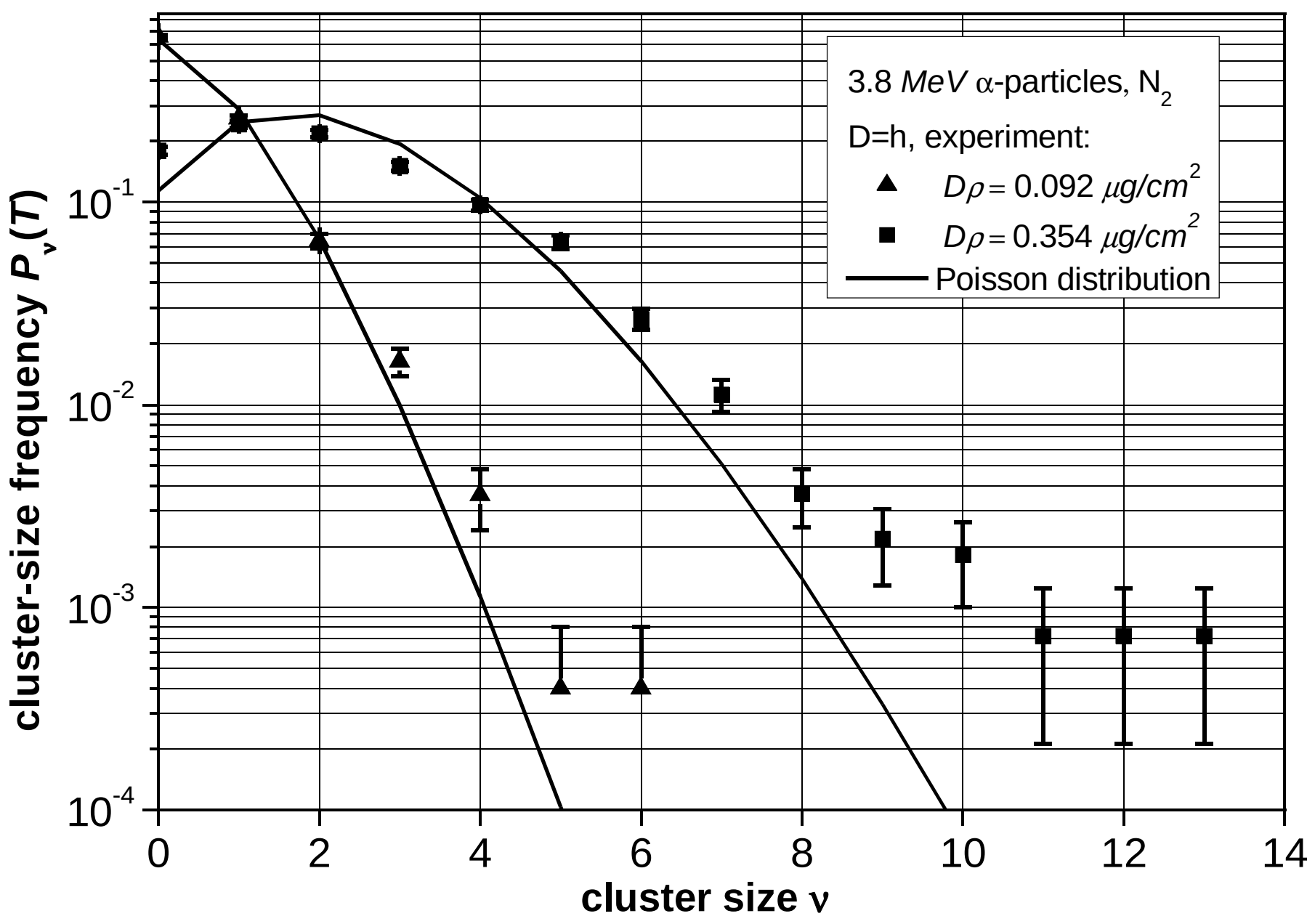

*Figure 65: Measured frequency distribution of ion cluster-size spectra for (▲) – 0.092 μg/cm² and (■) – 0.354 μg/cm² diameters of sensitive volume irradiated by 3.8 MeV α-particles. (–) – Poisson-like distribution based on the measured mean ion cluster size. Molecular nitrogen.*

## 6.3 Cluster-size distributions due to α-particles in propane gas

The experimental cluster-size spectra for cylinders with the following dimensions (diameter by height), 0.11 *μg/cm²* x 0.11 *μg/cm²*, 0.25 *μg/cm²* x 0.25 *μg/cm²* and 0.37 *μg/cm²* x 0.37 *μg/cm²* irradiated by a narrow beam of 4.6 *MeV* $\alpha$-particles entering the sensitive cavity through a 1 *mg/cm²* mylar foil (which degrades the energy to 3.8 *MeV*) were measured. The actual size of the SNS is derived from transmission measurements – see chapter 3.4.4.

Using (Eq.24), 0.11 *μg/cm²* of propane gas corresponds to a water cylinder. ($h = D$) of 0.138 *μg/cm²*, 0.25 *μg/cm²* to 0.313 *μg/cm²*, 0.37 *μg/cm²* to 0.463 *μg/cm²* respectively.

The cluster size frequency distribution was compared with that obtained from a Monte Carlo calculation.

The experimental and theoretical results are presented in Figures 66-68. The experimental distribution for 0.11 *μg/cm²* was measured for 60 *%* ions detection efficiency, for 0.25 *μg/cm²* – 40 *%* and 0.37 *μg/cm²* – 30 *%*, respectively. It must be added that no normalization procedure was applied to the experimental results. It can be seen that the agreement of measured and calculated cluster distributions is rather satisfactory apart from some small deviations for the higher cluster sizes, which may be the result of a higher contribution from delta electrons. These deviations increase with increasing dimensions of the SNS, as with the increasing ionization yield produced by secondary electrons. Also, the deviations from the calculations could be caused by the limited knowledge of the cross sections for propane gas. The influence of cross-section data on the Monte Carlo calculation is presented in [64, 82].

The de-convolution of the measured distributions for 0.11 *μg/cm²*, 0.25 *μg/cm²* and 0.37 *μg/cm²* to the true one (100 *%*) are presented in Figures 66-68. The agreement is very good only for 0.11 *μg/cm²* as the detection efficiency is rather high (60 *%*) and the low measuring statistics were enough for a good de-convolution. The de-convolution results for 0.25 *μg/cm²* and 0.37 *μg/cm²* do not look good but it is still possible to find some similarity.

Generally, experiments with propane give the same physical pattern as for molecular nitrogen, with some indication of a stronger influence on the stability of the ion detector (sparks, gain change).

The numerical values of all ionization cluster-size distribution spectra presented in Figures 66-68 together with the calculated $M_1$, $F_2$ (radiation descriptors on the nanometric scale) and $M_2$ (second moment – useful in (Eq.18) and (Eq.19)) values are tabulated in Tables 17-19 (Appendix C).

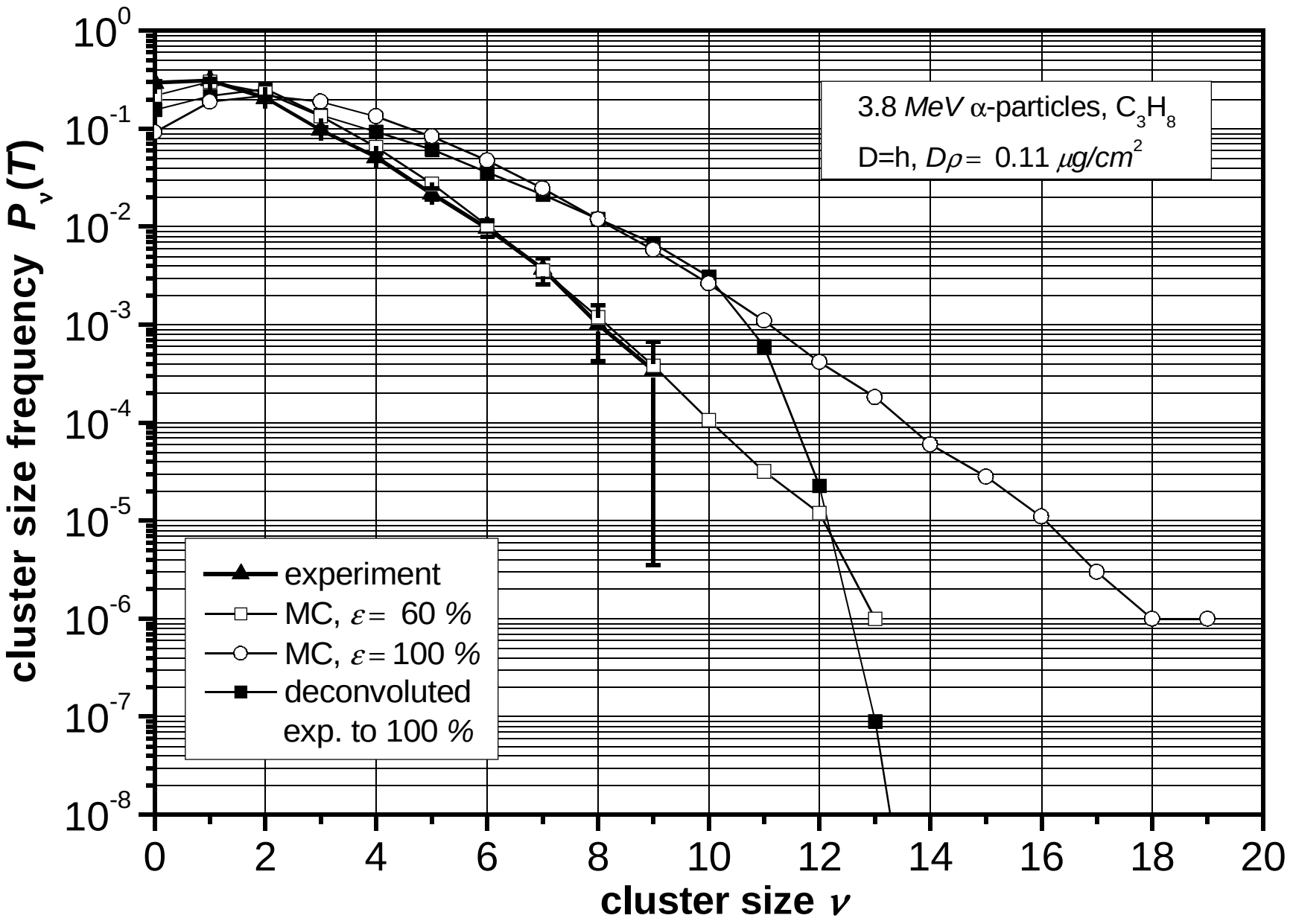

*Figure 66: Frequency distribution of ion cluster-size spectra for 0.11 μg/cm² diameters of sensitive volume irradiated by 3.8 MeV α-particles. (▲) – experimental spectra, (□) and (○) – Monte Carlo calculations for different ion detection efficiencies, (■) – deconvoluted experimental results to ε = 100 % with the assumption of an experimental single-ion detection efficiency ε = 30 %. Propane gas.*

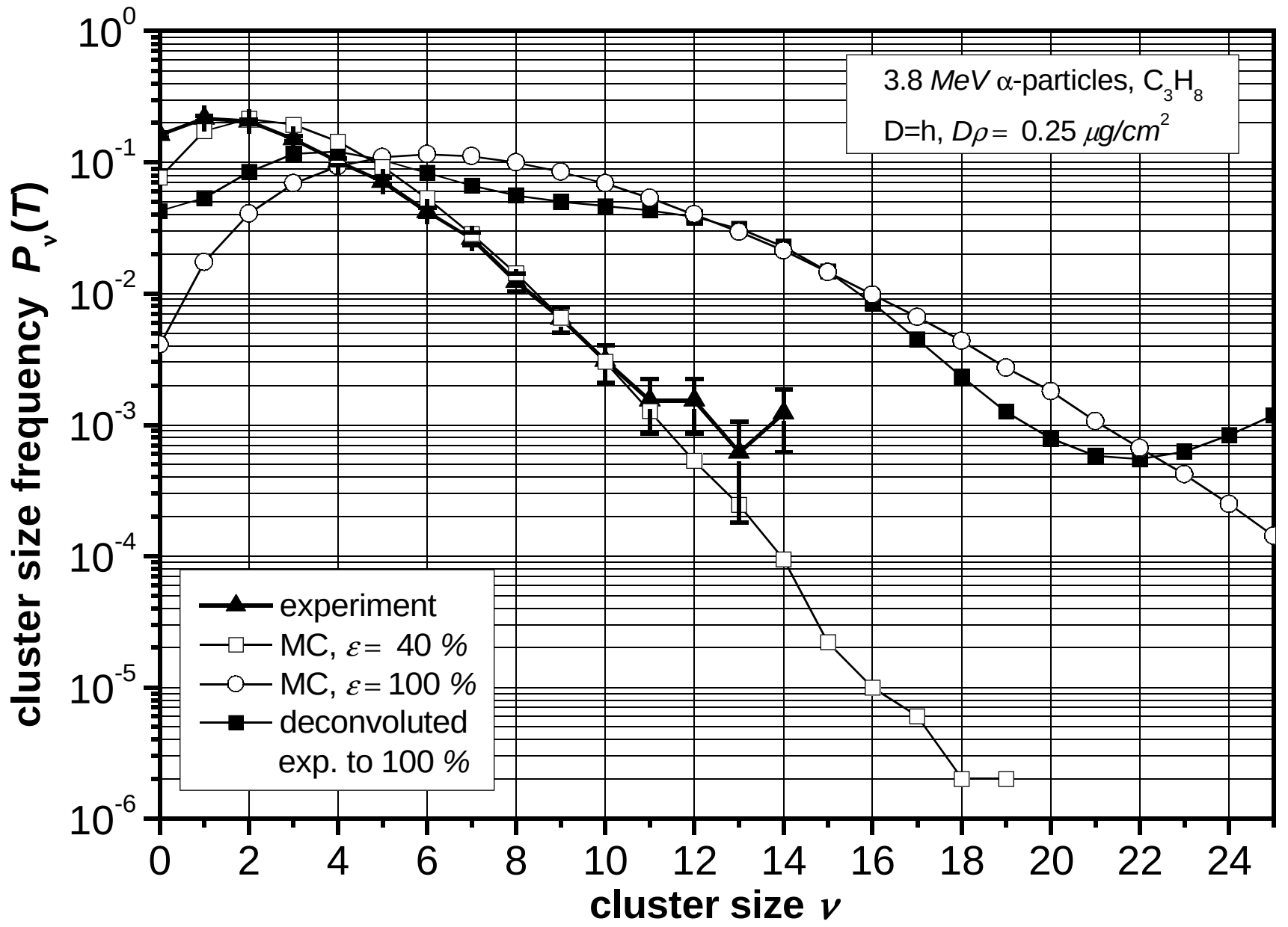

*Figure 67: The same as in Figure 66 for 0.25 μg/cm².*

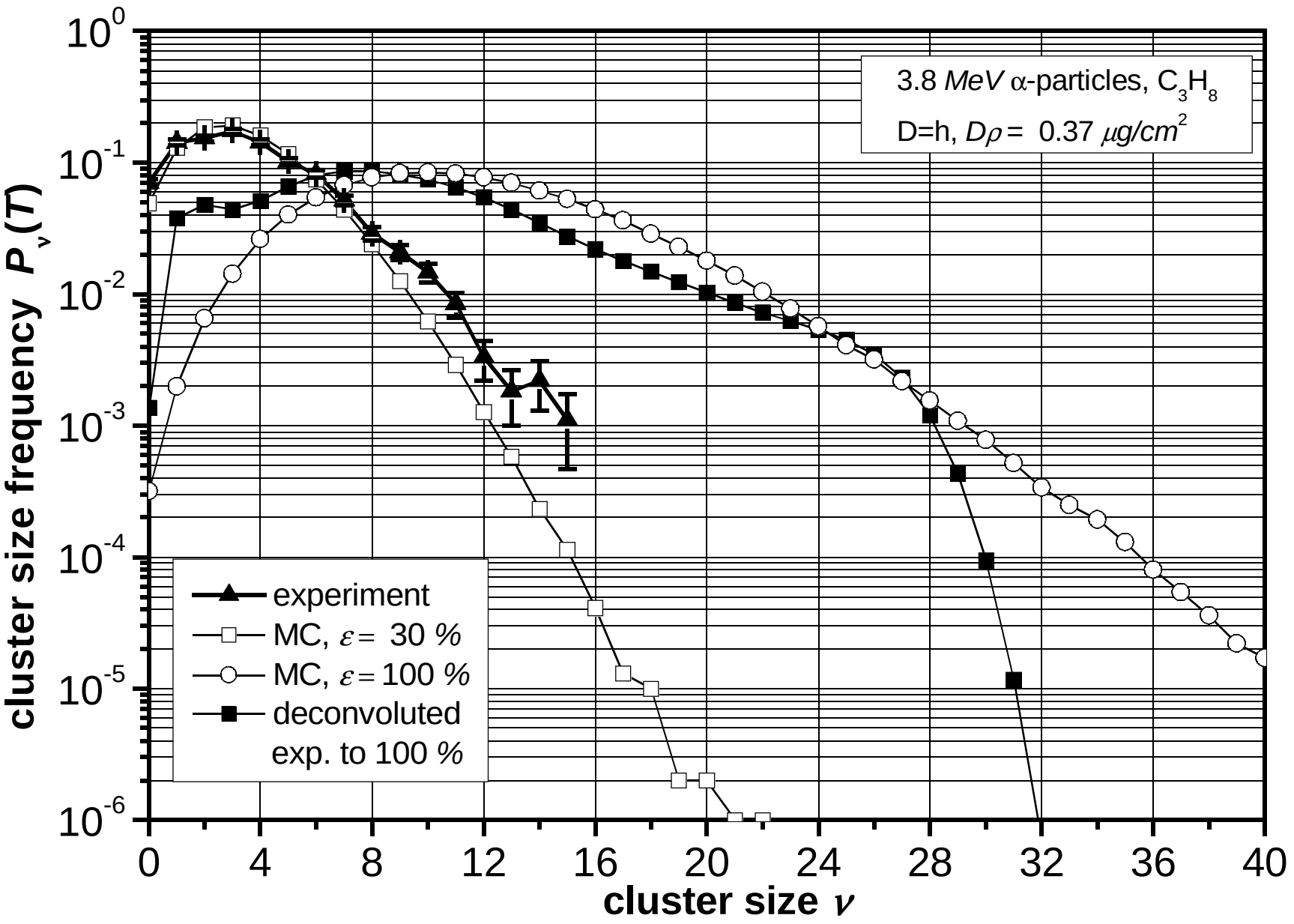

*Figure 68: The same as in Figure 66 for 0.37 μg/cm².*

## 6.4 Conclusions

It has been confirmed that the JC is a unique set-up with the ability to simulate cylindrical sensitive volumes at nanometre levels using propane gas or molecular nitrogen. Such target volumes are comparable in size to sub-cellular structures like segments of DNA or nucleosomes.

The counter has a rather high efficiency for the detection of single ions and represents the first measuring device based on single-ion counting which can be used to investigate ionisation-cluster formation in target volumes 0.1–5 *nm* in diameter at unit density.

The experimental data given by the JC are related to single particle tracks. The absolute (no normalization) discrete frequency distribution of cluster size for a given charged particle versus ion cluster size was derived experimentally for a wide range of *nm* site sizes. Using experimental frequency distributions, new descriptors of radiation quality for radiological protection were determined: the probability $P_1$ of creating cluster size $\nu = 1$, the first moment of the distribution $M_1$, the cumulative distribution $F_2$, are candidates for a description of radiation quality for radiation protection and radiobiology.

Because the JC can also be applied to other radiation qualities it is, at least in principle, one of the first measuring devices which might be used in the future for the development of a standard for the formation of ionization clusters in ‘‘nanometric’’ targets such as short DNA segments.

The results presented for propane gas and molecular nitrogen show the JC to be an efficient tool for the investigation of radiation quality for “single” electrons and $\alpha$-particles at the nanometre level.

In should be mentioned that our knowledge of the Jet Counter detector is good enough for a proper Monte Carlo simulation of cluster size formation in a simulated nanometer sensitive volume.

Also, the de-convolution procedure (see chapter 5) was presented and applied to experimental frequency distributions to de-convolute them to the true distributions. Comparisons of de-convoluted results with Monte Carlo simulations (100 % efficiency) show very good agreement for electrons (good measuring statistics). The agreements for $\alpha$-particles in molecular nitrogen is also acceptable. Only the comparison for $\alpha$-particles in propane gas is not so good due to low statistics. Nevertheless, the de-convolution procedure presented here works well and may be used in these kinds of measurements.

# 7 Nanodosimetric quantities – application approach

As mentioned in chapter 2, the metrological challenge of nanodosimetry is to replace or, at least, supplement the quantity absorbed dose by more appropriate quantities. These nanodosimetric quantities should be

- measurable (for application purposes),
- strongly correlated with the structure of the particle track and with the stochastics of particle interaction in nanometric volume comparable in size to short segments of DNA,
- correlated with radiobiological effects (SSB, DBS and so on).

In view of this result, it can be hoped that quantities which are based on the probability of forming clusters of multiple ionization due to particle interactions in volumes which are comparable in size with short segments of DNA can be used as candidates.

As presented in chapter 6, it is already possible to form clusters on a scale comparable with DNA and from these it is possible to derive statistical quantities ($P_1(Q)$, $M_1(Q)$ and $F_2(Q)$ – see Tables 4-19) which should be correlated with radiobiological effects caused by ionization radiation.

## 7.1 Mean cluster size – $M_1$

In the special case of a macroscopic target volume, where the initial particle energy $T$ is completely absorbed, $M_1(Q;d)$ (mean cluster size (Eq.18)) is equal to the mean number $N(T)$ of ion pairs formed, which is conventionally expressed by $N(T)=T/W(T)$, where $W(T)$ is the so-called *W*-value defined as the mean energy expended per ion pair formed upon the complete degradation of a charged particle [36]. Unfortunately, the latter condition is never fulfilled in nanometric volumes. Nevertheless, $M_1(Q;d)$ can be assumed to be equivalent to absorbed dose $D$ (for radiobiology) and to charge (for nanoelectronics).

p.s. W value is macroscopic parameter and not applicable for nano-volumes in principle.

## 7.2 Cluster size frequency $P_1$ and cumulative distribution function $F_2$

To investigate the validity of the nanodosimetric concept of defining radiation quality in terms of ionization cluster-size probabilities, the probability $P_1(Q)$ of forming a cluster size $\nu=1$, and the sum distribution function $F_2(Q)$ (Eq.19) of forming an ionization cluster size $\nu \geq 2$ must be compared, as a function of radiation quality, with radio-biological data regarding the formation of strand breaks of DNA. Here, use should be made of radio-biological measurements which were performed for different light ions over a wide range of radiation quality.

Both hypotheses were brilliantly checked by Grosswendt [18] using the radiobiological data of Taucher-Scholz and Kraft [83].

For checking the first hypothesis that the formation of ionization clusters of size $\nu = 1$ behaves, as a function of radiation quality, like the formation of single-strand breaks of DNA, Figure 69 shows the radio-biological cross sections of SV40 viral DNA with respect to SSB-formation as a function of LET. These data were measured by Taucher-Scholz and Kraft [83] in a low-scavenging buffer system for X-rays, $^{4}$He-ions, $^{12}$C-ions, $^{16}$O-ions, and $^{20}$Ne-ions over a wide LET range. The measured cross sections of SSB-formation are compared with the LET dependence of the calculated probabilities $P_1(Q)$, after normalization of the $P_1(Q)$ data to the experimental cross section at 50 $keV/\mu m$ and using for $Q$ the LET of the primary particles.

A Monte Carlo simulation was performed for a cylindrical target volume of water (DNA-like segment – 2.3 $nm$ in diameter and 3.4 $nm$ in height) homogeneously irradiated by ions. At first glance, it can be seen from Figure 69 that the normalized probabilities behave, as a function of LET, similarly to the radio-biological cross sections of SSB-formation. Hence, a measurement of the ionization cluster-size probability $P_1(Q)$ in an irradiation field of specified fluence could be directly related to the yield of single-strand breaks of SV40 viral DNA to be expected in this field.

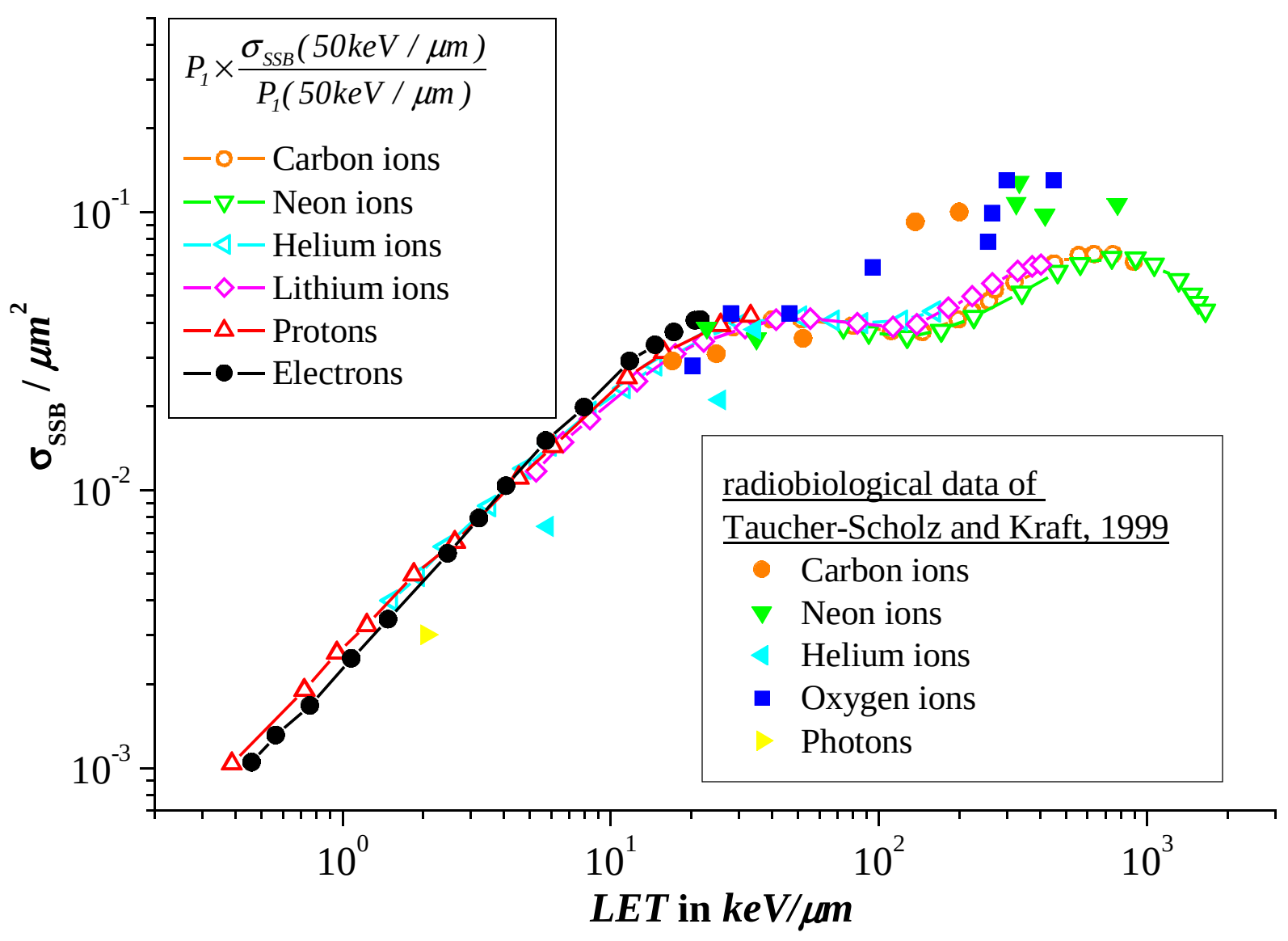

*Figure 69: Radio-biological cross section $\sigma_{SSB}$ for single-strand-break formation in SV40 viral DNA, as a function of LET: Comparison of experimental data of Taucher-Scholz and Kraft [83] with the cluster-size probability $P_1(Q)$ due to electrons or light ions normalized to the radio-biological data at an LET of 50 keV/μm; for the meaning of the symbols, see the insert. Ref.[18].*

In order to check the second hypothesis that the formation of ionization clusters of size $\nu \geq 2$ behaves, as a function of radiation quality, like the formation of DNA-double-strand breaks, Figure 70 shows the experimental results of Taucher-Scholz and Kraft [83] regarding the cross sections of SV40 viral DNA with respect to DSB-formation, again as a function of LET. These data are compared with the sum distribution function $F_2(Q)$ due to electrons or light ions after normalization to the DSB cross section measured at 50 $keV/\mu m$ and using again for $Q$ the LET of the primary particles.

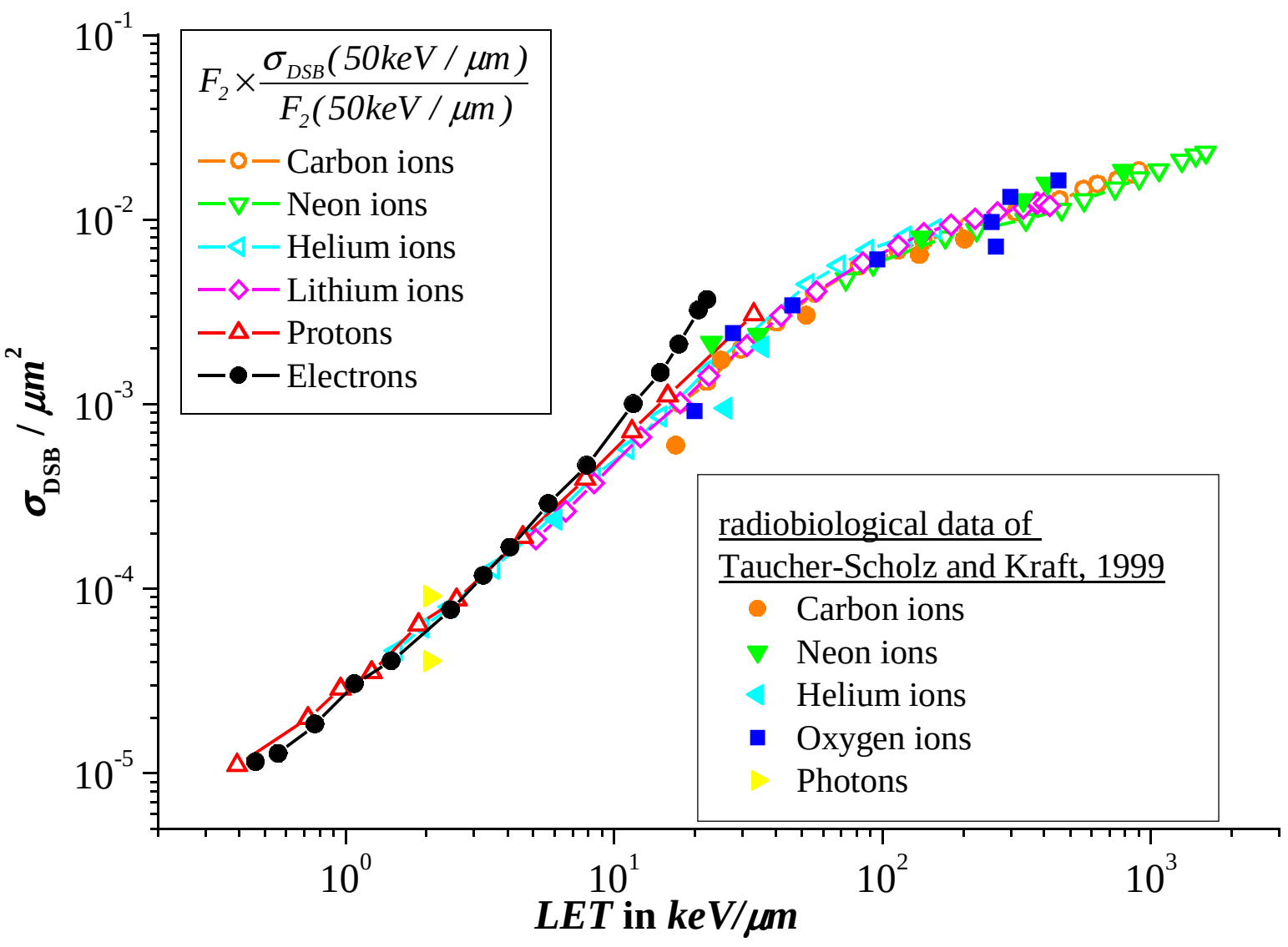

*Figure 70: Radio-biological cross section $\sigma_{DSB}$ for double-strand-break formation in SV40 viral DNA, as a function of LET: Comparison of the experimental data of Taucher-Scholz and Kraft [83] with the sum distribution function $F_2(Q)$ due to electrons or light ions normalized to the radio-biological data at an LET of 50 keV/μm; for the meaning of the symbols, see the insert. Ref.[18].*

## 7.3 Electrons – range and energy restricted LET

The experiments using "single" low-energy electrons interacting with a nitrogen jet of nanometer size, that is comparable to a short segment of DNA, show rather extended cluster-size distributions with values of cluster size $\nu$ as high as ten (see chapter 6.1). Consequently, it can be assumed, for instance, that in the field of nanoelectronics these electrons might be responsible for the formation of large charge clusters, and that in radiation biology they might be able to cause considerable DNA damage such as, for example, single or double-strand breaks (SSB, DSB) and even clustered base damage.

The first moment $M_1$ of the cluster-size distributions (the mean cluster size) can be used as a tool for a qualitative interpretation of biological endpoints observed, for instance, for delta electrons produced by charged particles or for monoenergetic electrons generated by photoelectric absorption of characteristic X rays of low-Z materials like carbon (280 *eV*) and aluminum (1.49 *keV*).

It should to be pointed out that the mean cluster size is directly related to the range restricted linear energy transfer $L_r$ by the following relation:

$$L_r = \frac{\omega(T)}{D\rho \times \varepsilon \, N_{mean}} M_{1E}(T, D\rho) \qquad (47)$$

where: $M_{1E}(T, D\rho)$ is the first moment of an experimental cluster-size distribution for a given energy $T$ and specified diameter $D\rho$; $N_{mean}$ is the mean value of the electron rate; $\omega(T)$ is the differential mean energy expended per ion pair formed in nitrogen. This differential value is related to the so-called $W$ value by

$$\omega(T) = \frac{W(T)}{1 - \frac{T}{W(T)} \times \frac{\mathrm{d}W}{\mathrm{d}T}} \qquad \textbf{(48)}$$

Since $dW/dT \leq 0$ for electrons, $\omega(T)$ is always less than or equal to $W(T)$ independent of energy $T$. $W(T)$ is the mean energy expended per ion pair formed in nitrogen for electrons at energy $T$ [64].

The values for $L_r$, based on (Eq.47), were derived using the experimental values for $M_{1\mathrm{E}}$ together with data for $\omega(T)$ and $W(T)$ (see Table 20). They are presented in Table 20 as $L_{rE}$. Other parameters used in the evaluation of $L_{rE}$ were $\varepsilon = 0.3$ and $N_{mean}$ as in Table 20. The results for $L_{rE}$ are in striking agreement with the corresponding values, $L_{rMC}$, which were determined from the first moments $M_1^{(\mathrm{MC})}$ of cluster-size distributions calculated by Monte Carlo simulations for single (one) electrons. A cumulanta $F_{2E}$ is also included in Table 20, together with values for $L_{100}(\mathrm{N_2})$ which were directly calculated from the cross sections of electron interaction in nitrogen (for the cross sections, see Ref. [64]). To test their applicability, the values of $L_r$ calculated according to (Eq.47) are compared with the data for energy restricted linear energy transfer $L_{100}$ in nitrogen. The results of the comparison are shown in Figure 71. As can be seen from the figure, the $L_{rE}$ derived from the $M_{1\mathrm{E}}$ of the measured cluster-size distributions in the energy range 100 *eV* to 2 *keV* follow $L_{rMC}$ and $L_{100}(\mathrm{N_2})$ for these energies. Larger discrepancies between $L_{rE}$ and $L_{rMC}$ or $L_{100}(\mathrm{N_2})$ for 100 *eV* electrons are most probably due to experimental errors. Here, it seems to be worth mentioning that the energy restricted linear energy transfer $L_{100}$ was suggested some time ago as a basic parameter to characterize the radiation quality of different types of ionizing radiation [10]. Up to now, data for $L_{100}$ for low energy electrons were based only on calculations due to the lack of adequate measuring methods. The results of this comparison indicate that the proposed system for measuring cluster size distributions for single low energy electrons has shown its practical feasibility. In addition, it should be mentioned that $L_r$ for nitrogen shows the same dependence on electron energy as $L_{100}$ for liquid water (see page 12 of Ref. [3]), see Table 20. It has been shown here that $L_r$ derived from cluster distributions for low energy electrons can replace $L_{100}$.

As far as the interpretation of biological endpoints is concerned, $M_1$ and $F_2$ represent new quantities. These quantities are based on the passage of a single particle through a sensitive target volume and are related to the fluence concept. With this, the derived quantities differ from the quantities used up to now, based on the absorbed-dose concept. The present experimental results are the first of their kind for low-energy electrons with energies ranging from 100 *eV* to 2000 *eV*. The immediate use of these results for the interpretation of the biological experiments of Virsik *et al.* [84] on the formation of chromosome aberrations in human lymphocytes, and of the experiments of de Lara *et al.* [85] on the yields of the formation of DSBs in Chinese Hamster V79-4 cells is now possible. In both experiments it was shown that monoenergetic K-X rays from Carbon (280 *eV*) are more effective than Aluminum K-X rays (1490 *eV*). The same fact is also evident from our experimental data presented in Figures 51-56 which show that low-energy electrons (100 *eV* – 300 *eV*) are much more efficient in producing larger cluster sizes in DNA-like target volumes than electrons at higher energies (1000 *eV* – 2000 *eV*).

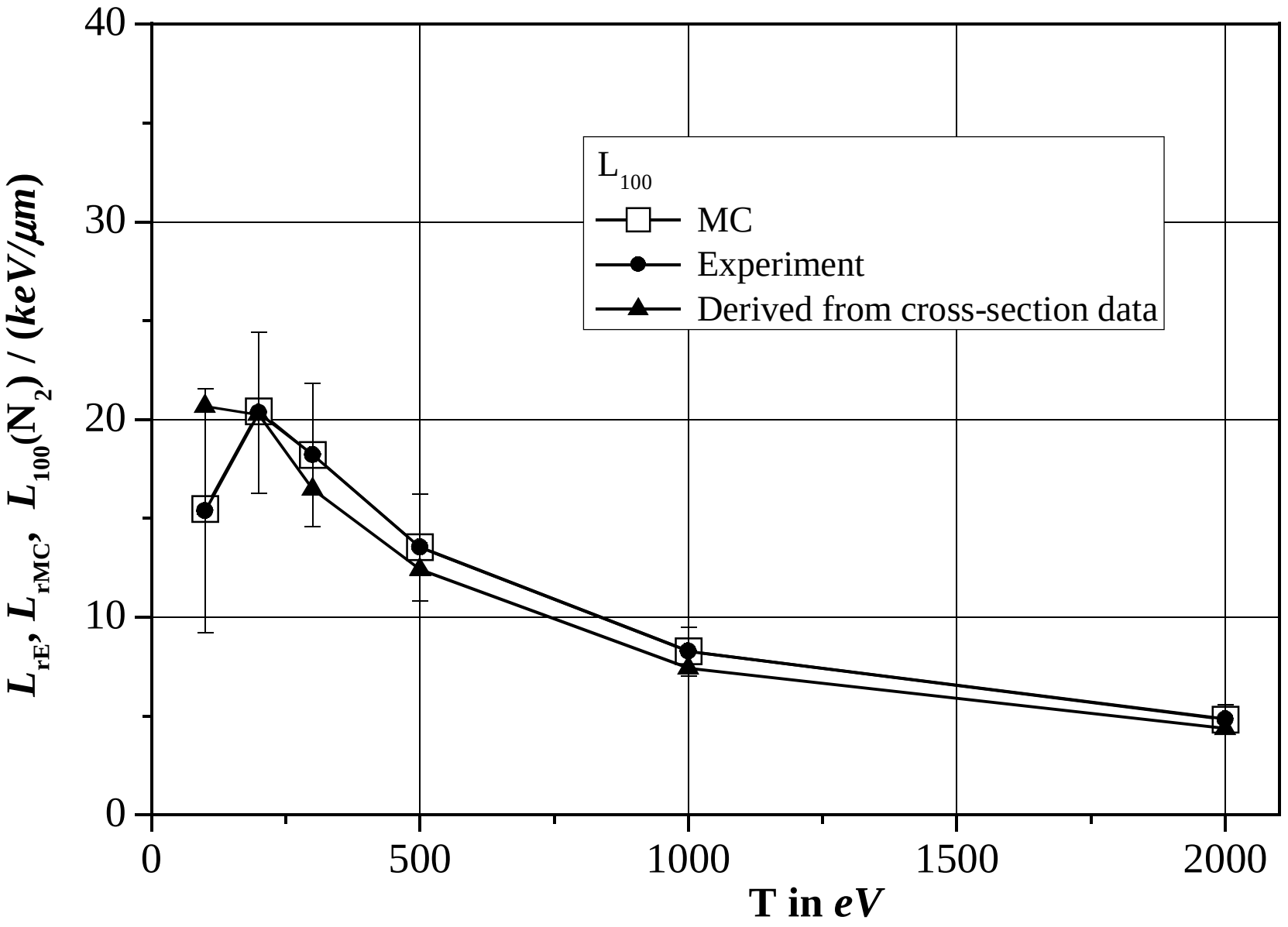

*Figure 71: (●) – range restricted linear energy transfer $L_{rE}$ derived from the first moments, $M_{1E}$, of the measured cluster-size distributions (see (Eq.47)); (□) – range restricted linear energy transfer $L_{rMC}$ derived from the first moments, $M_1^{(MC)}$, of cluster-size distribution calculated by Monte Carlo simulation; (▲) – energy restricted linear energy transfer $L_{100}(N_2)$ for nitrogen, as a function of energy of monoenergetic electrons.*

## 7.4 Conclusions

The experiments using single low-energy electrons (100 *eV* – 2 *keV*) and 3.8 *MeV* $\alpha$-particles interacting with a nitrogen or propane jet of nanometre size comparable to that of a short DNA segment, show discrete frequency distributions of cluster size with extended cluster sizes. These cluster-size distributions were determined for the first time for electrons. As a result, it has been shown that not only 3.8 *MeV* $\alpha$-particles but also low-energy electrons interacting with a DNA-like segment are able to create single and clustered damage (assuming that SSB formation is proportional to the frequency of a single ionization while DSB formation needs at least two ionizations within a short DNA segment). In nanoelectronics, they can generate charge clusters. Based on these distributions it has been shown that:

- the first moment or mean cluster size, $M_1$, of a cluster distribution can replace the restricted linear energy transfer $L_{100}$,
- the cumulative distribution function, $F_2$, can serve as a physical descriptor for the yields of DSBs and chromosome aberrations in radiobology,
- cluster-size frequency , $P_1(Q)$, can serve as a physical descriptor for the yields of SSBs in radiobiology.

In view of this, the three nanodosimetric quantities $M_1$, $F_2$, and $P_1(Q)$ can be used as new tools for the qualitative interpretation of observed biological endpoints due to charged particles. And of course, a fluence parameter (number of particles per unit area) is needed to totally characterize the whole radiation field on the nanometer scale.

In the future, experimental results for electrons may be used for radiobiological interpretation of monoenergetic electrons arising from the photoelectric effect of low-energy characteristic X rays, due to low-energy Auger electrons and delta electrons of charged particle tracks.

# 8 Nanodosimetry with the Jet Counter at present and in the future

## 8.1 Summary and conclusions

The summary of the main results achieved and presented in the thesis “Ionization cluster size distributions created by low energy electrons and alpha particles in nanometric track segment in gases” can be formulated as follows:

1. State of the art of the basics of experimental nanodosimetry followed by a review of track formation theory.
2. Description of the experimental set up called the Jet Counter with its improvement and adaptation to the experiments for the purpose of this thesis.
3. Detailed description of the components of the Jet Counter facility, namely:
   - method of simulation of nanometer size sites as well as the scaling procedure to related measurements in gases to data for other materials like liquid water;
   - single ion detection system;
   - method for measuring the frequency of cluster size spectra;
   - method of analyzing the experimental data;
   - approach to formulating a set of new quantities based on the experiments performed.
4. results of the experiments with low energy mono-energetic electrons from 100 *eV* to 2000 *eV*, namely:
   - frequency cluster size distributions produced by low energy electrons (100 *eV*, 200 *eV*, 300 *eV*, 500 *eV*, 1000 *eV* and 2000 *eV*) in molecular nitrogen with target area density 0.34 $\mu g/cm^2$.
5. Experiments with $\alpha$-particles:
   - frequency cluster size distributions produced by 3.8 *MeV* $\alpha$-particles in molecular nitrogen with target area density 0.092 $\mu g/cm^2$, 0.130 $\mu g/cm^2$, 0.187 $\mu g/cm^2$, 0.291 $\mu g/cm^2$, 0.354 $\mu g/cm^2$, 0.387 $\mu g/cm^2$ and 0.538 $\mu g/cm^2$;
   - frequency cluster size distributions produced by 3.8 *MeV* $\alpha$-particles in propane gas with target area density 0.11 $\mu g/cm^2$, 0.25 $\mu g/cm^2$ and 0.37 $\mu g/cm^2$.
6. Approach to formulate a set of new quantities based on the experiments performed, namely:
   - the first moment of the frequency distribution (mean cluster size), $M_1$ (Eq.18), i.e., the mean number of ions (ionizations) in a cluster for a given geometry of irradiation as well as for a given SNS;
   - the cumulative frequency, $F_2(Q;d)$ (Eq.19) – the frequency required to create a cluster-size equal to 2 or higher;
   - the parameters $P_1$, $M_1$ and $F_2$ directly describe the radiation quality of the ionizing radiation at a specific nanometer scale. Also, they are well correlated with radiobiological data and as a result these quantities may be used alone or to supplement traditional ones (absorbed dose, LET, …) in radiotherapy, radiation protection and other applications of ionizing radiations.

7. The Jet Counter is the first measuring facility based on single-ion counting which can be used to investigate ionization-cluster formation in nanometer target volumes (up to a few μg/cm2) for “single” low energy electrons and $\alpha$-particles.
8. Also, the Bayesian unfolding procedure presented here works well with the experimental results to unfold the measured cluster size distribution to the true one.

It should be mentioned that the experimental results with “single” electrons are the first results of this type in the world. These results may give a realistic impression of the role of δ-electrons in track structure formation by charged particles.

All experimental results are compared with sufficient Monte Carlo simulations. In most cases the agreement is striking, only the results with propane gas have some deviations for the highest cluster sizes.

For all measured frequency cluster size distributions the statistical values were calculated.

Finally, taking into consideration the experiments with $\alpha$-particles and electrons, the Jet Counter is well suited for studying the formation of ionization clusters by all kinds of charged particles. Because the Jet Counter can also be applied to other radiation qualities it is, at least in principle, one of the first measuring devices which might be used in the future for the development of a standard for the formation of ionization clusters in "nanometric" targets such as short DNA segments.

## 8.2 Perspectives

The future line of development of „nanodosimetry” with the Jet Counter:

1. In rather short horizon (5 years):
    - Improvements of ion counting system toward 100% efficiency of ion collection;
    - Increasing the ranges of the simulated sites up to 30 nm;
    - Experiments with neutrons;
    - Development of more compact „nanodevice” for environmental studies ( low dose range);
    - More close cooperation between radiobiological and physical experiments especially for targeted radiotherapy;
2. In longer horizon:
    - Searching for a new system of units especially for radiation protection purposes based on nanodosimetry concept.horizon:
    - Comparison of new radiological theories based on new system of units with biological systems;
    - Development of a new standard (nanodosimetry) from metrological point of view;
    - Development of the application of the experimental nanodosimetry for studying the radiation damage in nanoelectronic devices;
    - Searching for nanodosimetry system based on semiconductor detectors.

# Appendix A Electron cross sections for nitrogen gas

(After Grosswendt and Pszona [64])

## A.1 Treatment of elastic electron scattering

As proposed by Grosswendt and Waibel [86], the treatment of elastic electron scattering was based on integrated cross sections $\sigma_{el}(T)$ obtained by experiment and on Rutherford's differential cross section $(d\sigma/d\Omega)_{el}$, with respect to the solid angle, modified to take atomic screening effects into account:

$$\left(\frac{d\sigma(T)}{d\Omega}\right)_{el} = \frac{Z(Z+1)e^4}{(1-\cos\vartheta+2\eta)^2(4\pi\varepsilon_0)^2}\left[\frac{T+mc^2}{T(T+2mc^2)}\right]^2 \qquad \textbf{(49)}$$

The quantity $\vartheta$ is the polar angle of scattering relative to the initial electron direction, and $T$ the kinetic electron energy; $Z$=7 is the atomic number of nitrogen, $e$ the electron charge, $\varepsilon_0$ the permittivity of vacuum, $mc^2$ the electron rest energy, and $\eta$ the so-called screening parameter. If $\eta$ is known, the scattering angle $\vartheta$ can be sampled conventionally using (Eq.49), and interpreting the ratio $(d\sigma/d\Omega)_{el}/d_{el}\Omega(T)$ as the probability density with respect to $\vartheta$. This procedure is a satisfactory approximation to differential elastic scattering at energies greater than about 200 *eV.* At smaller energies, however, large angle scattering is greatly underestimated. A correction factor was used, therefore, at lower electron energies.

The following strategy was applied to determine $\eta$: as a first step, an analytical function was fitted to measured integral cross sections as a function of energy $T$ (supplemented by theoretical cross sections at higher electron energies). As a second step, the resulting cross section was set equal to the integral of $(d\sigma/d\Omega)_{el}$ over the solid angle [see (Eq.50)] which was afterwards solved for $\eta$:

$$\sigma_{el}(T) = \frac{Z(Z+1)e^4}{\eta(1+\eta)^2(4\pi\varepsilon_0)^2}\left[\frac{T+mc^2}{T(T+2mc^2)}\right]^2 \qquad \textbf{(50)}$$

The integrated elastic cross section $\sigma_{el}(T)$ was determined at electron energies between 10 *eV* and 100 *keV.* In the energy range between 10 *eV* and 10 *keV,* the analytical function given by (Eq.51) was used, modified by the factor $f(T)$ of (Eq.52) at energies $T \le 30\ eV$.

$$\sigma_{el}^{fit}(T) = \frac{c_1}{c_2 + c_3 t + c_4 t^2 + t^{c_5}} \qquad \textbf{(51)}$$

$$f(T) = \frac{1}{c_6 + c_7(\ln t)^{-1} + c_8 \ln t + c_9(\ln t)^2} \qquad \textbf{(52)}$$

The parameters $c_\nu$ for $\nu = 1, 2 \ldots 9$ are fitting constants and $t = T/(1\,eV)$. The values of the parameters $c_1 \div c_9$ are: $6.23 \cdot 10^{-16}$ $cm^2$, -1.165, 0.00493, $-9.42 \cdot 10^{-8}$, 0.1474, -30.33, 35.22, 0.749, and -1.0539.

At energies $T > 10\ keV$, the integrated elastic cross section was calculated on the basis of (Eq.50) normalized to $\sigma_{el}^{fit}(T)$ of (Eq.51) at 10 *keV,* using Moliére's screening parameter [87] which is given by (Eq. 53):

$$\eta = \left[1.13 + 3.76\left(\frac{Z}{137}\right)^2 \frac{mc^2}{2T}\right] \frac{1.7\times10^{-5} Z^{2/3} (mc^2)^2}{T(T+2mc^2)} \qquad \textbf{(53)}$$

where the parameters have the same meaning as those in (Eq.49).

## A.2 Treatment of electron impact ionization

Electron impact ionization was based on the binary-encounter-Bethe model of Kim and Rudd [88] , which combines the Mott cross section with the high-T behavior of the Bethe cross section. Within the framework of this model, the integrated partial ionization cross section $\sigma_{ion}^{(k)}(T)$ with respect to a subshell $k$ with electron binding energy $B_k^{ion}$, kinetic energy $U_k$ of an electron of the subshell, and electron occupation number $N_k$ is given by (Eq.54):

$$\sigma_{ion}^{(k)}(T) = \frac{S}{t+u+1}\left[\frac{\ln t}{2}\left(1-\frac{1}{t^2}\right) + \left(1-\frac{1}{t}-\frac{\ln t}{t+1}\right)\right] \qquad \textbf{(54)}$$

Here, $t = T/B_k$, $u = U_k/B_k$, and $S = 4\pi a_0{}^2 N_k R^2 / B_k{}^2$ where $a_0 = 0.5292\cdot10^{-8} cm$ is the Bohr radius and $R = 13.61 eV$ is the Rydberg constant. Equation (Eq.54) was applied to calculate the integrated partial ionization cross section $\sigma_{ion}^{(k)}(T)$ of electrons for four molecular subshells $k$ using again the data of Hwang *et al.* [69] for $B_k$, $U_k$, and $N_k$. The total integrated ionization cross section $\sigma_{ion}^{(tot)}(T)$ can be calculated from (Eq.54) by summation over all subshells. Since $\sigma_{ion}^{(tot)}(T)$ determined in this way agrees well with measurements which are not able to separate direct ionization and autoionization of highly excited states, it has been assumed that the auto-ionization is included in $\sigma_{ion}^{(tot)}(T)$.

The energy distribution of secondary electrons emitted after electron impact ionization was determined on the basis of a single-differential cross section $d\sigma(T)/d\varepsilon$ with respect to the kinetic electron energy $\varepsilon$ expressed by the Breit-Wigner formula, as proposed by Green and Sawada [89]. Integration of this formula over $\varepsilon$ leads to a simple analytical equation which can easily be solved for the upper integration limit and thus makes the Monte Carlo simulation very convenient. The sampling of $\varepsilon$ was, therefore, performed using the Breit-Wigner form of the differential cross section $d\sigma(T)/d\varepsilon$, after normalization to its integral over $\varepsilon$ within the limits $0 \le \varepsilon \le \varepsilon_{\max}(T)$, as the probability density with respect to the secondary electron energy. By convention, the faster electron after impact ionization is the primary one and as a result the maximum energy $\varepsilon_{\max}$ of the secondaries is given by $\varepsilon_{\max}(T) = (T-I)/2$ where $I$ is the ionization threshold energy assumed in the calculation. This procedure leads to the following expression for the secondary electron energy $\varepsilon$:

$$\varepsilon = \varepsilon_0(T) + \Gamma(T)\tan\left\{\xi\left[\arctan\left(\frac{\varepsilon_{\max}(T)-\varepsilon_0(T)}{\Gamma(T)}\right) + \arctan\left(\frac{\varepsilon_0(T)}{\Gamma(T)}\right) - \arctan\left(\frac{\varepsilon_0(T)}{\Gamma(T)}\right)\right]\right\} \quad \textbf{(55)}$$

Here, $\Gamma(T) = \Gamma_S T/(T+\Gamma_B)$ and $\varepsilon_0(T) = \varepsilon_S - \varepsilon_A/(T+\varepsilon_B)$, the quantities $\Gamma_S = 13.8\ eV$, $\Gamma_B = 15.6\ eV$, $\varepsilon_S = 4.71\ eV$, $\varepsilon_A = 1000\ eV^2$ or $\varepsilon_B = 2I$ are the fitting parameters of Green and Sawada [89] to the experimental data of Opal *et al.* [90], and $\xi$ is a pseudo-random number uniformly distributed between 0 and 1. The errors induced by this procedure due to the wrong shape of the energy distribution and due to the non-ideal behavior at high primary energies can be assumed to be acceptable for most applications.

The energy $T'$ of the primary electron after impact ionization was set equal to $T - \varepsilon - I(T)$, where $I(T)$ is the ionization threshold energy used at a specified electron energy $T$. This ionization threshold was not fixed at 15.58 *eV* but it was assumed that it depends on the electron energy $T$, to take into account, at least approximately, the contribution of subshells with binding energies greater than the lowest ionization threshold which can contribute to the electron degradation only if the electron energy is high enough. To determine $I(T)$, it was set equal to the average binding energy of the weakly-bound valence electrons of nitrogen, which was calculated as a function of electron energy on the basis of the partial electron ionization cross sections of (Eq.54) for different subshells. For practical reasons of Monte Carlo simulation, the calculated values of $I(T)$ were then fitted to the function given by (Eq.56) which is valid for $T \geq 18\,eV$. At lower energies, $I(T)$ was set equal to 15.58 *eV.*

$$I(T) = c_1 + c_2\left(\frac{T}{R}\right)^{c^3} \quad \textbf{(56)}$$

The quantity $R$ is again the Rydberg constant, and the three parameters $c_\nu$, $\nu = 1, 2$ and $3$, are fitting constants, which are equal to 19.83 *eV*, -4.158 *eV* and -0.4692, respectively.

The last aspect of ionization impact is the determination of the flight direction of the initial electron after scattering and of the liberated secondary electron. As complete sets of data are lacking, these directions were determined approximately using the kinematic equations proposed by Berger [91]. These equations are based on the conservation of momentum and energy and are in rather satisfactory agreement with the electron distributions measured by Opal *et al.* [90]. For details of the determination of the electron flight directions, in particular at lower electron energies, see the publication by Grosswendt and Waibel [86].

## A.3 Treatment of electron impact excitation

The treatment of excitation processes in nitrogen was based on the formula and cross section parameters of Porter *et al.* [92], which relate the excitation cross sections to generalized oscillator strengths and, by a distortion factor, take into account the fact that the Bethe formula is not valid at low electron energies.

For allowed discrete excitations and for the excitation of Rydberg states, the cross section $\sigma_{exc}^{(j)}(T)$ to a state *j* of electrons at energy $T$ is given by the following equation:

$$\sigma_{exc}^{(j)}(T) = 4\pi a_0^2 R^2 \frac{F_j}{W_j^2} \xi \Phi(\xi) \left[ \ln\left( \frac{4C_j(\tau+1)^2}{\xi} + e \right) - \frac{\tau(\tau+2)}{(\tau+1)^2} \right] \qquad \textbf{\textit{(57)}}$$

Here, $\xi = 2[W/mc^2](\tau+1)^2/[\tau(\tau+2)]$ and $\tau = T/(mc^2)$ where $mc^2$ is the electron energy at rest, $\Phi(\xi)$ is the distortion factor which is equal to 0 for $\xi > 1$ and equal to $\left[1-\xi^{\alpha_j}\right]^{\beta_j}$ for $\xi \leq 1$ and $F_j$, $W_j$, $C_j$, $\alpha_j$ *and* $\beta_j$ are fitting parameters. $W_j$ can be interpreted as the excitation energy, $F_j$ as the optical oscillator strength, and $C_j$ as a factor characterizing the shape of the generalized oscillator strength. The excitation of two different allowed states was taken into account in the present Monte Carlo model, using the cross section parameters of Porter *et al.* [92]. Their values of 0.666 and 0.321 for $F_j$ (see Table 1 of reference [92]), however, were replaced by 0.883 and 0.425, respectively, to give better agreement with the measured $\sigma_{tot}(T)$ total scattering cross section of Grosswendt *et al.* [93] and that calculated according to (Eq.34).

The cross section parameters in the case of the excitation of Rydberg states were also taken from the publication by Porter *et al.* [92] using the method of Green and Dutta [94] to calculate $W_j$ or $F_j$, and assuming a probability of 0.5 with respect to autoionization if the excitation energy $W_j$ of a Rydberg state is greater than the lowest ionization threshold of nitrogen at 15.58 *eV.* The secondary electrons produced by autoionization were assumed to be emitted isotropically.

In the case of optically forbidden excitations, the cross section $\sigma_{exc}^{(j)}(T)$ at energy $T$ is calculated according to (Eq.58):

$$\sigma_{exc}^{(j)}(T) = 4\pi a_0^2 R^2 \frac{F_j}{W_j^2} \Phi(\xi) \xi^{\Omega_j} \qquad \textbf{\textit{(58)}}$$

where $\xi$ or $\Phi(\xi)$ have the same meaning as in (Eq.57), and $F_j$, $W_j$, $\alpha_j$, $\beta_j$ or $\Omega_j$ are parameters which are characteristic of different excitation processes. The excitation of eight possible forbidden states was taken into account in the present calculation, again using the cross section parameters of Porter *et al.* [92].

To improve the agreement between the measured total electron scattering cross sections [93] and those calculated on the basis of (Eq.34), a missing excitation contribution was fitted, as a function of electron energy, to the cross section shape of (Eq.58) and treated as an additional excitation process. The cross section parameters of the additional excitation are $W_j = 11.79\,eV$, $F_j = 57.07$, $\Omega_j = 6.159$, $\alpha_j = 3.227$ and $\beta_j = 9.731$.

## A.4 Check of the model of electron degradation in nitrogen gas

To give an overview of the data set used in the present Monte Carlo model to simulate the degradation of electrons in nitrogen, Figure 72 shows the cross section for elastic scattering, the cross sections summed with respect to discrete allowed excitations, to the excitation of Rydberg states, or to optically forbidden excitations, the total cross section for direct ionization, and the cross section with respect to autoionization, as a function of electron energy $T$. At energies greater than 100 *eV*, the very similar energy dependence of the cross sections for impact ionization or excitation is obvious, apart from the cross section for excitation to optically forbidden states.

Tabulated values of $\sigma_{tot}(T)$, $\sigma_{el}(T)$, $\sigma_{ion}(T)$ and $\sigma_{exc}^{(tot)}(T)$ is given in Table 21 (Appendix C).

A glance at the figure reveals that the high values of the elastic cross section compared to those of other interactions at energies smaller than 100 *eV* are remarkable.

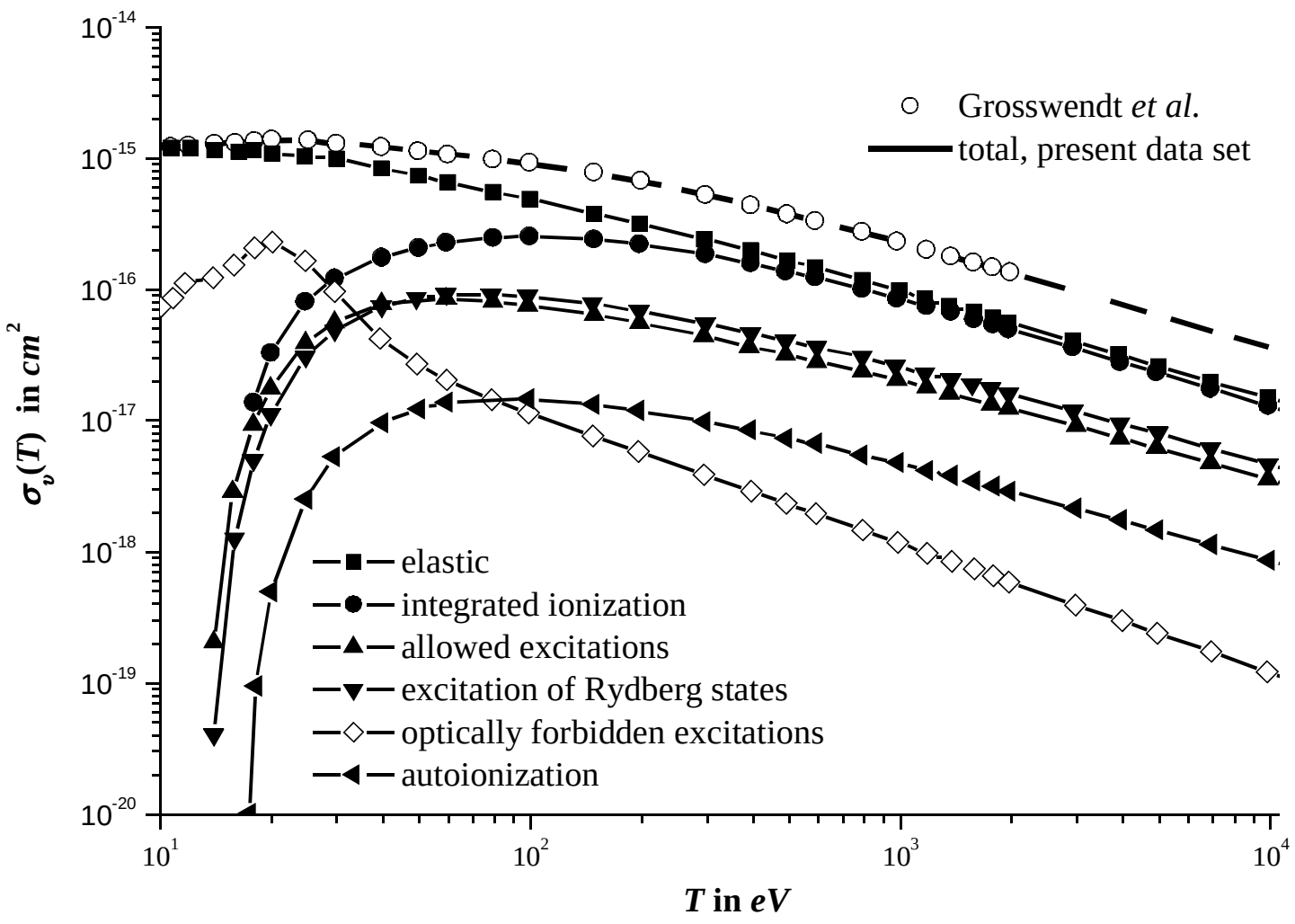

*Figure 72: Cross sections of electrons in molecular nitrogen as a function of energy T: (− −) – summed total scattering cross section using (Eq.34),present data set; (○) – measured total scattering cross section of Grosswendt et al. [93]; (■) – elastic scattering cross section; (●) – integrated ionization cross section; (▲) – summed cross sections with respect to discrete allowed excitations; (▼) – to the excitation of Rydberg states, (◇) – to optically forbidden excitations; and (◀) – cross section with respect to autoionization.*

As a first test of the cross section data set, the total scattering cross section $\sigma_{tot}(T)$ of electrons at energy $T$ in nitrogen calculated from (Eq.34) is also included in Figure 72 and can be compared with the measurements by Grosswendt *et al.* [93], which are denoted by the unfilled circles. The difference between measured or calculated cross sections is smaller than 2 % at energies greater than 15 *eV*.

For a second data check, Figure 73 shows the comparison between the mass stopping powers published by ICRU [20] and by Majeed and Strickland [95] for electrons in nitrogen and those calculated using the present set of scattering cross sections and the appropriate energy losses with respect to the different interaction processes. It can be seen at a glance that the agreement of the data in the overlapping energy region is very satisfactory.

To test the Monte Carlo model for the complete electron degradation, Figure 74 shows a comparison between W-values calculated for nitrogen and the experimental values obtained by Combecher [96] or Waibel and Grosswendt [97]. Good agreement between the calculation and experiment is obvious over the whole energy range and indicates that the simulation model for electrons should also be well suited for the calculation of ionization clusters in nitrogen.

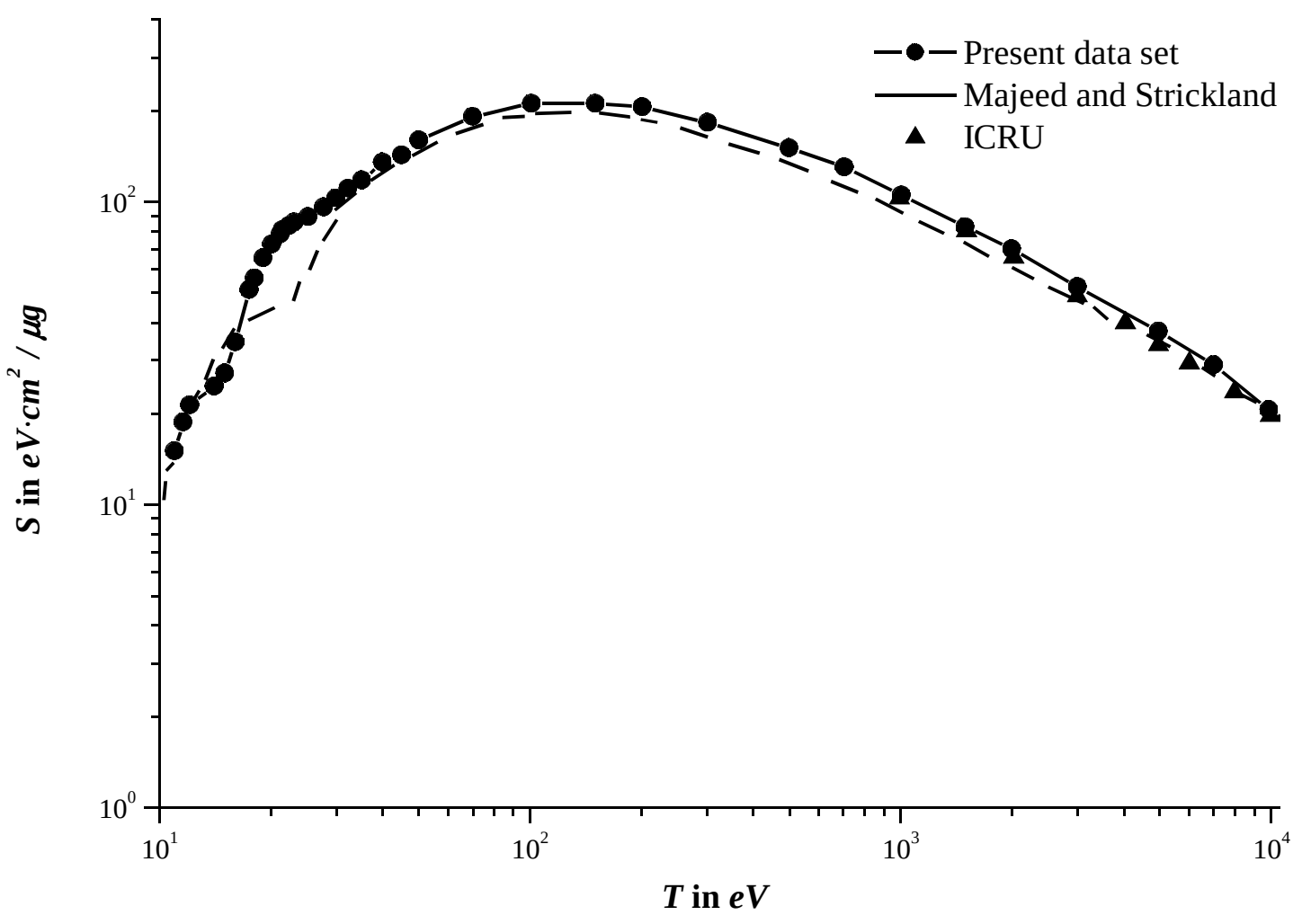

*Figure 73: Stopping power S of electrons in molecular nitrogen as a function of electron energy T: (●) – present data set, (– – ) – data of Majeed and Strickland [95], (▲) – data of ICRU [20].*

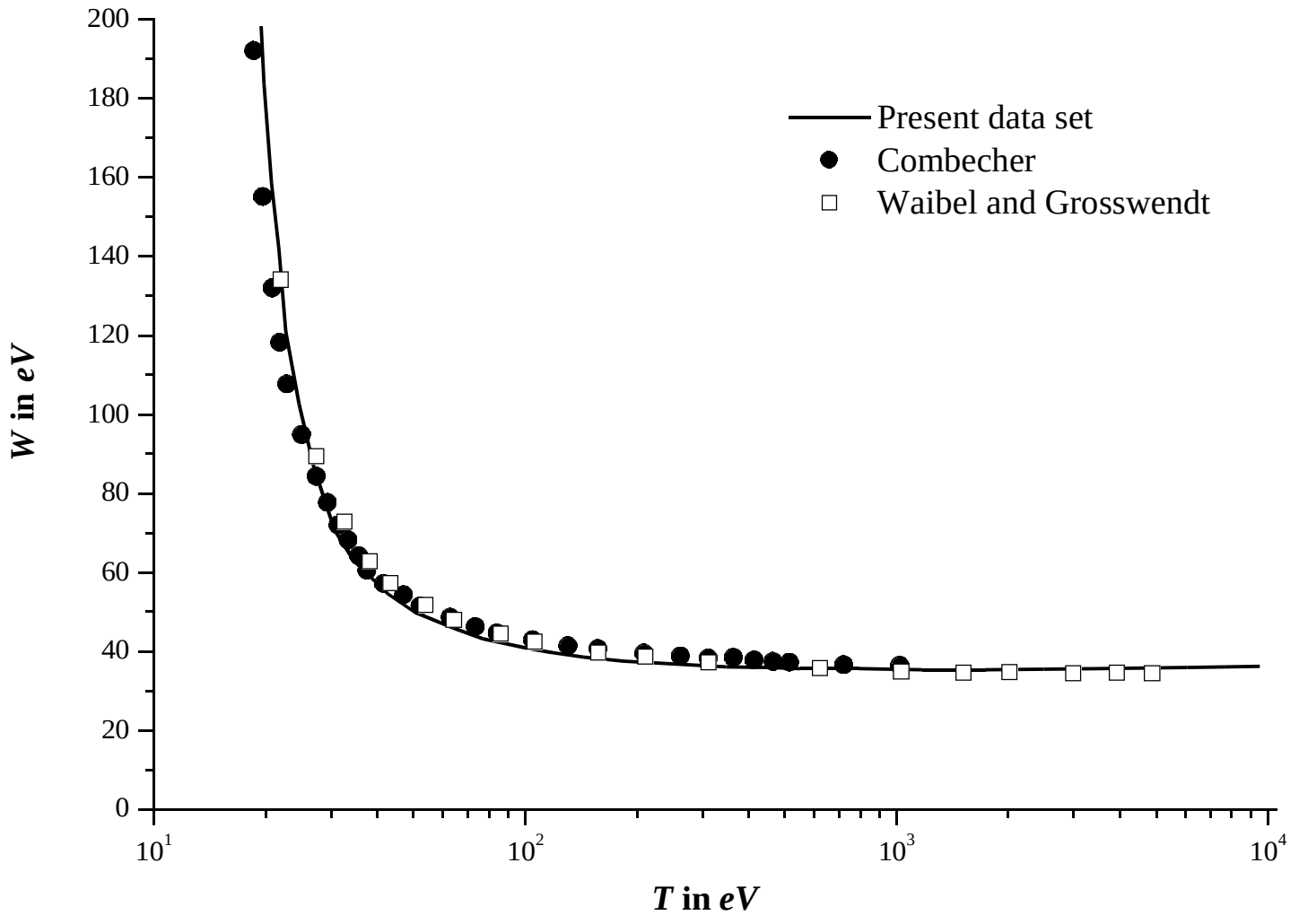

*Figure 74: Mean energy W required to produce an ion pair in the case of the complete slow-down of electrons in nitrogen as a function of energy T: (–) – Monte Carlo results, present data set, (●) – experimental data of Combecher [96], (□) – experimental data of Waibel and Grosswendt [97].*

# Appendix B Electron cross sections for propane gas

(After De Nardo *et al.* [34])

The preparation of a reliable set of electron cross sections in propane was based for the most part on the set of cross sections published by Chouki [98] which in part was derived from swarm experiments and in part from direct measurements. This data set includes the integral elastic scattering cross section, one integral ionization cross section, one discrete excitation cross section, a series of cross sections with respect to vibrational excitation, one cross section concerning dissociation, and one cross section with respect to electron attachment. These data were compared with other experimental data for elastic scattering, ionization and excitation (for details, see following subsections).

## B.1 Treatment of electron elastic scattering

Since no comprehensive experimental data on the differential elastic scattering of electrons in propane could be found in the literature, the treatment of electron elastic scattering was based on Rutherford's differential cross section $(d\sigma / d\Omega)_{el}$ with respect to the solid angle, modified to take atomic screening effects into account and calculated according to (Eq.49).

The integral atomic elastic scattering cross section $\sigma_{el}(T)$ with respect to an initial electron of kinetic energy $T$ is obtained by integration of (Eq.49) with respect to the solid angle and calculated according to (Eq.50):

As proposed by Grosswendt and Waibel [86], (Eq.50) was used to determine the screening parameter $\eta$ as a function of electron energy $T$, on the basis of integral cross sections $\sigma_{el}(T)$ derived from experiments. The polar scattering angle is then sampled conventionally, interpreting the ratio $(d\sigma / d\Omega)_{el} / \sigma_{el}(T)$ as the probability density with respect to $\vartheta$. This procedure is a satisfactory approximation of differential elastic scattering at energies greater than about 200 *eV*; at smaller energies, however, large angle scattering is greatly underestimated. As a result, an additional correction factor was applied at lower electron energies.

As directly measured values for the integral elastic cross section are also not available in the energy range of interest, $\sigma_{el}(T)$ was determined on the basis of (Eq.34) using the total scattering cross section $\sigma_{tot}(T)$ measured by Grosswendt *et al.* [93] and those for impact ionization or excitation described in the following subsections. To give an impression of the data obtained in this way, Figure 75 shows $\sigma_{el}(T)$ in propane as a function of electron energy $T$ in comparison with the comparable data of Boesten *et al.* [99] and Mark *et al.* [100], and with data calculated using the theoretical cross sections of Mayol and Salvat [101] for hydrogen and carbon atoms, assuming that the independent atom model is valid. As can be seen from the figure, the agreement between the present cross sections and those from Boesten *et al.* and Märk *et al.* is quite satisfactory. What is very significant is the resonance structure around 7 *eV* and the strong decrease in $\sigma_{el}(T)$ as a function of energy $T$.

## B.2 Treatment of impact ionization

Since only a few experimental data are available at present for the partial ionization cross sections of electrons in propane, we decided to calculate the ionization part of our Monte Carlo simulation of electron histories almost exclusively on the basis of the single integral ionization cross section used by Chouki [98] in his analysis of swarm data. This means that the sum of partial ionization cross sections of (Eq.34) is replaced by a single cross section.

To enhance the agreement between the ionization cross section and the measurements of Durić *et al.* [102] and Nishimura and Tawara [103] close to the ionization threshold, we repeated the fitting procedure of Chouki using the same semi-empirical function but in a somewhat restricted energy range (up to 10 *keV* instead of up to 2.7 *MeV*), as higher energies are of no importance for the present work. This fitting function, which is given by (Eq.59), shows a high-energy dependence that is consistent with the Bethe theory.

$$\sigma_{ion}(T) = 4\pi a_0^2 \frac{c_1}{(T/R)} \ln\left(1+\frac{T-I}{R}\right)\left[e^{-\frac{c_2}{(T/R)}} + c_3 e^{-\frac{c_4}{(T/R)^2}} + c_5 e^{-\frac{(T-I)c_6}{R(T/R)^2}}\right] \qquad (59)$$

Here, $a_0$ is the Bohr radius, $R = 13.61\,eV$ the Rydberg constant, $I = 11.08\,eV$ the lowest ionization threshold of propane, and $c_\nu$, $\nu = 1, 2..., 6$ are dimensionless fitting parameters. The resulting values for these six parameters are: 16.01, 3.369, -59.96, $8.858 \cdot 10^{-4}$, 59.84, and $-9.295 \cdot 10^{-3}$. Figure 76 shows the ionization cross section $\sigma_{ion}(T)$ given by (Eq.59) and the new fitting parameters $c_\nu$ as a function of electron energy $T$, compared with the data of Chouki [98] and those of Duric *et al.* [102], Nishimura and Tawara [103], Grill *et al.* [104], Hayashi [105], Kebarle and Godbole [106], and Schram *et al.* [107]. The steep increase in $\sigma_{ion}(T)$ with increasing energy near the ionization threshold, the maximum at around 80 *eV*, and the decrease at higher energies is typical of the ionization of atoms and molecules induced by electrons.

The energy distribution of secondary electrons emitted after electron impact ionization was determined on the basis of a single-differential cross section $d\sigma(T)d\tau$ with respect to the electron kinetic energy $\tau$ expressed by the Breit-Wigner formula, as proposed by Green and Sawada [89]. The integration of this formula over $\tau$ leads to a simple equation which can be easily solved for the upper integration limit and thus makes the Monte Carlo simulation very convenient. For the details of the sampling procedure for $\tau$, see Grosswendt and Waibel [86] or Grosswendt and Pszona [64]. The maximum energy $\tau_{\max}$ of the secondaries is assumed to be equal to $(T-I)/2$ where $I = 11.08\,eV$ is again the lowest ionization threshold energy. This assumption for $\tau_{\max}$ reflects the convention that the faster electron after impact ionization is the primary one. Since the necessary parameters of the Breit-Wigner form of the single-differential cross section $d\sigma(T)d\tau$ are lacking for propane, we applied the parameters of Green and Sawada [89] for methane. The errors induced by this procedure due to the wrong shape of the energy distribution, for slow electrons in particular, and the non-ideal behaviour at high energies can be assumed to be acceptable for most applications.

The energy $T$ of a primary electron after impact ionization is set equal to $T-\tau-I_{thr}(T)$ where $I_{thr}(T)$ is the ionization threshold energy used at a specified electron energy $T$. This ionization threshold was not fixed at 11.08 *eV* but was assumed to depend on the electron energy $T$, to take into account, at least approximately, the contribution of subshells with binding energies greater than the lowest ionization threshold which can contribute to the electron degradation only if the electron energy is high enough. To determine $I_{thr}(T)$, it was set equal to the average binding energy of the weakly bound valence electrons of propane which was calculated as a function of electron energy on the basis of the electron ionization cross sections of Hwang *et al.* [69] for the different subshells. For practical reasons of Monte Carlo simulation, the calculated values of $I_{thr}(T)$ were fitted afterwards to the function given by (Eq.60) which is valid for $T > 12.95\,eV$. At lower energies, $I_{thr}(T)$ is set equal to 11.08 *eV*.

$$I_{thr}(T) = c_1 + c_2\left(\frac{T}{R}\right)^{c_3} \tag{60}$$

Here *R* is the Rydberg constant and the three fitting parameters $c_\nu$, $\nu = 1, 2$ and $3$, are equal to 15.93 *eV*, -4.613 *eV* and -1.014, respectively.

The last aspect of the ionization impact is the determination of the flight direction of the initial electron after scattering and of that of the liberated secondary electron. As appropriate data are lacking, this is performed in an approximate way using the following kinematic equations which were proposed by Berger [91] on the basis of conservation of momentum and energy:

$$\sin^2\vartheta_p = \frac{\frac{\tau}{T}}{\left(1-\frac{\tau}{T}\right)\frac{T}{2mc^2}+1} \tag{61}$$

$$\sin^2\vartheta_s = \frac{1-\frac{\tau}{T}}{1+\frac{\tau}{2mc^2}} \tag{62}$$

Here, $\vartheta_p$ – is the polar scattering angle of the initial electron after ionization impact relative to its initial direction and

$\vartheta_s$ – is the corresponding polar angle of the secondary electron,

$T$ and $\tau$ – are the kinetic energies of the initial or secondary electron, and

$mc^2$ – is the electron rest energy.

The azimuthal angle $\varphi_p$ of the initial electron after scattering is assumed to be uniformly distributed between 0 and $2\pi$ and that of the secondary electron is set equal to $\varphi_s = \varphi_p - \pi$.

A comparison of $\vartheta_s$ with the angular distributions of secondary electrons measured by Opal *et al.* [90] for different atomic or molecular targets shows that (Eq.62) is a satisfactory approximation of the measurements, at least at energies greater than about 200 $eV$, whereas it is unsatisfactory at smaller energies. The following assumptions are, therefore, made which are more consistent with the experimental data than (Eq.61) and (Eq.62):

1. secondary electrons at energies smaller than 50 *eV* are emitted isotropically;
2. in the energy range between 50 *eV* and 200 *eV*, 90 *%* of the secondaries are emitted in the angular range between 45° and 90° whereas the remaining ones are emitted isotropically;
3. the scattering angle $\vartheta_p$ of primary electrons at energies greater than 100 *eV* after an ionization event is given by (Eq.61) whereas $\vartheta_p$ is uniformly distributed between 0° and 45° at smaller energies.

## B.3 Treatment of impact excitation

As mentioned above, the treatment of excitation processes in propane was based for the most part on the data set of Chouki [98], which contains one discrete excitation cross section with a threshold at 9.13 *eV*, a series of cross sections with respect to vibrational excitation, one cross section for molecular dissociation, and one with respect to electron attachment. These excitation cross sections were used in the Monte Carlo simulation but for practical reasons were fitted to empirical functions.

Since Chouki's cross section with respect to discrete excitation shows an energy dependence similar to that of impact ionization [see (Eq.59)], we applied a similar function to represent discrete excitation:

$$\sigma_{exc}(T) = 4\pi a_0^2 \frac{c_1}{T/R} \ln\left(1+\frac{T-c_2}{R}\right)\left[e^{-\frac{c_3}{(T/R)}} + c_4 e^{-\frac{c_5(T-c_2)}{R(T/R)^2}}\right] \qquad \textbf{(63)}$$

Here, the meaning of the different parameters is the same as for those in (Eq.59) except that the constants $c_\nu$ for $\nu = 1, 2...,5$ are now equal to 2.772, 8.938 *eV*, 0.1407, 0.5165 and 4.587.

Chouki's data for electron attachment, vibrational excitation and molecular dissociation were fitted to the formula given by (Eq.64) which was recommended, for instance, by Jackman *et al.* [108] for either forbidden or allowed excitations.

$$\sigma_{exc}^{(j)} = 4\pi a_0^2 R^2 \frac{f_j}{W_j^2}\left[1-\xi^{A_j}\right]^{B_j} \xi^{\Omega_j} \qquad \textbf{(64)}$$

where ζ=$W_j$/T and $f_j$, $W_j$, $A_j$, $B_j$ or $\Omega_j$ are parameters which are characteristic of different excitation processes; the other quantities are those of (Eq.59). For the parameters, see Table 22. It should be noted that, in addition to Chouki's data for impact excitation, another two excitation processes were added to explain the shape of the total cross section of Grosswendt *et al.* [93] close to 25 *eV* and 45 *eV*.

To give an impression of the excitation cross sections taken into account in our Monte Carlo model, Figure 77 shows $\sigma_{exc}(T)$ according to (Eq.63), and $\sigma_{exc}^{(j)}(T)$ according to (Eq.64) and the parameters of Table 22, as a function of electron energy $T$. Bearing in mind that the lower energy limit of the Monte Carlo simulations is 10 *eV*, it is obvious from the figure that:

1. only the higher-energy contributions of the three vibrational excitations ($j = 2, 4$ and $6$ of Table 22) are of importance,
2. the contribution of dissociation to electron degradation is of the order of only 1 *%*,
3. electron attachment can be completely neglected, and (iv) the energy loss by excitation is dominated by the single cross section of Chouki [98] with respect to discrete excitation, somewhat modified by the two contributions ($j = 9$ and $10$ of Table 22) which had to be added to fit the experimental total scattering cross section as mentioned above.

## B.4 Check of the model of electron degradation in propane gas

Figure 78 shows an overview of the electron cross sections used in our present Monte Carlo model for propane where, to avoid confusion, the cross sections with respect to excitation are bundled. At energies greater than 100 *eV*, the very similar energy dependence of the cross sections with respect to excitation and to impact ionization is obvious. As the figure shows, the high values of the elastic cross section compared with those of other interaction effects at energies smaller than 100 *eV* are remarkable. To perform a first check of the validity of our Monte Carlo model, the total scattering cross sections of electrons in propane calculated from (Eq.34) and represented by the unbroken curve are compared with the measured data of Grosswendt *et al.* [93] which are characterized by the open circles. The deviations between measured and calculated total cross sections at energies greater than 10 *eV* are smaller than 2.5 *%*. For a second data check, we compared the mass stopping powers published by ICRU [20] for electrons at energies smaller than 5 *keV* in propane with those calculated on the basis of the present set of scattering cross sections and energy losses taken into account in the Monte Carlo model. At 5 *keV*, for instance, the calculated mass stopping power is equal to 43.1 $MeV \cdot cm^2/g$, very close to the tentative value of 45.2 $MeV \cdot cm^2/g$ published by ICRU.

Tabulated values of $\sigma_{tot}(T)$, $\sigma_{el}(T)$, $\sigma_{ion}(T)$ and $\sigma_{exc}{}^{(tot)}(T)$ is given in Table 23 (Appendix C).

In addition to these direct data checks, we performed a Monte Carlo simulation of the complete electron slow down in propane and calculated the $W$-values of electrons at energies up to 5 *keV*. The results of these simulations showed the typical strong decrease of $W$ with increasing electron energy in the low-energy region up to about 100 *eV*, followed by convergence to a constant value of $W$ if the electron energy is further increased. At electron energies of 30 *eV*, 100 *eV*, and 200 *eV*, for instance, the $W$-value in propane is equal to 36.07 *eV*, 27.37 *eV* or 25.99 *eV* according to our Monte Carlo simulation. These data agree within 5 *%* with the corresponding experimental values of Combecher [96]. A similar satisfactory agreement between the results of the Monte Carlo simulation and measurements was also found at higher energies. At electron energies of 1 *keV* and 5 *keV*, for instance, the calculated $W$-values are equal to 25.09 *eV* or 25.06 *eV* and also agree within 5 % with the value of 25.9 *eV* measured by Krajcar Bronic *et al.* [109] at 1.2 *keV* and with the high-energy value of 24 *eV* recommended by ICRU [36].

The very satisfactory outcome of these comparisons indicates that the present Monte Carlo model should also be well suited for track structure calculations.

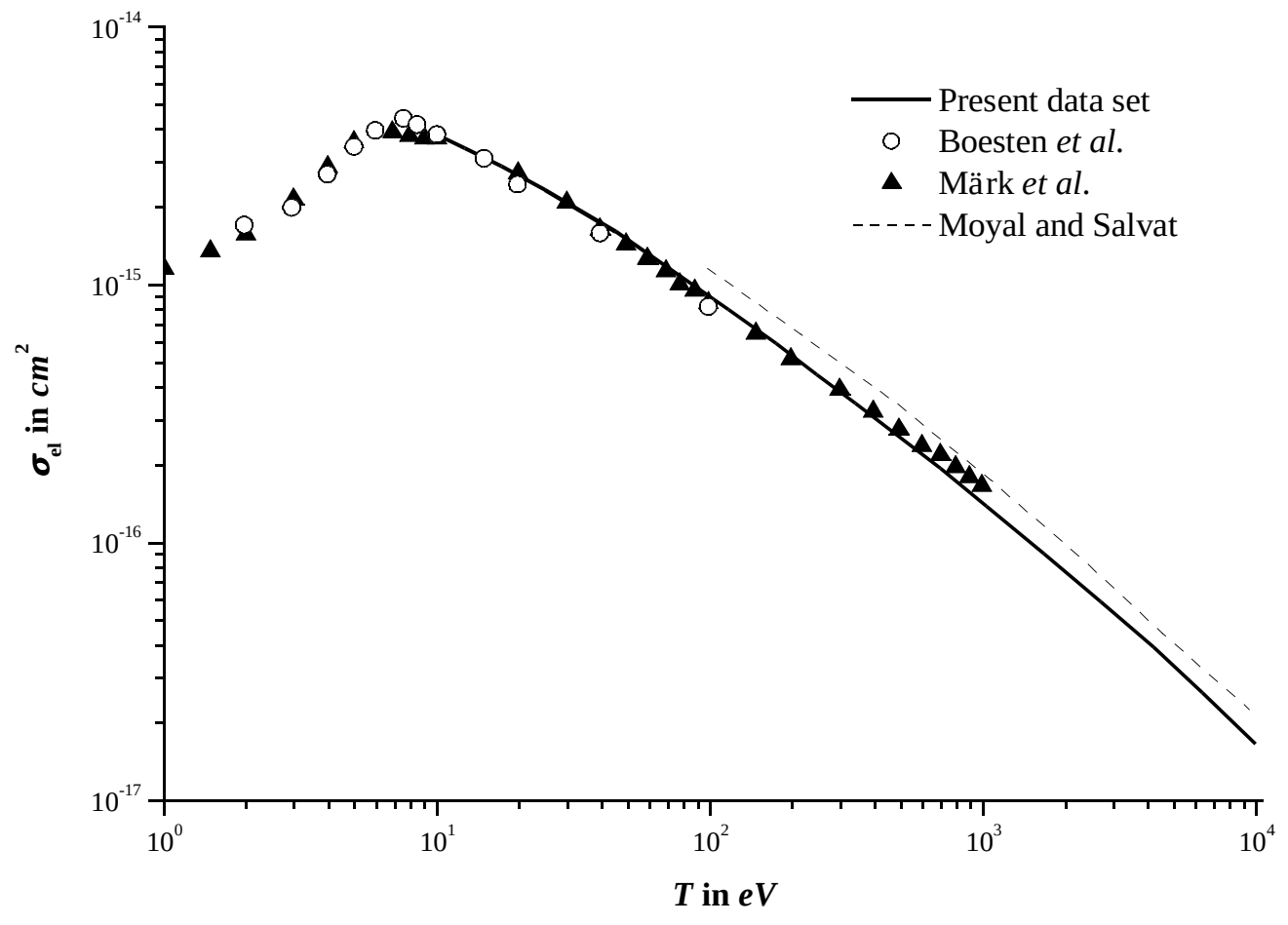

*Figure 75: Integral elastic scattering cross section $\sigma_{el}(T)$ of electrons in propane gas as a function of energy $T$ : (−) – represents values at energies greater than 10 eV, present data set, (○) – the data of Boesten et al. [99], (▲) – the data of Märk et al. [100], (---) – the calculated data using the theoretical cross sections of Mayol and Salvat [101] for hydrogen and carbon atoms within the framework of the independent atom model).*

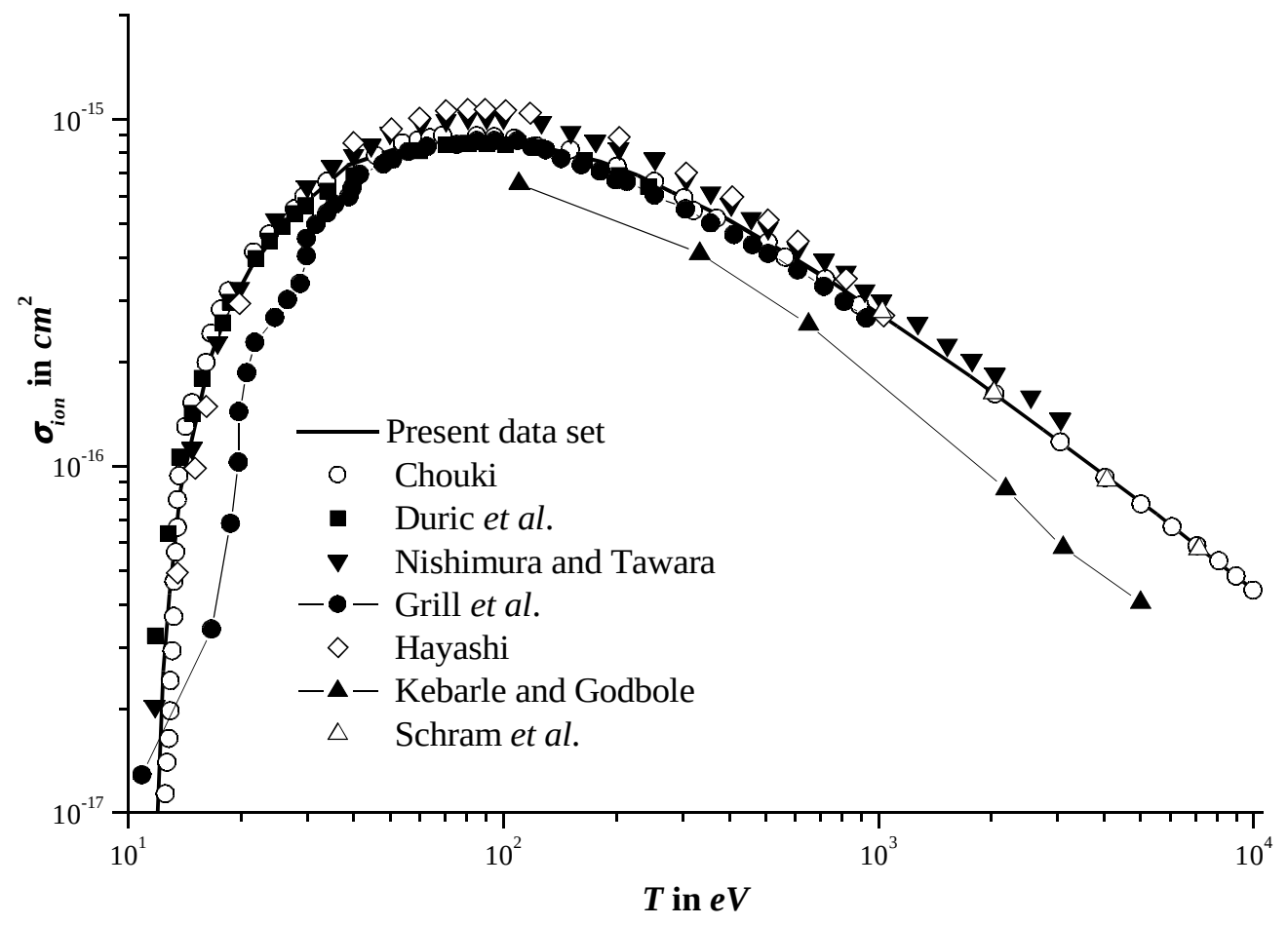

*Figure 76: Integral ionization cross section $\sigma_{ion}(T)$ of electrons in propane gas as a function of energy $T$: (−) – represents the present data set, (○) – the data of Chouki 98, (■) – Duric et al. [102], (▼) – Nishimura and Tawara [103], (●) – Grill et al. [104], (◇) – Hayashi [105], (▲) – Kebarle and Godbole [106], (Δ) – Schram et al. [107]).*

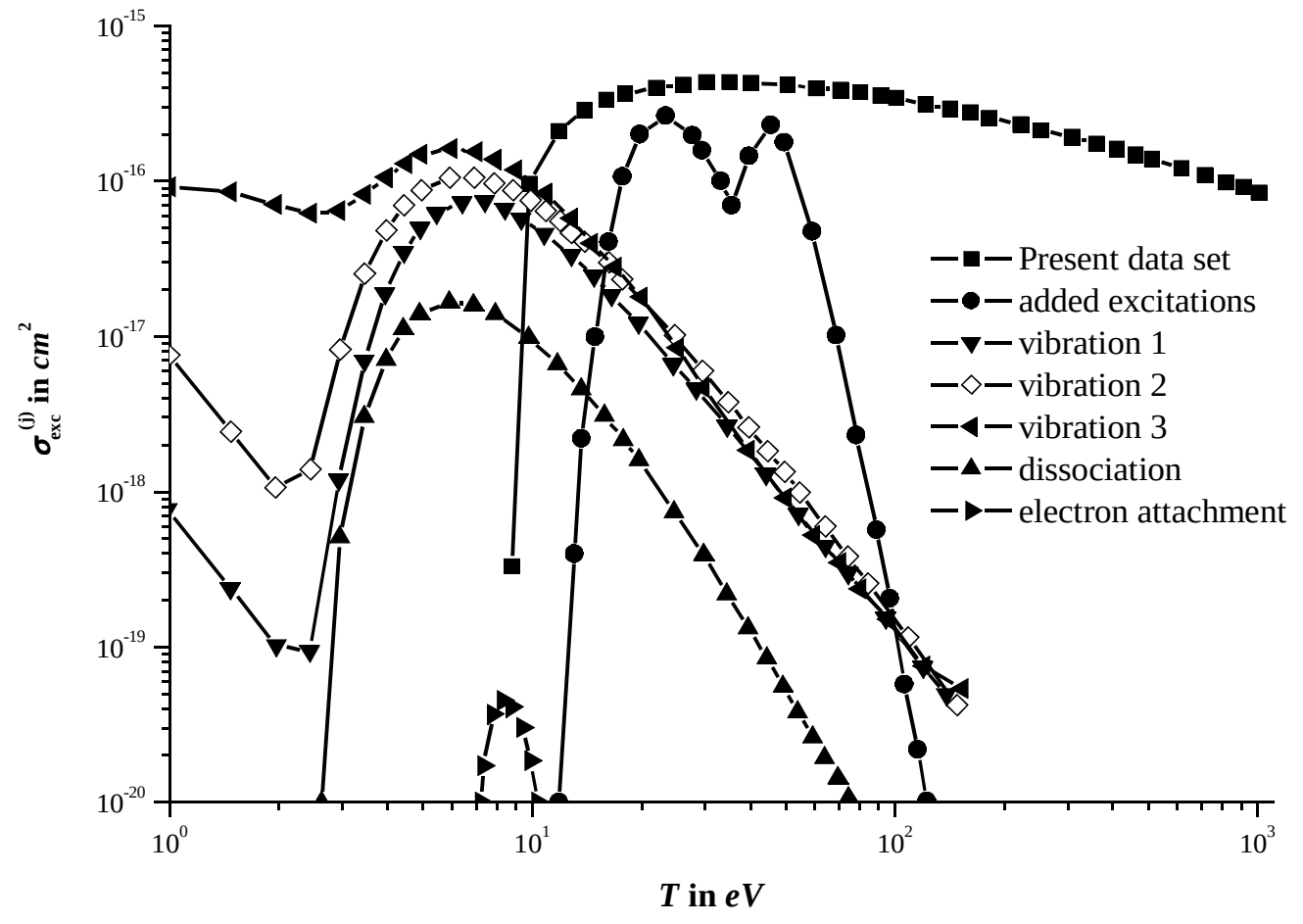

*Figure 77: Fig. 12 Overview of excitation cross sections used in the present Monte Carlo model: (▪) – cross section for discrete excitation $\sigma_{exc}(T)$ according to (Eq.63), present data set, (●) – added excitations $\sigma_{exc}^{(j)}(T)$, $j=9+10$ of Table 22, according to (Eq.64) in order to fit the measured total scattering cross section of Grosswendt et al. [93] using (Eq.60), (▼) – vibration 1, $j=1+2$; (◇) – vibration 2, $j=3+4$; (◀) – vibration 3, $j=5+6$; (▲) – dissociation, $j=7$; and (▶) – electron attachment, $j=8$).*

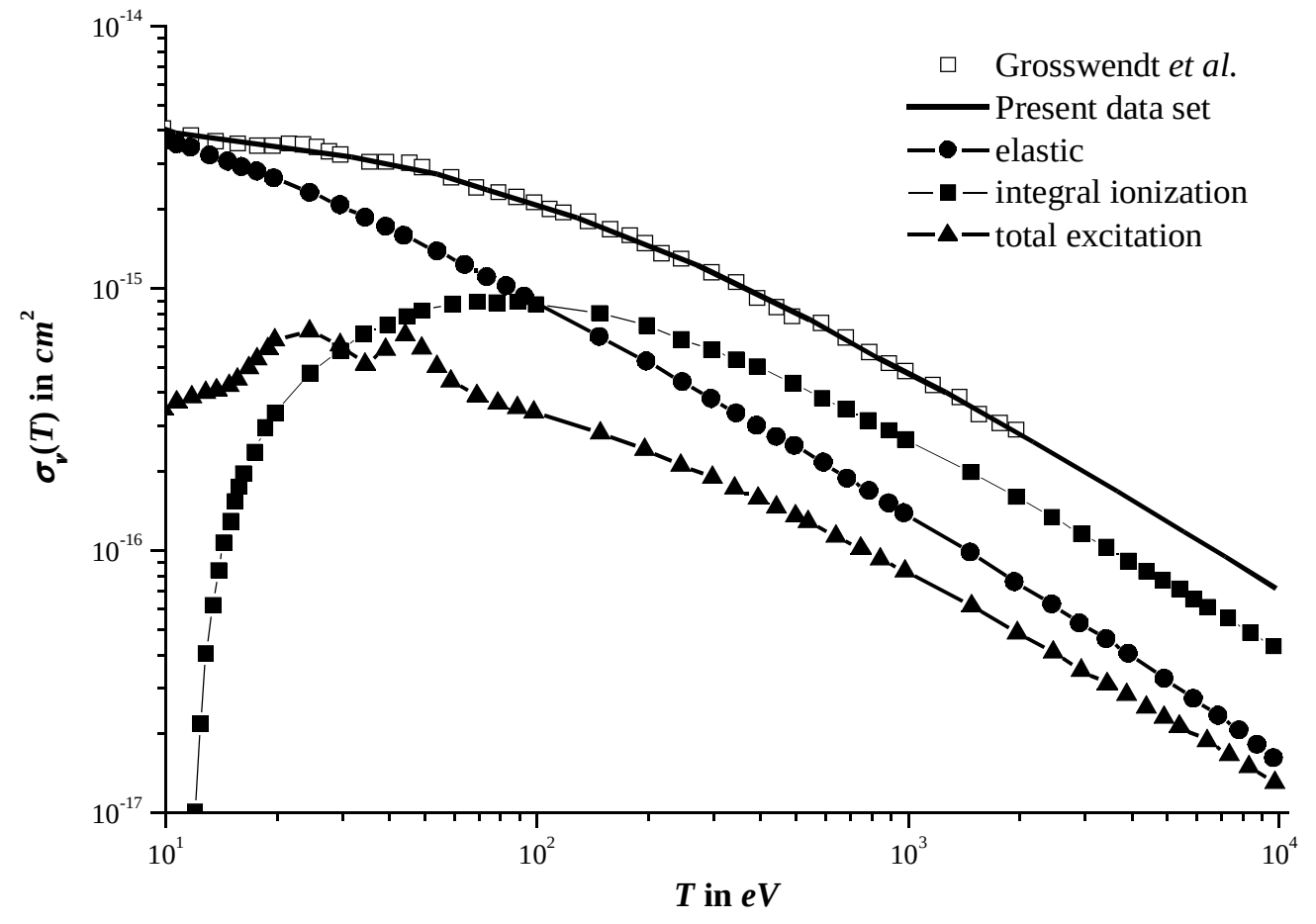

*Figure 78: Set of scattering cross section $\sigma_v(T)$ of electrons in propane gas as a function of energy $T$: (□) – total scattering cross section $\sigma_{tot}(T)$ measured by Grosswendt et al. [93]; (─) – calculated total scattering cross section $\sigma_{tot}(T)$ using (Eq.34), present data set; (●) – elastic scattering cross section $\sigma_{el}(T)$; (■) – integral ionization cross section $\sigma_{ion}(T)$; (▲) – total excitation cross section $\sigma_{exc}^{(tot)}(T)$.*

# Appendix C Tables

*Table 1:* ***Total scattering cross section for electrons $\sigma_{tot}(T)$ in molecular nitrogen and propane gas. Ref.[56].***

| Molecular nitrogen | |
|---|---|
| Electron energy *T*, *eV* | $\sigma_{tot}(T)$, $10^{-16}$ $cm^2$ |
| 20 | 13.74 |
| 30 | 12.77 |
| 40 | 11.97 |
| 50 | 11.16 |
| 60 | 10.59 |
| 80 | 9.54 |
| 100 | 8.83 |
| 150 | 7.37 |
| 200 | 6.49 |
| 300 | 5.08 |
| 400 | 4.37 |
| 500 | 3.74 |
| 600 | 3.26 |
| 800 | 2.67 |
| 1000 | 2.19 |
| 1200 | 1.91 |
| 1400 | 1.71 |
| 1600 | 1.58 |
| 1800 | 1.44 |
| 2000 | 1.39 |

| Propane gas | |
|---|---|
| Electron energy *T*, *eV* | $\sigma_{tot}(T)$, $10^{-16}$ $cm^2$ |
| 20 | 35.72 |
| 29.98 | 33.04 |
| 39.9 | 31.23 |
| 50 | 29.68 |
| 60 | 27.16 |
| 80 | 23.75 |
| 100.1 | 21.81 |
| 120 | 19.98 |
| 140.1 | 18.47 |
| 160.1 | 17.26 |
| 200 | 15.32 |
| 250 | 13.40 |
| 299.8 | 11.95 |
| 350.1 | 10.95 |
| 400.2 | 9.52 |
| 499.8 | 8.12 |
| 600 | 7.62 |
| 700 | 6.72 |
| 800 | 5.91 |
| 1000 | 4.97 |
| 1200 | 4.41 |
| 1400 | 3.98 |
| 1600 | 3.45 |
| 1800 | 3.17 |
| 2000 | 2.99 |

*Table 2:* ***Transmission of mono-energetic electrons (100 eV – 2000 eV) through molecular nitrogen vs. area density*** $D\rho$.

| Transmission | Electron energy, *eV* | | | | | |
|---|---|---|---|---|---|---|
| Area density $D\rho,\ \mu g/cm^2$ | 100 | 200 | 300 | 500 | 1000 | 2000 |
| 0.01 | 8.238E-01 | 8.676E-01 | 8.943E-01 | 9.229E-01 | 9.523E-01 | 9.718E-01 |
| 0.02 | 6.786E-01 | 7.527E-01 | 7.997E-01 | 8.518E-01 | 9.069E-01 | 9.444E-01 |
| 0.03 | 5.590E-01 | 6.530E-01 | 7.151E-01 | 7.861E-01 | 8.636E-01 | 9.178E-01 |
| 0.04 | 4.605E-01 | 5.665E-01 | 6.395E-01 | 7.255E-01 | 8.224E-01 | 8.919E-01 |
| 0.05 | 3.793E-01 | 4.915E-01 | 5.719E-01 | 6.696E-01 | 7.832E-01 | 8.668E-01 |
| 0.06 | 3.125E-01 | 4.264E-01 | 5.114E-01 | 6.180E-01 | 7.458E-01 | 8.423E-01 |
| 0.07 | 2.574E-01 | 3.700E-01 | 4.573E-01 | 5.704E-01 | 7.103E-01 | 8.186E-01 |
| 0.08 | 2.120E-01 | 3.210E-01 | 4.090E-01 | 5.264E-01 | 6.764E-01 | 7.955E-01 |
| 0.09 | 1.747E-01 | 2.785E-01 | 3.657E-01 | 4.858E-01 | 6.441E-01 | 7.731E-01 |
| 0.10 | 1.439E-01 | 2.416E-01 | 3.271E-01 | 4.484E-01 | 6.134E-01 | 7.513E-01 |
| 0.12 | 9.763E-02 | 1.818E-01 | 2.615E-01 | 3.819E-01 | 5.563E-01 | 7.095E-01 |
| 0.14 | 6.625E-02 | 1.369E-01 | 2.092E-01 | 3.253E-01 | 5.045E-01 | 6.701E-01 |
| 0.16 | 4.495E-02 | 1.030E-01 | 1.673E-01 | 2.771E-01 | 4.575E-01 | 6.329E-01 |
| 0.18 | 3.050E-02 | 7.755E-02 | 1.338E-01 | 2.360E-01 | 4.149E-01 | 5.977E-01 |
| 0.20 | 2.070E-02 | 5.837E-02 | 1.070E-01 | 2.010E-01 | 3.763E-01 | 5.645E-01 |
| 0.22 | 1.405E-02 | 4.393E-02 | 8.554E-02 | 1.712E-01 | 3.412E-01 | 5.331E-01 |
| 0.24 | 9.531E-03 | 3.307E-02 | 6.841E-02 | 1.459E-01 | 3.094E-01 | 5.035E-01 |
| 0.26 | 6.467E-03 | 2.489E-02 | 5.470E-02 | 1.242E-01 | 2.806E-01 | 4.755E-01 |
| 0.28 | 4.389E-03 | 1.873E-02 | 4.375E-02 | 1.058E-01 | 2.545E-01 | 4.490E-01 |
| 0.30 | 2.978E-03 | 1.410E-02 | 3.498E-02 | 9.014E-02 | 2.308E-01 | 4.241E-01 |
| 0.32 | 2.021E-03 | 1.061E-02 | 2.798E-02 | 7.678E-02 | 2.093E-01 | 4.005E-01 |
| 0.34 | 1.371E-03 | 7.989E-03 | 2.237E-02 | 6.540E-02 | 1.898E-01 | 3.783E-01 |
| 0.36 | 9.305E-04 | 6.013E-03 | 1.789E-02 | 5.571E-02 | 1.721E-01 | 3.572E-01 |
| 0.38 | 6.314E-04 | 4.526E-03 | 1.431E-02 | 4.745E-02 | 1.561E-01 | 3.374E-01 |
| 0.40 | 4.284E-04 | 3.407E-03 | 1.144E-02 | 4.042E-02 | 1.416E-01 | 3.186E-01 |
| 0.45 | 1.625E-04 | 1.675E-03 | 6.543E-03 | 2.706E-02 | 1.109E-01 | 2.762E-01 |
| 0.50 | 6.164E-05 | 8.231E-04 | 3.742E-03 | 1.812E-02 | 8.684E-02 | 2.394E-01 |
| 0.55 | 2.338E-05 | 4.046E-04 | 2.140E-03 | 1.214E-02 | 6.801E-02 | 2.075E-01 |
| 0.60 | 8.868E-06 | 1.988E-04 | 1.224E-03 | 8.126E-03 | 5.327E-02 | 1.798E-01 |
| 0.65 | 3.364E-06 | 9.774E-05 | 6.999E-04 | 5.441E-03 | 4.172E-02 | 1.559E-01 |
| 0.70 | 1.276E-06 | 4.804E-05 | 4.003E-04 | 3.643E-03 | 3.268E-02 | 1.351E-01 |
| 0.75 | 4.840E-07 | 2.361E-05 | 2.289E-04 | 2.440E-03 | 2.559E-02 | 1.171E-01 |
| 0.80 | 1.836E-07 | 1.161E-05 | 1.309E-04 | 1.634E-03 | 2.004E-02 | 1.015E-01 |
| 0.90 | 2.641E-08 | 2.804E-06 | 4.282E-05 | 7.325E-04 | 1.229E-02 | 7.627E-02 |
| 1.00 | 3.800E-09 | 6.774E-07 | 1.400E-05 | 3.284E-04 | 7.541E-03 | 5.730E-02 |
| 1.10 | 5.467E-10 | 1.637E-07 | 4.580E-06 | 1.473E-04 | 4.626E-03 | 4.305E-02 |
| 1.20 | 7.865E-11 | 3.954E-08 | 1.498E-06 | 6.603E-05 | 2.838E-03 | 3.235E-02 |
| 1.30 | 1.132E-11 | 9.553E-09 | 4.899E-07 | 2.961E-05 | 1.741E-03 | 2.430E-02 |
| 1.40 | 1.628E-12 | 2.308E-09 | 1.602E-07 | 1.327E-05 | 1.068E-03 | 1.826E-02 |
| 1.50 | 2.342E-13 | 5.576E-10 | 5.240E-08 | 5.952E-06 | 6.549E-04 | 1.372E-02 |
| 1.60 | 3.370E-14 | 1.347E-10 | 1.714E-08 | 2.669E-06 | 4.017E-04 | 1.031E-02 |
| 1.80 | 6.975E-16 | 7.862E-12 | 1.833E-09 | 5.365E-07 | 1.512E-04 | 5.817E-03 |
| 2.00 | 1.444E-17 | 4.589E-13 | 1.961E-10 | 1.079E-07 | 5.687E-05 | 3.284E-03 |
| 2.20 | 2.988E-19 | 2.679E-14 | 2.098E-11 | 2.169E-08 | 2.140E-05 | 1.853E-03 |
| 2.40 | 6.186E-21 | 1.563E-15 | 2.244E-12 | 4.360E-09 | 8.052E-06 | 1.046E-03 |
| 2.60 | 1.280E-22 | 9.125E-17 | 2.400E-13 | 8.765E-10 | 3.030E-06 | 5.906E-04 |
| 2.80 | 2.650E-24 | 5.326E-18 | 2.567E-14 | 1.762E-10 | 1.140E-06 | 3.333E-04 |
| 3.00 | 5.486E-26 | 3.109E-19 | 2.746E-15 | 3.543E-11 | 4.289E-07 | 1.882E-04 |

*Table 3:* ***Transmission of mono-energetic electrons (100 eV – 2000 eV) through propane gas vs. area density*** $D\rho$**.**

| Transmission | Electron energy, *eV* | | | | | |
|---|---|---|---|---|---|---|
| Area density $D\rho,\ \mu g / cm^2$ | 100 | 200 | 300 | 500 | 1000 | 2000 |
| 0.01 | 8.836E-01 | 9.134E-01 | 9.312E-01 | 9.501E-01 | 9.693E-01 | 9.819E-01 |
| 0.02 | 7.808E-01 | 8.342E-01 | 8.671E-01 | 9.027E-01 | 9.395E-01 | 9.642E-01 |
| 0.03 | 6.900E-01 | 7.619E-01 | 8.074E-01 | 8.577E-01 | 9.107E-01 | 9.467E-01 |
| 0.04 | 6.097E-01 | 6.959E-01 | 7.519E-01 | 8.149E-01 | 8.827E-01 | 9.296E-01 |
| 0.05 | 5.388E-01 | 6.356E-01 | 7.001E-01 | 7.742E-01 | 8.556E-01 | 9.128E-01 |
| 0.06 | 4.761E-01 | 5.806E-01 | 6.519E-01 | 7.356E-01 | 8.294E-01 | 8.963E-01 |
| 0.07 | 4.207E-01 | 5.303E-01 | 6.071E-01 | 6.989E-01 | 8.039E-01 | 8.801E-01 |
| 0.08 | 3.717E-01 | 4.843E-01 | 5.653E-01 | 6.640E-01 | 7.792E-01 | 8.642E-01 |
| 0.09 | 3.285E-01 | 4.424E-01 | 5.264E-01 | 6.309E-01 | 7.553E-01 | 8.486E-01 |
| 0.10 | 2.903E-01 | 4.040E-01 | 4.901E-01 | 5.994E-01 | 7.321E-01 | 8.332E-01 |
| 0.12 | 2.266E-01 | 3.370E-01 | 4.250E-01 | 5.411E-01 | 6.879E-01 | 8.034E-01 |
| 0.14 | 1.770E-01 | 2.812E-01 | 3.685E-01 | 4.885E-01 | 6.463E-01 | 7.746E-01 |
| 0.16 | 1.382E-01 | 2.346E-01 | 3.195E-01 | 4.410E-01 | 6.072E-01 | 7.469E-01 |
| 0.18 | 1.079E-01 | 1.957E-01 | 2.771E-01 | 3.981E-01 | 5.705E-01 | 7.201E-01 |
| 0.20 | 8.425E-02 | 1.632E-01 | 2.402E-01 | 3.593E-01 | 5.360E-01 | 6.943E-01 |
| 0.22 | 6.579E-02 | 1.362E-01 | 2.083E-01 | 3.244E-01 | 5.036E-01 | 6.694E-01 |
| 0.24 | 5.137E-02 | 1.136E-01 | 1.806E-01 | 2.928E-01 | 4.731E-01 | 6.454E-01 |
| 0.26 | 4.011E-02 | 9.477E-02 | 1.566E-01 | 2.643E-01 | 4.445E-01 | 6.223E-01 |
| 0.28 | 3.132E-02 | 7.906E-02 | 1.358E-01 | 2.386E-01 | 4.177E-01 | 6.000E-01 |
| 0.30 | 2.445E-02 | 6.595E-02 | 1.178E-01 | 2.154E-01 | 3.924E-01 | 5.785E-01 |
| 0.32 | 1.909E-02 | 5.502E-02 | 1.021E-01 | 1.944E-01 | 3.687E-01 | 5.578E-01 |
| 0.34 | 1.491E-02 | 4.590E-02 | 8.854E-02 | 1.755E-01 | 3.464E-01 | 5.378E-01 |
| 0.36 | 1.164E-02 | 3.829E-02 | 7.677E-02 | 1.585E-01 | 3.255E-01 | 5.185E-01 |
| 0.38 | 9.090E-03 | 3.194E-02 | 6.657E-02 | 1.430E-01 | 3.058E-01 | 5.000E-01 |
| 0.40 | 7.098E-03 | 2.665E-02 | 5.772E-02 | 1.291E-01 | 2.873E-01 | 4.820E-01 |
| 0.45 | 3.824E-03 | 1.694E-02 | 4.041E-02 | 9.997E-02 | 2.458E-01 | 4.400E-01 |
| 0.50 | 2.060E-03 | 1.077E-02 | 2.829E-02 | 7.740E-02 | 2.103E-01 | 4.017E-01 |
| 0.55 | 1.110E-03 | 6.843E-03 | 1.981E-02 | 5.993E-02 | 1.800E-01 | 3.666E-01 |
| 0.60 | 5.980E-04 | 4.350E-03 | 1.387E-02 | 4.640E-02 | 1.540E-01 | 3.347E-01 |
| 0.65 | 3.222E-04 | 2.765E-03 | 9.708E-03 | 3.592E-02 | 1.318E-01 | 3.055E-01 |
| 0.70 | 1.736E-04 | 1.757E-03 | 6.797E-03 | 2.781E-02 | 1.127E-01 | 2.789E-01 |
| 0.75 | 9.352E-05 | 1.117E-03 | 4.758E-03 | 2.153E-02 | 9.646E-02 | 2.546E-01 |
| 0.80 | 5.038E-05 | 7.100E-04 | 3.331E-03 | 1.667E-02 | 8.254E-02 | 2.324E-01 |
| 0.90 | 1.462E-05 | 2.869E-04 | 1.633E-03 | 9.994E-03 | 6.043E-02 | 1.936E-01 |
| 1.00 | 4.245E-06 | 1.159E-04 | 8.004E-04 | 5.991E-03 | 4.424E-02 | 1.613E-01 |
| 1.10 | 1.232E-06 | 4.683E-05 | 3.923E-04 | 3.591E-03 | 3.239E-02 | 1.344E-01 |
| 1.20 | 3.576E-07 | 1.892E-05 | 1.923E-04 | 2.153E-03 | 2.371E-02 | 1.120E-01 |
| 1.30 | 1.038E-07 | 7.644E-06 | 9.425E-05 | 1.290E-03 | 1.736E-02 | 9.333E-02 |
| 1.40 | 3.013E-08 | 3.088E-06 | 4.619E-05 | 7.735E-04 | 1.271E-02 | 7.777E-02 |
| 1.50 | 8.745E-09 | 1.248E-06 | 2.264E-05 | 4.637E-04 | 9.305E-03 | 6.480E-02 |
| 1.60 | 2.538E-09 | 5.041E-07 | 1.110E-05 | 2.780E-04 | 6.812E-03 | 5.399E-02 |
| 1.80 | 2.139E-10 | 8.229E-08 | 2.666E-06 | 9.988E-05 | 3.651E-03 | 3.749E-02 |
| 2.00 | 1.802E-11 | 1.343E-08 | 6.406E-07 | 3.589E-05 | 1.957E-03 | 2.603E-02 |
| 2.20 | 1.518E-12 | 2.193E-09 | 1.539E-07 | 1.290E-05 | 1.049E-03 | 1.807E-02 |
| 2.40 | 1.279E-13 | 3.579E-10 | 3.697E-08 | 4.634E-06 | 5.623E-04 | 1.255E-02 |
| 2.60 | 1.078E-14 | 5.842E-11 | 8.882E-09 | 1.665E-06 | 3.014E-04 | 8.711E-03 |
| 2.80 | 9.078E-16 | 9.537E-12 | 2.134E-09 | 5.983E-07 | 1.615E-04 | 6.048E-03 |
| 3.00 | 7.648E-17 | 1.557E-12 | 5.127E-10 | 2.150E-07 | 8.658E-05 | 4.199E-03 |

*Table 4:* ***Frequency distribution of ion cluster-size spectra due to 100 eV electrons with mean number of electrons*** $N_{mean}$ ***= 1.75 in molecular nitrogen in the case of a target volume with mass per area of the diameter of 0.34 µg/cm². Experimental results with statistical uncertainties, Monte Carlo simulation with detection efficiency 30 % and 100 %, deconvolution of experimental distribution with detection efficiency 30 % to true distribution with detection efficiency 100 % , Monte Carlo simulation for the case of a single (one) electron with detection efficiency 100 %. Calculated statistical parameters for all distributions: mean cluster size*** $M_1$***, second moment*** $M_2$ ***(Eq.18) and the cumulative frequency*** $F_2$ ***(Eq.19).***

| Energy, *eV* | 100 | | | | | |
|---|---|---|---|---|---|---|
| $N_{mean}$ | 1.75 | | | | | Single (one) *e* |
| Cluster size $\nu$ | Experiment | Stat.Error | MC 30% | MC 100% | Deconv 30% | MC 100% |
| 0 | 5.110E-01 | 2.70E-003 | 5.191E-01 | 2.534E-01 | 2.045E-01 | 2.138E-01 |
| 1 | 2.868E-01 | 2.02E-003 | 2.726E-01 | 1.217E-01 | 1.825E-01 | 2.756E-01 |
| 2 | 1.307E-01 | 1.37E-003 | 1.331E-01 | 1.665E-01 | 1.675E-01 | 3.088E-01 |
| 3 | 4.864E-02 | 8.34E-004 | 5.091E-02 | 1.484E-01 | 1.409E-01 | 1.771E-01 |
| 4 | 1.549E-02 | 4.70E-004 | 1.716E-02 | 1.012E-01 | 1.074E-01 | 2.452E-02 |
| 5 | 5.390E-03 | 2.77E-004 | 5.240E-03 | 7.685E-02 | 7.500E-02 | 1.700E-04 |
| 6 | 1.410E-03 | 1.42E-004 | 1.460E-03 | 5.226E-02 | 4.893E-02 | |
| 7 | 3.857E-04 | 7.42E-005 | 3.370E-04 | 3.294E-02 | 3.035E-02 | |
| 8 | 1.714E-04 | 4.95E-005 | 7.600E-05 | 2.028E-02 | 1.814E-02 | |
| 9 | 2.857E-05 | 2.02E-005 | 1.200E-05 | 1.206E-02 | 1.057E-02 | |
| 10 | | | 1.000E-06 | 6.720E-03 | 6.070E-03 | |
| 11 | | | 1.000E-06 | 3.720E-03 | 3.480E-03 | |
| 12 | | | | 1.970E-03 | 2.010E-03 | |
| 13 | | | | 1.070E-03 | 1.170E-03 | |
| 14 | | | | 4.900E-04 | 6.937E-04 | |
| 15 | | | | 2.310E-04 | 4.123E-04 | |
| 16 | | | | 1.000E-04 | 2.419E-04 | |
| 17 | | | | 5.000E-05 | 1.366E-04 | |
| 18 | | | | 1.400E-05 | 7.212E-05 | |
| 19 | | | | 3.000E-06 | | |
| 20 | | | | 0.000E+00 | | |
| 21 | | | | 1.000E-06 | | |
| 22 | | | | | | |
| 23 | | | | | | |
| 24 | | | | | | |
| 25 | | | | | | |
| 26 | | | | | | |
| | | | | | | |
| sum | 1.000E+00 | 7.957E-03 | 1.000E+00 | 1.000E+00 | 1.000E+00 | 1.000E+00 |
| $M_1$ | 7.958E-01 | 1.248E-02 | 7.982E-01 | 2.663E+00 | 2.653E+00 | 1.523E+00 |
| $M_2$ | 1.246E-01 | 2.194E-03 | 1.188E-01 | 2.361E-01 | 2.483E-01 | 3.632E-01 |
| $F_2$ | 2.022E-01 | 3.237E-03 | 2.083E-01 | 6.249E-01 | 6.131E-01 | 5.106E-01 |

*Table 5:* ***The same as in Table 4 for 200 eV electrons with*** $N_{mean}$ ***= 1.19.***

| Energy, *eV* | 200 | | | | | |
|---|---|---|---|---|---|---|
| $N_{mean}$ | 1.19 | | | | | Single (one) *e* |
| Cluster size $\nu$ | Experiment | Stat.Error | MC 30% | MC 100% | Deconv 30% | MC 100% |
| 0 | 5.758E-01 | 2.68E-003 | 6.007E-01 | 3.920E-01 | 3.075E-01 | 2.112E-01 |
| 1 | 2.444E-01 | 1.75E-003 | 2.120E-01 | 1.156E-01 | 1.719E-01 | 2.490E-01 |
| 2 | 1.115E-01 | 1.18E-003 | 1.093E-01 | 1.121E-01 | 1.424E-01 | 2.062E-01 |
| 3 | 4.390E-02 | 7.41E-004 | 4.816E-02 | 9.718E-02 | 1.119E-01 | 1.436E-01 |
| 4 | 1.661E-02 | 4.56E-004 | 1.904E-02 | 8.060E-02 | 8.426E-02 | 9.606E-02 |
| 5 | 5.440E-03 | 2.61E-004 | 7.150E-03 | 6.443E-02 | 6.120E-02 | 5.909E-02 |
| 6 | 1.680E-03 | 1.45E-004 | 2.440E-03 | 4.615E-02 | 4.295E-02 | 2.674E-02 |
| 7 | 4.000E-04 | 7.07E-005 | 8.490E-04 | 3.053E-02 | 2.912E-02 | 7.180E-03 |
| 8 | 2.875E-04 | 5.99E-005 | 2.610E-04 | 2.095E-02 | 1.905E-02 | 8.400E-04 |
| 9 | 5.000E-05 | 2.50E-005 | 8.300E-05 | 1.445E-02 | 1.203E-02 | 3.000E-05 |
| 10 | 0.000E+00 | 0.00E+000 | 2.400E-05 | 9.560E-03 | 7.360E-03 | |
| 11 | 0.000E+00 | 1.25E-005 | 9.000E-06 | 6.210E-03 | 4.390E-03 | |
| 12 | | | 2.000E-06 | 4.000E-03 | 2.570E-03 | |
| 13 | | | | 2.480E-03 | 1.500E-03 | |
| 14 | | | | 1.530E-03 | 8.731E-04 | |
| 15 | | | | 9.330E-04 | 5.102E-04 | |
| 16 | | | | 5.490E-04 | 2.985E-04 | |
| 17 | | | | 3.220E-04 | 1.737E-04 | |
| 18 | | | | 1.910E-04 | 9.935E-05 | |
| 19 | | | | 1.240E-04 | 5.509E-05 | |
| 20 | | | | 5.800E-05 | 2.915E-05 | |
| 21 | | | | 3.100E-05 | 1.450E-05 | |
| 22 | | | | 1.300E-05 | | |
| 23 | | | | 1.600E-05 | | |
| 24 | | | | 7.000E-06 | | |
| 25 | | | | | | |
| 26 | | | | | | |
| | | | | | | |
| sum | 1.000E+00 | 7.380E-03 | 1.000E+00 | 1.000E+00 | 1.000E+00 | 9.999E-01 |
| $M_1$ | 7.083E-01 | 1.166E-02 | 7.108E-01 | 2.366E+00 | 2.361E+00 | 1.990E+00 |
| $M_2$ | 9.163E-02 | 1.655E-03 | 7.754E-02 | 1.399E-01 | 1.769E-01 | 2.679E-01 |
| $F_2$ | 1.798E-01 | 2.950E-03 | 1.873E-01 | 4.924E-01 | 5.208E-01 | 5.397E-01 |

*Table 6:* ***The same as in Table 4 for 300 eV electrons*** $N_{mean}$ **= 1.02.**

| Energy, *eV* | 300 | | | | | |
|---|---|---|---|---|---|---|
| $N_{mean}$ | 1.02 | | | | | Single (one) *e* |
| Cluster size *ν* | Experiment | Stat.Error | MC 30% | MC 100% | Deconv 30% | MC 100% |
| 0 | 6.565E-01 | 3.06E-003 | 6.757E-01 | 4.680E-01 | 4.028E-01 | 2.559E-01 |
| 1 | 2.133E-01 | 1.75E-003 | 1.880E-01 | 1.318E-01 | 1.806E-01 | 2.767E-01 |
| 2 | 8.429E-02 | 1.10E-003 | 8.363E-02 | 1.139E-01 | 1.332E-01 | 2.005E-01 |
| 3 | 3.041E-02 | 6.59E-004 | 3.336E-02 | 8.685E-02 | 9.497E-02 | 1.212E-01 |
| 4 | 1.081E-02 | 3.93E-004 | 1.256E-02 | 6.305E-02 | 6.591E-02 | 6.925E-02 |
| 5 | 3.200E-03 | 2.14E-004 | 4.530E-03 | 4.471E-02 | 4.461E-02 | 3.793E-02 |
| 6 | 1.000E-03 | 1.20E-004 | 1.510E-03 | 3.118E-02 | 2.945E-02 | 1.959E-02 |
| 7 | 3.000E-04 | 6.55E-005 | 4.720E-04 | 2.152E-02 | 1.896E-02 | 1.076E-02 |
| 8 | 8.571E-05 | 3.50E-005 | 1.530E-04 | 1.443E-02 | 1.190E-02 | 5.110E-03 |
| 9 | 4.286E-05 | 2.47E-005 | 3.300E-05 | 9.330E-03 | 7.300E-03 | 2.330E-03 |
| 10 | 2.857E-05 | 2.02E-005 | 1.400E-05 | 5.730E-03 | 4.370E-03 | 5.700E-04 |
| 11 | | | 7.000E-06 | 3.750E-03 | 2.570E-03 | 9.000E-05 |
| 12 | | | 1.000E-06 | 2.290E-03 | 1.490E-03 | 2.000E-05 |
| 13 | | | | 1.390E-03 | 8.561E-04 | |
| 14 | | | | 8.170E-04 | 4.877E-04 | |
| 15 | | | | 5.180E-04 | 2.765E-04 | |
| 16 | | | | 3.185E-04 | 1.562E-04 | |
| 17 | | | | 1.545E-04 | 8.766E-05 | |
| 18 | | | | 9.700E-05 | 4.875E-05 | |
| 19 | | | | 5.600E-05 | 2.671E-05 | |
| 20 | | | | 2.700E-05 | 1.432E-05 | |
| 21 | | | | 2.000E-05 | 7.466E-06 | |
| 22 | | | | 7.000E-06 | | |
| 23 | | | | 5.000E-06 | | |
| 24 | | | | 1.000E-06 | | |
| 25 | | | | 1.000E-06 | | |
| 26 | | | | 2.000E-06 | | |
| | | | | | | |
| sum | 1.000E+00 | 7.441E-03 | 1.000E+00 | 1.000E+00 | 1.000E+00 | 1.000E+00 |
| $M_1$ | 5.418E-01 | 1.045E-02 | 5.423E-01 | 1.808E+00 | 1.807E+00 | 1.770E+00 |
| $M_2$ | 6.302E-02 | 1.280E-03 | 5.342E-02 | 1.040E-01 | 1.321E-01 | 2.308E-01 |
| $F_2$ | 1.302E-01 | 2.631E-03 | 1.363E-01 | 4.002E-01 | 4.167E-01 | 4.674E-01 |

*Table 7:* ***The same as in Table 4 for 500 eV electrons*** $N_{mean}$ ***= 0.93 .***

| Energy, *eV* | 500 | | | | | |
|---|---|---|---|---|---|---|
| $N_{mean}$ | 0.93 | | | | | Single (one) *e* |
| Cluster size $\nu$ | Experiment | Stat.Error | MC 30% | MC 100% | Deconv 30% | MC 100% |
| 0 | 7.457E-01 | 3.05E-003 | 7.545E-01 | 5.478E-01 | 5.167E-01 | 3.505E-01 |
| 1 | 1.750E-01 | 1.48E-003 | 1.634E-01 | 1.536E-01 | 1.803E-01 | 3.031E-01 |
| 2 | 5.625E-02 | 8.39E-004 | 5.654E-02 | 1.117E-01 | 1.186E-01 | 1.778E-01 |
| 3 | 1.684E-02 | 4.59E-004 | 1.791E-02 | 7.288E-02 | 7.555E-02 | 8.855E-02 |
| 4 | 4.560E-03 | 2.39E-004 | 5.410E-03 | 4.608E-02 | 4.644E-02 | 4.306E-02 |
| 5 | 1.190E-03 | 1.22E-004 | 1.580E-03 | 2.777E-02 | 2.752E-02 | 1.999E-02 |
| 6 | 3.750E-04 | 6.85E-005 | 4.820E-04 | 1.674E-02 | 1.574E-02 | 9.900E-03 |
| 7 | 1.500E-04 | 4.33E-005 | 1.280E-04 | 1.000E-02 | 8.760E-03 | 3.930E-03 |
| 8 | | | 3.900E-05 | 5.780E-03 | 4.790E-03 | 1.720E-03 |
| 9 | | | 8.000E-06 | 3.270E-03 | 2.600E-03 | 7.400E-04 |
| 10 | | | 3.000E-06 | 1.940E-03 | 1.420E-03 | 4.000E-04 |
| 11 | | | | 1.080E-03 | 7.750E-04 | 1.900E-04 |
| 12 | | | | 5.950E-04 | 4.252E-04 | 5.000E-05 |
| 13 | | | | 3.340E-04 | 4.655E-04 | 3.000E-05 |
| 14 | | | | 1.970E-04 | 1.080E-04 | 2.000E-05 |
| 15 | | | | 1.050E-04 | 7.082E-06 | 1.000E-05 |
| 16 | | | | 6.400E-05 | 8.673E-08 | |
| 17 | | | | 2.500E-05 | 1.390E-10 | |
| 18 | | | | 1.500E-05 | | |
| 19 | | | | 7.000E-06 | | |
| 20 | | | | 7.000E-06 | | |
| 21 | | | | 2.000E-06 | | |
| 22 | | | | 1.000E-06 | | |
| 23 | | | | 1.000E-06 | | |
| 24 | | | | | | |
| 25 | | | | | | |
| 26 | | | | | | |
| | | | | | | |
| sum | 1.000E+00 | 6.300E-03 | 1.000E+00 | 1.000E+00 | 1.000E+00 | 1.000E+00 |
| $M_1$ | 3.655E-01 | 6.812E-03 | 3.640E-01 | 1.214E+00 | 1.220E+00 | 1.311E+00 |
| $M_2$ | 3.789E-02 | 7.635E-04 | 3.419E-02 | 7.963E-02 | 9.248E-02 | 1.888E-01 |
| $F_2$ | 7.937E-02 | 1.770E-03 | 8.210E-02 | 2.986E-01 | 3.032E-01 | 3.464E-01 |

*Table 8:* ***The same as in Table 4 for 1000 eV electrons*** $N_{mean}$ **= 0.93 .**

| Energy, *eV* | 1000 | | | | | |
|---|---|---|---|---|---|---|
| $N_{mean}$ | 0.93 | | | | | Single (one) *e* |
| Cluster size $\nu$ | Experiment | Stat.Error | MC 30% | MC 100% | Deconv 30% | MC 100% |
| 0 | 8.289E-01 | 3.39E-003 | 8.306E-01 | 6.410E-01 | 6.318E-01 | 5.237E-01 |
| 1 | 1.317E-01 | 1.35E-003 | 1.289E-01 | 1.695E-01 | 1.794E-01 | 2.809E-01 |
| 2 | 3.058E-02 | 6.52E-004 | 3.098E-02 | 9.218E-02 | 9.489E-02 | 1.188E-01 |
| 3 | 6.820E-03 | 3.08E-004 | 7.380E-03 | 4.841E-02 | 4.818E-02 | 4.683E-02 |
| 4 | 1.510E-03 | 1.45E-004 | 1.710E-03 | 2.464E-02 | 2.363E-02 | 1.813E-02 |
| 5 | 4.028E-04 | 7.48E-005 | 3.630E-04 | 1.232E-02 | 1.135E-02 | 7.270E-03 |
| 6 | 9.722E-05 | 3.68E-005 | 7.200E-05 | 6.200E-03 | 5.450E-03 | 2.580E-03 |
| 7 | 1.389E-05 | 1.39E-005 | 1.900E-05 | 3.050E-03 | 2.650E-03 | 1.080E-03 |
| 8 | | | 2.000E-06 | 1.400E-03 | 1.320E-03 | 4.900E-04 |
| 9 | | | 1.000E-06 | 7.270E-04 | 6.771E-04 | 1.400E-04 |
| 10 | | | | 3.240E-04 | 3.509E-04 | 4.000E-05 |
| 11 | | | | 1.600E-04 | 1.806E-04 | 3.000E-05 |
| 12 | | | | 6.600E-05 | 9.017E-05 | |
| 13 | | | | 3.300E-05 | 4.251E-05 | |
| 14 | | | | 1.400E-05 | 1.847E-05 | |
| 15 | | | | 1.400E-05 | 7.242E-06 | |
| 16 | | | | 1.000E-06 | | |
| 17 | | | | 1.000E-06 | | |
| 18 | | | | 2.000E-06 | | |
| 19 | | | | | | |
| 20 | | | | | | |
| 21 | | | | | | |
| 22 | | | | | | |
| 23 | | | | | | |
| 24 | | | | | | |
| 25 | | | | | | |
| 26 | | | | | | |
| | | | | | | |
| sum | 1.000E+00 | 5.970E-03 | 1.000E+00 | 1.000E+00 | 1.000E+00 | 1.000E+00 |
| $M_1$ | 2.220E-01 | 4.848E-03 | 2.222E-01 | 7.422E-01 | 7.404E-01 | 7.968E-01 |
| $M_2$ | 1.935E-02 | 4.499E-04 | 1.871E-02 | 5.626E-02 | 6.028E-02 | 1.153E-01 |
| $F_2$ | 3.942E-02 | 1.230E-03 | 4.053E-02 | 1.895E-01 | 1.888E-01 | 1.954E-01 |

*Table 9:* ***The same as in Table 4 for 2000 eV electrons*** $N_{mean}$ ***= 1.06 .***

| Energy, *eV* | 2000 | | | | | |
|---|---|---|---|---|---|---|
| $N_{mean}$ | 1.06 | | | | | Single (one) *e* |
| Cluster size *ν* | Experiment | Stat.Error | MC 30% | MC 100% | Deconv 30% | MC 100% |
| 0 | 8.789E-01 | 3.540E-03 | 8.783E-01 | 7.141E-01 | 7.179E-01 | 6.859E-01 |
| 1 | 9.947E-02 | 1.190E-03 | 9.990E-02 | 1.658E-01 | 1.575E-01 | 2.156E-01 |
| 2 | 1.820E-02 | 5.099E-04 | 1.789E-02 | 6.964E-02 | 7.343E-02 | 6.687E-02 |
| 3 | 2.810E-03 | 2.005E-04 | 3.180E-03 | 2.921E-02 | 3.092E-02 | 2.097E-02 |
| 4 | 5.570E-04 | 8.920E-05 | 6.030E-04 | 1.241E-02 | 1.237E-02 | 6.880E-03 |
| 5 | 8.600E-05 | 3.500E-05 | 8.400E-05 | 5.230E-03 | 4.850E-03 | 2.670E-03 |
| 6 | | | 2.000E-05 | 2.160E-03 | 1.890E-03 | 7.500E-04 |
| 7 | | | 2.000E-06 | 8.660E-04 | 7.162E-04 | 2.500E-04 |
| 8 | | | | 3.560E-04 | 2.575E-04 | 8.000E-05 |
| 9 | | | | 1.300E-04 | 8.471E-05 | 3.000E-05 |
| 10 | | | | 7.200E-05 | 2.478E-05 | |
| 11 | | | | 2.900E-05 | 6.199E-06 | |
| 12 | | | | 1.000E-05 | 1.319E-06 | |
| 13 | | | | 2.000E-06 | | |
| 14 | | | | 2.000E-06 | | |
| 15 | | | | 2.000E-06 | | |
| 16 | | | | | | |
| 17 | | | | | | |
| 18 | | | | | | |
| 19 | | | | | | |
| 20 | | | | | | |
| 21 | | | | | | |
| 22 | | | | | | |
| 23 | | | | | | |
| 24 | | | | | | |
| 25 | | | | | | |
| 26 | | | | | | |
| | | | | | | |
| sum | 1.000E+00 | 5.565E-03 | 1.000E+00 | 1.000E+00 | 1.000E+00 | 1.000E+00 |
| $M_1$ | 1.470E-01 | 3.343E-03 | 1.482E-01 | 4.928E-01 | 4.904E-01 | 4.603E-01 |
| $M_2$ | 1.058E-02 | 2.777E-04 | 1.065E-02 | 4.054E-02 | 3.921E-02 | 5.697E-02 |
| $F_2$ | 2.165E-02 | 8.346E-04 | 2.178E-02 | 1.201E-01 | 1.246E-01 | 9.850E-02 |

*Table 10:* ***Frequency distribution of ion cluster-size spectra due to 3.8 MeV α-particles in molecular nitrogen in the case of a target volume with mass per area of the diameter of 0.092 µg/cm². Experimental results with statistical uncertainties, Monte Carlo simulation with detection efficiency 40 % and 100 %, deconvolution of experimental distribution with detection efficiency 40 % to true distribution with detection efficiency 100 %. Calculated statistical parameters for all distributions: mean cluster size $M_1$, second moment $M_2$ (Eq.18) and the cumulative frequency $F_2$ (Eq.19).***

| Area density $D\rho,\ \mu g/cm^2$ | 0.092 | | | | |
|---|---|---|---|---|---|
| Cluster size $\nu$ | Experiment | Stat.Error | MC 40% | MC 100% | Deconv 40% |
| 0 | 6.560E-01 | 1.620E-02 | 6.213E-01 | 3.227E-01 | 3.769E-01 |
| 1 | 2.588E-01 | 1.017E-02 | 2.867E-01 | 3.411E-01 | 3.347E-01 |
| 2 | 6.440E-02 | 5.080E-03 | 7.532E-02 | 2.023E-01 | 1.585E-01 |
| 3 | 1.640E-02 | 2.560E-03 | 1.424E-02 | 8.901E-02 | 7.194E-02 |
| 4 | 3.600E-03 | 1.200E-03 | 2.100E-03 | 3.190E-02 | 3.492E-02 |
| 5 | 4.000E-04 | 4.000E-04 | 3.100E-04 | 9.410E-03 | 1.414E-02 |
| 6 | 4.000E-04 | 4.000E-04 | 3.000E-05 | 2.810E-03 | 5.050E-03 |
| 7 | | | | 5.700E-04 | 2.020E-03 |
| 8 | | | | 1.600E-04 | 1.030E-03 |
| 9 | | | | 3.000E-05 | 5.959E-04 |
| 10 | | | | | 3.006E-04 |
| 11 | | | | | |
| 12 | | | | | |
| 13 | | | | | |
| 14 | | | | | |
| 15 | | | | | |
| 16 | | | | | |
| 17 | | | | | |
| 18 | | | | | |
| 19 | | | | | |
| 20 | | | | | |
| 21 | | | | | |
| 22 | | | | | |
| 23 | | | | | |
| 24 | | | | | |
| 25 | | | | | |
| 26 | | | | | |
| 27 | | | | | |
| 28 | | | | | |
| 29 | | | | | |
| | | | | | |
| Sum | 1.000E+00 | 3.601E-02 | 1.000E+00 | 1.000E+00 | 1.000E+00 |
| $M_1$ | 4.556E-01 | 3.721E-02 | 4.902E-01 | 1.210E+00 | 1.139E+00 |
| $M_2$ | 7.613E-02 | 6.863E-03 | 9.417E-02 | 2.265E-01 | 1.839E-01 |
| $F_2$ | 8.520E-02 | 9.640E-03 | 9.200E-02 | 3.362E-01 | 2.884E-01 |

*Table 11:* ***The same as in Table 10 for 0.130 μg/cm². Monte Carlo 40 % and 100 %. Deconvolution from 40 % to 100 %.***

| Area density $D\rho,\ \mu g/cm^2$ | 0.130 | | | | |
|---|---|---|---|---|---|
| Cluster size $\nu$ | Experiment | Stat.Error | MC 40% | MC 100% | Deconv 40% |
| 0 | 5.311E-01 | 1.232E-02 | 5.088E-01 | 1.997E-01 | 2.408E-01 |
| 1 | 3.031E-01 | 9.310E-03 | 3.291E-01 | 2.954E-01 | 3.134E-01 |
| 2 | 1.103E-01 | 5.610E-03 | 1.206E-01 | 2.395E-01 | 1.745E-01 |
| 3 | 4.000E-02 | 3.380E-03 | 3.246E-02 | 1.460E-01 | 1.157E-01 |
| 4 | 1.029E-02 | 1.710E-03 | 7.240E-03 | 7.144E-02 | 7.733E-02 |
| 5 | 3.430E-03 | 9.897E-04 | 1.520E-03 | 3.081E-02 | 4.536E-02 |
| 6 | 5.714E-04 | 4.041E-04 | 2.900E-04 | 1.139E-02 | 1.946E-02 |
| 7 | 8.571E-04 | 4.949E-04 | 3.000E-05 | 3.910E-03 | 5.920E-03 |
| 8 | 2.857E-04 | 2.857E-04 | 1.000E-05 | 1.370E-03 | 1.650E-03 |
| 9 | | | | 3.300E-04 | 6.561E-04 |
| 10 | | | | 1.200E-04 | 5.465E-04 |
| 11 | | | | 4.000E-05 | 1.050E-03 |
| 12 | | | | 2.000E-05 | 3.640E-03 |
| 13 | | | | | |
| 14 | | | | | |
| 15 | | | | | |
| 16 | | | | | |
| 17 | | | | | |
| 18 | | | | | |
| 19 | | | | | |
| 20 | | | | | |
| 21 | | | | | |
| 22 | | | | | |
| 23 | | | | | |
| 24 | | | | | |
| 25 | | | | | |
| 26 | | | | | |
| 27 | | | | | |
| 28 | | | | | |
| 29 | | | | | |
| | | | | | |
| Sum | 1.000E+00 | 3.450E-02 | 1.000E+00 | 1.000E+00 | 1.000E+00 |
| $M_1$ | 7.137E-01 | 5.063E-02 | 7.063E-01 | 1.764E+00 | 1.784E+00 |
| $M_2$ | 1.215E-01 | 9.115E-03 | 1.408E-01 | 2.920E-01 | 2.362E-01 |
| $F_2$ | 1.657E-01 | 1.287E-02 | 1.621E-01 | 5.049E-01 | 4.458E-01 |

*Table 12:* ***The same as in Table 10 for 0.187 µg/cm²*****. Monte Carlo 40 % and 100 %. Deconvolution from 40 % to 100 %.**

| Area density $D\rho,\ \mu g/cm^2$ | 0.187 | | | | |
|---|---|---|---|---|---|
| Cluster size $\nu$ | Experiment | Stat.Error | MC 40% | MC 100% | Deconv 40% |
| 0 | 3.450E-01 | 1.072E-02 | 3.693E-01 | 9.993E-02 | 1.226E-01 |
| 1 | 3.053E-01 | 1.009E-02 | 3.491E-01 | 2.033E-01 | 1.625E-01 |
| 2 | 1.897E-01 | 7.950E-03 | 1.829E-01 | 2.294E-01 | 1.762E-01 |
| 3 | 9.067E-02 | 5.500E-03 | 6.910E-02 | 1.905E-01 | 1.545E-01 |
| 4 | 4.767E-02 | 3.990E-03 | 2.172E-02 | 1.279E-01 | 1.252E-01 |
| 5 | 1.133E-02 | 1.940E-03 | 5.990E-03 | 7.438E-02 | 9.427E-02 |
| 6 | 7.330E-03 | 1.560E-03 | 1.460E-03 | 3.945E-02 | 6.465E-02 |
| 7 | 2.000E-03 | 8.165E-04 | 3.500E-04 | 1.964E-02 | 4.061E-02 |
| 8 | 1.000E-03 | 5.774E-04 | 1.100E-04 | 9.100E-03 | 2.425E-02 |
| 9 | | | | 3.910E-03 | 1.447E-02 |
| 10 | | | | 1.630E-03 | 8.940E-03 |
| 11 | | | | 5.600E-04 | 5.750E-03 |
| 12 | | | | 2.100E-04 | 3.750E-03 |
| 13 | | | | 8.000E-05 | 2.330E-03 |
| 14 | | | | 0.000E+00 | |
| 15 | | | | 1.000E-05 | |
| 16 | | | | 1.000E-05 | |
| 17 | | | | 1.000E-05 | |
| 18 | | | | | |
| 19 | | | | | |
| 20 | | | | | |
| 21 | | | | | |
| 22 | | | | | |
| 23 | | | | | |
| 24 | | | | | |
| 25 | | | | | |
| 26 | | | | | |
| 27 | | | | | |
| 28 | | | | | |
| 29 | | | | | |
| | | | | | |
| Sum | 1.000E+00 | 4.314E-02 | 1.000E+00 | 1.000E+00 | 1.000E+00 |
| $M_1$ | 1.270E+00 | 8.784E-02 | 1.051E+00 | 2.626E+00 | 3.175E+00 |
| $M_2$ | 1.999E-01 | 1.710E-02 | 2.052E-01 | 3.614E-01 | 3.119E-01 |
| $F_2$ | 3.497E-01 | 2.233E-02 | 2.816E-01 | 6.968E-01 | 7.149E-01 |

*Table 13:* ***The same as in Table 10 for 0.291 μg/cm². Monte Carlo 40 % and 100 %. Deconvolution from 40 % to 100 %.***

| Area density $D\rho,\ \mu g/cm^2$ | 0.291 | | | | |
|---|---|---|---|---|---|
| Cluster size $\nu$ | Experiment | Stat.Error | MC 40% | MC 100% | Deconv 40% |
| 0 | 2.123E-01 | 8.080E-03 | 2.030E-01 | 2.761E-02 | 3.126E-02 |
| 1 | 2.779E-01 | 9.250E-03 | 2.990E-01 | 8.380E-02 | 1.097E-01 |
| 2 | 2.077E-01 | 7.990E-03 | 2.396E-01 | 1.398E-01 | 1.578E-01 |
| 3 | 1.326E-01 | 6.390E-03 | 1.426E-01 | 1.648E-01 | 1.386E-01 |
| 4 | 8.492E-02 | 5.110E-03 | 6.875E-02 | 1.615E-01 | 1.079E-01 |
| 5 | 4.708E-02 | 3.810E-03 | 3.022E-02 | 1.360E-01 | 9.154E-02 |
| 6 | 1.846E-02 | 2.380E-03 | 1.112E-02 | 1.030E-01 | 8.401E-02 |
| 7 | 1.138E-02 | 1.870E-03 | 3.810E-03 | 7.076E-02 | 7.548E-02 |
| 8 | 5.540E-03 | 1.310E-03 | 1.330E-03 | 4.588E-02 | 6.179E-02 |
| 9 | 1.230E-03 | 6.154E-04 | 3.200E-04 | 2.957E-02 | 4.592E-02 |
| 10 | 3.077E-04 | 3.077E-04 | 1.700E-04 | 1.758E-02 | 3.212E-02 |
| 11 | 6.154E-04 | 4.351E-04 | 3.000E-05 | 9.690E-03 | 2.188E-02 |
| 12 | | | 1.000E-05 | 4.690E-03 | 1.473E-02 |
| 13 | | | 2.000E-05 | 2.570E-03 | 9.810E-03 |
| 14 | | | | 1.410E-03 | 6.510E-03 |
| 15 | | | | 7.400E-04 | 4.360E-03 |
| 16 | | | | 4.000E-04 | 3.000E-03 |
| 17 | | | | 1.600E-04 | 2.130E-03 |
| 18 | | | | 5.000E-05 | 1.520E-03 |
| 19 | | | | 4.000E-05 | |
| 20 | | | | 1.000E-05 | |
| 21 | | | | 0.000E+00 | |
| 22 | | | | 1.000E-05 | |
| 23 | | | | | |
| 24 | | | | | |
| 25 | | | | | |
| 26 | | | | | |
| 27 | | | | | |
| 28 | | | | | |
| 29 | | | | | |
| | | | | | |
| Sum | 1.000E+00 | 4.755E-02 | 1.000E+00 | 1.000E+00 | 1.000E+00 |
| $M_1$ | 1.922E+00 | 1.351E-01 | 1.741E+00 | 4.344E+00 | 4.805E+00 |
| $M_2$ | 2.594E-01 | 2.309E-02 | 2.896E-01 | 4.523E-01 | 3.603E-01 |
| $F_2$ | 5.098E-01 | 3.022E-02 | 4.980E-01 | 8.886E-01 | 8.590E-01 |

*Table 14:* ***The same as in Table 10 for 0.354 µg/cm². Monte Carlo 40 % and 100 %. Deconvolution from 40 % to 100 %.***

| Area density $D\rho,\ \mu g/cm^2$ | 0.354 | | | | |
|---|---|---|---|---|---|
| Cluster size $\nu$ | Experiment | Stat.Error | MC 40% | MC 100% | Deconv 40% |
| 0 | 1.800E-01 | 8.090E-03 | 1.403E-01 | 1.232E-02 | 5.370E-02 |
| 1 | 2.429E-01 | 9.400E-03 | 2.507E-01 | 4.563E-02 | 3.102E-02 |
| 2 | 2.182E-01 | 8.910E-03 | 2.427E-01 | 9.009E-02 | 9.414E-02 |
| 3 | 1.506E-01 | 7.400E-03 | 1.744E-01 | 1.273E-01 | 2.061E-01 |
| 4 | 9.709E-02 | 5.940E-03 | 1.020E-01 | 1.418E-01 | 1.391E-01 |
| 5 | 6.364E-02 | 4.810E-03 | 5.190E-02 | 1.398E-01 | 6.229E-02 |
| 6 | 2.655E-02 | 3.110E-03 | 2.202E-02 | 1.229E-01 | 4.026E-02 |
| 7 | 1.127E-02 | 2.020E-03 | 9.830E-03 | 9.860E-02 | 4.762E-02 |
| 8 | 3.640E-03 | 1.150E-03 | 3.930E-03 | 7.367E-02 | 8.047E-02 |
| 9 | 2.180E-03 | 8.907E-04 | 1.460E-03 | 5.332E-02 | 1.177E-01 |
| 10 | 1.820E-03 | 8.131E-04 | 4.900E-04 | 3.593E-02 | 8.900E-02 |
| 11 | 7.273E-04 | 5.143E-04 | 2.200E-04 | 2.366E-02 | 2.448E-02 |
| 12 | 7.273E-04 | 5.143E-04 | 4.000E-05 | 1.414E-02 | 2.200E-03 |
| 13 | 7.273E-04 | 5.143E-04 | 0.000E+00 | 9.260E-03 | 7.565E-05 |
| 14 | | | 1.000E-05 | 5.020E-03 | 1.441E-06 |
| 15 | | | | 2.840E-03 | 2.695E-08 |
| 16 | | | | 1.670E-03 | 1.015E-09 |
| 17 | | | | 9.600E-04 | 1.696E-10 |
| 18 | | | | 4.700E-04 | 2.615E-10 |
| 19 | | | | 2.700E-04 | 6.167E-09 |
| 20 | | | | 1.500E-04 | 2.492E-06 |
| 21 | | | | 8.000E-05 | 1.184E-02 |
| 22 | | | | 7.000E-05 | |
| 23 | | | | 0.000E+00 | |
| 24 | | | | 2.000E-05 | |
| 25 | | | | 1.000E-05 | |
| 26 | | | | | |
| 27 | | | | | |
| 28 | | | | | |
| 29 | | | | | |
| | | | | | |
| Sum | 1.000E+00 | 5.408E-02 | 1.000E+00 | 1.000E+00 | 1.000E+00 |
| $M_1$ | 2.169E+00 | 1.739E-01 | 2.180E+00 | 5.457E+00 | 5.419E+00 |
| $M_2$ | 2.855E-01 | 2.817E-02 | 3.307E-01 | 4.959E-01 | 5.338E-01 |
| $F_2$ | 5.771E-01 | 3.659E-02 | 6.090E-01 | 9.421E-01 | 9.153E-01 |

*Table 15:* ***The same as in Table 10 for 0.387 µg/cm$^2$. Monte Carlo 30 % and 100 %. Deconvolution from 30 % to 100 %.***

| Area density $D\rho,\ \mu g/cm^2$ | 0.387 | | | | |
|---|---|---|---|---|---|
| Cluster size $\nu$ | Experiment | Stat.Error | MC 30% | MC 100% | Deconv 30% |
| 0 | 2.084E-01 | 1.872E-02 | 1.898E-01 | 8.400E-03 | 6.650E-02 |
| 1 | 2.437E-01 | 2.024E-02 | 2.898E-01 | 3.336E-02 | 3.174E-02 |
| 2 | 2.420E-01 | 2.017E-02 | 2.426E-01 | 6.967E-02 | 4.071E-02 |
| 3 | 1.496E-01 | 1.586E-02 | 1.504E-01 | 1.051E-01 | 6.195E-02 |
| 4 | 9.244E-02 | 1.246E-02 | 7.575E-02 | 1.296E-01 | 8.902E-02 |
| 5 | 3.697E-02 | 7.880E-03 | 3.238E-02 | 1.328E-01 | 1.152E-01 |
| 6 | 1.345E-02 | 4.750E-03 | 1.269E-02 | 1.276E-01 | 1.310E-01 |
| 7 | 8.400E-03 | 3.760E-03 | 4.350E-03 | 1.062E-01 | 1.272E-01 |
| 8 | 5.040E-03 | 2.910E-03 | 1.500E-03 | 8.798E-02 | 1.042E-01 |
| 9 | | | 5.300E-04 | 6.462E-02 | 7.348E-02 |
| 10 | | | 1.500E-04 | 4.662E-02 | 4.725E-02 |
| 11 | | | 3.000E-05 | 3.198E-02 | 2.999E-02 |
| 12 | | | | 2.127E-02 | 2.025E-02 |
| 13 | | | | 1.351E-02 | 1.523E-02 |
| 14 | | | | 8.740E-03 | 1.275E-02 |
| 15 | | | | 5.480E-03 | 1.131E-02 |
| 16 | | | | 2.950E-03 | 9.680E-03 |
| 17 | | | | 1.840E-03 | 7.060E-03 |
| 18 | | | | 1.050E-03 | 3.840E-03 |
| 19 | | | | 4.700E-04 | 1.360E-03 |
| 20 | | | | 3.800E-04 | 2.794E-04 |
| 21 | | | | 1.100E-04 | 3.007E-05 |
| 22 | | | | 1.500E-04 | |
| 23 | | | | 1.000E-04 | |
| 24 | | | | 5.000E-05 | |
| 25 | | | | 0.000E+00 | |
| 26 | | | | 0.000E+00 | |
| 27 | | | | 1.000E-05 | |
| 28 | | | | | |
| 29 | | | | | |
| | | | | | |
| Sum | 1.000E+00 | 1.068E-01 | 1.000E+00 | 1.000E+00 | 1.000E+00 |
| $M_1$ | 1.911E+00 | 2.755E-01 | 1.816E+00 | 6.041E+00 | 6.370E+00 |
| $M_2$ | 2.865E-01 | 5.720E-02 | 2.989E-01 | 5.180E-01 | 5.126E-01 |
| $F_2$ | 5.479E-01 | 6.779E-02 | 5.204E-01 | 9.582E-01 | 9.018E-01 |

*Table 16*: ***The same as in Table 10 for 0.538 μg/cm². Monte Carlo 25 % and 100 %. Deconvolution from 25 % to 100 %.***

| Area density $D\rho,\ \mu g/cm^2$ | 0.538 | | | | |
|---|---|---|---|---|---|
| Cluster size $\nu$ | Experiment | Stat.Error | MC 25% | MC 100% | Deconv 25% |
| 0 | 2.033E-01 | 8.600E-03 | 1.317E-01 | 1.540E-03 | 2.224E-02 |
| 1 | 2.593E-01 | 9.710E-03 | 2.448E-01 | 7.090E-03 | 3.505E-02 |
| 2 | 2.240E-01 | 9.030E-03 | 2.481E-01 | 1.971E-02 | 5.049E-02 |
| 3 | 1.466E-01 | 7.300E-03 | 1.785E-01 | 3.858E-02 | 6.631E-02 |
| 4 | 8.509E-02 | 5.560E-03 | 1.055E-01 | 5.972E-02 | 7.986E-02 |
| 5 | 4.400E-02 | 4.000E-03 | 5.267E-02 | 8.117E-02 | 8.903E-02 |
| 6 | 1.855E-02 | 2.600E-03 | 2.376E-02 | 9.585E-02 | 9.278E-02 |
| 7 | 9.450E-03 | 1.850E-03 | 9.210E-03 | 1.033E-01 | 9.111E-02 |
| 8 | 6.180E-03 | 1.500E-03 | 3.780E-03 | 1.013E-01 | 8.497E-02 |
| 9 | 2.180E-03 | 8.907E-04 | 1.310E-03 | 9.511E-02 | 7.565E-02 |
| 10 | 0.000E+00 | 0.000E+00 | 4.100E-04 | 8.443E-02 | 6.542E-02 |
| 11 | 1.090E-03 | 6.298E-04 | 1.100E-04 | 7.231E-02 | 5.319E-02 |
| 12 | 0.000E+00 | 0.000E+00 | 7.000E-05 | 6.073E-02 | 4.290E-02 |
| 13 | 3.636E-04 | 3.636E-04 | 1.000E-05 | 4.828E-02 | 3.386E-02 |
| 14 | | | 2.000E-05 | 3.747E-02 | 2.656E-02 |
| 15 | | | | 2.760E-02 | 2.082E-02 |
| 16 | | | | 1.997E-02 | 1.638E-02 |
| 17 | | | | 1.474E-02 | 1.297E-02 |
| 18 | | | | 1.031E-02 | 1.033E-02 |
| 19 | | | | 6.780E-03 | 8.250E-03 |
| 20 | | | | 5.220E-03 | 6.550E-03 |
| 21 | | | | 3.230E-03 | 5.120E-03 |
| 22 | | | | 2.080E-03 | 3.910E-03 |
| 23 | | | | 1.340E-03 | 2.870E-03 |
| 24 | | | | 7.600E-04 | 2.010E-03 |
| 25 | | | | 5.300E-04 | 1.320E-03 |
| 26 | | | | 4.200E-04 | |
| 27 | | | | 2.300E-04 | |
| 28 | | | | 1.600E-04 | |
| 29 | | | | 2.000E-05 | |
| | | | | | |
| Sum | 1.000E+00 | 5.203E-02 | 1.000E+00 | 1.000E+00 | 1.000E+00 |
| $M_1$ | 1.971E+00 | 1.521E-01 | 2.217E+00 | 8.871E+00 | 7.830E+00 |
| $M_2$ | 2.737E-01 | 2.611E-02 | 3.411E-01 | 5.941E-01 | 4.434E-01 |
| $F_2$ | 5.375E-01 | 3.372E-02 | 6.235E-01 | 9.914E-01 | 9.427E-01 |

*Table 17:* ***Frequency distribution of ion cluster-size spectra due to 3.8 MeV α-particles in propane gas in the case of a target volume with mass per area of the diameter of 0.110 µg/cm². Experimental results with statistical uncertainties, Monte Carlo simulation with detection efficiency 60 % and 100 %, deconvolution of experimental distribution with detection efficiency 60 % to true distribution with detection efficiency 100 %. Calculated statistical parameters for all distributions: mean cluster size $M_1$, second moment $M_2$ (Eq.18) and the cumulative frequency $F_2$ (Eq.19).***

| Area density $D\rho,\ \mu g/cm^2$ | 0.110 | | | | |
|---|---|---|---|---|---|
| Cluster size $\nu$ | Experiment | Stat.Error | MC 60% | MC 100% | Deconv 60% |
| 0 | 2.954E-01 | 9.970E-03 | 2.191E-01 | 9.184E-02 | 1.559E-01 |
| 1 | 3.126E-01 | 1.026E-02 | 3.017E-01 | 1.884E-01 | 2.166E-01 |
| 2 | 2.059E-01 | 8.320E-03 | 2.350E-01 | 2.181E-01 | 2.551E-01 |
| 3 | 9.791E-02 | 5.740E-03 | 1.359E-01 | 1.887E-01 | 1.377E-01 |
| 4 | 5.148E-02 | 4.160E-03 | 6.516E-02 | 1.346E-01 | 9.330E-02 |
| 5 | 2.187E-02 | 2.710E-03 | 2.754E-02 | 8.402E-02 | 6.140E-02 |
| 6 | 9.760E-03 | 1.810E-03 | 1.034E-02 | 4.754E-02 | 3.613E-02 |
| 7 | 3.700E-03 | 1.120E-03 | 3.570E-03 | 2.467E-02 | 2.128E-02 |
| 8 | 1.010E-03 | 5.828E-04 | 1.200E-03 | 1.190E-02 | 1.205E-02 |
| 9 | 3.400E-04 | 3.365E-04 | 3.810E-04 | 5.810E-03 | 6.810E-03 |
| 10 | | | 1.070E-04 | 2.634E-03 | 3.110E-03 |
| 11 | | | 3.200E-05 | 1.102E-03 | 6.029E-04 |
| 12 | | | 1.200E-05 | 4.170E-04 | 2.268E-05 |
| 13 | | | 1.000E-06 | 1.820E-04 | 9.042E-08 |
| 14 | | | | 6.00E-005 | 2.79E-011 |
| 15 | | | | 2.80E-005 | 6.21E-016 |
| 16 | | | | 1.10E-005 | 1.08E-021 |
| 17 | | | | 3.00E-006 | 1.70E-028 |
| 18 | | | | 1.00E-006 | 2.94E-036 |
| 19 | | | | 1.00E-006 | 6.73E-045 |
| 20 | | | | | |
| | | | | | |
| Sum | 1.000E+00 | 4.501E-02 | 1.000E+00 | 1.000E+00 | 1.000E+00 |
| $M_1$ | 1.429E+00 | 1.007E-01 | 1.679E+00 | 2.802E+00 | 2.382E+00 |
| $M_2$ | 2.249E-01 | 1.923E-02 | 2.784E-01 | 3.646E-01 | 3.003E-01 |
| $F_2$ | 3.920E-01 | 2.478E-02 | 4.792E-01 | 7.198E-01 | 6.275E-01 |

*Table 18:* ***The same as in Table 17 for 0.250 μg/cm²*. Monte Carlo 40 % and 100 %. Deconvolution from 40 % to 100 %.**

| Area density $D\rho,\ \mu g/cm^2$ | 0.250 | | | | |
|---|---|---|---|---|---|
| Cluster size $\nu$ | Experiment | Stat.Error | MC 40% | MC 100% | Deconv 40% |
| 0 | 1.619E-01 | 7.060E-03 | 7.678E-02 | 4.080E-03 | 4.265E-02 |
| 1 | 2.157E-01 | 8.150E-03 | 1.727E-01 | 1.733E-02 | 5.336E-02 |
| 2 | 2.046E-01 | 7.930E-03 | 2.139E-01 | 4.065E-02 | 8.448E-02 |
| 3 | 1.505E-01 | 6.800E-03 | 1.934E-01 | 6.899E-02 | 1.156E-01 |
| 4 | 1.009E-01 | 5.570E-03 | 1.437E-01 | 9.342E-02 | 1.209E-01 |
| 5 | 7.138E-02 | 4.690E-03 | 9.224E-02 | 1.091E-01 | 1.043E-01 |
| 6 | 4.185E-02 | 3.590E-03 | 5.303E-02 | 1.152E-01 | 8.312E-02 |
| 7 | 2.615E-02 | 2.840E-03 | 2.841E-02 | 1.113E-01 | 6.661E-02 |
| 8 | 1.231E-02 | 1.950E-03 | 1.423E-02 | 9.982E-02 | 5.612E-02 |
| 9 | 6.460E-03 | 1.410E-03 | 6.550E-03 | 8.454E-02 | 5.009E-02 |
| 10 | 3.080E-03 | 9.730E-04 | 3.010E-03 | 6.902E-02 | 4.641E-02 |
| 11 | 1.540E-03 | 6.880E-04 | 1.270E-03 | 5.338E-02 | 4.302E-02 |
| 12 | 1.540E-03 | 6.880E-04 | 5.300E-04 | 4.006E-02 | 3.822E-02 |
| 13 | 8.754E-04 | 4.351E-04 | 2.460E-04 | 2.942E-02 | 3.123E-02 |
| 14 | 1.230E-03 | 6.154E-04 | 9.400E-05 | 2.118E-02 | 2.280E-02 |
| 15 | | | 2.200E-05 | 1.453E-02 | 1.468E-02 |
| 16 | | | 1.000E-05 | 9.790E-03 | 8.430E-03 |
| 17 | | | 6.000E-06 | 6.630E-03 | 4.470E-03 |
| 18 | | | 2.000E-06 | 4.350E-03 | 2.310E-03 |
| 19 | | | 2.000E-06 | 2.720E-03 | 1.360E-03 |
| 20 | | | | 1.800E-03 | 7.815E-04 |
| 21 | | | | 1.070E-03 | 6.842E-04 |
| 22 | | | | 6.640E-04 | 5.441E-04 |
| 23 | | | | 4.200E-04 | 6.248E-04 |
| 24 | | | | 2.480E-04 | 8.362E-04 |
| 25 | | | | 1.420E-04 | 1.280E-03 |
| 26 | | | | 7.500E-05 | 1.560E-03 |
| 27 | | | | 6.300E-05 | 1.660E-03 |
| 28 | | | | 2.500E-05 | 1.250E-03 |
| 29 | | | | 2.100E-05 | 5.793E-04 |
| 30 | | | | 1.300E-05 | |
| 31 | | | | 2.000E-06 | |
| 32 | | | | 1.000E-06 | |
| 33 | | | | 1.000E-06 | |
| 34 | | | | | |
| | | | | | |
| Sum | 1.000E+00 | 5.339E-02 | 1.000E+00 | 1.000E+00 | 1.000E+00 |
| $M_1$ | 2.522E+00 | 1.997E-01 | 1.997E-01 | 2.962E+00 | 7.406E+00 |
| $M_2$ | 2.815E-01 | 2.752E-02 | 7.752E-04 | 3.833E-01 | 5.445E-01 |
| $F_2$ | 6.225E-01 | 3.818E-02 | 9.848E-01 | 7.505E-01 | 9.786E-01 |

*Table 19:* ***The same as in Table 17 for 0.370 μg/cm². Monte Carlo 30 % and 100 %. Deconvolution from 30 % to 100 %.***

| Area density $D\rho,\ \mu g/cm^2$ | 0.370 | | | | |
|---|---|---|---|---|---|
| Cluster size $\nu$ | Experiment | Stat.Error | MC 30% | MC 100% | Deconv 30% |
| 0 | 7.038E-02 | 5.080E-03 | 4.927E-02 | 3.170E-04 | 1.360E-03 |
| 1 | 1.422E-01 | 7.220E-03 | 1.294E-01 | 1.980E-03 | 3.789E-02 |
| 2 | 1.532E-01 | 7.490E-03 | 1.862E-01 | 6.530E-03 | 4.789E-02 |
| 3 | 1.734E-01 | 7.970E-03 | 1.917E-01 | 1.427E-02 | 4.386E-02 |
| 4 | 1.430E-01 | 7.240E-03 | 1.609E-01 | 2.620E-02 | 5.103E-02 |
| 5 | 1.019E-01 | 6.110E-03 | 1.159E-01 | 4.007E-02 | 6.593E-02 |
| 6 | 8.211E-02 | 5.490E-03 | 7.488E-02 | 5.432E-02 | 7.988E-02 |
| 7 | 5.169E-02 | 4.350E-03 | 4.381E-02 | 6.738E-02 | 8.652E-02 |
| 8 | 2.896E-02 | 3.260E-03 | 2.411E-02 | 7.755E-02 | 8.609E-02 |
| 9 | 2.089E-02 | 2.770E-03 | 1.259E-02 | 8.295E-02 | 8.150E-02 |
| 10 | 1.466E-02 | 2.320E-03 | 6.220E-03 | 8.398E-02 | 7.436E-02 |
| 11 | 8.430E-03 | 1.760E-03 | 2.890E-03 | 8.233E-02 | 6.515E-02 |
| 12 | 3.300E-03 | 1.100E-03 | 1.270E-03 | 7.714E-02 | 5.459E-02 |
| 13 | 1.830E-03 | 8.197E-04 | 5.780E-04 | 6.992E-02 | 4.402E-02 |
| 14 | 2.200E-03 | 8.979E-04 | 2.330E-04 | 6.105E-02 | 3.477E-02 |
| 15 | 1.100E-03 | 6.349E-04 | 1.140E-04 | 5.267E-02 | 2.747E-02 |
| 16 | | | 4.100E-05 | 4.385E-02 | 2.221E-02 |
| 17 | | | 1.300E-05 | 3.625E-02 | 1.796E-02 |
| 18 | | | 1.000E-05 | 2.869E-02 | 1.485E-02 |
| 19 | | | 2.000E-06 | 2.283E-02 | 1.256E-02 |
| 20 | | | 2.000E-06 | 1.793E-02 | 1.039E-02 |
| 21 | | | 1.000E-06 | 1.376E-02 | 8.610E-03 |
| 22 | | | 1.000E-06 | 1.041E-02 | 7.270E-03 |
| 23 | | | | 7.740E-03 | 6.320E-03 |
| 24 | | | | 5.700E-03 | 5.450E-03 |
| 25 | | | | 4.090E-03 | 4.490E-03 |
| 26 | | | | 3.170E-03 | 3.490E-03 |
| 27 | | | | 2.180E-03 | 2.320E-03 |
| 28 | | | | 1.540E-03 | 1.200E-03 |
| 29 | | | | 1.080E-03 | 4.287E-04 |
| 30 | | | | 7.790E-04 | 9.504E-05 |
| 31 | | | | 5.190E-04 | 1.168E-05 |
| 32 | | | | 3.390E-04 | 7.205E-07 |
| 33 | | | | 2.490E-04 | 2.048E-08 |
| 34 | | | | 1.930E-04 | 2.504E-10 |
| | | | | | |
| Sum | 9.993E-01 | 6.451E-02 | 1.000E+00 | 1.000E+00 | 1.000E+00 |
| $M_1$ | 3.674E+00 | 3.085E-01 | 3.485E+00 | 1.155E+01 | 9.198E+00 |
| $M_2$ | 3.640E-01 | 4.176E-02 | 4.207E-01 | 6.385E-01 | 4.727E-01 |
| $F_2$ | 7.874E-01 | 5.221E-02 | 8.213E-01 | 9.977E-01 | 9.608E-01 |

*Table 20:* ***The Basic physical and nanodosimetric data used or derived from the present measurements for electrons, as a function of electron energy. For an explanation of the different quantities, see chapter 7.3 , page 77.***

| Energy | *eV* | 100 | 200 | 300 | 500 | 1000 | 2000 |
|---|---|---|---|---|---|---|---|
| $M_{1E}$ | | 0.796 | 0.708 | 0.542 | 0.366 | 0.222 | 0.147 |
| $M_1$ | | 2.65 | 2.36 | 1.81 | 1.22 | 0.74 | 0.49 |
| $N_{mean}$ | | 1.75 | 1.19 | 1.02 | 0.93 | 0.93 | 1.06 |
| $M_1^{(MC)}$ | | 1.52 | 1.99 | 1.77 | 1.31 | 0.80 | 0.46 |
| $W$(T) | *eV* | 40.86 | 37.64 | 36.71 | 36.11 | 35.83 | 35.82 |
| $\omega$(T) | *eV* | 34.51 | 34.88 | 34.97 | 35.13 | 35.37 | 35.60 |
| $L_{rE}$ | *keV/µm* | 15.39 | 20.35 | 18.21 | 13.53 | 8.28 | 4.84 |
| $L_{rMC}$ | *keV/µm* | 15.46 | 20.41 | 18.21 | 13.54 | 8.29 | 4.82 |
| $L_{100}(N_2)$ | *keV/µm* | 20.65 | 20.25 | 16.46 | 12.41 | 7.43 | 4.36 |
| $L_{100}(H_2O)$ | *keV/µm* | 27.78 | 22.15 | 19.33 | 14.76 | 9.23 | 5.46 |
| $F_{2E}$ | | 0.202 | 0.180 | 0.130 | 0.079 | 0.039 | 0.022 |
| $F_{2E}/M_{1E}$ | | 0.254 | 0.254 | 0.240 | 0.217 | 0.178 | 0.147 |

*Table 21:* ***Set of scattering cross section $\sigma_v(T)$ of electrons in molecular nitrogen as a function of energy $T$: calculated total scattering cross section $\sigma_{tot}(T)$ using (Eq.34), present data set; elastic scattering cross section $\sigma_{el}(T)$; integral ionization cross section $\sigma_{ion}(T)$; total excitation cross section $\sigma_{exc}^{(tot)}(T)$.***

| Energy, *eV* | $\sigma_{exc}^{(tot)}, cm^2$ | $\sigma_{ion}, cm^2$ | $\sigma_{el}, cm^2$ | $\sigma_{tot}, cm^2$ |
|---|---|---|---|---|
| 1.00E+01 | 7.6038E-17 | 0.0000E+00 | 1.1798E-15 | 1.2559E-15 |
| 1.05E+01 | 8.1971E-17 | 0.0000E+00 | 1.1828E-15 | 1.2648E-15 |
| 1.10E+01 | 8.4465E-17 | 0.0000E+00 | 1.1841E-15 | 1.2686E-15 |
| 1.15E+01 | 9.9754E-17 | 0.0000E+00 | 1.1839E-15 | 1.2836E-15 |
| 1.20E+01 | 1.0991E-16 | 0.0000E+00 | 1.1822E-15 | 1.2921E-15 |
| 1.25E+01 | 1.1706E-16 | 0.0000E+00 | 1.1794E-15 | 1.2965E-15 |
| 1.30E+01 | 1.2058E-16 | 0.0000E+00 | 1.1756E-15 | 1.2962E-15 |
| 1.35E+01 | 1.2179E-16 | 0.0000E+00 | 1.1710E-15 | 1.2927E-15 |
| 1.40E+01 | 1.2289E-16 | 0.0000E+00 | 1.1656E-15 | 1.2885E-15 |
| 1.45E+01 | 1.2624E-16 | 0.0000E+00 | 1.1596E-15 | 1.2859E-15 |
| 1.50E+01 | 1.3358E-16 | 0.0000E+00 | 1.1532E-15 | 1.2868E-15 |
| 1.55E+01 | 1.4550E-16 | 0.0000E+00 | 1.1465E-15 | 1.2920E-15 |
| 1.60E+01 | 1.6139E-16 | 1.3358E-18 | 1.1395E-15 | 1.3022E-15 |
| 1.65E+01 | 1.7989E-16 | 2.9801E-18 | 1.1323E-15 | 1.3152E-15 |
| 1.70E+01 | 1.9935E-16 | 4.6622E-18 | 1.1250E-15 | 1.3290E-15 |
| 1.75E+01 | 2.1819E-16 | 8.8592E-18 | 1.1177E-15 | 1.3447E-15 |
| 1.80E+01 | 2.3522E-16 | 1.3556E-17 | 1.1103E-15 | 1.3591E-15 |
| 1.90E+01 | 2.6098E-16 | 2.3090E-17 | 1.0958E-15 | 1.3799E-15 |
| 2.00E+01 | 2.7450E-16 | 3.2603E-17 | 1.0818E-15 | 1.3889E-15 |
| 2.50E+01 | 2.4071E-16 | 8.1761E-17 | 1.0229E-15 | 1.3454E-15 |
| 3.00E+01 | 1.9771E-16 | 1.2215E-16 | 9.8307E-16 | 1.3029E-15 |
| 3.50E+01 | 1.8281E-16 | 1.5281E-16 | 8.9475E-16 | 1.2304E-15 |
| 4.00E+01 | 1.8073E-16 | 1.7598E-16 | 8.2576E-16 | 1.1825E-15 |
| 4.50E+01 | 1.8225E-16 | 1.9439E-16 | 7.6987E-16 | 1.1465E-15 |
| 5.00E+01 | 1.8400E-16 | 2.0903E-16 | 7.2336E-16 | 1.1164E-15 |
| 5.50E+01 | 1.8496E-16 | 2.2040E-16 | 6.8384E-16 | 1.0892E-15 |
| 6.00E+01 | 1.8500E-16 | 2.2918E-16 | 6.4968E-16 | 1.0639E-15 |
| 6.50E+01 | 1.8423E-16 | 2.3590E-16 | 6.1977E-16 | 1.0399E-15 |
| 7.00E+01 | 1.8283E-16 | 2.4098E-16 | 5.9327E-16 | 1.0171E-15 |
| 7.50E+01 | 1.8096E-16 | 2.4474E-16 | 5.6956E-16 | 9.9526E-16 |
| 8.00E+01 | 1.7875E-16 | 2.4743E-16 | 5.4820E-16 | 9.7438E-16 |
| 8.50E+01 | 1.7632E-16 | 2.4924E-16 | 5.2880E-16 | 9.5436E-16 |
| 9.00E+01 | 1.7375E-16 | 2.5035E-16 | 5.1108E-16 | 9.3518E-16 |
| 9.50E+01 | 1.7110E-16 | 2.5087E-16 | 4.9480E-16 | 9.1677E-16 |
| 1.00E+02 | 1.6840E-16 | 2.5091E-16 | 4.7979E-16 | 8.9910E-16 |
| 1.50E+02 | 1.4345E-16 | 2.3795E-16 | 3.7413E-16 | 7.5552E-16 |
| 2.00E+02 | 1.2456E-16 | 2.1817E-16 | 3.1139E-16 | 6.5413E-16 |
| 2.50E+02 | 1.1046E-16 | 1.9940E-16 | 2.6876E-16 | 5.7861E-16 |
| 3.00E+02 | 9.9591E-17 | 1.8295E-16 | 2.3747E-16 | 5.2001E-16 |
| 3.50E+02 | 9.0946E-17 | 1.6880E-16 | 2.1334E-16 | 4.7309E-16 |

Table 21: **(continued)**

| Energy, $eV$ | $\sigma_{exc}^{(tot)}, cm^2$ | $\sigma_{ion}, cm^2$ | $\sigma_{el}, cm^2$ | $\sigma_{tot}, cm^2$ |
|---|---|---|---|---|
| 4.00E+02 | 8.3887E-17 | 1.5662E-16 | 1.9406E-16 | 4.3457E-16 |
| 4.50E+02 | 7.7999E-17 | 1.4609E-16 | 1.7825E-16 | 4.0233E-16 |
| 5.00E+02 | 7.3001E-17 | 1.3691E-16 | 1.6500E-16 | 3.7491E-16 |
| 5.50E+02 | 6.8696E-17 | 1.2885E-16 | 1.5373E-16 | 3.5127E-16 |
| 6.00E+02 | 6.4943E-17 | 1.2172E-16 | 1.4400E-16 | 3.3066E-16 |
| 6.50E+02 | 6.1638E-17 | 1.1537E-16 | 1.3551E-16 | 3.1252E-16 |
| 7.00E+02 | 5.8700E-17 | 1.0969E-16 | 1.2803E-16 | 2.9641E-16 |
| 7.50E+02 | 5.6069E-17 | 1.0456E-16 | 1.2138E-16 | 2.8201E-16 |
| 8.00E+02 | 5.3697E-17 | 9.9921E-17 | 1.1543E-16 | 2.6905E-16 |
| 8.50E+02 | 5.1545E-17 | 9.5696E-17 | 1.1007E-16 | 2.5731E-16 |
| 9.00E+02 | 4.9583E-17 | 9.1834E-17 | 1.0522E-16 | 2.4663E-16 |
| 9.50E+02 | 4.7786E-17 | 8.8290E-17 | 1.0080E-16 | 2.3687E-16 |
| 1.00E+03 | 4.6132E-17 | 8.5025E-17 | 9.6756E-17 | 2.2791E-16 |
| 1.50E+03 | 3.4704E-17 | 6.2474E-17 | 6.9557E-17 | 1.6673E-16 |
| 2.00E+03 | 2.8184E-17 | 4.9742E-17 | 5.4725E-17 | 1.3265E-16 |
| 2.50E+03 | 2.3909E-17 | 4.1508E-17 | 4.5343E-17 | 1.1076E-16 |
| 3.00E+03 | 2.0864E-17 | 3.5722E-17 | 3.8860E-17 | 9.5445E-17 |
| 3.50E+03 | 1.8573E-17 | 3.1419E-17 | 3.4108E-17 | 8.4100E-17 |
| 4.00E+03 | 1.6780E-17 | 2.8087E-17 | 3.0474E-17 | 7.5341E-17 |
| 4.50E+03 | 1.5334E-17 | 2.5426E-17 | 2.7606E-17 | 6.8365E-17 |
| 5.00E+03 | 1.4141E-17 | 2.3248E-17 | 2.5285E-17 | 6.2673E-17 |
| 5.50E+03 | 1.3138E-17 | 2.1432E-17 | 2.3369E-17 | 5.7938E-17 |
| 6.00E+03 | 1.2281E-17 | 1.9892E-17 | 2.1761E-17 | 5.3934E-17 |
| 6.50E+03 | 1.1541E-17 | 1.8569E-17 | 2.0394E-17 | 5.0504E-17 |
| 7.00E+03 | 1.0894E-17 | 1.7419E-17 | 1.9218E-17 | 4.7532E-17 |
| 7.50E+03 | 1.0323E-17 | 1.6411E-17 | 1.8197E-17 | 4.4931E-17 |
| 8.00E+03 | 9.8151E-18 | 1.5518E-17 | 1.7302E-17 | 4.2636E-17 |
| 8.50E+03 | 9.3603E-18 | 1.4722E-17 | 1.6513E-17 | 4.0595E-17 |
| 9.00E+03 | 8.9504E-18 | 1.4008E-17 | 1.5812E-17 | 3.8770E-17 |
| 9.50E+03 | 8.5788E-18 | 1.3363E-17 | 1.5186E-17 | 3.7128E-17 |
| 1.00E+04 | 8.2403E-18 | 1.2778E-17 | 1.4624E-17 | 3.5643E-17 |

*Table 22:* ***Parameters of (Eq.64) used to calculate the excitation cross sections $\sigma_{exc}^{(j)}(T)$ of electrons at energy $T$ for $j$ from 1 to 10: The vibrational excitation 1 of Chouki [98] corresponds to the sum of $\sigma_{exc}^{(j)}(T)$ for $j=1$ and 2; vibrational excitation 2 to the sum for $j=3$ and 4; vibrational excitation 3 to the sum for $j=5$ and 6. The cross sections for $j=7$ and 8. represent molecular dissociation and electron attachment; and the sum of cross sections for $j=9$ and 10 was introduced to obtain better agreement between the total scattering cross section calculated using (Eq.34) and experimental data of Grosswendt et al. [93]. $\Delta T_j$ is the energy loss assumed in the calculation.***

| $j$ | $f_j$ | $W_j$, eV | $A_j$ | $B_j$ | $\Omega_j$ | $\Delta T_j$, eV |
|---|---|---|---|---|---|---|
| 1 | 0.989 | 0.083 | 3.628 | 0.01078 | 2.0 | 0.08 |
| 2 | 1.304 | 1.014 | 2.906 | 2.0 | 66.02 | 1.0 |
| 3 | 2.34 | 0.175 | 3.882 | 0.03071 | 2.241 | 0.17 |
| 4 | 4.733 | 1.539 | 3.245 | 1.22 | 2.241 | 1.5 |
| 5 | 40.0 | 0.139 | 2.698 | 0.1433 | 6.322 | 0.13 |
| 6 | 1903 | 1.274 | 4.913 | 0.7776 | 15.61 | 1.27 |
| 7 | 220 | 2.3 | 5.0 | 0.3731 | 6.0 | 2.3 |
| 8 | 3581 | 5.214 | 24.08 | 2.933 | 26.62 | 5.2 |
| 9 | 1.546E+6 | 9.0 | 10.1 | 1.552 | 22.21 | 1.5 |
| 10 | 9.344E+4 | 34.4 | 14.96 | 1.263 | 4.85 | 1.5 |

*Table 23:* ***Set of scattering cross section $\sigma_v(T)$ of electrons in propane gas as a function of energy $T$: calculated total scattering cross section $\sigma_{tot}(T)$ using (Eq.34), present data set; elastic scattering cross section $\sigma_{el}(T)$; integral ionization cross section $\sigma_{ion}(T)$; total excitation cross section $\sigma_{exc}^{(tot)}(T)$.***

| Energy, *eV* | $\sigma_{exc}^{(tot)}, cm^2$ | $\sigma_{ion}, cm^2$ | $\sigma_{el}, cm^2$ | $\sigma_{tot}, cm^2$ |
|---|---|---|---|---|
| 1.00E+01 | 3.4900E-16 | 0.0000E+00 | 3.7837E-15 | 4.1327E-15 |
| 1.05E+01 | 3.6459E-16 | 0.0000E+00 | 3.7023E-15 | 4.0669E-15 |
| 1.10E+01 | 3.7620E-16 | 0.0000E+00 | 3.6249E-15 | 4.0011E-15 |
| 1.15E+01 | 3.8536E-16 | 4.6902E-19 | 3.5512E-15 | 3.9371E-15 |
| 1.20E+01 | 3.9289E-16 | 7.5319E-18 | 3.4810E-15 | 3.8814E-15 |
| 1.25E+01 | 3.9930E-16 | 2.2366E-17 | 3.4139E-15 | 3.8356E-15 |
| 1.30E+01 | 4.0503E-16 | 4.1324E-17 | 3.3498E-15 | 3.7962E-15 |
| 1.35E+01 | 4.1054E-16 | 6.2745E-17 | 3.2885E-15 | 3.7617E-15 |
| 1.40E+01 | 4.1644E-16 | 8.5505E-17 | 3.2296E-15 | 3.7316E-15 |
| 1.45E+01 | 4.2346E-16 | 1.0885E-16 | 3.1732E-15 | 3.7055E-15 |
| 1.50E+01 | 4.3241E-16 | 1.3227E-16 | 3.1190E-15 | 3.6837E-15 |
| 1.55E+01 | 4.4396E-16 | 1.5545E-16 | 3.0670E-15 | 3.6664E-15 |
| 1.60E+01 | 4.5858E-16 | 1.7817E-16 | 3.0169E-15 | 3.6536E-15 |
| 1.65E+01 | 4.7637E-16 | 2.0030E-16 | 2.9686E-15 | 3.6453E-15 |
| 1.70E+01 | 4.9710E-16 | 2.2179E-16 | 2.9221E-15 | 3.6410E-15 |
| 1.75E+01 | 5.2022E-16 | 2.4259E-16 | 2.8772E-15 | 3.6400E-15 |
| 1.80E+01 | 5.4492E-16 | 2.6269E-16 | 2.8339E-15 | 3.6415E-15 |
| 1.90E+01 | 5.9542E-16 | 3.0086E-16 | 2.7516E-15 | 3.6479E-15 |
| 2.00E+01 | 6.4149E-16 | 3.3646E-16 | 2.6746E-15 | 3.6526E-15 |
| 2.50E+01 | 7.0629E-16 | 4.8361E-16 | 2.3529E-15 | 3.5428E-15 |
| 3.00E+01 | 6.1120E-16 | 5.9388E-16 | 2.1074E-15 | 3.3125E-15 |
| 3.50E+01 | 5.2714E-16 | 6.7850E-16 | 1.9131E-15 | 3.1187E-15 |
| 4.00E+01 | 5.9141E-16 | 7.4357E-16 | 1.7549E-15 | 3.0899E-15 |
| 4.50E+01 | 6.7249E-16 | 7.9325E-16 | 1.6234E-15 | 3.0891E-15 |
| 5.00E+01 | 6.0157E-16 | 8.3072E-16 | 1.5120E-15 | 2.9443E-15 |
| 5.50E+01 | 5.1400E-16 | 8.5851E-16 | 1.4163E-15 | 2.7888E-15 |
| 6.00E+01 | 4.5615E-16 | 8.7860E-16 | 1.3332E-15 | 2.6679E-15 |
| 6.50E+01 | 4.2249E-16 | 8.9259E-16 | 1.2601E-15 | 2.5752E-15 |
| 7.00E+01 | 4.0212E-16 | 9.0172E-16 | 1.1954E-15 | 2.4993E-15 |
| 7.50E+01 | 3.8829E-16 | 9.0699E-16 | 1.1376E-15 | 2.4329E-15 |
| 8.00E+01 | 3.7761E-16 | 9.0919E-16 | 1.0857E-15 | 2.3725E-15 |
| 8.50E+01 | 3.6852E-16 | 9.0894E-16 | 1.0387E-15 | 2.3162E-15 |
| 9.00E+01 | 3.6033E-16 | 9.0675E-16 | 9.9601E-16 | 2.2631E-15 |
| 9.50E+01 | 3.5269E-16 | 9.0300E-16 | 9.5700E-16 | 2.2127E-15 |
| 1.00E+02 | 3.4547E-16 | 8.9802E-16 | 9.2120E-16 | 2.1647E-15 |
| 1.50E+02 | 2.8761E-16 | 8.1924E-16 | 6.7697E-16 | 1.7838E-15 |
| 2.00E+02 | 2.4721E-16 | 7.3547E-16 | 5.4065E-16 | 1.5233E-15 |
| 2.50E+02 | 2.1751E-16 | 6.6367E-16 | 4.5274E-16 | 1.3339E-15 |
| 3.00E+02 | 1.9473E-16 | 6.0402E-16 | 3.9095E-16 | 1.1897E-15 |
| 3.50E+02 | 1.7666E-16 | 5.5435E-16 | 3.4493E-16 | 1.0759E-15 |

Table 23: **(continued)**

| Energy, *eV* | $\sigma_{exc}^{(tot)}, cm^2$ | $\sigma_{ion}, cm^2$ | $\sigma_{el}, cm^2$ | $\sigma_{tot}, cm^2$ |
|---|---|---|---|---|
| 4.00E+02 | 1.6195E-16 | 5.1256E-16 | 3.0923E-16 | 9.8374E-16 |
| 4.50E+02 | 1.4972E-16 | 4.7698E-16 | 2.8066E-16 | 9.0736E-16 |
| 5.00E+02 | 1.3938E-16 | 4.4634E-16 | 2.5722E-16 | 8.4294E-16 |
| 5.50E+02 | 1.3050E-16 | 4.1969E-16 | 2.3763E-16 | 7.8782E-16 |
| 6.00E+02 | 1.2279E-16 | 3.9628E-16 | 2.2098E-16 | 7.4005E-16 |
| 6.50E+02 | 1.1603E-16 | 3.7555E-16 | 2.0665E-16 | 6.9823E-16 |
| 7.00E+02 | 1.1004E-16 | 3.5706E-16 | 1.9417E-16 | 6.6127E-16 |
| 7.50E+02 | 1.0469E-16 | 3.4046E-16 | 1.8320E-16 | 6.2835E-16 |
| 8.00E+02 | 9.9891E-17 | 3.2547E-16 | 1.7346E-16 | 5.9882E-16 |
| 8.50E+02 | 9.5552E-17 | 3.1185E-16 | 1.6477E-16 | 5.7217E-16 |
| 9.00E+02 | 9.1610E-17 | 2.9942E-16 | 1.5695E-16 | 5.4798E-16 |
| 9.50E+02 | 8.8010E-17 | 2.8804E-16 | 1.4988E-16 | 5.2593E-16 |
| 1.00E+03 | 8.4708E-17 | 2.7756E-16 | 1.4345E-16 | 5.0572E-16 |
| 1.50E+03 | 6.2219E-17 | 2.0539E-16 | 1.0104E-16 | 3.6865E-16 |
| 2.00E+03 | 4.9690E-17 | 1.6463E-16 | 7.8425E-17 | 2.9275E-16 |
| 2.50E+03 | 4.1616E-17 | 1.3819E-16 | 6.4217E-17 | 2.4402E-16 |
| 3.00E+03 | 3.5943E-17 | 1.1952E-16 | 5.4404E-17 | 2.0987E-16 |
| 3.50E+03 | 3.1720E-17 | 1.0559E-16 | 4.7189E-17 | 1.8450E-16 |
| 4.00E+03 | 2.8445E-17 | 9.4759E-17 | 4.1645E-17 | 1.6485E-16 |
| 4.50E+03 | 2.5823E-17 | 8.6079E-17 | 3.7243E-17 | 1.4915E-16 |
| 5.00E+03 | 2.3674E-17 | 7.8953E-17 | 3.3656E-17 | 1.3628E-16 |
| 5.50E+03 | 2.1878E-17 | 7.2991E-17 | 3.0675E-17 | 1.2554E-16 |
| 6.00E+03 | 2.0352E-17 | 6.7922E-17 | 2.8156E-17 | 1.1643E-16 |
| 6.50E+03 | 1.9039E-17 | 6.3557E-17 | 2.5997E-17 | 1.0859E-16 |
| 7.00E+03 | 1.7895E-17 | 5.9754E-17 | 2.4125E-17 | 1.0178E-16 |
| 7.50E+03 | 1.6890E-17 | 5.6410E-17 | 2.2487E-17 | 9.5787E-17 |
| 8.00E+03 | 1.6000E-17 | 5.3444E-17 | 2.1040E-17 | 9.0483E-17 |
| 8.50E+03 | 1.5204E-17 | 5.0795E-17 | 1.9752E-17 | 8.5751E-17 |
| 9.00E+03 | 1.4489E-17 | 4.8412E-17 | 1.8599E-17 | 8.1501E-17 |
| 9.50E+03 | 1.3842E-17 | 4.6258E-17 | 1.7561E-17 | 7.7661E-17 |
| 1.00E+04 | 1.3255E-17 | 4.4299E-17 | 1.6621E-17 | 7.4174E-17 |